\documentclass[11pt,a4paper]{article}

\def\paperauthors{F. Abudinen, G. Morgante, and P. Skands} 
\def\paperasciititle{A framework for ME-corrected QED antenna showers in hadron decays} 
\def\papertitle{A framework for matrix-element-corrected QED antenna showers in hadron decays}
\def\paperkeywords{{High-energy Physics, Simulation generator, QED, Final-state radiation}}
\newif\ifEnableSectionTOCLinks
\EnableSectionTOCLinksfalse 

\usepackage[top=1in, bottom=1.25in, left=1in, right=1in]{geometry}

\usepackage{microtype}
\usepackage{lineno}  
\usepackage{xspace} 
\usepackage{caption} 

\usepackage{graphicx}  
\usepackage{subcaption} 
\usepackage{color}
\usepackage{colortbl}
\graphicspath{{./figs/}} 
\usepackage{float}

\usepackage{amsmath} 
\usepackage{amssymb}
\usepackage{amsfonts}
\usepackage{upgreek} 

\newcommand*\patchAmsMathEnvironmentForLineno[1]{%
\expandafter\let\csname old#1\expandafter\endcsname\csname #1\endcsname
\expandafter\let\csname oldend#1\expandafter\endcsname\csname
end#1\endcsname
 \renewenvironment{#1}%
   {\linenomath\csname old#1\endcsname}%
   {\csname oldend#1\endcsname\endlinenomath}%
}
\newcommand*\patchBothAmsMathEnvironmentsForLineno[1]{%
  \patchAmsMathEnvironmentForLineno{#1}%
  \patchAmsMathEnvironmentForLineno{#1*}%
}
\AtBeginDocument{%
\patchBothAmsMathEnvironmentsForLineno{equation}%
\patchBothAmsMathEnvironmentsForLineno{align}%
\patchBothAmsMathEnvironmentsForLineno{flalign}%
\patchBothAmsMathEnvironmentsForLineno{alignat}%
\patchBothAmsMathEnvironmentsForLineno{gather}%
\patchBothAmsMathEnvironmentsForLineno{multline}%
\patchBothAmsMathEnvironmentsForLineno{eqnarray}%
}

\usepackage[pdftex,
            pdfauthor={\paperauthors},
            pdftitle={\paperasciititle},
            pdfkeywords={\paperkeywords},
            allcolors=blue]{hyperref}

\usepackage{hyperxmp}
\usepackage[all]{hypcap} 

\hypersetup{
  colorlinks   = true, 
  urlcolor     = blue, 
  linkcolor    = blue, 
  citecolor    = purple   
}

\usepackage[colorinlistoftodos,textsize=scriptsize]{todonotes}

\usepackage[bottom,flushmargin,hang,multiple]{footmisc}

\ifEnableSectionTOCLinks
    \usepackage[explicit]{titlesec} 
    
    \let\oldcontentsline\contentsline
    \renewcommand

    \titleformat{\section}{\normalfont\Large\bf}{\hyperlink{tocsection.\thesection}{{\thesection} \parbox[t]{\dimexpr\textwidth-1pc}{#1}}}{1pc}{}

    \titleformat{\subsection}{\normalfont\bf}{\hyperlink{tocsubsection.\thesubsection}{{\thesubsection} \parbox[t]{\dimexpr\textwidth-1pc}{#1}}}{1pc}{}

    \titleformat{name=\section,numberless}[display]{}{}{0pt}{\normalfont\Huge\bfseries #1}
\fi

\usepackage{cite} 
\usepackage{mciteplus}

\usepackage{xcolor}
\definecolor{vir_purp}{HTML}{440154}
\definecolor{vir_blue}{HTML}{39568C}
\definecolor{vir_azzr}{HTML}{1F968B}
\definecolor{vir_gree}{HTML}{73D055}
\definecolor{vir_yell}{HTML}{FDE725}

\usepackage{booktabs}
\usepackage{cleveref}
\usepackage{authblk}

\usepackage{enumitem} 
\usepackage[export]{adjustbox} 
\usepackage{xspace}

\usepackage{ifthen} 

\definecolor{orange1}{HTML}{FFE7D8}
\definecolor{orange2}{HTML}{FECBA5}
\definecolor{orange3}{HTML}{FF9E55}
\definecolor{orange4}{HTML}{8D3E43}
\definecolor{cyan1}{HTML}{CCFFFF}
\definecolor{cyan2}{HTML}{79FCFC}
\definecolor{cyan3}{HTML}{00FFFF}
\definecolor{cyan4}{HTML}{CC00FF}
\definecolor{PHOTOS}{RGB}{253, 231, 37}
\definecolor{PHOTOSmec}{RGB}{173, 220, 48}
\definecolor{SherpaPhotons1}{RGB}{94, 201, 98}
\definecolor{SherpaPhotons20}{RGB}{40, 174, 128}
\definecolor{SherpaPhotons21}{RGB}{33, 145, 140}
\definecolor{VinciaQEDmec}{RGB}{44, 114, 142}
\definecolor{VinciaQED}{RGB}{59, 82, 139}
\definecolor{VinciaQEDdip}{RGB}{71, 45, 123}
\definecolor{PythiaQED}{RGB}{68, 1, 84}

\newboolean{pdflatex}
\setboolean{pdflatex}{true} 

\newboolean{articletitles}
\setboolean{articletitles}{true} 

\usepackage{microtype}
\usepackage[defaultlines=2,all]{nowidow}
\renewcommand{\d}{\ensuremath{\mathrm{d}}}

\newboolean{uprightparticles}
\setboolean{uprightparticles}{false} 

\usepackage{xspace} 
\usepackage{upgreek}

\def\MagUp {\mbox{\em Mag\kern -0.05em Up}\xspace}

\ifthenelse{\boolean{uprightparticles}}%
{
 
 \def\Pgamma      {\ensuremath{\upgamma}\xspace}

 \def\Pmu         {\ensuremath{\upmu}\xspace}                 
 \def\Pnu         {\ensuremath{\upnu}\xspace}                 
                  
 \def\Ppi         {\ensuremath{\uppi}\xspace}                 
                  
 \def\Prho        {\ensuremath{\uprho}\xspace}

 \def\Ppsi        {\ensuremath{\uppsi}\xspace}

 \def\PDelta      {\ensuremath{\Delta}\xspace}                 
 \def\PXi         {\ensuremath{\Xi}\xspace}                 
 \def\PLambda     {\ensuremath{\Lambda}\xspace}                 
 \def\PSigma      {\ensuremath{\Sigma}\xspace}                 
 \def\POmega      {\ensuremath{\Omega}\xspace}                 
 \def\PUpsilon    {\ensuremath{\Upsilon}\xspace}
 \let\oldPi\Pi
 \def\PPi         {\ensuremath{\oldPi}\xspace}

 \def\PB      {\ensuremath{\mathrm{B}}\xspace}                 
 \def\PD      {\ensuremath{\mathrm{D}}\xspace}                 
 \def\PH      {\ensuremath{\mathrm{H}}\xspace}                 
 \def\PJ      {\ensuremath{\mathrm{J}}\xspace}                 
 \def\PK      {\ensuremath{\mathrm{K}}\xspace}                 
 \def\PW      {\ensuremath{\mathrm{W}}\xspace}                 
 \def\PZ      {\ensuremath{\mathrm{Z}}\xspace}                 

 \def\Pe      {\ensuremath{\mathrm{e}}\xspace}                 

 \def\Pq      {\ensuremath{\mathrm{q}}\xspace}                 
 \def\Ps      {\ensuremath{\mathrm{s}}\xspace}

 \def\thebaroffset{0.0em}
}
{
 
 \def\Pgamma      {\ensuremath{\gamma}\xspace}

 \def\Pmu         {\ensuremath{\mu}\xspace}                 
 \def\Pnu         {\ensuremath{\nu}\xspace}                 
                  
 \def\Ppi         {\ensuremath{\pi}\xspace}                 
                  
 \def\Prho        {\ensuremath{\rho}\xspace}

 \def\Ppsi        {\ensuremath{\psi}\xspace}                 
                  
 \mathchardef\PDelta="7101
 \mathchardef\PXi="7104
 \mathchardef\PLambda="7103
 \mathchardef\PSigma="7106
 \mathchardef\POmega="710A
 \mathchardef\PUpsilon="7107
 \mathchardef\PPi="7105
 \def\PB      {\ensuremath{B}\xspace}                 
 \def\PD      {\ensuremath{D}\xspace}                 
 \def\PH      {\ensuremath{H}\xspace}                 
 \def\PJ      {\ensuremath{J}\xspace}                 
 \def\PK      {\ensuremath{K}\xspace}                 
 \def\PW      {\ensuremath{W}\xspace}                 
 \def\PZ      {\ensuremath{Z}\xspace}                 

 \def\Pe      {\ensuremath{e}\xspace}                 

 \def\Pq      {\ensuremath{q}\xspace}                 
 \def\Ps      {\ensuremath{s}\xspace}

 \def\thebaroffset{0.18em}
}
\newcommand{\offsetoverline}[2][\thebaroffset]{\kern #1\overline{\kern -#1 #2}}%

\makeatletter
\ifcase \@ptsize \relax
  \newcommand{\miniscule}{\@setfontsize\miniscule{4}{5}}
\or
  \newcommand{\miniscule}{\@setfontsize\miniscule{5}{6}}
\or
  \newcommand{\miniscule}{\@setfontsize\miniscule{5}{6}}
\fi
\makeatother

\DeclareRobustCommand{\optbar}[1]{\shortstack{{\miniscule (\rule[.5ex]{1.25em}{.18mm})}
  \\ [-.7ex] $#1$}}

\def\en         {{\ensuremath{\Pe^-}}\xspace}   
\def\ep         {{\ensuremath{\Pe^+}}\xspace}

\def\mup        {{\ensuremath{\Pmu^+}}\xspace}
\def\mun        {{\ensuremath{\Pmu^-}}\xspace} 

\def\mumu       {{\ensuremath{\Pmu^+\Pmu^-}}\xspace}

\def\ellp       {{\ensuremath{\ell^+}}\xspace}

\def\ellell     {\ensuremath{\ell^+ \ell^-}\xspace}

\def\neu        {{\ensuremath{\Pnu}}\xspace}
\def\neub       {{\ensuremath{\overline{\Pnu}}}\xspace}
\def\neue       {{\ensuremath{\neu_e}}\xspace}
\def\neueb      {{\ensuremath{\neub_e}}\xspace}
\def\neum       {{\ensuremath{\neu_\mu}}\xspace}
\def\neumb      {{\ensuremath{\neub_\mu}}\xspace}

\def\g      {{\ensuremath{\Pgamma}}\xspace}

\def\Wp     {{\ensuremath{\PW^+}}\xspace}

\def\Z      {{\ensuremath{\PZ}}\xspace}

\def\quark     {{\ensuremath{\Pq}}\xspace}
\def\quarkbar  {{\ensuremath{\overline \quark}}\xspace}
\def\qqbar     {{\ensuremath{\quark\quarkbar}}\xspace}

\def\squark    {{\ensuremath{\Ps}}\xspace}

\def\pion   {{\ensuremath{\Ppi}}\xspace}
\def\piz    {{\ensuremath{\pion^0}}\xspace}
\def\pip    {{\ensuremath{\pion^+}}\xspace}
\def\pim    {{\ensuremath{\pion^-}}\xspace}
\def\pipm   {{\ensuremath{\pion^\pm}}\xspace}
\def\pimp   {{\ensuremath{\pion^\mp}}\xspace}

\def\kaon    {{\ensuremath{\PK}}\xspace}
\def\Kbar    {{\ensuremath{\offsetoverline{\PK}}}\xspace}

\def\KorKbar {\kern \thebaroffset\optbar{\kern -\thebaroffset \PK}{}\xspace}
\def\Kz      {{\ensuremath{\kaon^0}}\xspace}
\def\Kzb     {{\ensuremath{\Kbar{}^0}}\xspace}
\def\Kp      {{\ensuremath{\kaon^+}}\xspace}
\def\Km      {{\ensuremath{\kaon^-}}\xspace}
\def\Kpm     {{\ensuremath{\kaon^\pm}}\xspace}

\def\KS      {{\ensuremath{\kaon^0_{\mathrm{S}}}}\xspace}

\def\KL      {{\ensuremath{\kaon^0_{\mathrm{L}}}}\xspace}

\def\Dbar    {{\ensuremath{\offsetoverline{\PD}}}\xspace}
\def\D       {{\ensuremath{\PD}}\xspace}

\def\DorDbar {\kern \thebaroffset\optbar{\kern -\thebaroffset \PD}\xspace}
\def\Dz      {{\ensuremath{\D^0}}\xspace}
\def\Dzb     {{\ensuremath{\Dbar{}^0}}\xspace}
\def\Dp      {{\ensuremath{\D^+}}\xspace}
\def\Dm      {{\ensuremath{\D^-}}\xspace}
\def\Dpm     {{\ensuremath{\D^\pm}}\xspace}
\def\Dmp     {{\ensuremath{\D^\mp}}\xspace}
\def\DpDm    {\ensuremath{\Dp {\kern -0.16em \Dm}}\xspace}

\def\Ds      {{\ensuremath{\D^+_\squark}}\xspace}

\def\B       {{\ensuremath{\PB}}\xspace}
\def\Bbar    {{\ensuremath{\offsetoverline{\PB}}}\xspace}

\def\BorBbar {\kern \thebaroffset\optbar{\kern -\thebaroffset \PB}\xspace}
\def\Bz      {{\ensuremath{\B^0}}\xspace}
\def\Bzb     {{\ensuremath{\Bbar{}^0}}\xspace}
\def\Bd      {{\ensuremath{\B^0}}\xspace}

\def\BdorBdbar {\kern \thebaroffset\optbar{\kern -\thebaroffset \Bd}\xspace}
\def\Bu      {{\ensuremath{\B^+}}\xspace}

\def\Bp      {{\ensuremath{\Bu}}\xspace}

\def\Bpm     {{\ensuremath{\B^\pm}}\xspace}

\def\Bs      {{\ensuremath{\B^0_\squark}}\xspace}

\def\BsorBsbar {\kern \thebaroffset\optbar{\kern -\thebaroffset \Bs}\xspace}

\def\jpsi     {{\ensuremath{{\PJ\mskip -3mu/\mskip -2mu\Ppsi}}}\xspace}

\def\Y#1S{\ensuremath{\PUpsilon{(#1S)}}\xspace}

\def\LorLbar     {\kern \thebaroffset\optbar{\kern -\thebaroffset \PLambda}\xspace}

\newcommand{\decay}[2]{\ensuremath{#1\!\to #2}\xspace} 

\def\to                 {\ensuremath{\rightarrow}\xspace}

\def\AT#1     {\ensuremath{A_{\mathrm{T}}^{#1}}\xspace}           

\def\C#1      {\ensuremath{\mathcal{C}_{#1}}\xspace}                       
\def\Cp#1     {\ensuremath{\mathcal{C}_{#1}^{'}}\xspace}                    
\def\Ceff#1   {\ensuremath{\mathcal{C}_{#1}^{\mathrm{(eff)}}}\xspace}        
\def\Cpeff#1  {\ensuremath{\mathcal{C}_{#1}^{'\mathrm{(eff)}}}\xspace}       
\def\Ope#1    {\ensuremath{\mathcal{O}_{#1}}\xspace}                       
\def\Opep#1   {\ensuremath{\mathcal{O}_{#1}^{'}}\xspace}                    

\newcommand{\aunit}[1]{\ensuremath{\text{\,#1}}}       

\newcommand{\tev}{\aunit{Te\kern -0.1em V}\xspace}
\newcommand{\gev}{\aunit{Ge\kern -0.1em V}\xspace}
\newcommand{\mev}{\aunit{Me\kern -0.1em V}\xspace}
\newcommand{\kev}{\aunit{ke\kern -0.1em V}\xspace}
\newcommand{\ev}{\aunit{e\kern -0.1em V}\xspace}
 
\newcommand{\mevc}{\ensuremath{\aunit{Me\kern -0.1em V\!/}c}\xspace}
\newcommand{\gevc}{\ensuremath{\aunit{Ge\kern -0.1em V\!/}c}\xspace}
\newcommand{\mevcc}{\ensuremath{\aunit{Me\kern -0.1em V\!/}c^2}\xspace}
\newcommand{\gevcc}{\ensuremath{\aunit{Ge\kern -0.1em V\!/}c^2}\xspace}

\def\gsim{{~\raise.15em\hbox{$>$}\kern-.85em
          \lower.35em\hbox{$\sim$}~}\xspace}
\def\lsim{{~\raise.15em\hbox{$<$}\kern-.85em
          \lower.35em\hbox{$\sim$}~}\xspace}

\def\cpp        {\mbox{\textsc{C\raisebox{0.1em}{{\footnotesize{++}}}}}\xspace}

\def\tell1  {TELL1\xspace}
\def\ukl1   {UKL1\xspace}

\newcommand{\eg}{\mbox{\itshape e.g.}\xspace}

\newcommand{\cf}{\mbox{\itshape cf.}\xspace}

\newcommand{\orcid}[1]{\href{https://orcid.org/#1}{\hspace*{0.1em}\raisebox{-0.45ex}{\includegraphics[width=1em]{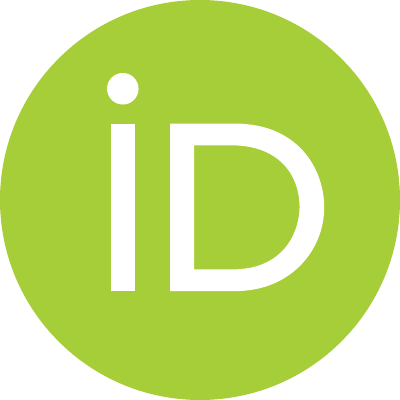}}}}

\newcommand{\pT}[1][]{{\ensuremath {p_{\perp{#1}}}}\xspace}
\newcommand{\pTs}[1][]{{\ensuremath p_{\perp{#1}}^2}\xspace}

\newcommand{\code}[1]{\textsc{#1}\xspace}

\newcommand{\evtgen}{\code{Evtgen}}
\newcommand{\feynrules}{\code{FeynRules}}
\newcommand{\herwig}{\code{Herwig}}
\newcommand{\madgraph}{\code{MadGraph}}
\newcommand{\pythia}{\code{Pythia}}
\newcommand{\photons}{\code{Photons}}
\newcommand{\photonspp}{\code{Photons\raisebox{0.1em}{{\footnotesize{++}}}}}
\newcommand{\photos}{\code{Photos}}
\newcommand{\sherpa}{\code{Sherpa}}
\newcommand{\vincia}{\code{Vincia}}
\newcommand{\ariadne}{\code{Ariadne}}

\newcommand{\appRef}[1]{app.~\ref{#1}\xspace}

\newcommand{\eqRef}[1]{eq.~(\ref{#1})\xspace}
\newcommand{\EqRef}[1]{Eq.~(\ref{#1})\xspace}
\xspace

\newcommand{\figRef}[1]{fig.~\ref{#1}\xspace}
\newcommand{\figsRef}[1]{figs.~\ref{#1}\xspace}
\newcommand{\FigRef}[1]{Fig.~\ref{#1}\xspace}
\newcommand{\FigsRef}[1]{Figs.~\ref{#1}\xspace}
\newcommand{\tabRef}[1]{tab.~\ref{#1}\xspace}
\newcommand{\TabRef}[1]{Tab.~\ref{#1}\xspace}
\newcommand{\tabsRef}[1]{tabs.~\ref{#1}\xspace}

\newcommand{\secRef}[1]{sec.~\ref{#1}\xspace}

\newcommand{\secsRef}[1]{secs.~\ref{#1}\xspace}

\title{\bf\papertitle}

\author{F.~Abudin{\'e}n$^{1,2}$\orcid{0000-0002-6737-3528}, G.~Morgante$^3$\orcid{0009-0007-4141-3767}, and P.~Skands$^3$\orcid{0000-0003-0024-3822}}

\affil{
\footnotesize
\it $^1$School of Physics and Astronomy, University of Edinburgh, Edinburgh, United Kingdom\\
\it $^2$Department of Physics, University of Warwick, Coventry, United Kingdom \\
\it $^3$School of Physics and Astronomy, Monash University, Melbourne, Australia\\
}  

\date{November 2026}

\begin{document}
\maketitle

\begin{abstract}
We present a general framework for the description of QED parton showers in hadron decays.
The formalism is constructed in terms of coherent QED antennae with full dependence on particle masses and extends beyond the YFS approach by systematically incorporating spin-dependent (quasi)collinear single-unresolved limits to all orders. 
In addition, we incorporate corrections derived from tree-level matrix elements up to \(\mathcal{O}(\alpha^2)\) and describe a generic procedure to obtain such matrix elements using \textsc{FeynRules} and \textsc{MadGraph}.
We provide an explicit implementation of the method for decays involving pseudoscalar mesons and for decays of vectors and pseudoscalars into fermion-antifermion pairs, 
within the combined \textsc{Pythia}+\textsc{Vincia} parton-shower Monte Carlo framework, 
and compare the resulting QED emission spectra to those generated with \textsc{Photos} and \textsc{Sherpa}+\textsc{Photons\raisebox{0.1em}{{\footnotesize{++}}}} through the \textsc{EvtGen} generator.  
\end{abstract}

\vspace*{2.0cm}

\vspace{\fill}

\clearpage

\vspace*{1.2cm}
\setcounter{tocdepth}{2}
\tableofcontents
\clearpage

\section{Introduction}
\label{sec:intro}

Understanding the effect of Quantum Electrodynamics~(QED) radiation in particle decays is essential for precise measurements of particle decay properties.
This is particularly relevant for the evaluation of reconstruction and selection efficiencies in experimental analyses and for the modelling of kinematic distributions. 
A notable example is the reconstruction of resonance structures in invariant mass spectra, where QED radiation leads to characteristic low-mass tails arising from “missing” (i.e., unreconstructed) photons.
Simulations based on Monte Carlo event generators~\cite{Buckley:2011ms,ParticleDataGroup:2024cfk,vanBeekveld:2026uxl} play a central role in quantifying these effects and their associated uncertainties, and furthermore enable the embedding of decays into fully realistic collider-event environments. 
For QED radiation in hadron decays, state-of-the-art generators currently include the \sherpa's \photonspp module~\cite{Schonherr:2008av,Sherpa:2019gpd,Sherpa:2024mfk}, the \herwig~7 generator~\cite{Hamilton:2006xz,Bahr:2008pv,Bellm:2025pcw}, and the \photos package~\cite{Barberio:1990ms,Barberio:1993qi,Golonka:2005pn,Nanava:2006vv,davidson2015photos}. 
The latter is commonly used in combination with the \pythia~\cite{Sjostrand:2006za,Bierlich:2022pfr} and/or \evtgen generator~\cite{Lange:2001uf,Abudinen:2024wpc,evtgen:2025}. 

QED radiation generators account for the four‑momentum “lost” to QED radiation by modifying the four‑momenta of the decay products in decay chains that are initially generated without any final‑state radiation (i.e., taking each complete decay chain, including all its particles, as input). 
The four‑momenta are adjusted in such a way that the QED factorisation properties are preserved in the infrared regime, specifically in the soft and/or collinear limits.
In practice, this typically requires adopting a formalism based on the soft-photon approximation, as realised in the YFS approach~\cite{Yennie:1961ad} employed by \photonspp and \herwig, and implemented as a correction weight in \photos, or on the collinear approximation, as described by DGLAP evolution~\cite{Gribov:1972ri,Dokshitzer:1977sg,Altarelli:1977zs}, which is used in \pythia and \photos and can be incorporated through higher-order corrections in \photonspp.

An alternative and comparatively less explored framework for QED radiation is the antenna formalism~\cite{Azimov:1985zta,Gustafson:1987rq,Georgi:1988ba,Kosower:1997zr,Kosower:2003bh,Gehrmann-DeRidder:2005btv,Larkoski:2009ah,Kleiss:2017iir}. 
This formalism combines exact phase-space factorisations~\cite{Kosower:1997zr} with (multipole) QED antenna functions~\cite{Kleiss:2017iir} that have YFS-style eikonal factors as their soft limits and DGLAP splitting kernels as their (quasi)collinear limits.
From a theoretical standpoint, this offers a unified and accurate description of both soft and collinear limits, without requiring a prioritisation of one over the other.  
It provides the basis for the treatment of QED radiation in the \vincia parton-shower module~\cite{Skands:2020lkd} of \pythia~8~\cite{Bierlich:2022pfr}, but so far has only been developed for QED radiation from elementary particles such as quarks, leptons, and \PW~bosons.

In this paper, we present the initial developments towards a new framework for the simulation of QED radiation in generic hadron decays within the \vincia shower module. The algorithm is based on the sector-antenna approximation~\cite{Kosower:2003bh,Larkoski:2009ah,Lopez-Villarejo:2011pwr,Kleiss:2017iir,Brooks:2020upa,Skands:2020lkd}, implemented with full mass dependence~\cite{Gehrmann-DeRidder:2011gkt,Brooks:2019xso,Skands:2020lkd}, and extends beyond the YFS formalism by incorporating spin-dependent (quasi)collinear single-unresolved limits to all orders. 
In its present form, the framework includes the relevant spin-dependent contributions for photon emission from spin-$0$, spin-$1/2$, and spin-$1$ particles. 
In addition, secondary \decay{\gamma}{\ellell} and \decay{\gamma}{\qqbar} splittings can be activated, although in that case the user should take care of ensuring consistency between inclusive branching fractions and those in which explicit lepton pairs are generated.
We note that in the present work, we set all hadronic form factors to unity\footnote{We note that our implementation in \pythia/\vincia will allow for case-by-case form factors to be provided, via a \texttt{UserHooks} style interface.} and treat electromagnetic radiation from spin-3/2 particles identically to that from spin-$1/2$ particles. 

\begin{figure}[tp]
\centering
\includegraphics[width=0.95\textwidth]{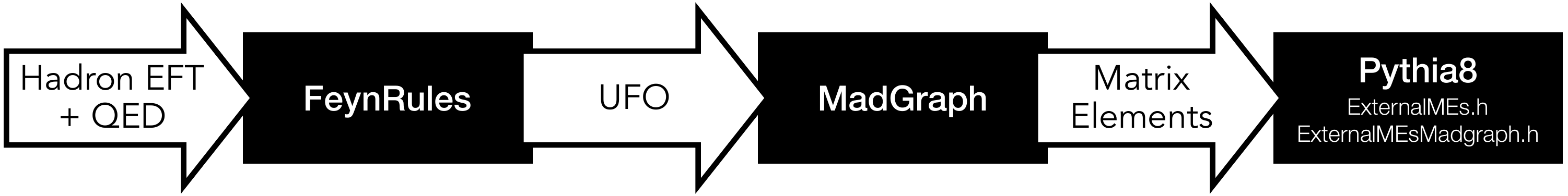}
\caption{Illustration of the tool chain from an effective field theory (EFT) + QED for the hadron sector to matrix elements that can be interfaced from \pythia~8.\label{fig:meChain}}
\end{figure}

An additional important development is the inclusion of a framework that incorporates matrix-element~(ME) corrections~\cite{Bengtsson:1986et,Seymour:1994df,Corcella:1998rs,Norrbin:2000uu,Giele:2011cb,Lopez-Villarejo:2011pwr} directly into the QED~parton-shower evolution via an interface to ME libraries.
It can be created semi-automatically with \feynrules~\cite{Degrande:2011ua,Alloul:2013bka} and \madgraph~\cite{Alwall:2011uj,Alwall:2014hca}, enabling modern effective field theory (EFT) and ME-matching methods for QED~radiative corrections in hadronic decays.  
We demonstrate the tool chain~(\figRef{fig:meChain}) for QED~corrections involving pseudoscalar- and vector-meson decays into fermion pairs. 
Extensions to additional classes of decays (for example, further vector-meson and baryonic decay modes) will be the subject of future studies.

We compare the obtained results with those from other well-established algorithms for a representative set of benchmark decay channels. 
For these comparisons, we simulate decays using the \evtgen generator, which is widely employed in numerous experiments to model the decays of unstable heavy hadrons. 
\evtgen relies on specialised external packages to incorporate final-state QED radiative corrections into the simulated decays, using \photos in its default configuration. 
Recently, the \evtgen generator has been upgraded to support both \photonspp and \vincia as alternative options to \photos, enabling comparisons between QED radiation generators within a single framework.

The outline of this paper is as follows.
In \secRef{sec:generators} we discuss the underlying theory approaches of common general-purpose QED-correction generators.
In \secRef{sec:vincia} we detail the \vincia QED shower formalism.
In \secRef{sec:deadConeEffect} we discuss the dead-cone effect in the centre of mass frame.
In \secRef{sec:IR} we look at the effects of changing the infrared cutoff between generators.
In \secRef{sec:MEs} we look in more detail at the size of the ME corrections.
In \secRef{sec:results} we discuss results for benchmark decay modes and compare the results obtained with \vincia with the results obtained using other generators.
In \secRef{sec:performance} we compare the performance of the different generators in both space and time.
Finally, in \secRef{sec:summary}, we provide a summary and an outlook.

\section{Approaches for QED radiation in current generators}
\label{sec:generators}

In this section, we summarise the methodologies currently implemented in the event generators that are used as benchmarks for our comparison with the \vincia simulations. In general, QED radiation generators exploit the factorisation properties of QED splitting kernels such that the differential cross section factorises into the Born-level (leading-order) cross section multiplied by  appropriate QED splitting kernel(s) and --- beyond tree level --- Sudakov factors. 

The contributions from real-photon emission are, by themselves, infrared divergent. However, these divergences are cancelled order by order in perturbation theory by virtual corrections of opposite sign. 
According to the KLN theorem~\cite{Kinoshita:1962ur,Lee:1964is}, the combined sum of the virtual and bremsstrahlung contributions is free of infrared divergences. 
In practical implementations, 
\begin{itemize}
  \item Fixed-order generators numerically reconstruct virtual corrections from the real-photon contributions following the slicing approach~\cite{Giele:1991vf}, which partitions the phase space into a resolved and unresolved region based on a cutoff parameter.
In the unresolved region, the real-emission matrix element factorises and can be analytically integrated up to the cutoff by exploiting its universal infrared limits. The resolved region, being finite, is computed numerically. 
\item Shower generators, on the other hand, implement exact unitarity by construction. Their probability-conserving Sudakov factors embody the KLN theorem explicitly, 
\begin{equation}
\mathrm{Sing}\left[\mathrm{Virtual}\right] ~ = ~ - \mathrm{Sing}\left[\int \mathrm{Real}\right]~, \label{eq:unitarity}
\end{equation}
where the notation  $\mathrm{Sing[X]}$ means the singular part of $\mathrm{X}$ (typically cast as explicit poles in a regularisation parameter), 
order by order in the infinite-order sum generated by the shower. The infrared cutoff of the QED shower (which is typically user-definable, with a low default value $\ll1\mev$) then determines what one defines as the softest photon that could in principle produce  resolvable effects. 
\end{itemize}
Note that different applications may choose different functional forms for the slicing or cutoff variable. E.g., in one algorithm it may be a photon energy (in the decay rest frame) or energy fraction, while in another it may be a transverse-momentum type variable. This is important to keep in mind when comparing the numerical sizes of slicing or cutoff values to each other, and will be important when comparing results between different QED generators below.

Standard QED factorisation strategies effectively reduce to two main classes depending on the baseline approximation for the QED emission kernels: those that choose the soft–emission limit as their baseline, as formulated in the YFS framework, and subsequently incorporate corrections for the collinear and hard regimes, and those that instead start from the collinear limit, as formulated in the DGLAP framework, and then introduce corrections to account for the soft and hard regimes. 

Algorithms for QED corrections also differ in whether they operate as showers or as ``$N$-generators''. 
Showers use Sudakov-style evolution equations to sweep over phase space in an ordered manner, generating branchings iteratively as they go along, with the first emission always being the hardest (according to the given shower's evolution variable) and the last being the softest (closest to the infrared cutoff of the shower). 
Vetoes may be applied during one or more of these steps to implement corrections to the baseline shower kernels. 
$N$-generators begin from a prescribed target distribution (e.g., a Poissonian distribution with a specific mean) for the total number of QED branchings, sample a value of $N$ from this distribution, and subsequently generate those $N$ branchings in an unordered manner. 
In other words, no a priori ordering in hardness is imposed, and any of the generated branchings may correspond to the hardest one.
Vetoes may also be applied during or after this procedure, to adjust the final branching distribution relative to the initially prescribed target distribution. 

\TabRef{tab:overview} gives an overview of the type ($N$-generator or shower) and starting limit~(collinear, soft, or both) of each of the current Monte Carlo approaches to QED radiation in hadron decays.
\begin{table}[bp]
\centering
\includegraphics*[width=0.9\textwidth]{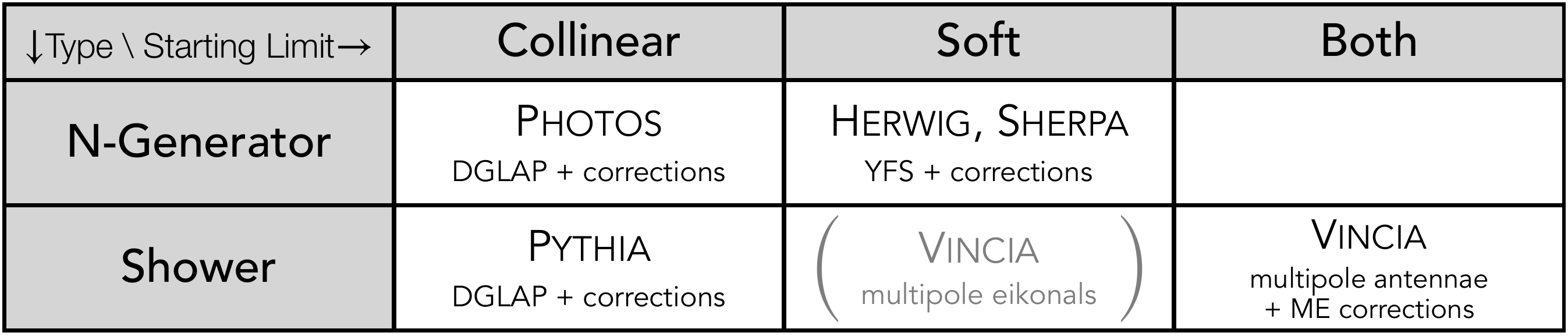}
\caption{Type and starting limit of each of the current Monte Carlo approaches to QED radiation in hadron decays.
The default starting limit of the \vincia event generator is based on the multipole antenna formalism. 
This configuration can be converted to a soft starting limit based on multipole eikonals (indicated with grey parenthesis) by setting the flag \texttt{Vincia:useSpinsQEDHadDec=\string{off,off,off\string}}, where the three entries are for spin-$1/2$, spin-$1$, and spin-$3/2$ particles respectively. 
}\label{tab:overview}
\end{table}

\subsection{PHOTOS}
\label{sec:photos}

As indicated in \tabRef{tab:overview}, the \photos generator~\cite{Barberio:1990ms,Barberio:1993qi,Golonka:2005pn,Nanava:2006vv,davidson2015photos} follows an approach based on the collinear approximation: the QED splitting kernel of a charged decay product into itself and a photon, \decay{X}{X\gamma}, is approximated with the DGLAP splitting function for a spin-$1/2$ particle. 
The QED splitting kernel incorporates the sum of Born-level, virtual, and real photon contributions, where the latter are integrated up to a cutoff energy.
To control systematic errors, the algorithm covers the full phase space without approximations, imposes four-momentum conservation, and makes use of the explicit analytical form of the approximate one-photon matrix element. 

If the exact matrix element for a certain channel is known, \photos can correct for possible deviations with the help of a weight, corresponding to the ratio between the exact matrix element and the approximation.
Otherwise, \photos employs generalised weighting factors to account for the soft-photon emission regime, as well as for discrepancies arising from the DGLAP evolution kernels associated with potentially different spins of the decay products.
In the case of a (pseudo)scalar meson decaying into two (pseudo)scalar particles, \photos has an implemented calculation of the NLO matrix element~\cite{Nanava:2006vv} that must be activated by the user to be effective (we will refer to this configuration as \photos~MEC in \secRef{sec:results}).
In the case of decays with multiple charged particles, \photos applies an additional correction weight to account for interference effects, ensuring proper infrared behaviour.

\subsection{PHOTONS++}
\label{sec:photons}

The approach followed by the \photonspp module~\cite{Schonherr:2008av} (henceforth referred to as \photons), which is embedded in the \sherpa generator~\cite{Sherpa:2019gpd,Sherpa:2024mfk}, is based on the soft-photon approximation following the YFS formalism~\cite{Yennie:1961ad}.
The algorithm separates and exponentiates the universal, process-independent components of the QED corrections, while the remaining process-dependent and spin-dependent contributions are incorporated via subtraction terms. 

The algorithm supports decays into multi-particle final states and implements a dedicated momentum-mapping procedure to relate the original momenta prior to QED radiation to those after radiation. This is required because the eikonal factors entering the YFS splitting kernel are formulated in terms of the already radiation-corrected final-state momenta, rather than the original pre-radiation kinematics. 
In this mapping, all final-state momenta are treated symmetrically: their three-momenta are uniformly rescaled by a common factor, the photon momenta are distributed accordingly, and the energy components of all four-vectors are assigned such that overall momentum conservation and all on-shell constraints are simultaneously satisfied.

As an optional feature, the \photons generator can account for the collinear regime by employing dipole splitting functions~\cite{Dittmaier:1999mb,Catani:2002hc}, generalised to consistently include polarisation degrees of freedom. 
These dipole splitting functions are used as universal multiplicative correction factors, ensuring the correct reproduction of the collinear limit. 
Since they smoothly interpolate between the soft and collinear regions, they inherently contain the soft limit. 
In constructing the correction weights, the soft contributions are explicitly subtracted, as these are already resummed in the original YFS eikonal factors.

To achieve higher precision, the \photons generator incorporates exact first-order, full matrix elements. 
Substantial deviations from approximate treatments of the matrix elements occur in nonsingular regions characterised by comparably hard, wide-angle emissions.  
Accordingly, the algorithm employs infrared-subtracted one-photon real-emission matrix elements, which are subsequently evaluated at the generated momentum configurations.  
Exact first-order matrix elements obtained in this manner are available for a number of relevant processes, such as the decay of a neutral vector boson into a pair of charged fermions.

In the comparative analyses presented in \secRef{sec:results}, the uncorrected configuration employing the YFS splitting kernels will be denoted by \photons~1. 
The configuration applying approximate corrections derived from dipole splitting functions will be denoted by \photons~20, while the configuration incorporating exact next-to-leading-order (NLO) ME corrections will be indicated by \photons~21.

\subsection{PYTHIA}
\label{sec:pythia}

The default parton shower in the \pythia generator~\cite{Bierlich:2022pfr} implements a fully interleaved treatment of final-state QCD+QED radiation~\cite{Sjostrand:2004ef,Corke:2010yf}. 
The shower evolution is ordered in decreasing transverse-momentum (\pT)~scales, in close analogy to the scheme employed in \vincia (see \secRef{sec:vincia}), but differs in that its starting point is the collinear approximation, using DGLAP splitting kernels for \decay{f}{f\gamma} and \decay{\gamma^{(*)}}{f\bar{f}}, with $f$ an elementary spin-$1/2$ charged fermion. 
The precise definitions of $\pT$ also differ between \pythia and \vincia: while they coincide in the soft-collinear limit, they deviate outside this regime. 
Specifically, for a final-state branching \decay{XR}{xr\,\gamma}, where $X$ denoted the radiating charged particle, $R$ its dipole recoiler, and uppercase (lowercase) letters label pre- (post-)branching momenta. 
In \pythia, the evolution variable $\pT$ is defined as~\cite{Sjostrand:2004ef}
\begin{equation}
p_{\perp\mathrm{E}}^2 = z (1-z) Q^2\,,
\label{eq:pypt}
\end{equation}
where the subscript $\mathrm{E}$ stands for ``evolution'', $z = z_x\in[0,1]$ is the energy fraction retained by the emitting particle, and 
\begin{equation*}
    Q^2 = s_{x\gamma} \equiv (p_x + p_\gamma)^2 - m_x^2
\end{equation*}
represents the virtuality (offshellness) of the branching. 
In the expression above, $p_x$ and $p_\gamma$ are the four-momenta of the particles, and $s$ is a Lorentz-invariant quantity defined in general as\footnote{Following the convention of~\cite{Gehrmann-DeRidder:2011gkt}.}
\begin{equation}
    s_{ij} \equiv 2\, p_i \cdot p_j = (p_i+p_j)^2 - m_i^2 - m_j^2\,,
    \label{eq:sij}
\end{equation}
for particles $i$ and $j$ with four-momenta $p_i$, $p_j$ and masses $m_i$, $m_j$.
For comparison, in \vincia the branching is assigned the resolution scale~\cite{Gustafson:1987rq,Giele:2007di}
\begin{equation}
p_{\perp\mathrm{A}}^2 = \frac{s_{x\gamma} s_{r\gamma}}{s_{x\gamma}  +s_{r\gamma}+s_{xr}} ~\stackrel{\gamma \parallel x}\to~ (1-z)\,Q^2\,,
\label{eq:vcpt}
\end{equation}
where the subscript $\mathrm{A}$ stands for both \textsc{Ariadne} (which used the same evolution variable~\cite{Lonnblad:1992tz}) and antenna. 
The rightmost expression corresponds to the $\gamma \parallel x$ collinear limit. 

It is readily observed that \eqRef{eq:pypt} and the $\gamma \parallel x$ limit of \eqRef{eq:vcpt} both vanish in the same way for $z \to 1$ (i.e., in the soft-collinear limit), whereas they differ when considered away from this kinematic limit.
This difference means that the two algorithms will order the emissions that occur ``early'' and ``late'' in the respective showers differently. 
In particular, hard-collinear branchings with $Q^2 \to 0$ and $x_\gamma \to 1 \Rightarrow z \to 0$, will occur quite late in \pythia's evolution, since $p^2_{\perp\mathrm{E}}$ vanishes as the product of $Q^2$ and $z$ in that limit. 
The absence of the factor $z$ in $p_{\perp\mathrm{A}}^2$ implies that those same emissions will occur earlier in the \vincia algorithm. 
Since both algorithms are iterative, ordered showers, this has an impact on how much phase space remains accessible when a given emission is produced and on the degree of associated Sudakov suppression. 
Emissions occurring at ``later'' stages in the evolution are subject to stronger Sudakov suppression than those generated at ``earlier'' stages. 
The difference in the evolution variable also affects the interpretation of the infrared cutoff of the showers: 
for an identical numerical value of the infrared cutoff (for instance, $1\mev$), \pythia will classify more hard-collinear emissions as unresolved (i.e., lying below the infrared cutoff) than \vincia. 
This concludes our comparison of the evolution variables in \pythia and \vincia.

The shower evolution in \pythia is performed by identifying radiation dipoles associated with the decay process. 
In configurations with three or more charged particles, a principle of ``maximal screening'' is used to determine an optimal set of such dipoles. 
This does not capture the full multipole radiation pattern but should provide a reasonable approximation to it for the typical case of higher multipole moments being suppressed relative to the leading dipole contributions.

For simple ($2 \to 1$) hard processes and most $1 \to 2$ resonance decays, \pythia's parton shower incorporates NLO ME corrections; 
systematic matching and merging procedures can also be used to consistently embed further matrix elements into the shower evolution. 
For QED showers in hadron decays, the description is much less advanced. 
Prior to this study, the only available option was to enable QED showers in decays to a pair of leptons, with ME corrections available for vector-meson \decay{V^0}{\ellell} and \decay{V^+}{\ellp\nu} decays, based on those for \decay{\Z}{\ellell} and \decay{\Wp}{\ellp\nu}, respectively. 
In the course of this work, we have enabled \pythia to also generate QED showers for more general hadron decays, although it still has the limitation that it will always use the spin-$1/2$ DGLAP kernels regardless of the actual spins of the involved particles. This option is therefore mostly intended for comparison and will produce the wrong collinear limits for mesons.

The \pythia parton shower can be configured to evolve exclusively via its QED component. 
In \secRef{sec:results}, we use this restricted parton shower to obtain an additional reference sample to study QED radiation in the decay of a vector meson into a pair of charged leptons, for which \pythia's description should be accurate and includes the NLO ME correction. 
We note that no form factor for the decaying meson is included in this description, but such form factors are not included in any of the other approaches discussed here. 

\section{The VINCIA antenna-shower approach}
\label{sec:vincia}

The \vincia generator~\cite{Brooks:2020upa} implements algorithms for QED and QCD parton showers with an interleaved~\cite{Sjostrand:2004ef,Brooks:2021kji}, \pT-ordered evolution~\cite{Giele:2007di,Skands:2020lkd} based on sector antennae~\cite{Kleiss:2017iir,Lopez-Villarejo:2011pwr,Skands:2020lkd,Brooks:2020upa}. 

In the present study, we restrict our attention to \vincia's QED antenna showers which are detailed in~\cite{Kleiss:2017iir,Skands:2020lkd}.  
Conceptually, these are closely analogous to their QCD counterparts~\cite{Brooks:2020upa}, except of course that the photon is uncharged, electric charges are not confined, and there is no ``colour flow'' or $1/N_C$ expansion. 
QCD showers exploit the latter, in particular, to set up a system of radiating QCD dipoles which is unique at ``leading colour'' (lowest order in $1/N_C$). 
QED does not offer a similar expansion parameter, but the simpler Abelian structure of QED makes other approaches viable, such as the principle of maximal screening employed in \pythia, cf.~\secRef{sec:pythia}. 
In \vincia, the default option is to include the full multipole structure of soft QED radiation~\cite{Kleiss:2017iir,Skands:2020lkd}, although a simpler (and somewhat faster) option analogous to \pythia's maximal-screening principle is also available. 
Comparisons between these options are included in some of the validation plots presented in \secRef{sec:results}.

The antenna formalism provides a unified description of both the soft and collinear divergent behaviour of radiative matrix elements. 
In contrast to shower formalisms based on Catani-Seymour~(CS) dipoles~\cite{Catani:1996jh,Catani:1996vz}, in which the antenna function is effectively partitioned into two separately radiating ``dipole ends'', the antenna framework treats the radiation pattern as arising coherently from the full dipole pair\footnote{This is also the case for the Lund-dipole formalism~\cite{Gustafson:1987rq}, implemented, e.g., in the \ariadne~\cite{Lonnblad:1992tz} generator.}. 
In the dipole formalism, one parton is designated as the radiator for each emission, while the other acts as a purely longitudinal recoiler. 
In the antenna framework, both dipole ends act simultaneously as radiators for each emission, sharing the recoil between them, with a smooth interpolation between the two collinear limits. 

Compared to approaches based on YFS exponentiation, such as \photonspp, or to algorithms based on collinear logarithms that are included only order by order, such as \photos, and which are used as “afterburners” to generate coherent soft-photon radiation in particle decays, the antenna parton-shower approach not only reproduces coherent QED radiation but can also be systematically interleaved with QCD radiation and with resonance decays~\cite{Brooks:2021kji}. 
In the present work, however, we restrict ourselves to using \vincia as an afterburner and defer a detailed study of its interleaving with hadronic decays to future investigations.

The default QED antenna functions implemented in \vincia \cite{Kleiss:2017iir,Skands:2020lkd} incorporate fully coherent multipole soft-interference contributions in addition to the standard collinear DGLAP structures. 
Mass effects are systematically taken into account~\cite{Brooks:2020upa,Gehrmann-DeRidder:2011gkt} both through mass-dependent corrections to the antenna functions, ensuring that all relevant (quasi-)collinear limits are correctly reproduced, and through the application of exact phase-space factorisations for massive particles. Hadronic form factors are neglected, as are possible effects arising from photon mixing with hadronic vector resonances (for example, through vector meson dominance) and effective photon splitting into pairs of charged particles. 
Treatment of these aspects is deferred to future work.

\subsection{The QED multipole shower formalism in VINCIA}

\label{sec:multipole}

The default QED parton shower in \vincia employs a fully coherent multipole formalism for the simulation of photon radiation. 
In its original implementation, the algorithm was designed to describe radiation from systems of elementary charged fermions, vector bosons, and scalar particles, as well as photon splitting into pairs of charged fermions~\cite{Kleiss:2017iir,Skands:2020lkd}. 

The principal extension introduced in the present work is a generalisation of the antenna functions to allow for the description of radiation emitted from charged hadrons, incorporating an explicit dependence on the hadron spin. 
In the current implementation, this includes dedicated antenna functions for spin-$0$, spin-$1/2$, and spin-$1$ hadrons.
Hadrons with higher integer spins, if provided, are effectively treated as scalar particles, whereas higher-spin fermions are treated the same as spin-$1/2$ particles.

To introduce the underlying antenna formalism, we consider \decay{IK}{ik\gamma} branchings\footnote{In \secRef{sec:pythia} we used the notation \decay{XR}{xr\,\gamma}, where $X$ is the radiating charged particle and $R$ its dipole recoiler. That notation reflects that, in \pythia (as in any DGLAP or partitioned-dipole shower), the roles of radiator and recoiler are distinct. 
In the antenna language, however, the photon is emitted coherently from the pair, so there is no unique radiator or recoiler; singling one out would be artificial. 
We therefore label them $I$ and $K$ here, to emphasise that they are equivalent in a way that they are not in the DGLAP/dipole picture.} 
with charged emitters $I$ and $K$ carrying four-momenta $p_I$ and $p_K$.
After the branching, the recoiling particles carry the four-momenta $p_i$ and $p_k$, while the emitted photon has four-momentum $p_j$. 
The corresponding branching process is therefore
\begin{equation*}
I(p_I) \,+\, K(p_K) ~\mapsto~ i(p_i) \,  + \,k(p_k)\,+\, \gamma(p_j)\,.
\end{equation*}
If both $I$ and $K$ are final-state particles, the antenna is called a final-final (FF) antenna. If particle $I$ is the decaying hadron, the antenna is called an initial-final (IF) antenna\footnote{Also called a resonance-final (RF) antenna~\cite{Brooks:2019xso}.}.

The emission probability densities are described by (unpolarised) antenna functions, whose leading (eikonal) terms are universal, i.e., independent of the total spins of the emitting particles and the same for both FF and IF antennae.
We therefore express the photon-emission dipole antenna functions as
{\addtolength{\abovedisplayskip}{4pt}
\addtolength{\belowdisplayskip}{4pt}
\begin{equation*}
a_\mathrm{emit}^{ik}(s_{ij},s_{jk},s_{ik},m_i^2,m_k^2)~=~
a_\mathrm{scalar}^{ik}\,+\,a_\mathrm{spin}^{i}\,+\,a_\mathrm{spin}^{k}\,,
\end{equation*}}
where the spin-independent eikonal term 
 (presented here without the corresponding electromagnetic charge and coupling factors) is:
\begin{align}
    a_{\rm scalar}^{ik} \,&=\, 4\frac{s_{ik}}{s_{ij}s_{jk}} \,-\, 4\frac{m_i^2}{s_{ij}^2} \,-\, 4\frac{m_k^2}{s_{jk}^2}\,,
    \label{eq:AntennaScalar}
\end{align}
with the Lorentz-invariant quantity $s_{xy}\equiv 2p_{x\mu} p_{y}^{\mu}$, \cf \eqRef{eq:sij}, and rest masses $m_i=m_I$ and $m_k=m_K$.

The hard-collinear terms for spin-1/2 emitters are also independent of whether the antenna is FF or IF:
\begin{align}
 a^{i}_{\text{spin-}\frac{1}{2}} \,&=\,  \frac{2}{s_{IK}} \frac{s_{jk}}{s_{ij}}\,,
   \\[2mm]  a^{k}_{\text{spin-}\frac{1}{2}} \,&=\,  \frac{2}{s_{IK}} \frac{s_{ij}}{s_{jk}}\,,
\end{align}
while for spin-$1$ emitters, the signs of some of the collinear invariants depend on whether the antenna is FF or IF (related by crossing),
\begin{align}
    \mbox{FF :} & & a^{\mathrm{FF},i}_{\text{spin-}1} \,&=\,  \frac{8}{3} \frac{1}{s_{ij}}\left( \frac{s_{jk}}{s_{IK}} + \frac{s_{jk}}{s_{IK}-s_{jk}} - \frac{s_{jk}^2}{s^2_{IK}} \right) 
    \,,  \\[2mm]
     & & a^{\mathrm{FF},k}_{\text{spin-}1} \,&=\, \frac{8}{3} \frac{1}{s_{jk}}\left( \frac{s_{ij}}{s_{IK}} + \frac{s_{ij}}{s_{IK}-s_{ij}} - \frac{s_{ij}^2}{s^2_{IK}} \right)\,,\\
    \mbox{IF :} & & a^{\mathrm{IF},i}_{\text{spin-}1}
    \,&=\,  
    \frac{8}{3} \frac{1}{s_{ij}}\left( \frac{s_{jk}}{s_{IK}} 
    +\frac{s_{jk}}{s_{IK}+s_{jk}} +\frac{s_{jk}^2}{s_{IK}^2} \right) 
    \,,  \\[2mm]
     & & a^{\mathrm{IF},k}_{\text{spin-}1} \,&=\, 
      \frac{8}{3} \frac{1}{s_{jk}}\left( \frac{s_{ij}}{s_{ik}+s_{kj}} + \frac{s_{ij}}{s_{IK}+s_{jk}} - \frac{s_{ij}^2}{(s_{IK}+s_{jk})^2} \right)\,.
\end{align}
The full dipole antenna function, for a pair of charged particles $i$ and $k$, is thus,
\begin{align}
        a_{\rm emit}^{ik} (s_{ij},s_{jk},s_{ik},m_i^2,m_k^2) \,=&\, a_{\rm scalar}^{ik} \nonumber\\[2mm] &\,+\,  \delta^{i}_{\text{spin-}\frac{1}{2}}a^{i}_{\text{spin-}\frac{1}{2}} 
        \,+\,\delta^{k}_{\text{spin-}\frac{1}{2}}a^{k}_{\text{spin-}\frac{1}{2}} 
        \nonumber\\[2mm]
        &\,+\, \delta^{i}_{\text{spin-}1}a^{i}_{\text{spin-}1} 
        \,+\, \delta^{k}_{\text{spin-}1}a^{k}_{\text{spin-}1}        
    \label{eq:FFAntenna}\,,
\end{align}
where $\delta^x_{\text{spin-}S}=1$ for particle $x$ having (total) spin $S$ and zero otherwise, and the spin-$1$ terms use either the FF or IF forms given above, according to the type of dipole antenna.

The fully coherent (multipole) antenna function, in which all eikonal contributions are incorporated simultaneously within a single splitting kernel, is given by
\begin{equation}
       a^{\rm multipole}_{\rm emit}(\{p\},p_{j}) \;=\; -\sum_{\{i < k\}}\sigma_{i}Q_{i}\,\sigma_{k}Q_{k}\,a_{\rm emit}^{ik} (s_{ij},s_{jk},s_{ik},m_i^2,m_k^2)\,,
  \label{eq:QEDAntenna}
\end{equation}
where the sum over \(\{i <k\}\) runs over all unordered pairs\footnote{I.e., the pair $\{i,k\}$ appears only once, not twice as would be the case if $\{i,k\}$ was considered distinct from $\{k,i\}$ (ordered pairs).} of charged particles  with electromagnetic charges \(Q_i\) and \(Q_k\) (including contributions from both equal and opposite signs). 
The factors \(\sigma_i\) and \(\sigma_k\) denote sign factors (a.k.a., crossing factors) that are defined as \(+1\) for final-state particles and \(-1\) for initial-state particles. 

This splitting kernel contains all soft multipole contributions and also reproduces the correct collinear limits~\cite{Kleiss:2017iir}, although its singularity structure is quite intricate. 
In the \vincia sector shower, double counting of singular configurations is avoided by allowing only a single such splitting kernel to contribute at any given phase-space point. 
This is achieved by partitioning the phase space into sectors, following the formalism of ref.~\cite{Lopez-Villarejo:2011pwr}, with the modification that the antenna function itself is employed as the sector-defining variable.

Consequently, the shower approximation to the NLO matrix element can be written as
\begin{equation}
    |M^\mathrm{PS}_{n+1}(\{p\},p_{j})|^{2} \;=\; a^{\rm multipole}_{\rm emit}(\{p\},p_{j})\,4\pi\alpha
    \sum_{\{i <k\}} \Theta_j\big(a_{\rm emit}^{ik}\big)\,
    |M_{n}(\{p\})|^{2}\,,
    \label{eq:CoherentFactorisation}
\end{equation}
where the superscript $\mathrm{PS}$ stands for ``Parton Shower'', $\alpha$ is the (renormalised) electromagnetic coupling constant, and 
the Heaviside function \(\Theta\) is defined as
\begin{equation*} \label{eq:vincia-sector-resolution}
        \Theta_{j} \;=\; 
        \begin{cases}
                1 & \text{if } a_{\rm emit}^{ik}(s_{ij},s_{jk},s_{ik},m_i^2,m_k^2) \,\geq\,
                a_{\rm emit}^{vw}(s_{vj},s_{jw},s_{vw},m_v^2,m_w^2)
                ~\forall~\text{pairs } \{v,w\}\,,\\[8pt]
                0 & \text{otherwise}\,,
        \end{cases}
\end{equation*}
which effectively selects the sector corresponding to the pair of charged particles \(i\) and \(k\) in the state $\{p\}$ that yields the largest value of the antenna function for the emission of photon \(j\). Note that 
the sectorisation implies that, for each post-branching phase-space point, $(\{p\},p_j)$, only a single term inside the sum over $\{i<k\}$ in \eqRef{eq:CoherentFactorisation} is ever non-zero. 
A technical description of the implementation of the \vincia parton shower, as well as its interface within \pythia, is provided in~\appRef{app:QEDshower}. 

\subsection{The simple dipole-shower option}

\label{sec:dipole}

The \vincia generator also provides a simpler and computationally more efficient alternative in which the complete multipole sum is replaced by a set of individual dipole contributions, constructed according to a principle of maximal screening analogous to that used in \pythia's QED shower (see \secRef{sec:pythia}). 

In this formulation, radiation is generated solely from a collection of dipoles, with each charged particle only appearing in exactly one dipole, and the corresponding dipole shower adopts a faster, unsectorised scheme. 
The antenna function in \eqRef{eq:QEDAntenna} is no longer evaluated as a sum over all charged pairs, but instead only over a specific set of pairs which each satisfy $\sigma_i Q_i \sigma_k Q_k < 0$, i.e., each pair is defined by two opposite-sign charged hadrons (modulo crossings). 

Following the principle of maximal screening~\cite{Skands:2020lkd}, the selection of pairs $\{i,k\}$ to form within a given multi-particle configuration is defined by minimising $\sum_{\{i<k\}} \left[(p_i+p_k)^2 - (m_i+m_k)^2\right]$.
In \vincia, the combinatorial optimisation problem of finding the optimal pairings under this constraint is solved with $\mathcal{O}(n^3)$ time complexity using the so-called Hungarian algorithm~\cite{kuhn1955hungarian,munkres1957algorithms,jonker1987shortest}. 

Although this algorithmic strategy is generally more efficient, it only generates a (summed) dipole approximation to the full multipole structure. 
In particular, for configurations involving three or more charged particles, the exact soft-photon multipole radiation pattern is not fully captured.

\subsection{Matrix-element corrections}

\label{sec:mecs}

In the presence of hard, wide-angle photons, the nonsingular contributions in the matrix-element factorisation become phenomenologically significant. 
These are not guaranteed to be captured by shower kernels (irrespective of whether they are based on DGLAP, dipoles/antennae, or YFS).  
To incorporate these effects, the antenna function in \eqRef{eq:QEDAntenna} can be replaced by the exact tree-level matrix-element ratio, up to a given (fixed) tree-level order. 
In most current descriptions, such ME corrections are only implemented to first order. 
We have implemented corrections for up to two-photon emissions in \vincia by default, but it is fairly straightforward to generate and interface corrections for more photons if so desired. 
Instructions for how to generate and implement further ones will be given in the \pythia's HTML manual~\cite{pythia_html}.

\vincia currently supports generic QED ME corrections for two-body decays of a scalar into two leptons or two scalars, and for three-body decays of a scalar into all-scalar final states. 
We plan to extend these to processes involving generic fermions and spin-$1$ hadrons in future work.
The ME corrections are implemented by employing a combination of \feynrules~\cite{Degrande:2011ua,Alloul:2013bka} and \madgraph~\cite{Alwall:2011uj,Alwall:2014hca}.
Details of the implementation and a summary of the ME corrections included with \vincia in the next release of \pythia (tentatively, \pythia~8.319) are given in~\appRef{app:MEs}.

Currently, matrix elements are evaluated only for one photon at a time. 
That is, the QED shower proceeds by iterated single emissions, which can include iterated ME corrections but does not incorporate so-called direct multiple-emission contributions such as the direct $2 \to 4$ branching processes in \cite{Li:2016yez,El-Menoufi:2024sys}. 
Thus, at each step, the ME-corrected antenna corresponds to the ratio 
\begin{equation*}
    a^{\rm QED}_{\rm emit}(\{p\},p_{j}) \;=\; \frac{|M_{n+1}(\{p\},p_{j})|^{2}}{|M_{n}(\{p\}/p_{j})|^{2}}\,,
\end{equation*}
where the notation $\{p\}/p_j$ in the argument of the matrix element in the denominator means $p_j$ clustered into the state $\{p\}$ (i.e., the pre-branching configuration) while the numerator is evaluated on the unclustered momenta (i.e., the post-branching configuration).
In the next iteration of the shower, one can then either use an ME-corrected ratio again (if the new numerator ME is available) or the antenna approximation. 
When ME corrections are switched on and available for the considered decay topology, the current default setting is to include ME corrections for the first two photon emissions. 

We note that ME corrections for branchings involving \decay{\gamma}{\ellell} splitting have not been implemented at this point. 
They do remain available to switch on in the shower (without ME corrections), although we emphasise that enabling such branchings in hadron decays should be accompanied by a careful consideration of the relative decay branching ratios to inclusive vs. exclusive channels with/without additional \ellell pairs. 

The treatment of virtual corrections proceeds analogously to their implementation in standard parton-shower algorithms, namely through the enforcement of unitarity via Sudakov form factors.
In parton showers, one exploits the fact that, at the level of singular contributions, the virtual correction is exactly equal in magnitude and opposite in sign to the phase-space integral of the corresponding real-emission correction, cf.~\eqRef{eq:unitarity}. 
The nonsingular remainder is neglected, which is consistent with the approximation that the correction being implemented is dominated by the infrared and collinear singular structures.

When employing tree-level matrix elements for real emissions instead of their antenna approximations, the integral over real emissions is modified by a nonsingular correction. Since the corresponding virtual corrections are still incorporated via~\eqRef{eq:unitarity}, this also induces a change in the effective virtual correction. 
The formal perturbative accuracy of the calculation is neither improved nor degraded by this. 
To include explicit virtual corrections in a controlled manner, one would have to perform a loop-level matching to fixed-order calculations, such as in \textsc{Mc@Nlo}~\cite{Frixione:2002ik}, \textsc{Powheg}~\cite{Nason:2004rx}, or equivalent strategies. 

At the level of inclusive decay rates, it is sufficient to normalise the total decay width to the known NLO prediction (or to experimentally measured total widths) in order to effectively incorporate the NLO virtual corrections. 
Virtual photon contributions induce relative corrections of \(\mathcal{O}(\alpha)\) with respect to the tree-level result; 
however, they do not modify the event-by-event kinematics, only the overall rates.
Therefore, provided that the total rate is normalised to a theoretical or experimental reference value that already includes the virtual contributions, these virtual corrections are effectively taken into account.

\section{The dead-cone effect}
\label{sec:deadConeEffect}

The dead-cone effect is a well-established phenomenon in both classical and quantum gauge field theories, manifesting itself as a strong suppression of gauge-boson emissions within a polar angle of $\sim m/E = 1/\gamma$ of a radiator with mass $m$ and energy $E$. 
See, e.g., \cite{ALICE:2021aqk,LHCb:2025mcq,CMS:2026lkj} for recent demonstrations of this effect for QCD radiation off charm and bottom quarks. 

\begin{figure}[t]
\centering
\includegraphics*[width=0.995\linewidth]{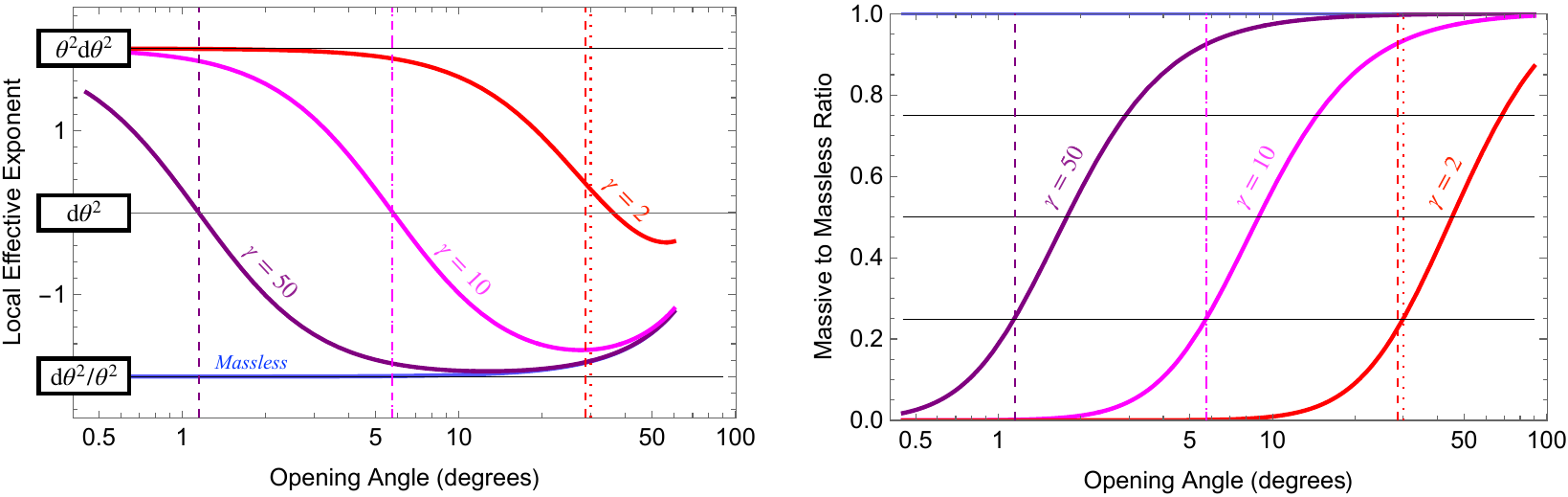}
\caption{The (left)~local effective exponent  and (right)~massive-to-massless antenna ratio for soft photon emissions, as functions of emission angle (note the logarithmic axes), for a few representative Lorentz factors $\gamma=2$, $10$, and $50$, with vertical lines indicating the standard dead-cone
$\theta = 1/\gamma$ (vertical dashed lines) and, for the small $\gamma$ factors, the slightly larger $\theta = \cos^{-1}(\beta)$ (vertical dotted lines).
\label{fig:schematic}}
\end{figure}

Briefly stated, the differential photon emission probability for bremsstrahlung off a \emph{massless} charged particle at small opening angles $\theta$ in the soft-photon limit exhibits the characteristic double-logarithmic form
\begin{equation}
\mathrm{d}{\cal P} \,\propto\, \frac{\mathrm{d}E}{E}\, \frac{\mathrm{d}\theta^2}{\theta^2}\,.
\label{eq:bremsprob}
\end{equation}
Plotted logarithmically in photon energy and opening angle, the emission density is roughly constant (and proportional to $\alpha$). 
For radiation off a \emph{massive} particle, the emission probability has the same $\mathrm{d}\theta^2_\gamma/\theta^2_\gamma$ behaviour as \eqRef{eq:bremsprob} for angles above $\theta_\mathrm{dead} \sim 1/\gamma = m/E$.
For angles below this threshold, the emission probability is suppressed, tending to $\theta^2_\gamma \mathrm{d}\theta^2_\gamma$ for $\theta_\gamma \to 0$.
This is illustrated in \figRef{fig:schematic}, where we show the local effective exponent of the soft-limit antenna function defined as $p_\mathrm{eff}(\theta_\gamma)=\mathrm{d}\ln(\mathrm{ant})/\mathrm{d}\ln \theta_\gamma$, and the ratio of the massive to the massless antenna functions in the same limit for Lorentz factors of $\gamma = 2$, $10$, and $50$.
Also shown are the conventional dead-cone definitions at $\theta = 1/\gamma$ and, for the small $\gamma$ factors, the slightly larger $\theta = \cos^{-1}(\beta)$ that we advocate below. 
In this section, we show how the dead cone arises from our antenna functions and phase-space constraints. 
We also present the full angular radiation pattern and study how the dead cone changes beyond the soft-photon and relativistic-radiator limits, and under different laboratory~(LAB) and centre-of-mass~(CM) definitions of opening angles and rapidities. 
This matters because experiments observe hadron decays with a broad LAB-frame four-momentum spectrum, whereas theory discussions are usually spectrum-independent and are formulated in the hadron CM frame.

\subsection{Notation and basics}
\label{sec:deadConeBasics}

Consider a QED dipole-antenna spanned between a pair of oppositely charged scalar particles (see \eqRef{eq:AntennaScalar}). 
For simplicity, consider the case with only $m_I=m_i \neq 0$ and $m_K=m_k=0$. The corresponding antenna function (without making any soft-photon approximation at this point and using $p_j=p_\gamma$) is proportional to~\cite{Kleiss:2016esx}
\begin{equation*}
a(p_i,p_\gamma,p_k) ~\propto~   \frac{s_{ik}}{s_{i\gamma}s_{k\gamma}} \,-\, \frac{m_i^2}{s_{i\gamma}^2}\,. 
\end{equation*}
Requiring the antenna function to be non-negative yields 
\begin{equation}
    s_{i\gamma}s_{ik} \,-\, m_i^2 s_{k\gamma} ~\ge~ 0\,.
    \label{eq:gram}
\end{equation}
This is exactly equal to the expression one obtains from requiring a positive Gram determinant, which defines the accessible phase space for massive kinematics. 
That is, \eqRef{eq:gram} represents both the physical boundary of the massive phase space as well as the contour along which the (negative) mass correction to the scalar antenna function becomes equal to the (positive) massless eikonal term.  

\subsection{Parametric rapidity and dead-cone angle}

Before discussing the dead cone in terms of angular variables, we first note that the constraint in \eqRef{eq:gram} implies hard boundaries on an invariant quantity, which we call the parametric rapidity of the photon,
\begin{eqnarray}
    y_\gamma^* &~\equiv~& \frac12\ln\left(\frac{s_{k\gamma}}{s_{i\gamma }}\right) ~ \le~ \frac12\ln\left( \frac{m_{IK}^2\left(1 - x_\gamma\right)-m_I^2}{m_I^2}\right)\,, 
    \label{eq:yStarMax}
\end{eqnarray}
where $m_{IK}$ is the total CM energy of the $IK$ system (equal to the mass of the decaying hadron if $I$ and $K$ were produced in a two-body decay), and 
$x_\gamma = 2E_\gamma/m_{IK}$ is the energy fraction of the emitted photon.
For a massive recoiler, $m_K \neq 0$, the bound is a bit more involved,
\begin{eqnarray}
y_\gamma^*&~\stackrel{m_K\neq 0}{\le}~&\,
 \frac12\ln\left( \frac{m_{IK}^2\left(1 - x_\gamma\right)-m_I^2-m_K^2}{m_I^2}\right) \nonumber \\ & & +\,
\frac12\ln\left(\frac12 + \frac12\sqrt{1 - \frac{4m_I^2 m_K^2}{(m_{IK}^2(1-x_\gamma) - m_I^2 - m_K^2)^2}}\right)\,.
\label{eq:yStarMaxmK}
\end{eqnarray}
We note that this bound is always smaller than that of \eqRef{eq:yStarMax}. 
In the soft-photon ($x_\gamma \to 0$) limit, a useful overestimate of the bounds on $y^*_\gamma$ for general masses is:
\begin{equation}
-\ln\left(\frac{m_{IK}-m_K^2}{m_K^2}\right) \,+\, \frac{m_I^2}{2m_{IK}^2} ~\le~ y^*_\gamma ~\le~ \frac12\ln\left(\frac{m_{IK}^2-m_I^2}{m_I^2}\right) \,-\, \frac{m_K^2}{2m_{IK}^2}\,,
\label{eq:yStarBounds}
\end{equation}
where terms quartic in $m/m_{IK}$ have been dropped.

The quantity $y^*_\gamma$ can be reconstructed experimentally by noting that, in an arbitrary frame, it can be expressed as 
\begin{equation}
y^*_\gamma = \frac12\ln\left(\frac{s_{k\gamma}}{s_{i\gamma}}\right)~=~ \frac12\ln\left(\frac{E_k-|\vec{p}_k|\cos\theta_{k\gamma}}{E_i-|\vec{p}_i|\cos\theta_{i\gamma}}\right)~=~\frac12\ln\left(\frac{E_k}{E_i}\right) \,+\, \frac12\ln\left(\frac{1-\beta_k\cos\theta_{k\gamma}}{1-\beta_i\cos\theta_{i\gamma}}\right)\,, 
\label{eq:yStarRec}
\end{equation}
with $E_{i,k}=\sqrt{|\vec{p}_{i,k}|^2+m_{i,k}^2}$. 
In the soft-photon limit, this becomes
\begin{equation*}
y^*_\gamma ~\stackrel{x_\gamma \to 0}\to 
\frac12\ln\left(\frac{m_{IK}^2 + m_K^2-m_I^2}{m_{IK}^2+m_I^2-m_K^2}\right) \,+\,\frac12\ln\left(\frac{1-\beta_K \cos\theta_{k\gamma}}{1-\beta_I\cos\theta_{i\gamma}}\right)\,,
\end{equation*}
where, in the CM-frame, $\cos{\theta_{k\gamma}}\sim -\cos{\theta_{i\gamma}}$.
We emphasise that the fact that $y_\gamma^*$ remains kinematically bounded in the soft-photon limit does not imply that there is any opening angle (or physical photon rapidity) which is kinematically inaccessible. 

Moving on to the conventional dead-cone angle for soft radiation around a massive emitter, we note that this does not correspond to an actual zero of the massive antenna function but to the ratio of the massive to massless antenna functions becoming small. 
To derive it, we start from this ratio,
\begin{equation*}
 \frac{\mathrm{ant}(m_I\neq 0)}{\mathrm{ant}(m_I=0)} ~=~
 \frac{(1-\cos\theta_{ij})(1-\beta_i \cos\theta_{ik})}
 {(1-\beta_i \cos\theta_{ij})(1-\cos\theta_{ik})}
 -\frac{(1-\cos\theta_{ij})(1-\cos\theta_{jk})}{2(1-\cos\theta_{ik})(1-\beta_i \cos\theta_{ij})^2\gamma_i^2}\,,
 \label{eq:antRatio}
\end{equation*}
where we have not yet made any approximations apart from setting $m_K=0$ (which does not affect the derivation to the accuracy we present), and $\gamma_i$
is the Lorentz factor of particle $i$. 

In the soft-photon limit, we adopt the no-recoil approximation in the center-of-mass~(CM) frame, i.e. $\cos\theta_{ik} \to -1$ and $\cos\theta_{jk} \to -\cos\theta_{ij}$. 
We then determine the condition under which the ratio of massive to massless antenna function is equal $1/4$ (chosen as an arbitrary but reasonable criterion quantifying the suppression of massive relative to massless emission). 
This procedure yields
\begin{equation}
\cos(\theta_\mathrm{dead}) ~=~\beta_i\,. 
\label{eq:deadExact}
\end{equation}
Performing an expansion about the relativistic limit $\beta_i \to 1$, one obtains the following expression for the dead-cone angle:
\begin{equation}
\theta_\mathrm{dead} ~=~
\frac{1}{\gamma_i}
\,+\,\frac{1}{6\gamma_i^3}
 \,+\,{\cal O}(\gamma_i^{-5})\,+\,{\cal O}(\gamma_i^{-3}\gamma_k^{-2})\,,
 \label{eq:deadExpand}
\end{equation}
where the first term corresponds to the standard dead-cone angle definition, $\gamma_i^{-1}=m_i/E_i$, and the first non-relativistic correction is independent of $m_k$.
The reason we include a non-relativistic correction is that the emitters in hadron decays are not always highly relativistic. 
In \tabRef{tab:boosts} we tabulate some (CM-frame) velocities and Lorentz factors for a selection of two-body hadron decays, \decay{X}{1 + 2}, along with the sizes of the dead-cone angles using just the first term in \eqRef{eq:deadExpand} or the unexpanded form in \eqRef{eq:deadExact}. 
\begin{table}[tp]
\centering
\begin{tabular}{l|cc|cc|cc|cc}
\toprule
\decay{X}{1+2} & $\beta_1^*$ & $\beta_2^*$ & $\gamma_1^*$ & $\gamma_2^*$ & 
\multicolumn{2}{c|}{$1/\gamma^*$ in deg.}
& \multicolumn{2}{c}{$\cos^{-1}(\beta^*)$}
 \\
\midrule
\decay{\Bd}{\Ds\Dm} & 0.677 & 0.696
& 1.36 & 1.39 
& $42.2^\circ$ & $41.2^\circ$
& $47.4^\circ$ & $45.9^\circ$ 
\\
\decay{\Bd}{\Dp\Km} & 0.773 & 0.977 & 1.58 & 4.72  & $36.3^\circ$ & $12.1^\circ$ 
& $39.4^\circ$ & $12.2^\circ$
\\
\decay{\Bd}{\Dp\pim} & 0.777 & 0.998 & 1.59 & 16.55 & $36.1^\circ$ & $3.5^\circ$ 
& $39.0^\circ$ & $3.5^\circ$
\\
\decay{\Bd}{\Kp\pim} & 0.983 & 0.999 & 5.39 & 18.76 & $10.6^\circ$ & $3.1^\circ$ 
& $10.7^\circ$ & $3.1^\circ$
\\
\decay{\jpsi}{\mumu} & \multicolumn{2}{c|}{0.998} & \multicolumn{2}{c|}{14.61} 
& \multicolumn{2}{c|}{$3.9^\circ$}  
& \multicolumn{2}{c}{$3.9^\circ$} 
\\
\decay{\Dz}{\Kp\Km} & \multicolumn{2}{c|}{0.848}
& \multicolumn{2}{c|}{1.89}
& \multicolumn{2}{c|}{$30.4^\circ$} 
& \multicolumn{2}{c}{$32.0^\circ$} 
\\
\decay{\Dz}{\Km\pip} & 0.867 & 0.987 & 2.01 & 6.25
& $28.5^\circ$ & $9.2^\circ$ 
& $29.8^\circ$ & $9.2^\circ$ 
\\
\decay{\Kz}{\pip\pim} & \multicolumn{2}{c|}{0.828} & \multicolumn{2}{c|}{1.78}
& \multicolumn{2}{c|}{$32.1^\circ$} 
& \multicolumn{2}{c}{$34.1^\circ$} 
\\
\bottomrule
\end{tabular}
\caption{CM-frame velocities $\beta^*$, Lorentz factors $\gamma^*$, and dead-cone angles $1/\gamma^*$ and $\cos^{-1}\beta^*$ for a selection of two-body hadron decays, with mass values from~\cite{ParticleDataGroup:2024cfk}.
\label{tab:boosts}}
\end{table}

\begin{figure}[tb]
    \centering
    \begin{subfigure}{0.49\textwidth}
        \includegraphics[width=\linewidth]{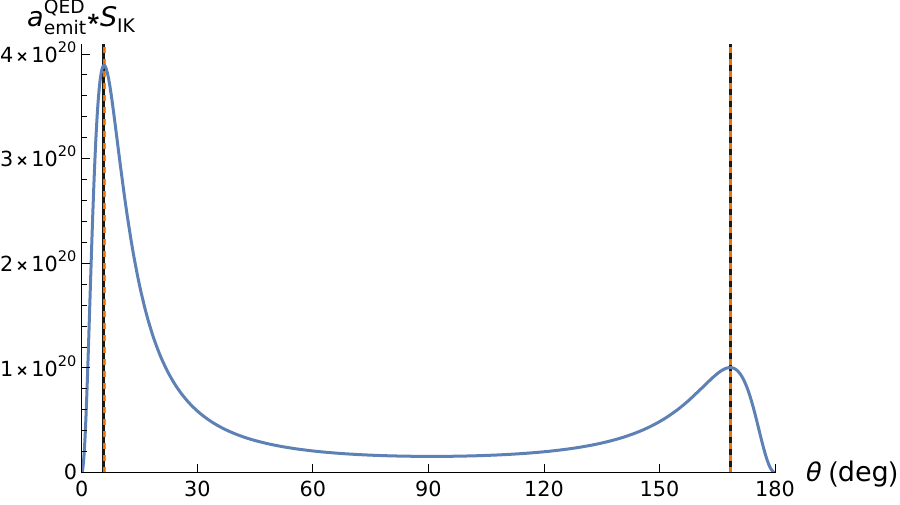}
        \caption{
        $(m_{\rm X}, m_{1}, m_{2}) = (20, 1, 2)\gev$,
                 $E_\gamma = 10^{-8}$\gev.
                 }
        \label{fig:stationaryRel}
    \end{subfigure}
    \hfill
    \begin{subfigure}{0.49\textwidth}
        \includegraphics[width=\linewidth]{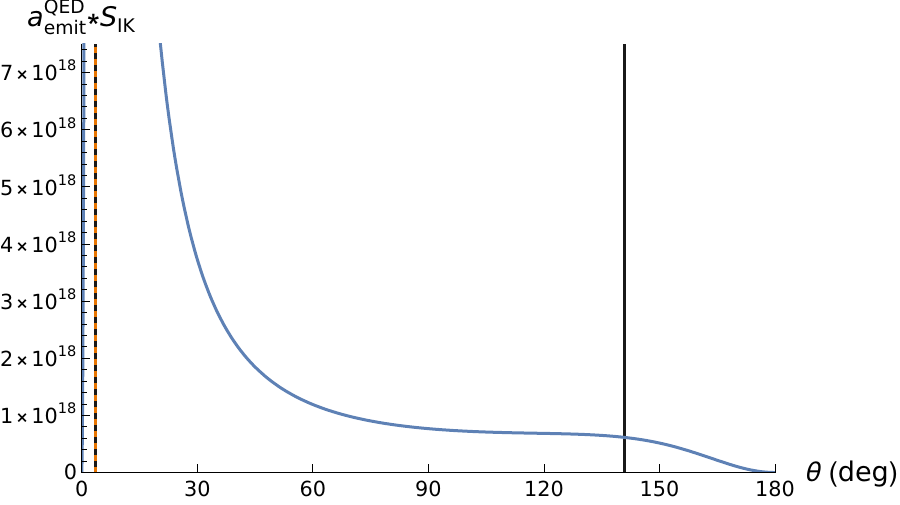}
        \caption{
        \decay{\Bz}{\Dp\pim}, $E_\gamma = 10^{-8}$\gev.
        \\
        }
        \label{fig:stationaryBad}
    \end{subfigure}
    \caption{Angular radiation patterns for massive antenna functions. 
             The dead-cone angles $\cos^{-1}(\beta)$ (black vertical lines) 
             and the stationary points of the antenna function (orange dashed 
             lines) are shown for comparison.
             In the figure on the left, the wide-angle radiation from the light (pion) emitter dominates, and the dipole function has no well-defined stationary point in the \D-meson hemisphere.}
    \label{fig:stationary}
\end{figure}

An alternative definition of the dead-cone angle can be derived by identifying the stationary point of the massive antenna function. 
At this point, the nominal divergence $1/\theta^2$ is regulated by the negative mass term (see the left-hand pane of \figRef{fig:schematic}).
For relativistic emitters, this procedure yields the same dead-cone angle as obtained by imposing that the ratio of the massive-to-massless antenna functions equals $1/4$.
This is illustrated in \figRef{fig:stationaryRel}, where we plot the massive antenna function for a hypothetical \decay{X^0}{1^+ 2^-} decay with highly relativistic decay products $(m_{\rm X}, m_{1}, m_{2}) = (20, 1, 2)\gev$ emitting a very soft photon with an energy of $10^{-8}\gev$. 
The black vertical lines indicate our dead-cone definition, $\cos^{-1}(\beta)$, and the orange dashed lines (which coincide almost exactly with the black solid ones) indicate the angle at which the antenna function ``turns over'', $\mathrm{d}(\mathrm{ant})/\mathrm{d}\theta = 0$. 
We also confirm, as  was noted above, that the dead-cone angle does not correspond to a total vanishing of radiation, but to the $1/\theta^2$ divergence one would have for a massless emitter being regulated, with a finite (but not divergent) radiation probability across the full range of polar angles, $\theta \in [0^\circ,180^\circ]$. 

However, when one or both emitters are non-relativistic, the stationary-point definition becomes more complicated and requires a partial-fractioning of the dipole antenna pattern to separate it into two pieces, one for each emitter. 
An illustration of such a case is shown in \figRef{fig:stationaryBad} corresponding to the \decay{\Bd}{\Dp\pim} decay mode.
In this case, the full dipole antenna pattern does not have an unambiguous stationary point on the \D-meson side, but the $1/4$ definition still gives a sensible result for the dead-cone angle. 
One could still use the stationary point of the corresponding massive DGLAP kernels, but here we have instead used the ratio = $1/4$ as a simple and numerically stable alternative with a clear physical meaning. 

\begin{figure}[htb]
    \centering
    \begin{subfigure}{0.4\textwidth}
  \centering
  \includegraphics[width=\linewidth]{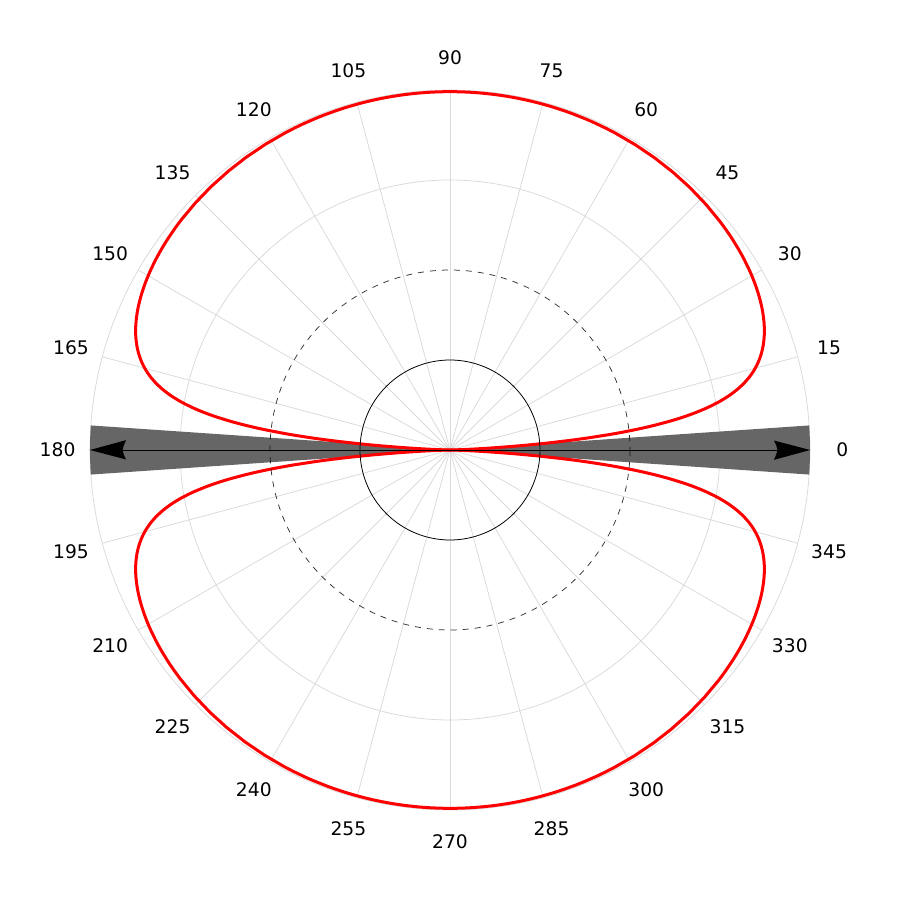}
  \caption{\decay{\jpsi}{\mumu}}
  \label{fig:JPsiMuMu}
\end{subfigure}
\hfill
\begin{subfigure}{0.4\textwidth}
  \centering
  \includegraphics[width=\linewidth]{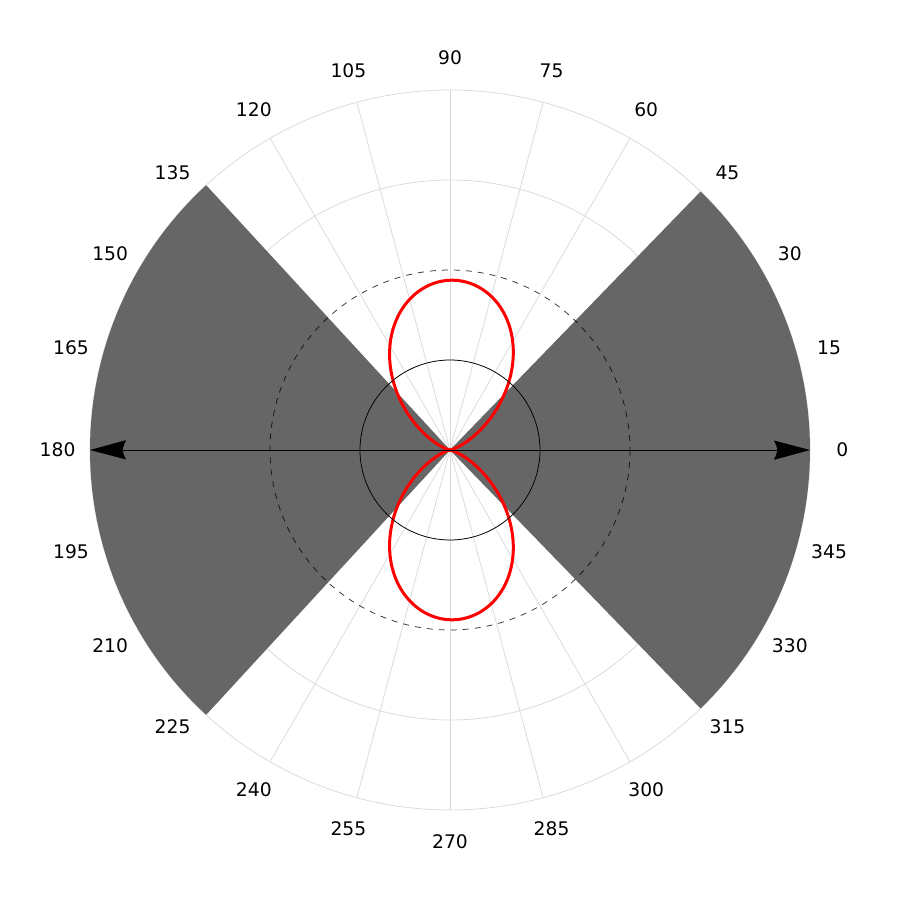}
  \caption{\decay{\Bz}{\Ds\Dm}}
  \label{fig:D0t}
\end{subfigure}
\begin{subfigure}{0.4\textwidth}
  \centering
  \includegraphics[width=\linewidth]{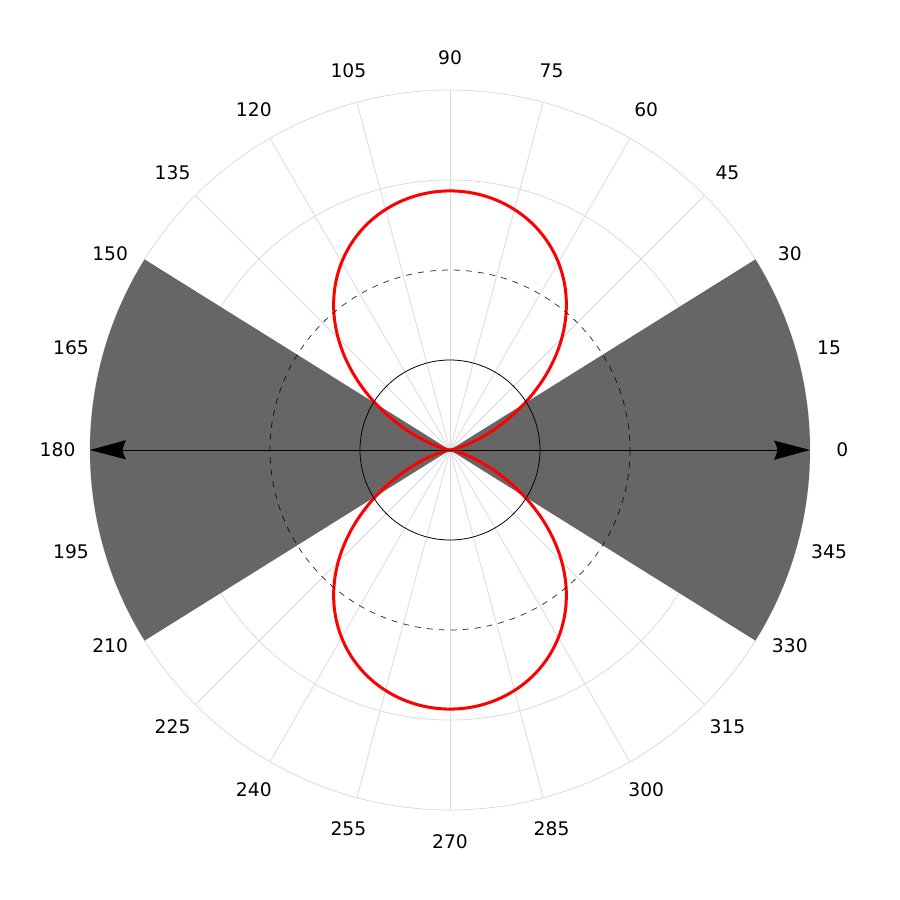}
  \caption{\decay{\Dz}{\Kp\Km}}
  \label{fig:D0toKK}
\end{subfigure}
\hfill
\begin{subfigure}{0.4\textwidth}
  \centering
  \includegraphics[width=\linewidth]{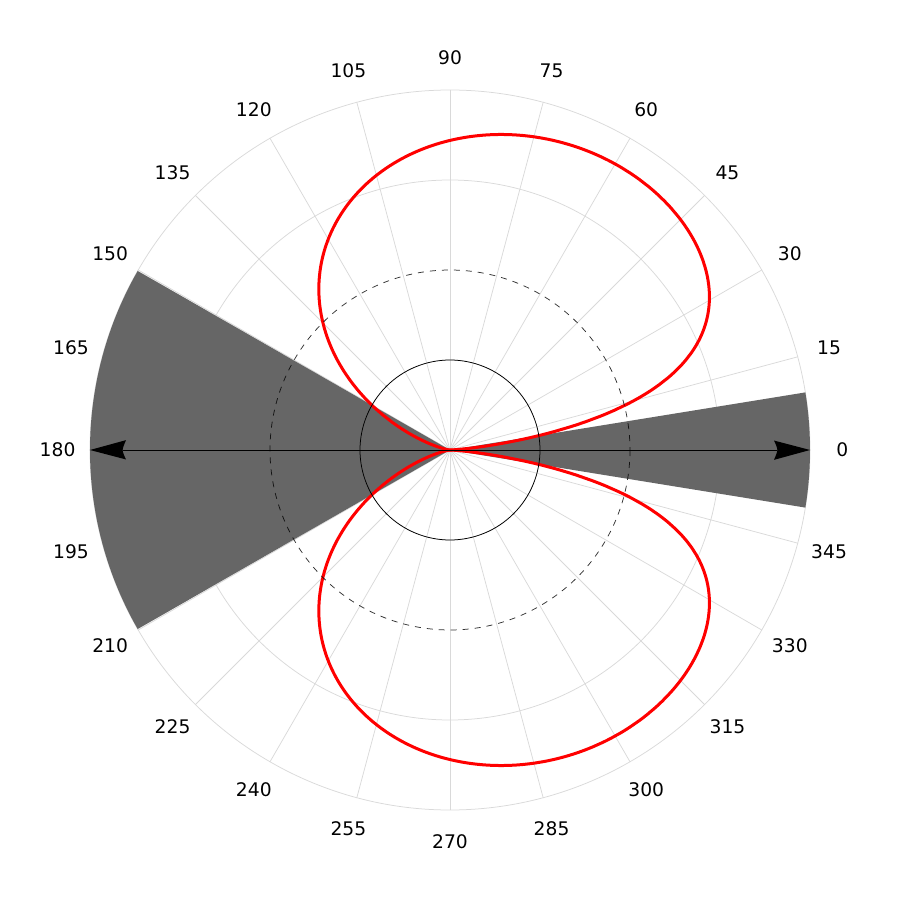}
  \caption{\decay{\Dz}{\Km\pip}}
  \label{fig:D0toKpi}
\end{subfigure}
    \caption{Massive-to-massless antenna ratio for soft photon emission vs. photon polar angle relative to the CM-frame decay axis. 
    The ratio is the radius of the red curve. 
    Dead-cone angles, $\cos^{-1}(\beta^*)$, from \tabRef{tab:boosts} are shown as dark grey regions. Circles mark ratios of 1 (outermost), $1/2$ (dashed), and $1/4$ (solid).}
    \label{fig:B0toDStarDs}
\end{figure}

We now examine the full angular dependence of the ratio of the massive to the massless antenna functions in the soft limit ($x_\gamma \to 0$), and compare it to the dead-cone result for a few representative two-body decays \decay{X^0}{1^+ 2^-}. 
In \figRef{fig:B0toDStarDs}, we show polar plots with this ratio on the radial axis (red lines), as a function of the polar emission angle of the photon, relative to the decay axis in the CM frame. 
Where the red line touches the outermost circle, the ratio is unity, i.e., massive = massless. 
Where the red line intersects the dashed black circle, the ratio is $1/2$, and where it intersects the solid black circle, the ratio is $1/4$, which is what we used in our definition of the dead cone above. 
The dark grey shaded areas indicate the size of the dead cones, as given by the $\cos^{-1}(\beta^*)$ columns in \tabRef{tab:boosts}. 
We consider the following four representative scenarios:
\begin{enumerate}[label=\alph*)]
\item \decay{\jpsi}{\mumu} (highly relativistic decay products): the massive-to-massless antenna ratio is approximately unity over most emission angles, and becomes smaller only within narrow cones around the dead-cone angles $\sim \pm 4^\circ$, cf.\ \tabRef{tab:boosts};
\item \decay{\Bz}{\Ds \Dm} (highly non-relativistic decay products): the massive-to-massless antenna ratio is substantially below unity for all angles, and remains below $1/2$ throughout;
\item \decay{\Dz}{\Kp\Km} (intermediate kinematics): the massive-to-massless ratio stays below unity over the full angular range, with comparatively large dead-cone angles $\sim 30^\circ$ (see \tabRef{tab:boosts});
\item \decay{\Dz}{\Km\pip}: a case with a pronounced mass difference between the two decay products, leading to an asymmetric antenna-ratio pattern, with a larger dead cone on the side of the more massive decay product.
\end{enumerate}

\subsection{Phase-space distributions for finite photon energies}

\label{ssec:phsp_photons}

\begin{figure}[tp]
    \centering
    \includegraphics[width=0.5\linewidth]{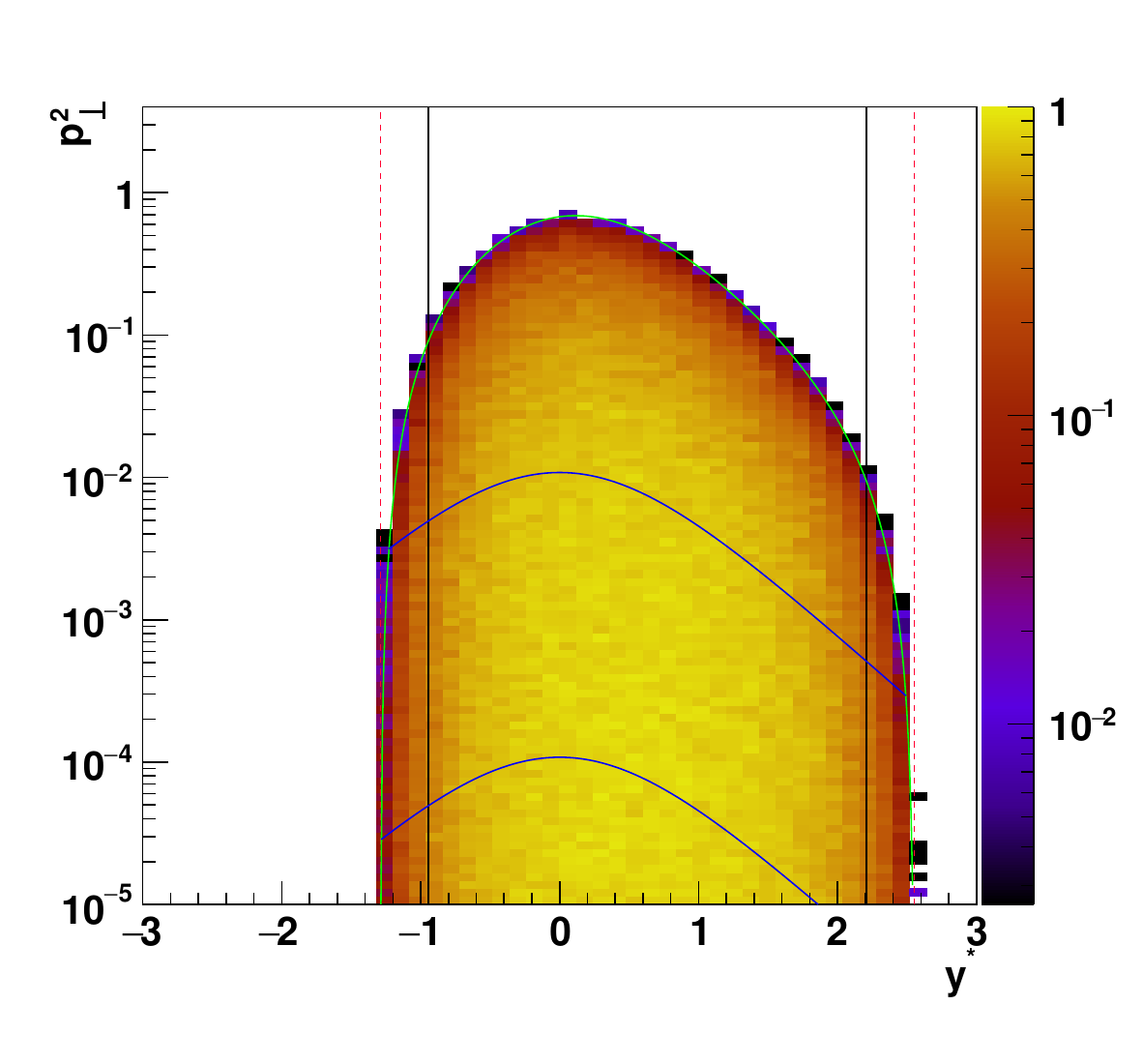}
    \caption{
Distribution of $\pTs$ vs. $y^*$, for the first photon emission in \mbox{\decay{\Dz}{\Km\pip}} decays.
The dashed red line corresponds to the value of $y^*$ for $\theta=0$ and the solid black line is the value of $y^*$ at the dead-cone angle $\theta=\cos^{-1}(\beta)$, where $\theta$ is the angle between the emitted photon and the post-branching emitter in the CM~frame. 
The solid green line is the boundary of the physical phase space, defined by the Gram determinant equal to $0$. 
The two solid blue lines are contours of emitted photon energy equalling $100\mev$ and $10\mev$. 
The plot is normalised such that the largest bin is set to $1$. 
The direction of the pion momentum is taken as reference axis, so that positive values of $y^*$ correspond to emissions closer to the pion while  negative values indicate emissions closer to the kaon. 
    }
    \label{fig:yStar}
\end{figure}

\begin{figure}[tp]
    \centering
    \begin{subfigure}{0.45\textwidth}
        \includegraphics[width=\linewidth]{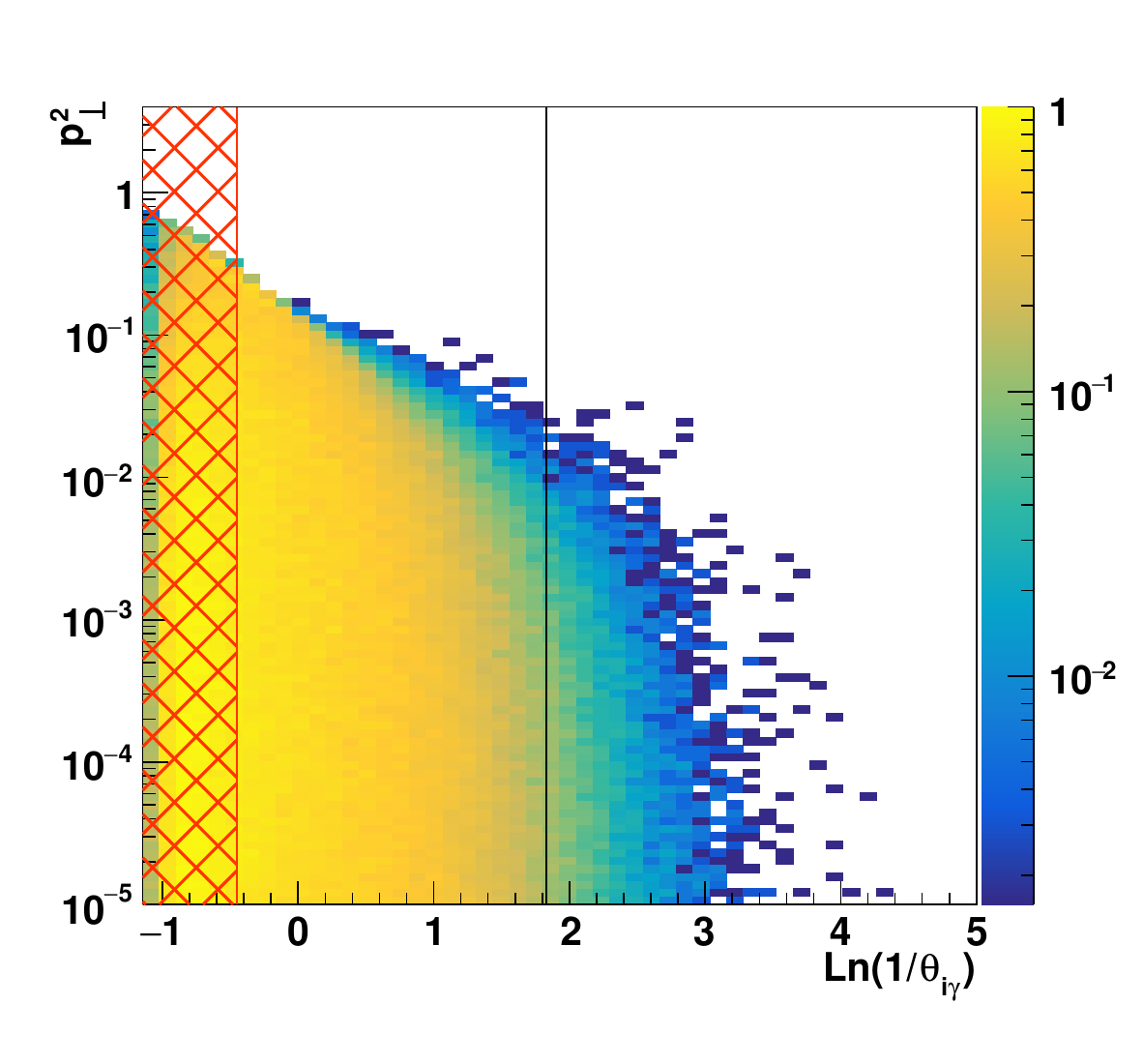}
    \caption{$\pTs$ vs. $\ln(1/\theta_{\pi\gamma})$}
    \label{fig:logTheta1}
    \end{subfigure}
    \hfill
    \begin{subfigure}{0.45\textwidth}
        \includegraphics[width=\linewidth]{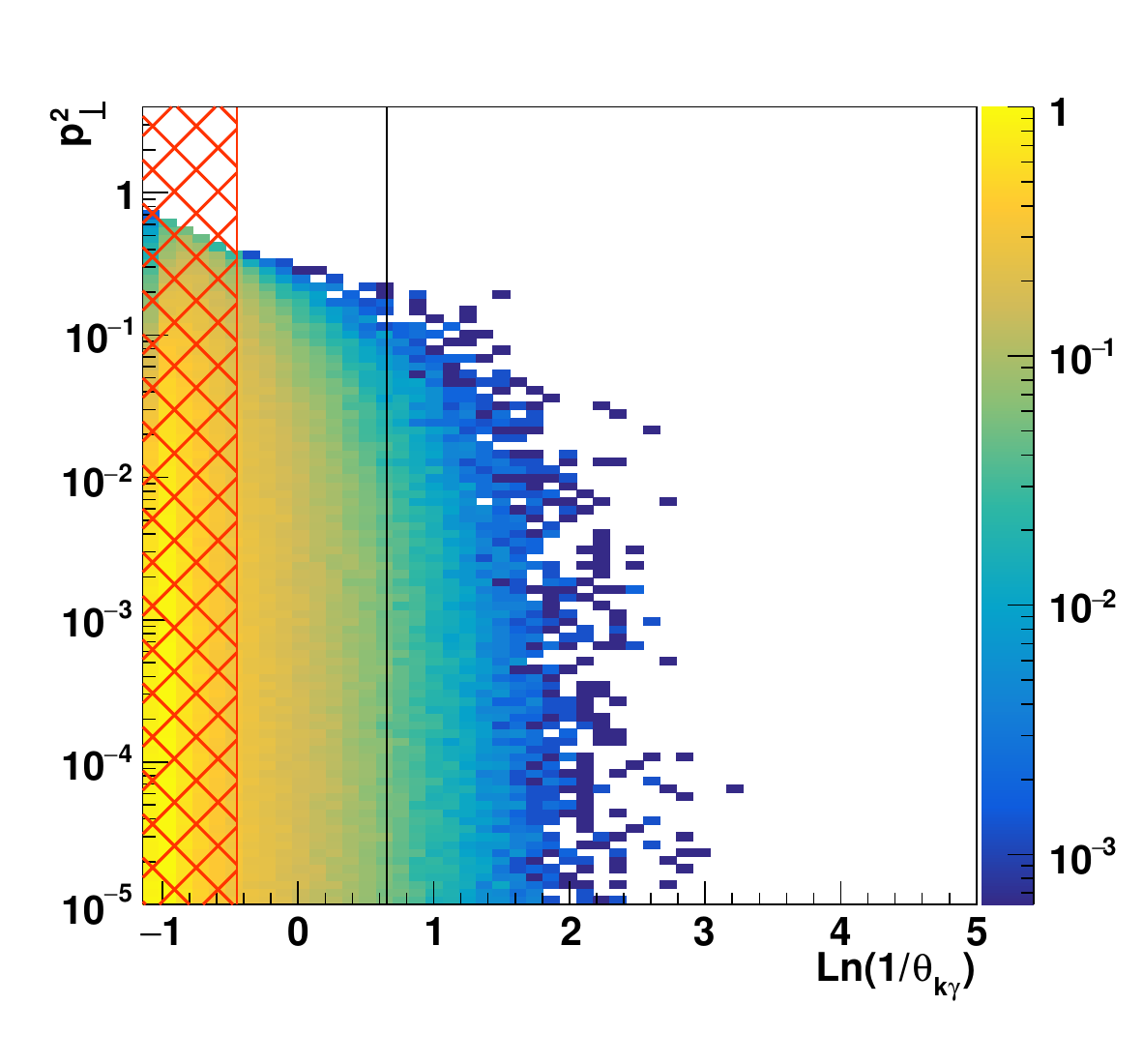}
    \caption{$\pTs$ vs. $\ln(1/\theta_{K\gamma})$}
    \label{fig:logTheta2}
    \end{subfigure}
    \caption{Distributions of $\pTs$ vs. $\ln(1/\theta)$, where $\theta$ is the angle between the emitted photon and the post-branching final-state (left)~pion and (right)~kaon in the CM~frame. 
    The solid black lines represent the deadcone angle $\theta=\cos^{-1}(\beta)$ and the hatched orange region marks emissions in the backwards direction~($\theta \geq \frac{\pi}{2}$).}
    \label{fig:logTheta}
\end{figure}

As emphasised above, the parametric rapidity, $y_\gamma^*$, defined in \eqRef{eq:yStarRec}, is kinematically bounded for all photon energies~(see \eqRef{eq:yStarBounds}).
In contrast, for opening angles $\theta_{i\gamma}$---or equivalently, physical photon rapidities $\eta_\gamma$---no strict kinematic bound arises in the soft limit; rather, there exists a characteristic angular scale at which the nominal $1/\theta^2$ collinear divergence is effectively regulated.
This is illustrated explicitly in \figsRef{fig:yStar} and~\ref{fig:logTheta}. 
Both figures present the distribution of the first photon emission for a representative decay, \decay{\Dz}{\Km\pip}, shown as a function of the photon $p_\perp^2$ (vertical axis), versus the parametric photon rapidity $y^*_\gamma$ in \figRef{fig:yStar} and versus $\ln(1/\theta)$ in \figRef{fig:logTheta} (horizontal axes).

In \figRef{fig:yStar}, a pure massless eikonal would give a constant density in these phase-space variables. 
The thin green line marks where the Gram determinant vanishes, i.e., the physical phase-space boundary for full kinematics. 
The pink dashed lines show the soft-photon-limit asymptotes of $y^*$ where the Gram determinant vanishes on the kaon (left) and pion (right) sides. 
The black lines give the $y^*$ values defining the dead-cone angle, beyond which emission is suppressed relative to the near-constant wide-angle rate. 
The solid blue lines are constant photon-energy contours at $10$ and $100\mev$.

In contrast, \figRef{fig:logTheta} shows the radiation pattern around a single emitter. 
The pion (\figRef{fig:logTheta1}) radiates at angles smaller than those of the kaon (\figRef{fig:logTheta2}) for a given $\pT$. 
The dead-cone angle (solid black line) shows a sharp photon drop-off, with adjacent bins differing by $\sim 2$. 
The most energetic photons are confined to $\theta \to \pi$, due to hard, high-recoil photons emitted at large angles relative to both the kaon and pion.

\begin{figure}[tp]
    \centering
    \begin{subfigure}{0.45\textwidth}
        \includegraphics[width=\linewidth]{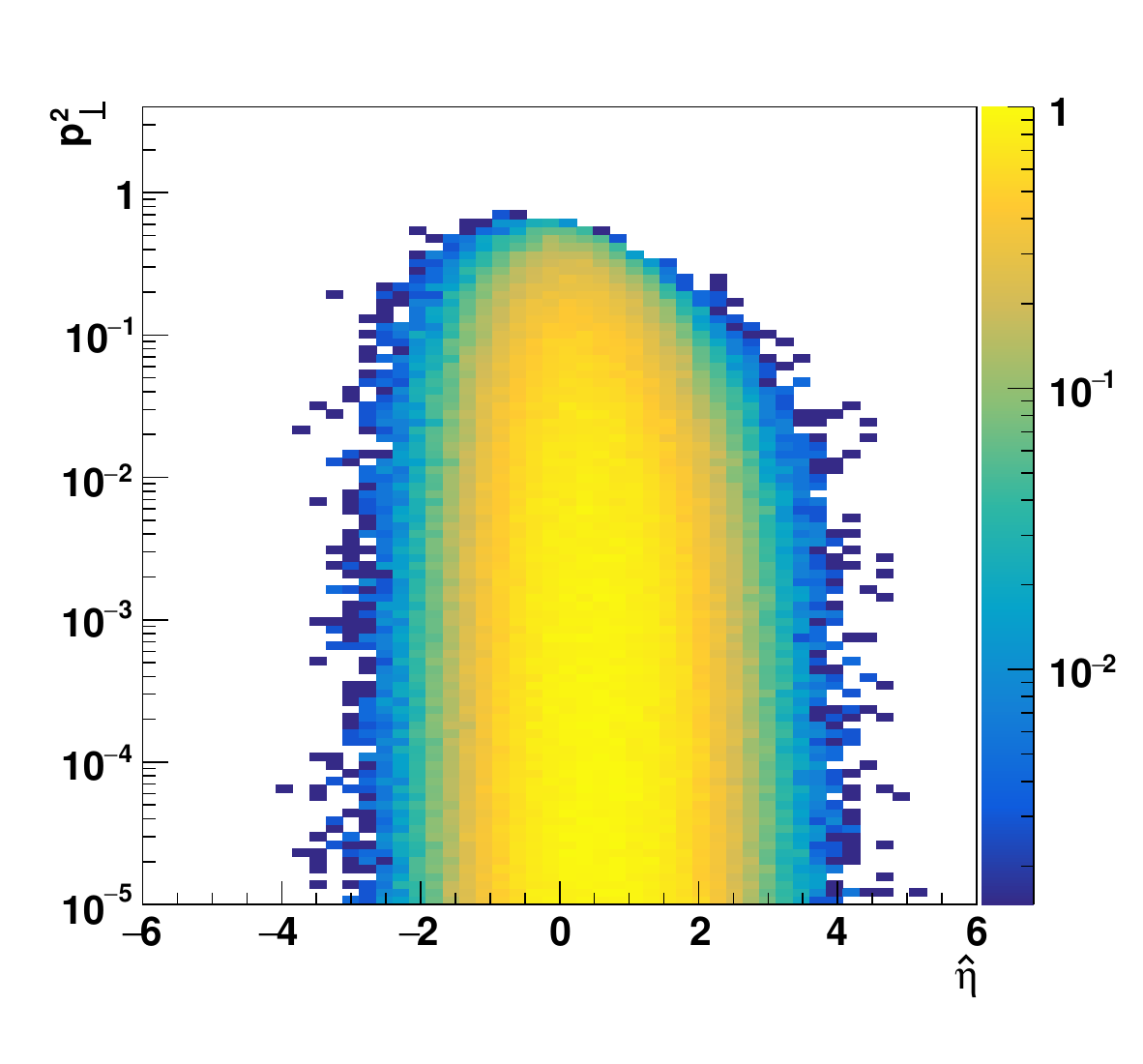}
    \caption{$\pTs$ vs. $\hat{\eta}$}
    \label{fig:deadConeD0kPiEta}
    \end{subfigure}
    \hfill
    \begin{subfigure}{0.45\textwidth}
        \includegraphics[width=\linewidth]{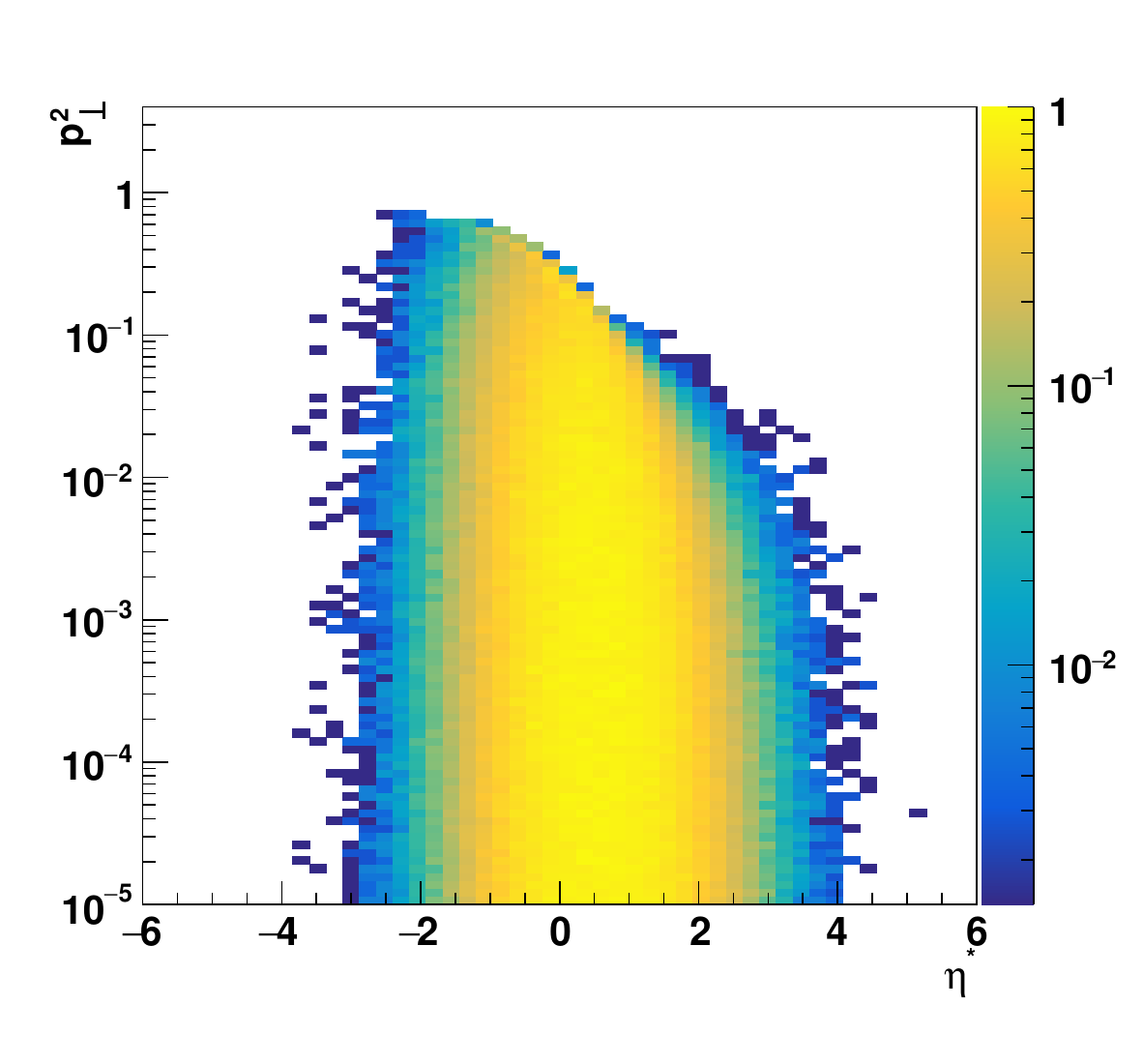}
    \caption{$\pTs$ vs. $\eta$}
    \label{fig:deadConeD0kPiEtaStar}
    \end{subfigure}
    \caption{Distributions of $\pTs$ as a function of (left) the pseudorapidity $\hat{\eta}$ with respect to the pre-branching Born-level decay axis, and (right) the post-branching pseudorapidity $\eta$ (see definitions in the text) for the first emitted photon in \mbox{\decay{\Dz}{\Km\pip}} decays.
    The direction of the pion momentum is taken as reference axis for the calculation of the angles.}
    \label{fig:etaVsEtaStar}
\end{figure}

Using $\log(1/\theta)$ on the x-axis obscures information about emissions in the $\theta \geq \pi/2$ region (hatched orange area in \figRef{fig:logTheta}). 
To show both limits ($\theta \to 0$ and $\theta\to \pi$) on the same plot, we re-express the same kinematics in a variable closely related to $y^*$ by using the photon pseudorapidity,
\begin{equation*}
\hat{\eta} = \frac{1}{2}\log\!\left(\frac{1+\cos{\hat{\theta}}}{1-\cos{\hat{\theta}}}\right)
~~~~~\mbox{or}~~~~~
\eta = \frac{1}{2}\log\!\left(\frac{1+\cos{\theta}}{1-\cos{\theta}}\right)\,,
\end{equation*}
where $\hat{\theta}$ is the opening angle between the photon and the pre-branching two-body decay axis in the CM~frame, and $\theta$ is the corresponding angle between the photon and the emitter after radiation (post-branching). 
The resulting plots are shown in \figRef{fig:etaVsEtaStar}.
A comparison of \figsRef{fig:deadConeD0kPiEta} and~\ref{fig:deadConeD0kPiEtaStar} shows that the distributions for low \pT are essentially identical between the pre- and post-branching pseudorapidity definitions (\ref{fig:deadConeD0kPiEta} and \ref{fig:deadConeD0kPiEtaStar}, respectively), while at higher \pT,  the recoil imparted by the first emission  makes an appreciable difference.

Recoil effects have received much attention lately in the context of QCD showers~\cite{Dasgupta:2020fwr}, where so-called \emph{local} recoil schemes (in which the recoil from each emission is distributed among only two partons, which are often referred to as the radiator and the recoiler) have been shown to yield only leading-logarithmic (LL) accuracy in the soft-collinear limit, while \emph{global} recoil schemes (in which the recoil is distributed among multiple partons) can reach higher accuracies. 
Among the QED generators considered here, \pythia employs a local dipole-recoil scheme, in which the emitter absorbs the full transverse recoil while the recoiler recoils only longitudinally in the rest frame of the emitting dipole, \vincia uses a local antenna-recoil scheme in which both parents share the transverse recoil (though for resonance-final emissions its recoil strategy does have global properties~\cite{Brooks:2019xso}), and \sherpa and \photos both use variants of global recoil schemes. 
These differences in recoil strategies imply that the generators will typically differ more from each other in the $\hat{\theta}$ variable (emission angle with respect to the Born-level decay axis) than in the post-branching opening angle $\theta$. 
Since the latter is more directly related to the experimentally accessible final state, we choose to plot $\theta$ rather than $\hat{\theta}$ for the validation plots shown in \secRef{sec:results}.

\begin{figure}[tp]
    \centering
    \begin{subfigure}{0.45\textwidth}
        \includegraphics[width=\linewidth]{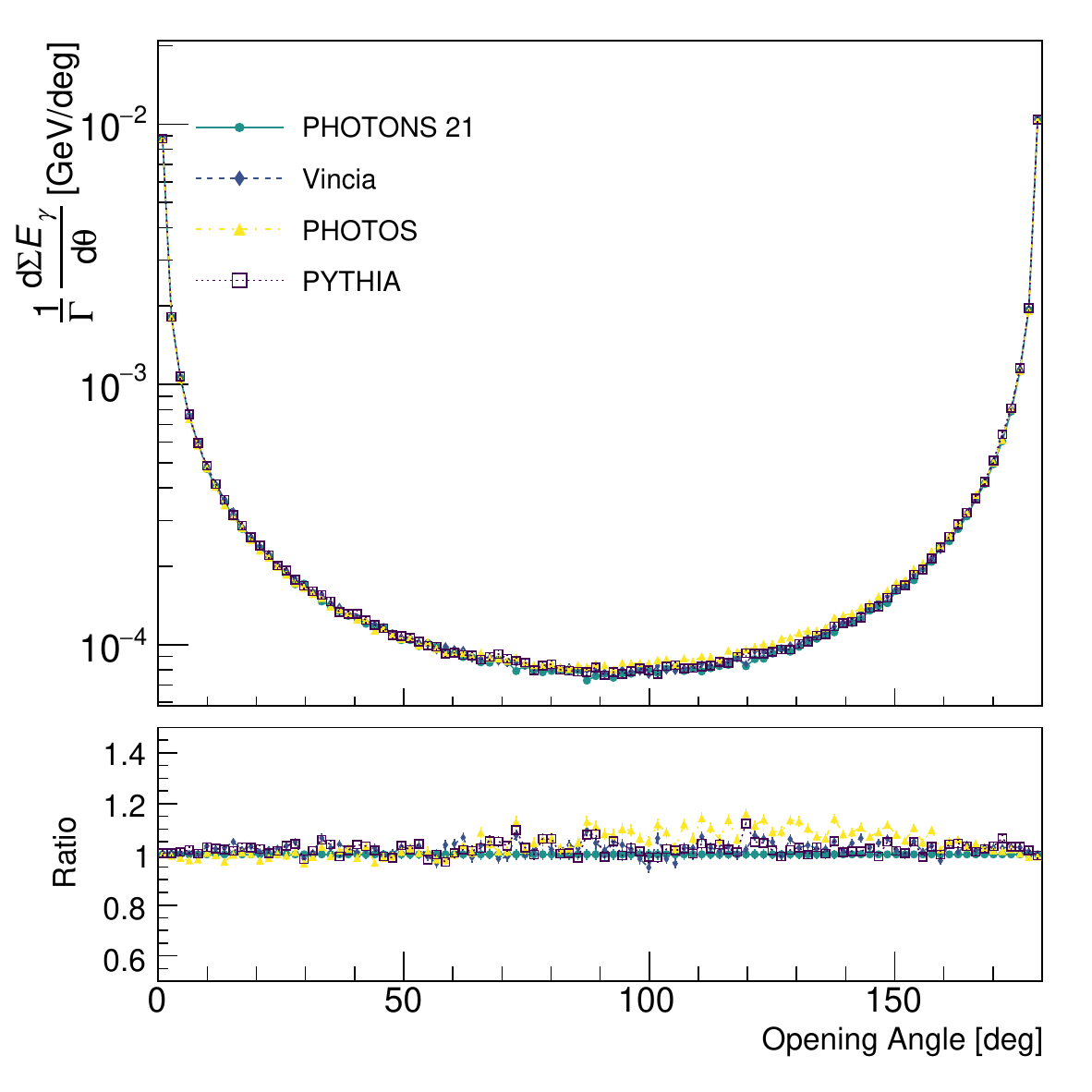}
    \caption{The x-axis is the opening angle between the photon and the post-branching final-state electron.}
    \label{fig:Jpsi_ee_Angle_Final}
    \end{subfigure}
    \hfill
    \begin{subfigure}{0.45\textwidth}
        \includegraphics[width=\linewidth]{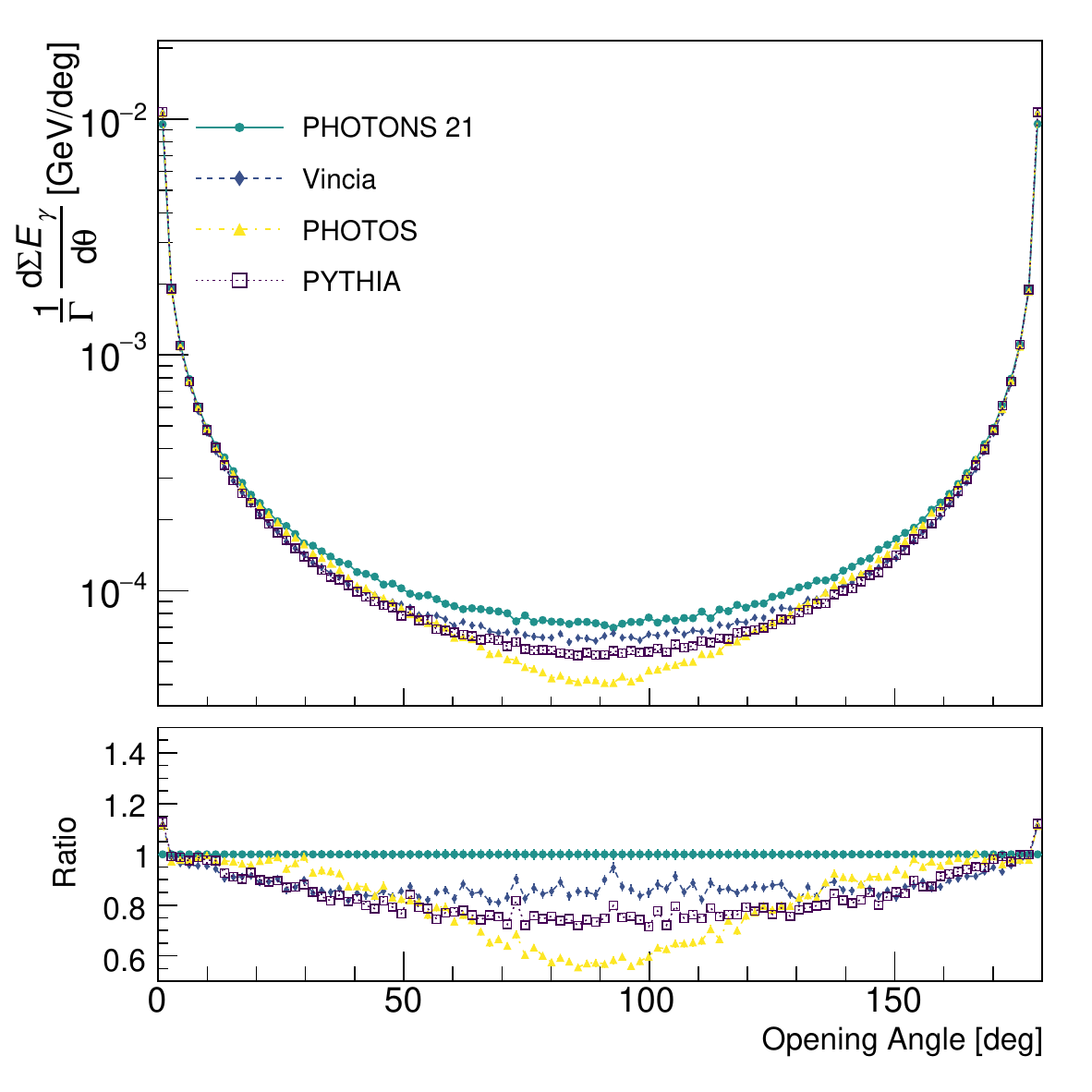}
    \caption{The x-axis is the opening angle between the photon and the pre-branching born-level electron.}
    \label{fig:Jpsi_ee_Angle_Born}
    \end{subfigure}
    \caption{Angular distribution of photon energy in \decay{\jpsi}{\ep\en} decays for (left)~post-branching and (right)~pre-branching opening angles.
    }
    \label{fig:Jpsi_ee_Angle_Comparisons}
\end{figure}

\begin{figure}[tp]
    \centering
    \begin{subfigure}{0.45\textwidth}
        \includegraphics[width=\linewidth]{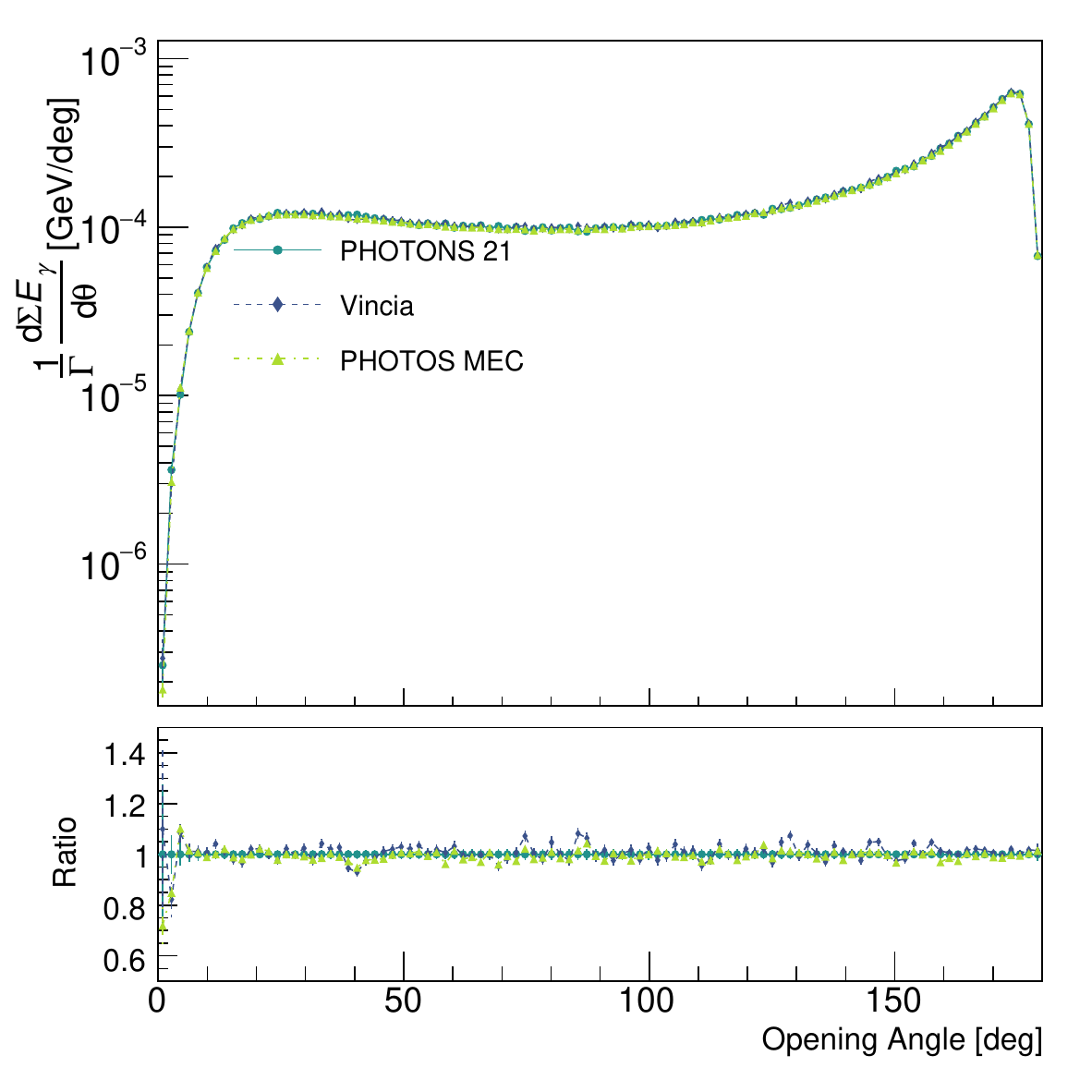}
    \caption{The x-axis is the opening angle between the photon and the post-branching final-state kaon.}
    \label{fig:Bd_Kpi_Angle_Final}
    \end{subfigure}
    \hfill
    \begin{subfigure}{0.45\textwidth}
        \includegraphics[width=\linewidth]{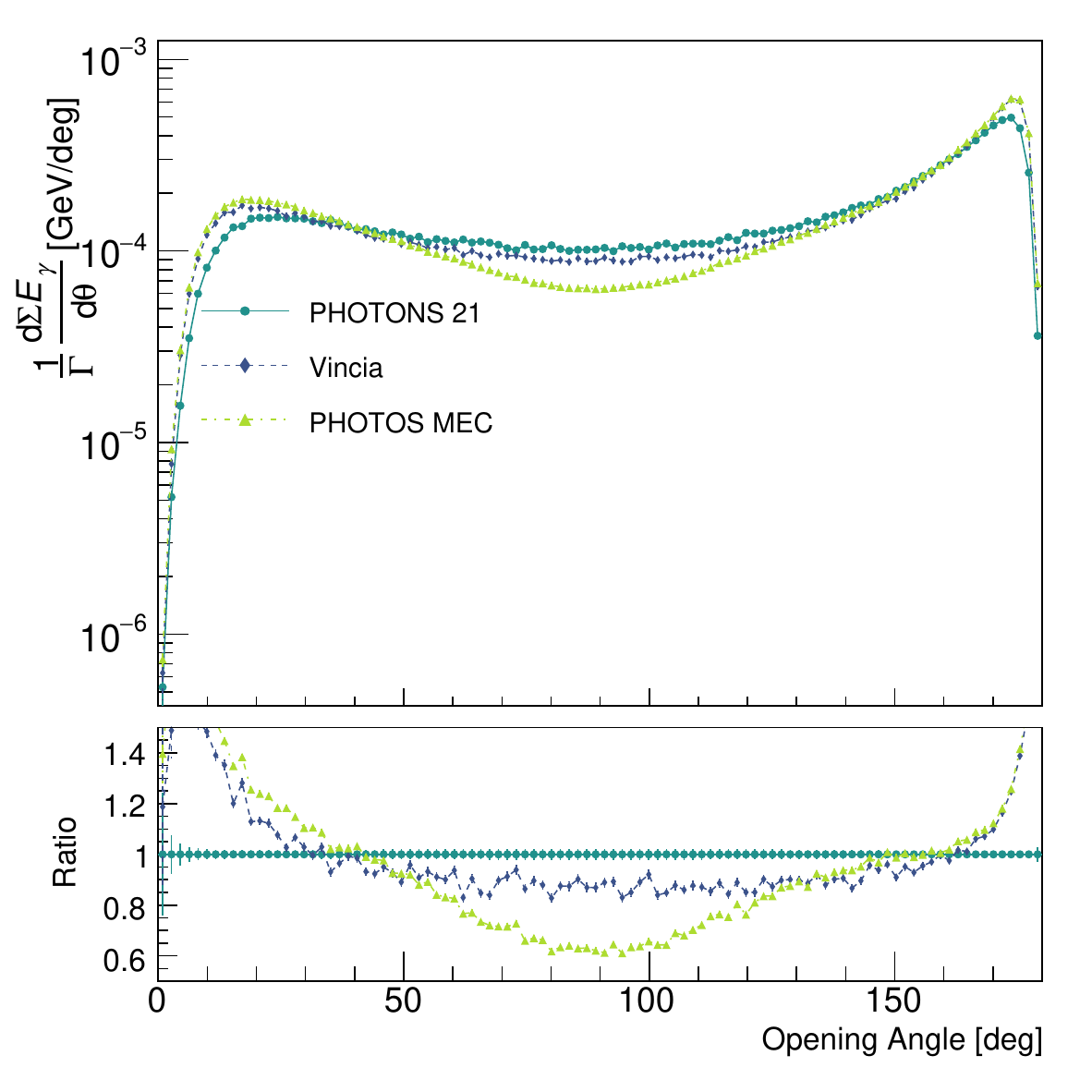}
    \caption{The x-axis is the opening angle between the photon and the pre-branching born-level kaon.}
    \label{fig:Bd_Kpi_Angle_Born}
    \end{subfigure}
    \caption{Angular distribution of photon energy in \decay{\Bd}{\Kp\pim} decays for (left)~post-branching and (right)~pre-branching opening angles.
    }
    \label{fig:Bd_Kpi_Angle_Comparisons}
\end{figure}

\begin{figure}[htb]
    \centering
    \begin{subfigure}{0.45\textwidth}
        \includegraphics[width=\linewidth]{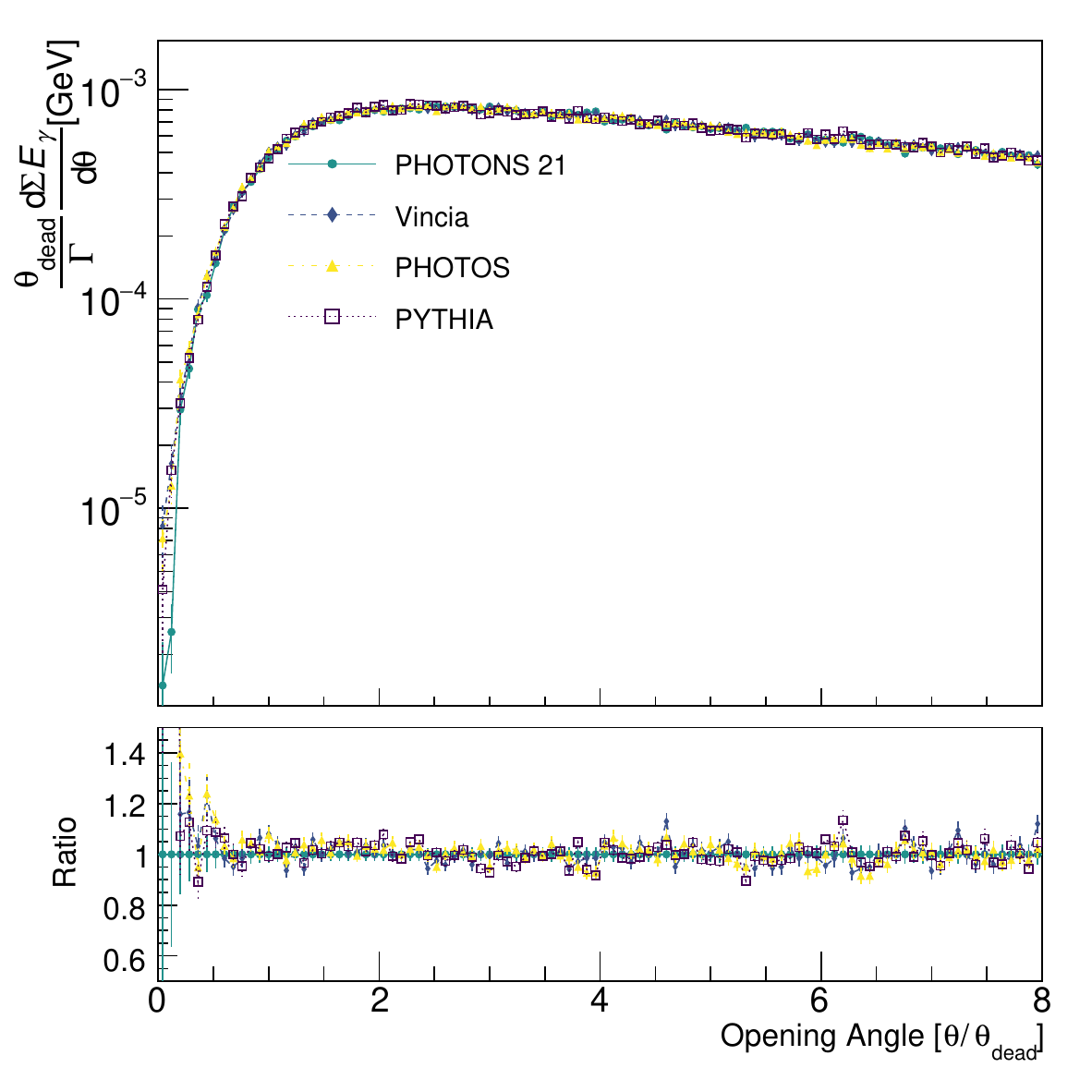}
    \caption{\decay{\jpsi}{\ep\en}}
    \label{fig:deadConeJpsi_ee_Final}
    \end{subfigure}
    \hfill
    \begin{subfigure}{0.45\textwidth}
        \includegraphics[width=\linewidth]{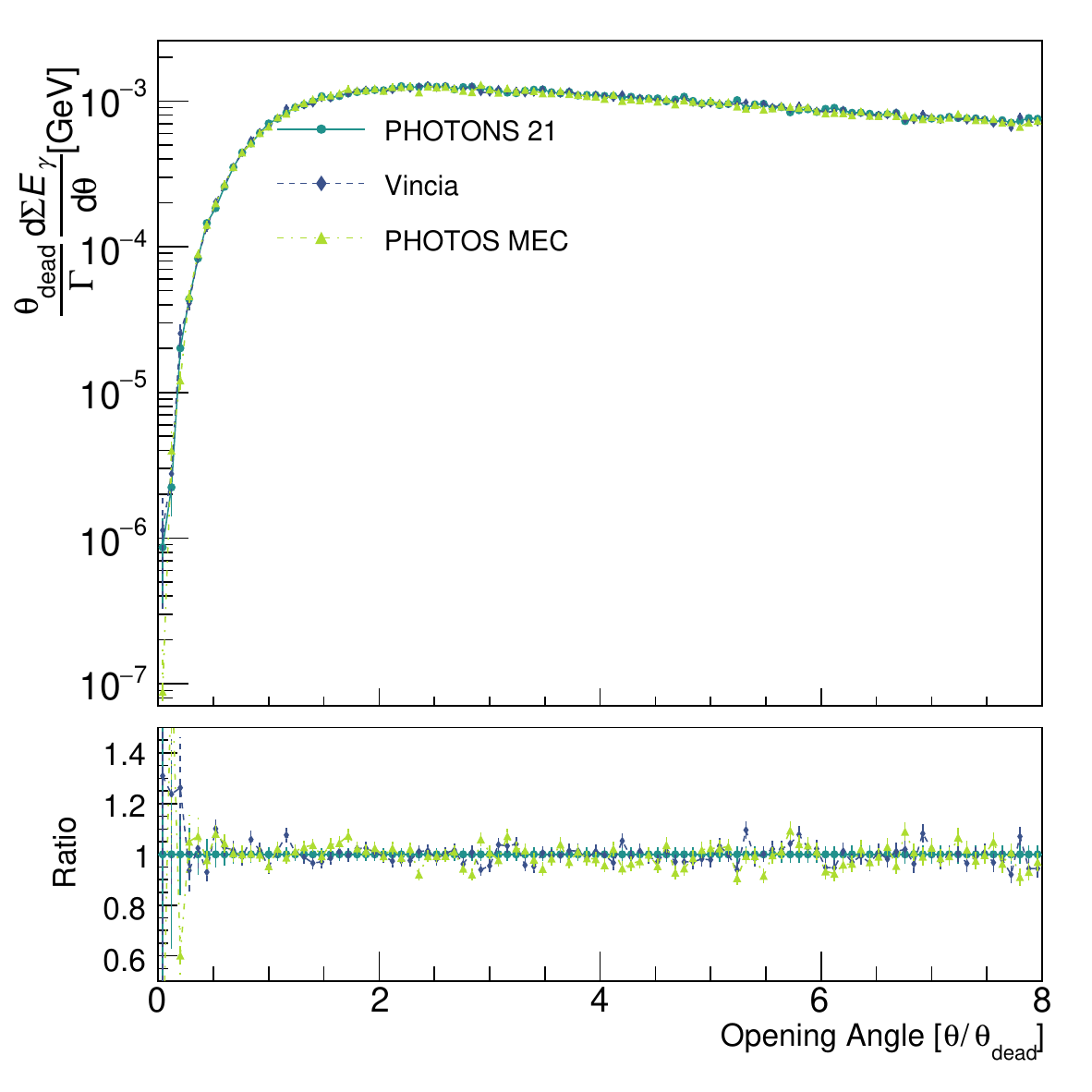}
    \caption{\decay{\Bd}{\pip\pim}}
    \label{fig:deadConeBd_pipi_Final}
    \end{subfigure}
    \caption{Angular distribution of photon energy in units of the dead-cone angle $\theta_{i\gamma}/\theta_\mathrm{dead} = \theta_{i\gamma} (E_i/m_i)$ in (left)~\mbox{\decay{\jpsi}{\ep\en}} and (right)~\decay{\Bd}{\pip\pim} decays.
    The angle $\theta_{i\gamma}$ is the opening angle between the photon and the post-branching final-state electron (or negative pion). }
    \label{fig:deadCone_Generator_Comparisons}
\end{figure}

To illustrate the differences between casting angular distributions in terms of $\theta$ or $\hat{\theta}$, \figsRef{fig:Jpsi_ee_Angle_Comparisons} and \ref{fig:Bd_Kpi_Angle_Comparisons} show the angular distributions of the total emitted photon energy in \decay{\jpsi}{\ep\en}  (\figRef{fig:Jpsi_ee_Angle_Comparisons}) and \decay{\Dz}{\Km\pip} decays (\figRef{fig:Bd_Kpi_Angle_Comparisons}) for various QED generators\footnote{Details on generator usage, configuration, and decay simulation are given in \secsRef{sec:IR} and~\ref{sec:results}.}. 
For \decay{\jpsi}{\ep\en} decays, the post-branching opening-angle distributions are asymmetric because $\theta$ 
is defined relative to the momentum of the final-state electron after 
branching: hard wide-angle emissions from the positron appear at larger 
$\theta$ because the reference axis itself has recoiled away from the photon, whereas the situation is fully symmetric relative to the (fixed) Born-level decay axis.
The distributions of the post-branching opening-angle show a better agreement between the generators compared to the pre-branching opening-angle distributions. 
\FigRef{fig:deadCone_Generator_Comparisons} shows that this agreement persists at small angles around the size of the dead cone (using the standard $m_i/E_i$ definition). 

\clearpage

\section{Infrared cutoff dependence}
\label{sec:IR}

Both parton showers and $N$-generators (cf., \tabRef{tab:overview}) impose a lower (“infrared”) cutoff on the phase space of photon emissions. 
To illustrate why this is needed and how it matters in practice, consider the leading-order photon-emission probability density for a highly relativistic particle as a function of photon energy and emission angle, 
which has the familiar
double-logarithmic form of \eqRef{eq:bremsprob} in the soft small-angle limit, with singularities for
$E_\gamma \to 0$ (soft) and $\theta_\gamma \to 0$ (collinear).
Physically, a charged particle is accompanied by infinitely many soft photons, and if massless, also infinitely many collinear photons (some carrying large energy fractions). 
These are the soft and collinear divergences that are also familiar in QCD.

The concept of a ``bare'' charged particle is therefore not quantum mechanically meaningful: 
only the concept of a charged particle \emph{plus} some amount of unresolved (very soft/collinear) radiation is. 
We emphasise that this does not, 
a priori, present a problem for experimental measurements. 
However, it \emph{does} become important when comparing experiment to theory. 
It cannot be emphasised enough that the \emph{number} of photons that are emitted, in a given theoretical calculation, completely depends on the infrared cutoff it uses. 
It would not be meaningful at all to compare this number between two calculations that do not use the same infrared cutoffs. 
This is related not only to the numerical value of the cutoff but also to the specific kinematic variable that defines it, which varies from generator to generator.
What can be meaningfully compared is the number of photons that are resolved with respect to some common definition of what a resolved photon means, e.g., in terms of photon energy, transverse momentum, and/or emission angle. 
This definition of a resolved photon must be \emph{infrared safe}: 
it must be insensitive to soft- and collinear photon emissions. 
For example, the number of photons with energies above some minimum value \emph{and} opening angles above some minimum emission angle would be well defined.
The cumulative energy carried away by the photons above the minimum emission angle is likewise meaningful. 

Considering first the soft singularity ($E \to 0$), 
experiments always have a minimum detectable photon energy set by the apparatus, so infinitely soft photons are never resolved. 
Theoretically, the infrared cutoff must therefore be below the lowest photon energy that the experiment can resolve such that all experimentally visible photons are resolved in the calculation. 
The cutoff must also be low enough so that the cumulative energy lost by each charged particle to unresolved radiation stays below the experimental energy resolution for reconstructed particles; 
otherwise the generator must account for the cumulative energy loss of charged particles to unresolved photons in another way.

The collinear singularity can be more subtle and, especially for highly relativistic emitters like electrons\footnote{And, less severely, muons or charged pions in the decays of very massive particles like $\PUpsilon$ or \PB~mesons, or in higher-energy contexts, e.g., leptonic decays of \PZ and \PW bosons~\cite{ATLAS:2014alx}.}, dealing with it accurately can be crucial. 
For any charged particle that has a non-zero rest mass, the collinear singularity is actually regulated:  
mathematically, at the dead-cone angle $\theta \sim m/E$, the angular dependence changes from $\d\theta^2/\theta^2$ to $\theta^2\d\theta^2$, \cf the discussion in \secRef{sec:deadConeEffect}. 
However, for highly relativistic emitters, the $\d\theta^2/\theta^2$ part can still result in high rates of very collinear photons; 
this is the same physics as the forward peaking of synchrotron radiation.  
The collinear photons can carry appreciable energy fractions and \emph{can} be resolvable experimentally, significantly affecting observable quantities such as the cumulative energy loss of the emitter. 

Experimentally, what happens if an electron emits a hard photon at almost zero emission angle? 
In the absence of magnetic fields in the detector, the electron and the photon would propagate together and hit the same calorimeter cell, which would only record their combined energy (disregarding possible 
information from the cluster shape). 
In such a setup, the effective experimental cutoff on photon emission angle would simply correspond to the size of the calorimeter cells. 
To reproduce this theoretically, the infrared cutoff must go deep enough into the collinear region to resolve all detectable photons emitted at angles larger than the angular resolution of the calorimeter, while photons at smaller angles should be summed back with their parent emitters, mimicking the sum done by the calorimeter.
In practice, however, the magnetic fields in the detector deflect the emitting electron away from any photons it was originally collinear with. 
To make the experimental electron energy \emph{insensitive} to photon emission, one must extrapolate the charged-particle trajectory back to its original direction and sum its energy together with that of any reconstructed photons collinear with it. 
This defines a “dressed” electron: the electron track plus any calorimeter clusters consistent with the chosen definition of a photon that should be added to the electron.\footnote{E.g., $\Delta R_{e\gamma} = \sqrt{\Delta\eta_{e\gamma}^2 + \Delta\phi_{e\gamma}^2}< 0.1$ in the study of \decay{\Z}{\ep\en} in \cite{ATLAS:2014alx}.}. 

For hadron decays, the default infrared cutoffs in event generators are typically small enough that they effectively satisfy the conditions above. 
A ``worst-case'' scenario is the \decay{\jpsi}{\ep\en} decay, for which the electrons are highly relativistic, with dead-cone angles of only $\sim0.02^\circ$. 
On the theory side, the infrared cutoff should be low enough to resolve even highly collinear photons, and on the experiment side, observed electrons should be dressed by adding back photons that are highly collinear with their original directions.
Note also that in practice, charged-particle reconstruction is further affected 
by bremsstrahlung emitted in the detector's magnetic field.
Accurate simulation of the 
decay bremsstrahlung is nevertheless essential to ensure that its contribution is 
correctly accounted for.

\begin{table}[tp]
\centering
\renewcommand{\arraystretch}{1.5}
\begin{tabular}{|c|c|c|c|c|} \hline
\cellcolor[HTML]{CCCCCC} \begin{tabular}[c]{@{}l@{}}\textbf{Generator} $\rightarrow$ \\[-2mm] $\downarrow$ \textbf{Cutoff} \end{tabular} & \cellcolor[HTML]{CCCCCC} \textbf{PHOTOS} & \cellcolor[HTML]{CCCCCC} \textbf{SHERPA} & \cellcolor[HTML]{CCCCCC} \textbf{VINCIA} & \cellcolor[HTML]{CCCCCC} \textbf{PYTHIA} \\ \hline
 \textbf{Variable}                          & Energy Fraction            & Energy [\gev]                  & $p_{\perp A}$ [\gev]                        & $p_{\perp\mathrm{evol}}$ [\gev]                        \\ \hline
\textbf{Default Value}                 & $10^{-7}$                     & $10^{-3}$                     & $10^{-6}$                     & $10^{-6}$                     \\ \hline
\textbf{Chosen Value}                     & $\frac{2}{M}10^{-5}$          & $10^{-5}$                     & $10^{-6}$                     & $10^{-6}$ \\ \hline
\end{tabular}
\renewcommand{\arraystretch}{1.0}
\caption{Infrared cutoff values used for the QED generators under comparison, along with their defaults. 
Since \photos uses the energy fraction as the cutoff variable, we divide the chosen energy cutoff by $M/2$, where $M$ is the mass of the decaying resonance.
The quoted \photos default is the one for the exponentiated mode.}\label{tab:cutoffs}
\end{table}

In \tabRef{tab:cutoffs}, we compare the four QED generators used in this study in terms of their cutoff variables and default values and provide the cutoff values used in this study: 
\begin{itemize}
\item \photos defines the infrared cutoff in terms of the energy fraction of a resolved photon,
\begin{equation*}
z_\gamma = \frac{2E_\gamma}{M_\mathrm{mother}}\,,
\end{equation*}
with a default value of $z_\gamma^\mathrm{min} = 10^{-7}$. 
We use $z_\gamma^\mathrm{min} = 2\times 10^{-5}/M_\mathrm{mother}$ to effectively translate it into a cutoff on photon energy of $E_\gamma^\mathrm{min} = 10^{-5}\gev=10\kev$. 
We did not find an explicit angular cutoff in \photos, so presume that it relies on particle masses to self-regulate the collinear divergence.
\item In \sherpa's \photons module, the cutoff corresponds to the minimum photon energy with default value 
$E_\gamma^\mathrm{min} = 10^{-3}\gev=1\mev$.\footnote{
In the original paper, the default cutoff in \sherpa was given as $0.1\gev$~\cite{Schonherr:2008av}. 
However, the current online manual~\cite{Sherpa:2024mfk} gives the default value for the \texttt{IR\_CUTOFF} parameter as $10^{-3}\,(\gev)$. We here quote the latter value.} 
To cover more of the soft wide-angle phase space and match the \sherpa cutoff to the one of \photos\footnote{Assuming they are evaluated in the same Lorentz frame.}, we lowered it to $E_\gamma^\mathrm{min} = 10^{-5}\gev = 10\kev$. 
As with \photos, we are not aware of an explicit angular cutoff in \photons, so we assume it self-regulates the collinear divergence using particle masses.
\item In \vincia, the infrared cutoff is applied to a Lorentz-invariant transverse-momentum measure, rather than energies or angles. 
This is the \textsc{Ariadne} \pT variable~\cite{Gustafson:1987rq,Lonnblad:1992tz,Giele:2007di,Brooks:2020upa} introduced in \secRef{sec:pythia}. 
In terms of the photon energy fraction $z_\gamma$ (in the Born-level rest frame) and parametric rapidity $y_\gamma^*$ (see \eqRef{eq:yStarMax}),
\begin{equation*}
p_{\perp \mathrm{A}}~=~\frac{z_\gamma}{e^{y_\gamma^*} + e^{-y_\gamma^*}}\,M_\mathrm{mother}~.
\end{equation*}
For relativistic emitters in the soft/collinear limit, \pT-like measures approach
\begin{equation*}
p_{\perp\gamma} ~\to~E_\gamma\,\theta_\gamma~,
\end{equation*}
so the cutoff effectively constrains a combination of energy and angle and is not directly comparable to an energy cutoff. Since at fixed $E_\gamma$ one has smaller $p_{\perp\gamma}$ at small angles, we use $p_{\perp \gamma}^\mathrm{min}=10^{-6}\gev=1\kev$, which is also the default QED infrared cutoff in \vincia.
\item In \pythia, QED parton showers are cut off in $p_{\perp\mathrm{E}}$~\cite{Sjostrand:2004ef} (see \eqRef{eq:pypt}), which has the same soft-collinear limit as the \vincia evolution variable~(see discussion in \secRef{sec:pythia}).
We therefore use $p_{\perp\mathrm{evol}}^\mathrm{min}\,=\,10^{-6}\gev\,=\,1\kev$, which is also \pythia's default. 
\end{itemize}

\begin{figure}[tp]
\centering
\begin{subfigure}{0.47\textwidth}
  \centering
  \includegraphics[width=\linewidth]{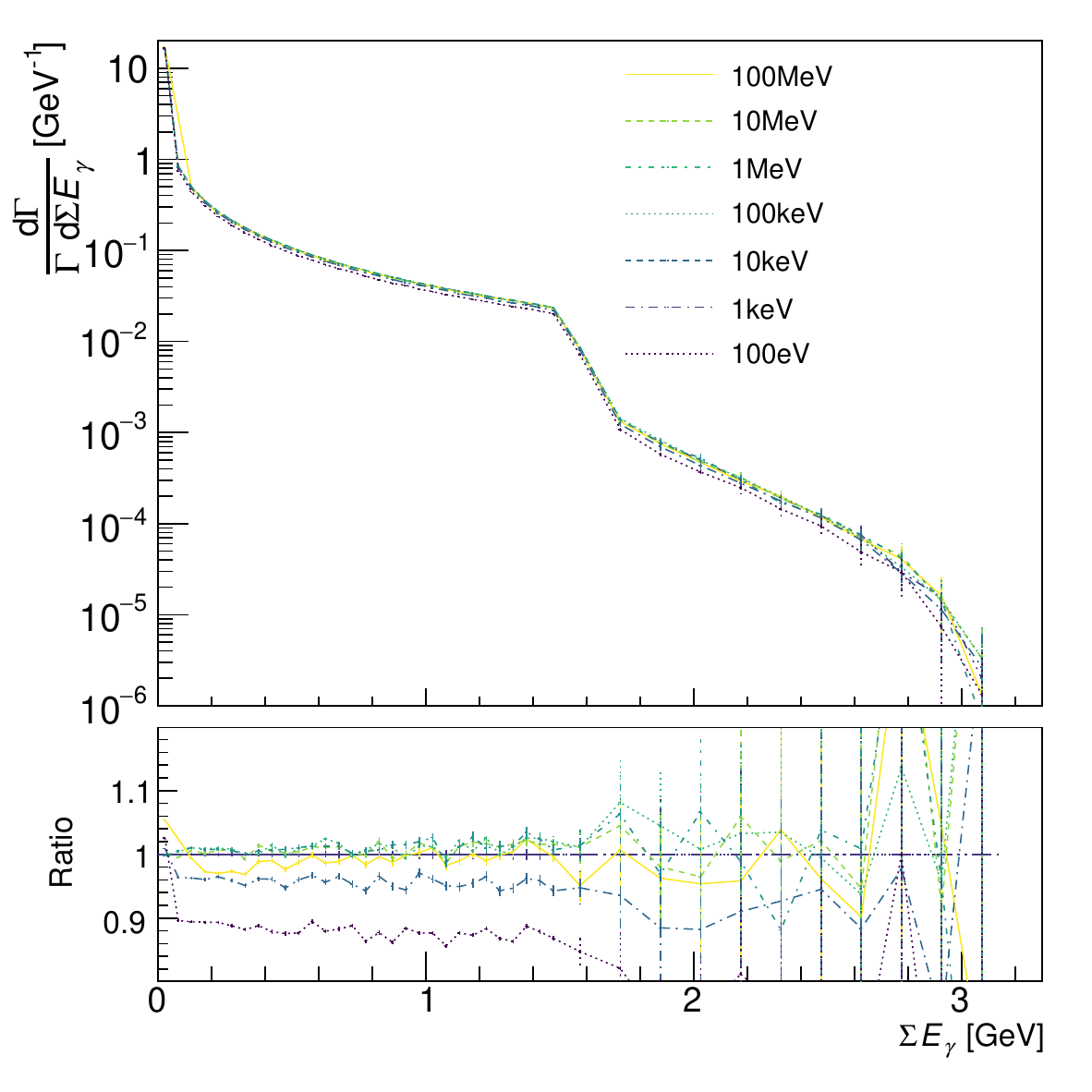}\vspace*{-5mm}
  \caption{\photos.}
  \label{fig:cutoffPHOTOS}
\end{subfigure}
\hfill
\begin{subfigure}{0.47\textwidth}
  \centering
  \includegraphics[width=\linewidth]{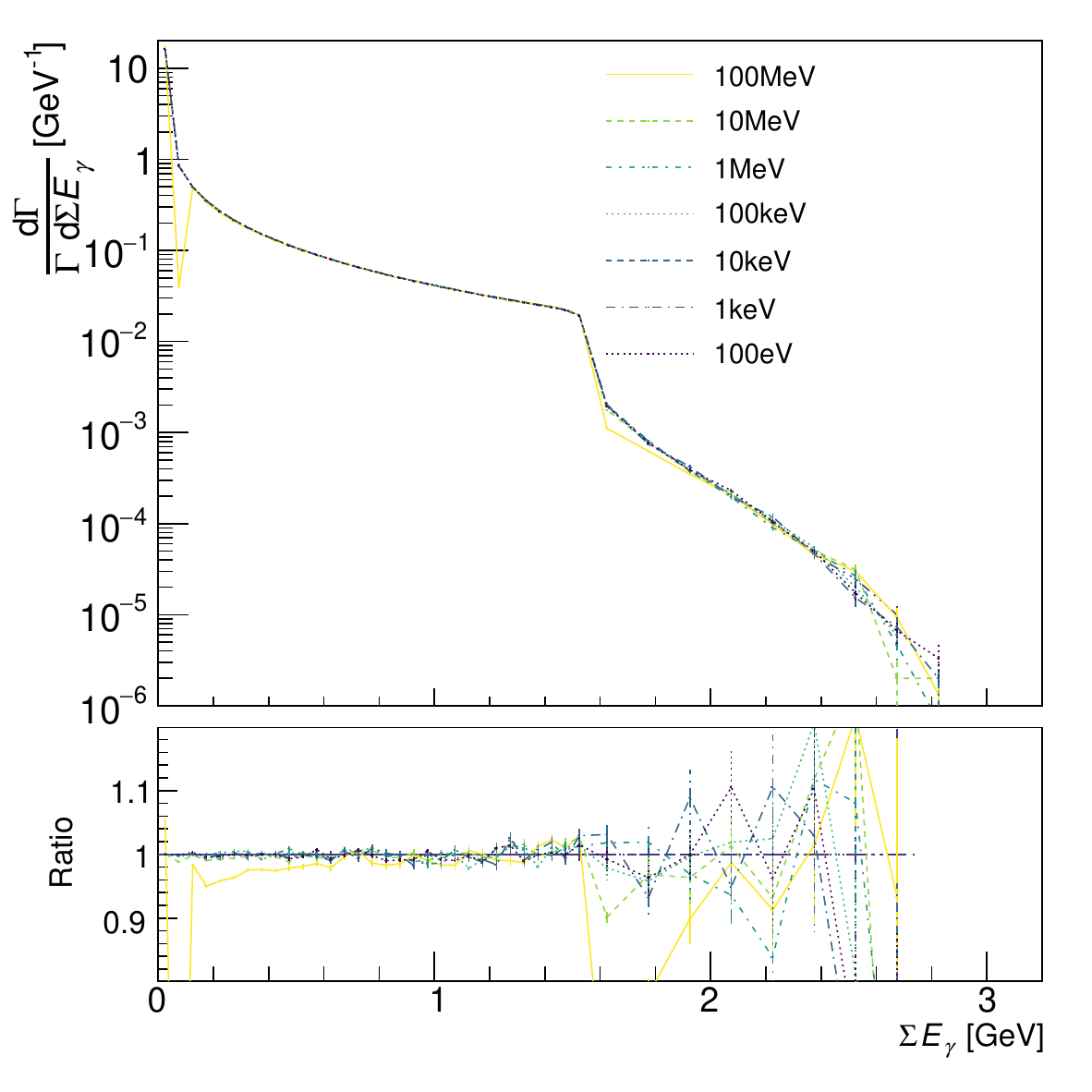}\vspace*{-5mm}
  \caption{\photons~21.}
  \label{fig:cutoffSherpaPhotons21}
\end{subfigure}
\begin{subfigure}{0.47\textwidth}
  \centering
  \includegraphics[width=\linewidth]{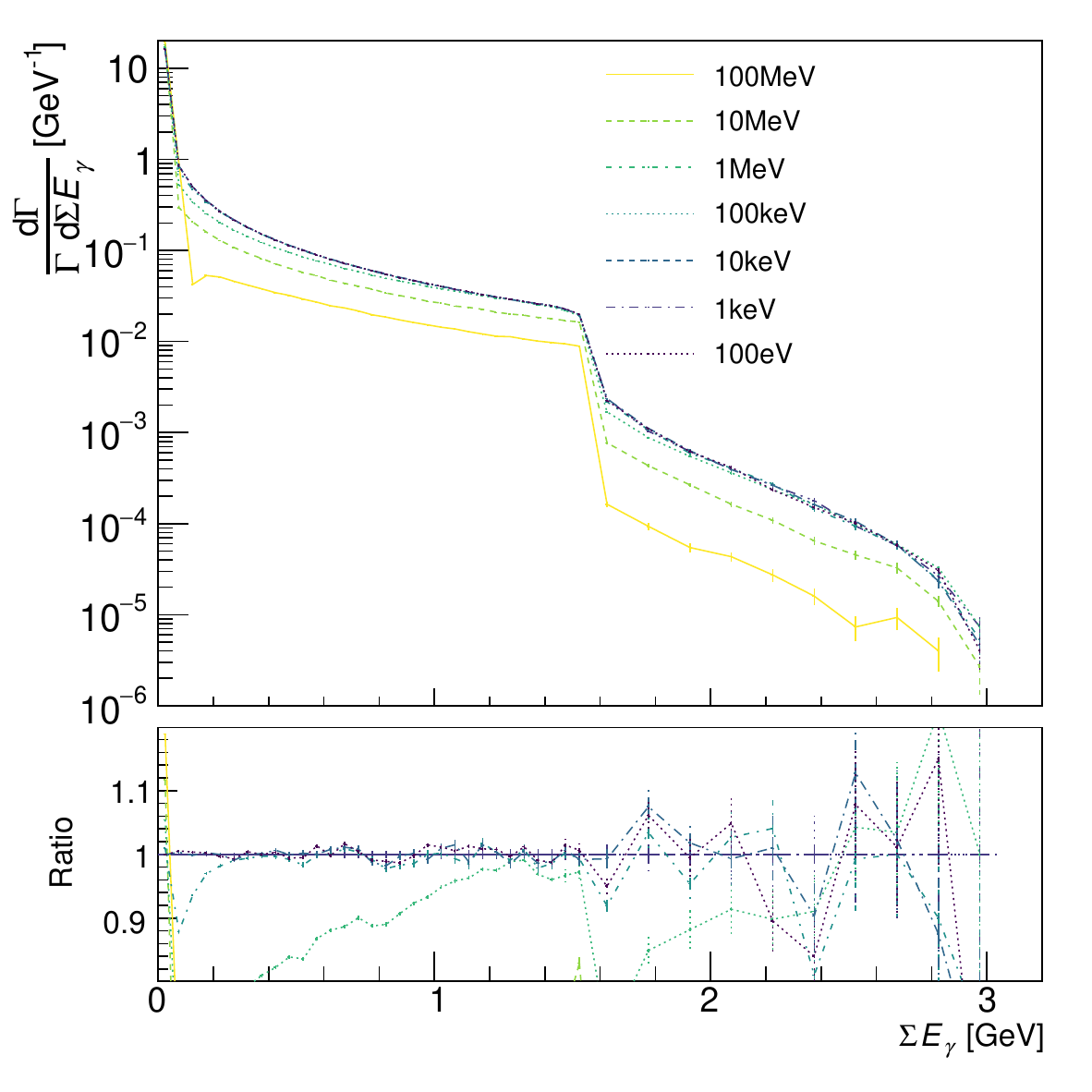}\vspace*{-5mm}
  \caption{\vincia.}
  \label{fig:cutoffVinciaQED}
\end{subfigure}
\hfill
\begin{subfigure}{0.47\textwidth}
  \centering
  \includegraphics[width=\linewidth]{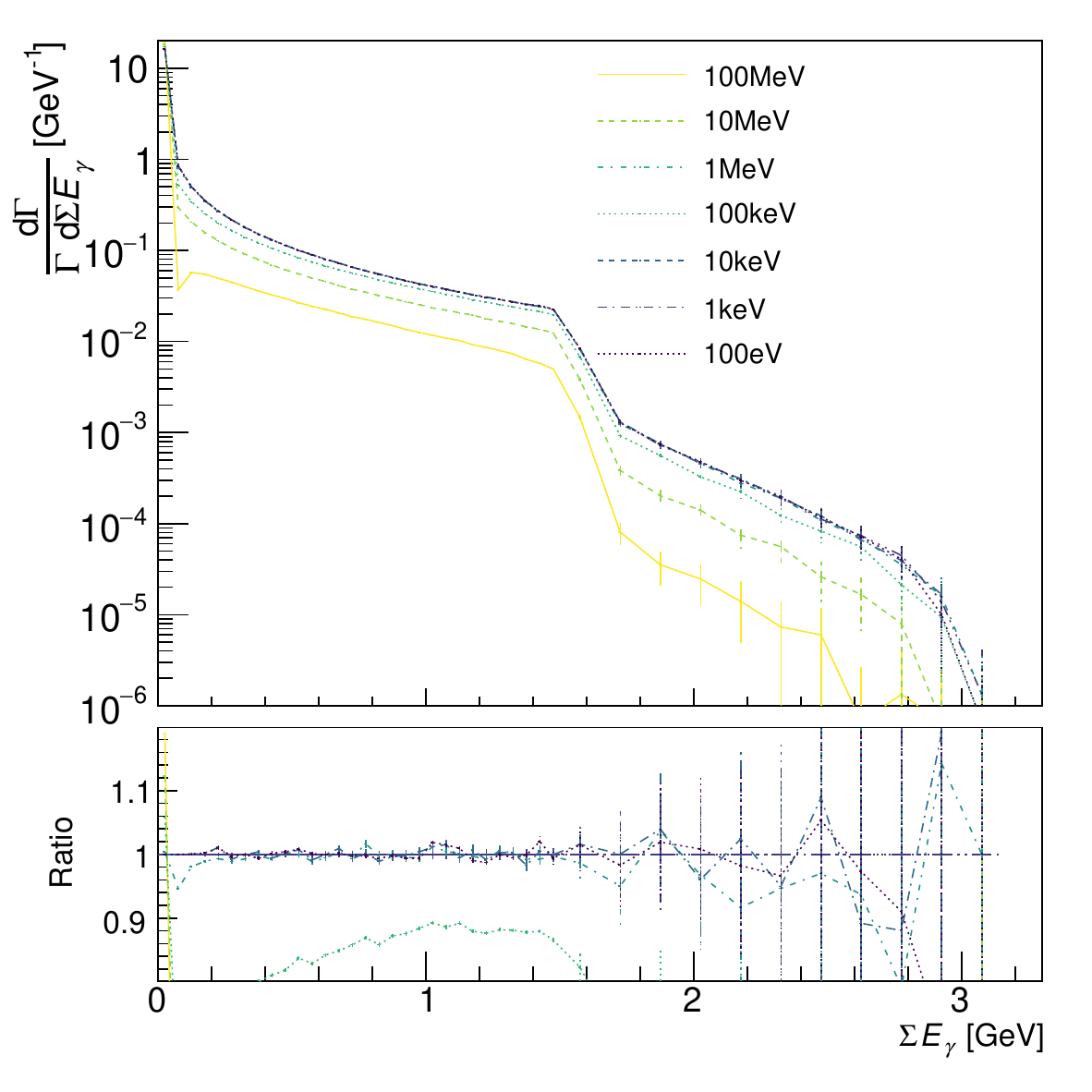}\vspace*{-5mm}
  \caption{\pythia.}
  \label{fig:cutoffPythiaQED}
\end{subfigure}    \caption{Total photon energy radiated in \decay{\jpsi}{\ep\en} decays for the four considered generators using different infrared cutoffs. 
The ratio pane uses the events generated with a $10\kev$ cutoff as the baseline. 
(Not shown: \photons~20 exhibits the same behaviour as \photons~21.)}
    \label{fig:cut-offComparisons}
\end{figure}

We study how the spectrum of total emitted photon energy, $\sum E_{\gamma}$, varies depending on the cutoff value for the \decay{\jpsi}{\ep\en} scenario in \photos, \sherpa, \vincia, and \pythia~(\figRef{fig:cut-offComparisons}). 
In each case, the cutoff is varied from $100\mev$ (solid yellow) down to $100\ev$ (dotted purple), with ratios shown relative to $10\kev$.
The photon-energy sum corresponds closely to individual photon energies; deviations arise only from multi-photon emission, which is suppressed by an extra power of $\alpha$. 
Accordingly, in \sherpa (which uses an energy cutoff), this observable is highly stable against the cutoff (see \figRef{fig:cutoffSherpaPhotons21}). 
Only for the largest cutoff, $E^\mathrm{min}_\gamma=100\mev$, do we observe a significant dip at $\sum E_\gamma=100\mev$, with a mild undershoot at slightly larger $\sum E_\gamma$ from multi-photon emissions. 
For $\sum E_\gamma > \frac12 m_{\jpsi} \approx 1.55\gev$, only events with two or more photons can populate the phase space. 
In this region, the $100\mev$ cutoff causes a clear dip relative to the other distributions. 
For \sherpa cutoffs $\le 1\mev$, the \sherpa distributions are cutoff-independent at the scale (and bin widths) shown.

We expected a similar behaviour in \photos, but \figRef{fig:cutoffPHOTOS} shows a more complex cutoff dependence than \sherpa. 
For cutoff values between $10\mev$ and $10\kev$, the results are stable within a few percent, motivating our default choice of $10^{-5}\gev\,=\,10\kev$. 
For the lowest cutoff values, $1\kev$ and $100\ev$, the energy sum is unexpectedly suppressed at large values. 
We do not know the cause, but note it here and advise against using cutoff values in \photos below $10\kev$ for this decay. 

In \vincia and \pythia, the cutoff is defined in terms of \pT-like variables, so we expect a different behaviour, as shown in \figsRef{fig:cutoffVinciaQED} and \ref{fig:cutoffPythiaQED}. 
If we instead plotted a \pT-like variable (or an angular-weighted energy sum), \vincia and \pythia would presumably exhibit sharper behaviour, while the one of \photos and \sherpa would appear more smeared. 
This simply reflects that different cutoff variables produce different behaviours. 
The $1\kev$ cutoffs used here are stable on the scales and bin widths shown.

\begin{figure}[tp]
\centering
\begin{subfigure}{0.47\textwidth}
  \centering
  \includegraphics[width=\linewidth]{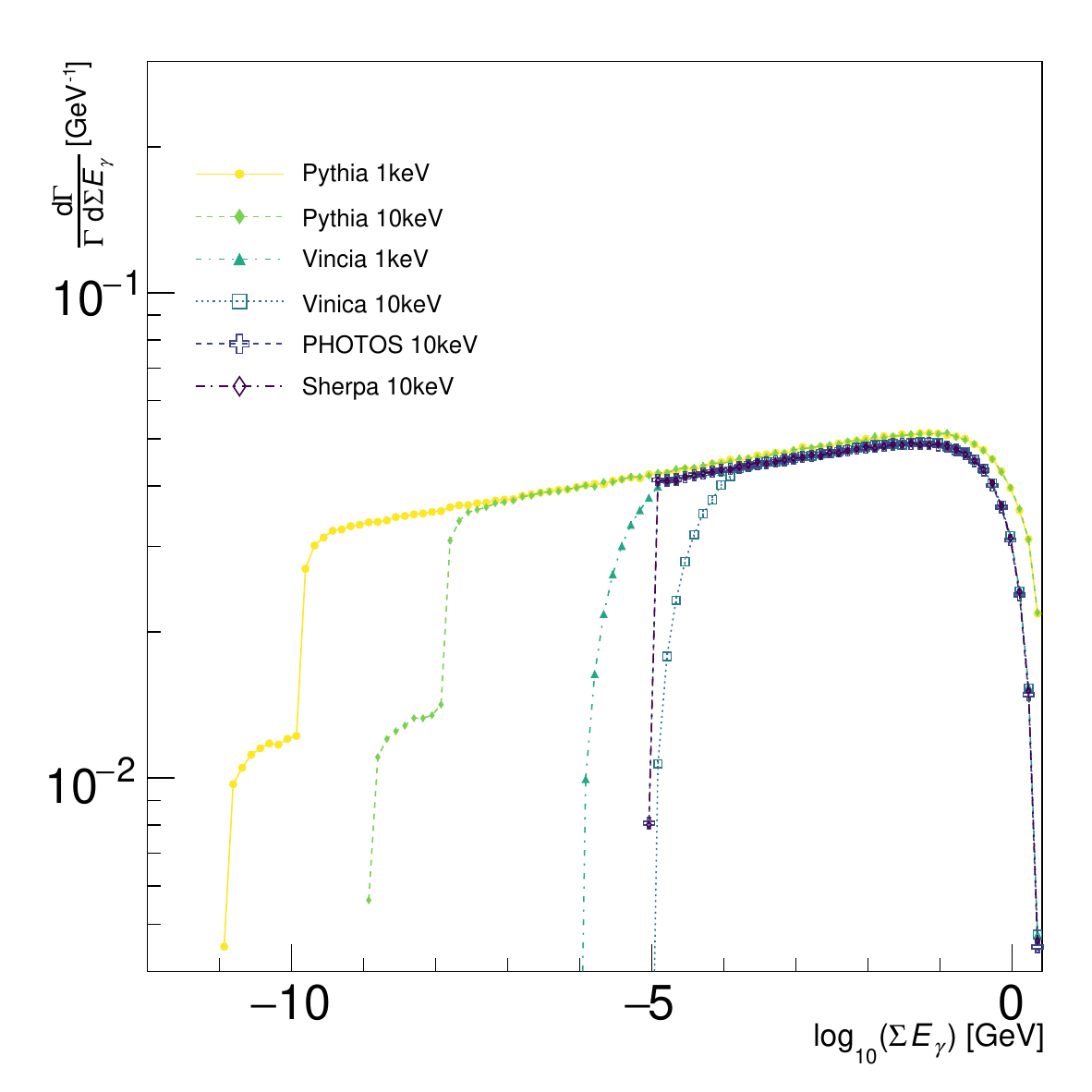}\vspace*{-5mm}
  \caption{\decay{\Bz}{\Kp\pim}.}
  \label{fig:deltaELogBKpi}
\end{subfigure}
\hfill
\begin{subfigure}{0.47\textwidth}
  \centering
  \includegraphics[width=\linewidth]{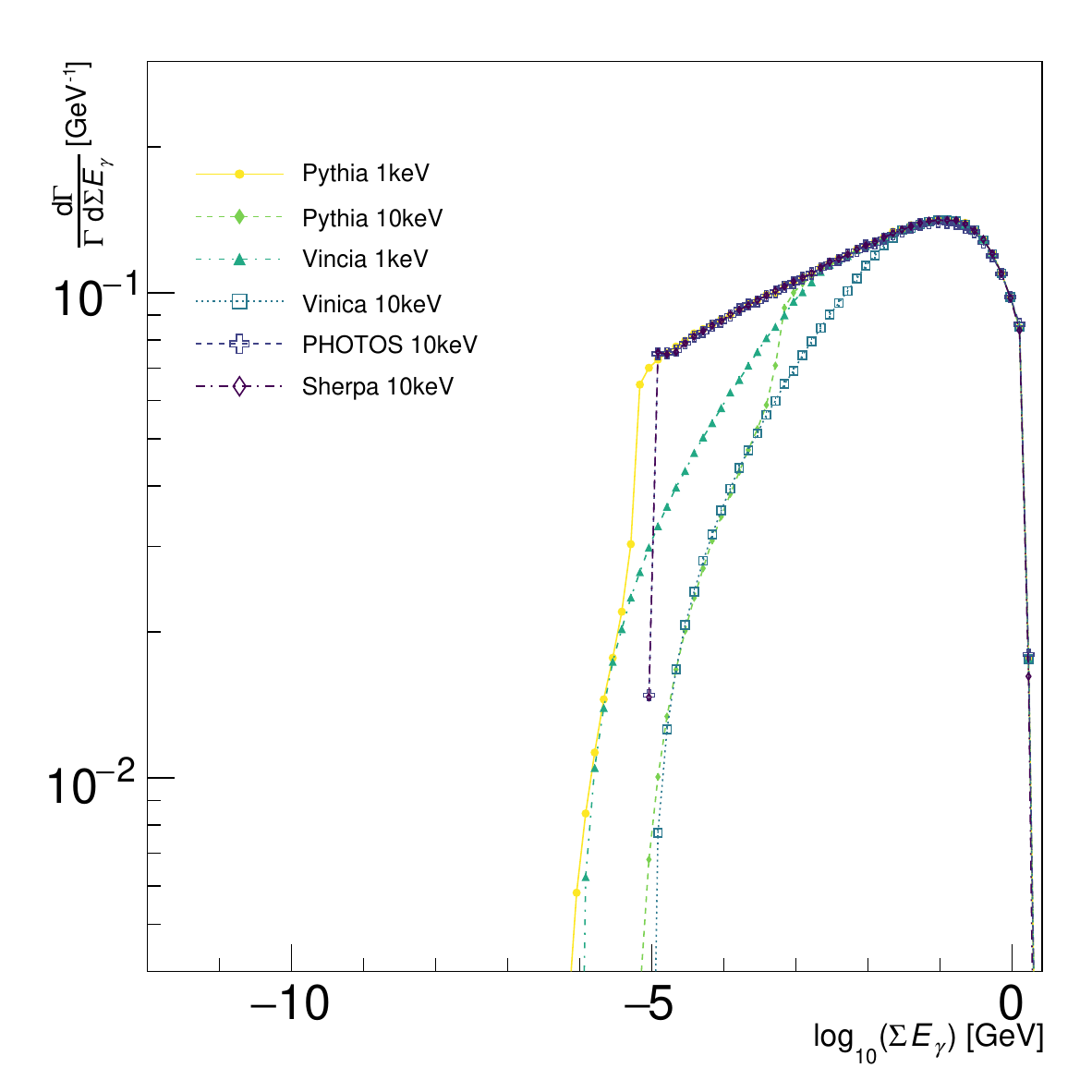}\vspace*{-5mm}
  \caption{\decay{\jpsi}{\ep\en}.}
  \label{fig:deltaELogJpsiee}
\end{subfigure}
   \caption{Base-10 logarithm of the total emitted photon energy in (left)~\decay{\Bz}{\Kp\pim} and (right)~\mbox{\decay{\jpsi}{\ep\en}} decays, shown for the four event generators under study and for several choices of the infrared cutoff value. 
   The two panels employ identical $x$- and $y$-axis ranges. }
    \label{fig:deltaELog}
\end{figure}

\FigRef{fig:deltaELog} compares $\pT$-like and energy cutoffs. 
\sherpa and \photos show step-function cutoffs, while \pythia and \vincia show smooth falloffs, as expected from their respective energy and $\pT$-like cutoff choices. 
In \decay{\Bz}{\Kp\pim} (\figRef{fig:deltaELogBKpi}), setting the \vincia $\pT$ cutoff equal to the energy cutoffs reduces the rate starting about an order of magnitude above the cutoff. 
Lowering the \vincia cutoff to $1\kev$ matches the energy-cutoff generators down to $10\kev$, then extends one order of magnitude lower with a much reduced rate.
For \decay{\jpsi}{\ep\en} (\figRef{fig:deltaELogJpsiee}), lowering the \vincia cutoff has a similar effect, bringing it in line with the other generators for another order of magnitude. 
This further motivates our chosen cutoff values for \vincia.
However, in the extreme case of \decay{\jpsi}{\ep\en}~decays, the smooth suppression starts above the energy cutoff of the other generators.
\pythia shows odd behaviour in both processes. 
In \figRef{fig:deltaELogBKpi} it follows the generators down to their energy cutoffs, but then continues well below its expected cutoff. 
This likely stems from \pythia’s handling of emitters whose masses are large relative to the available phase space, producing kinks in the spectral tails when only the kaon radiates. 
In \decay{\jpsi}{\ep\en} decays, around $\log_{10}(\sum E_\gamma) \sim -3.5$ and $\sim -5.5$, a similar two-component structure appears in \pythia's distributions.

In many real-world physics applications, one could likely use cutoffs about ten times larger than those used in these studies, unless resolving very soft and/or collinear photons is required. 
We instead use low cutoffs to ensure that we resolve all contributions to the observables studied.

\section{Quantifying matrix-element corrections}

\label{sec:MEs}

The first-order real-emission QED corrections for vector-to-dilepton decays such as \decay{\jpsi}{\mumu} and scalar-to-dilepton decays such as \decay{\Bs}{\mumu} differ only by nonsingular terms. 
Thus, from the point of view of pure bremsstrahlung descriptions (e.g., parton showers), they are typically described by the same (singular) radiation kernels.
The actual matrix-element ratio for a neutral hadron decay into a pair of charged leptons is~\cite{Gehrmann-DeRidder:2011gkt}:
\begin{equation*}
\frac{|M(\decay{X^0}{f\bar{f}\gamma})|^2}{|M(\decay{X^0}{f\bar{f}})|^2}
\,=\, \frac{8\pi\alpha Q_f^2}{M_V^2}\left(
  \frac{2y_{f\bar{f}}}{y_{f\gamma}y_{\bar{f}\gamma}}
 \,-\,\frac{2\mu_{f}^2}{y_{f\gamma}^2}
 \,-\,\frac{2\mu_{\bar{f}}^2}{y_{\bar{f}\gamma}^2}
 \,+\,\frac{y_{\bar{f}\gamma}}{y_{f\gamma}}
 \,+\,\frac{y_{f\gamma}}{y_{\bar{f}\gamma}} 
 \,+\,c^\mathrm{X}_\mathrm{NS}
\right)\,,
\label{eq:antVecFF}
\end{equation*}
where $f$ denotes a generic charged fermion with rest mass $m_f^2$, the dimensionless invariants are $\mu_f^2\equiv m_f^2/m_{X}^2$ and $y_{ij} \equiv s_{ij}/m_{X}^2=2(p_i\cdot p_j)/m_{X}^2$, and the first-order nonsingular matrix-element coefficients are~\cite{Gehrmann-DeRidder:2011gkt}:
\begin{eqnarray*}
\mbox{for X = Vector : }~~~~ c_{\mathrm{NS}} & = & ~0\,, \\[2mm]
\mbox{for X = Scalar : }~~~~ c_\mathrm{NS} & = & +2\,. \end{eqnarray*}

In \vincia, the baseline QED antenna function for fermion-antifermion 
dipoles~\cite{Skands:2020lkd} is chosen to have $c_\mathrm{NS}=0$, so that
\begin{equation*}
\mathrm{ant}^\mathrm{def}_{f\bar{f}\mapsto f\bar{f}\gamma}~=~
\frac{|M(\decay{X^0}{f\bar{f}\gamma})|^2}{|M(\decay{X^0}{f\bar{f}})|^2}
\Bigg|_{c^\mathrm{X}_\mathrm{NS}=0}\,.
\end{equation*}
Thus, for the vector \decay{\jpsi}{\mumu} decay, we expect the first-order shower-to-matrix-element ratio, 
\begin{equation}
R^\mathrm{PS}_\mathrm{ME} ~=~
\frac{\mbox{Expanded Shower}}{\mbox{Matrix Element}}
~=~\frac{\mathrm{ant}_{f\bar{f}\mapsto f\bar{f}\gamma}\,|M(\decay{X^0}{f\bar{f}})|^2}{|M(\decay{X^0}{f\bar{f}\gamma})|^2}
\label{eq:RPSME}
\end{equation}
to be unity over all phase space.
For the pseudoscalar \decay{\Bs}{\mumu} decay, we expect this ratio to be well below unity for hard wide-angle photons, since the uncorrected shower has $c_\mathrm{NS}=0$, while the scalar first-order matrix element has $c_\mathrm{NS}=2$. 
The ME correction factors that we apply to the shower are simply the inverse of this ratio,
\begin{equation*}
P^\mathrm{MEC} ~=~(R^\mathrm{PS}_\mathrm{ME})^{-1}\,,
\end{equation*}
generalised so that it can be applied to multiple emissions, not just the first, following the same strategy as for (sector) QCD showers~\cite{Lopez-Villarejo:2011pwr}. 
Unless otherwise stated, results labelled as ``\vincia MEC'' below include tree-level ME corrections for up to two photon emissions.

If we manually add a value of $+2$ to \vincia's baseline antenna function, we expect $R^\mathrm{PS}_\mathrm{ME}$ to be unity over all phase space for \decay{\Bs}{\mumu} decays. 
For \decay{\jpsi}{\mumu} decays, it would exceed unity, indicating that the baseline shower would overestimate the matrix element in the hard wide-angle region. 
We illustrate this by varying the coefficient $c_\mathrm{NS}$ in the antenna function\footnote{Analogous variations of $c_\mathrm{NS}$ for QCD splitting kernels are shown in~\cite{Mrenna:2016sih}}.

\begin{figure}[tp]
\centering
\begin{subfigure}{0.45\textwidth}
  \centering
  \includegraphics[width=\linewidth]{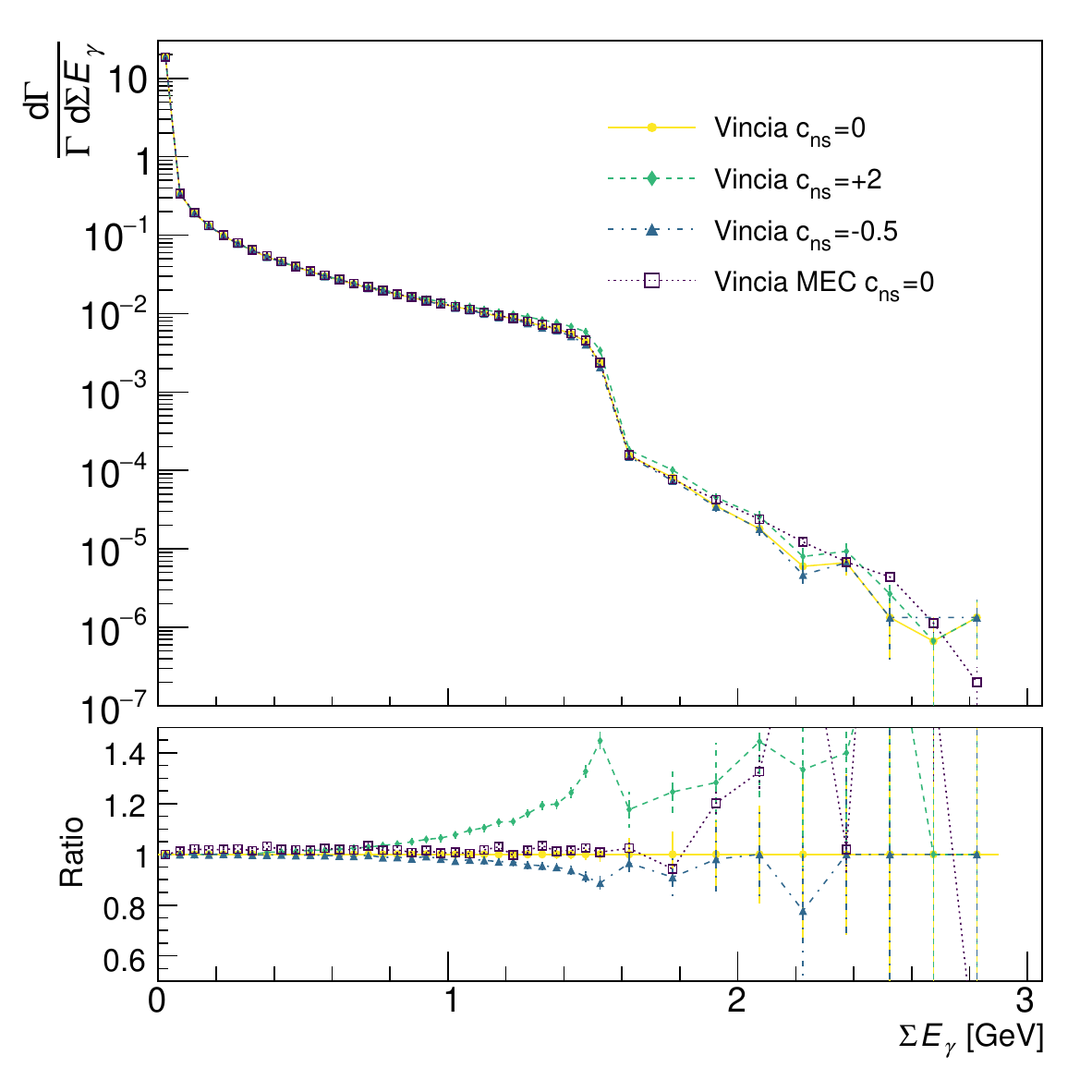}
  \caption{\decay{\jpsi}{\mumu}}
  \label{fig:JpsiMuMusize}
\end{subfigure}
\hfill
\begin{subfigure}{0.45\textwidth}
  \centering
  \includegraphics[width=\linewidth]{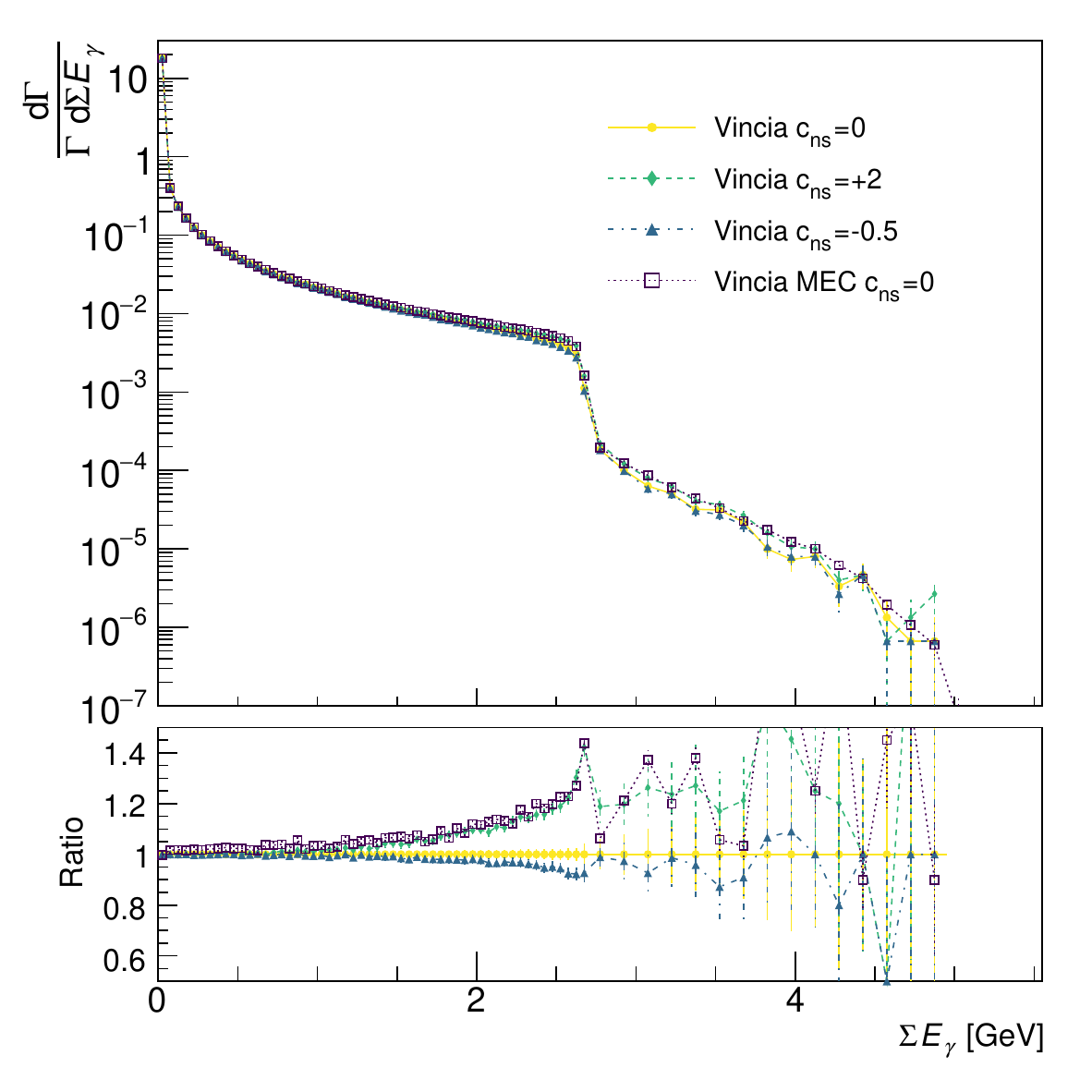}
  \caption{\decay{\Bs}{\mumu}}
  \label{fig:BsmumuMECSize}
\end{subfigure}
    \caption{Total photon energy radiated in (left)~\decay{\jpsi}{\mumu} and (right)~\decay{\Bs}{\mumu} decays for the baseline \vincia QED shower (solid yellow), the shower with manually added non singular contributions $+2$ (dashed green) and $-0.5$ (dash dotted blue) to the standard antenna function, and the shower including second-order tree-level ME corrections (dotted purple). 
    The ratio panel is normalised to the baseline shower prediction.}
    \label{fig:MECsize}
\end{figure}

In \figRef{fig:MECsize}, we compare the total emitted photon energy for \decay{\jpsi}{\mumu} and \decay{\Bs}{\mumu} decays, for \vincia antenna functions in the default configuration, with manually added $+2$ and $-0.5$ nonsingular terms, and with ME corrections through ${\cal O}(\alpha^2)$. 
For \decay{\jpsi}{\mumu} decays (\figRef{fig:JpsiMuMusize}), the effect of nonsingular terms and ME corrections in the soft region (low $\sum E_\gamma$) is irrelevant, as expected. 
The differences start to be noticeable above $\sim m_{\jpsi}/4 \sim 800\mev$. We confirm that, for this decay, the baseline shower and the ME-corrected shower are essentially identical.
The ME-corrected shower radiates slightly more because the corrections affect the second emission.
For \decay{\Bs}{\mumu} decays (\figRef{fig:BsmumuMECSize}), we find strong agreement in the soft region up to $m_{\Bs}/4 \sim 1.3\gev$. 
Above this, the showers with positive nonsingular term and with ME correction deviate from the baseline shower and are nearly identical, as expected from the nonsingular $+2$ term matching the exact \decay{\PH}{\quark\g\quarkbar} antenna~\cite{Gehrmann-DeRidder:2011gkt}.
The important point is that, regardless of the baseline antenna function chosen, the framework for ME corrections developed in this work imposes the correct tree-level matrix elements in both decays.

\begin{figure}[tp]
\centering
\begin{subfigure}{0.45\textwidth}
  \centering
  \includegraphics[width=\linewidth]{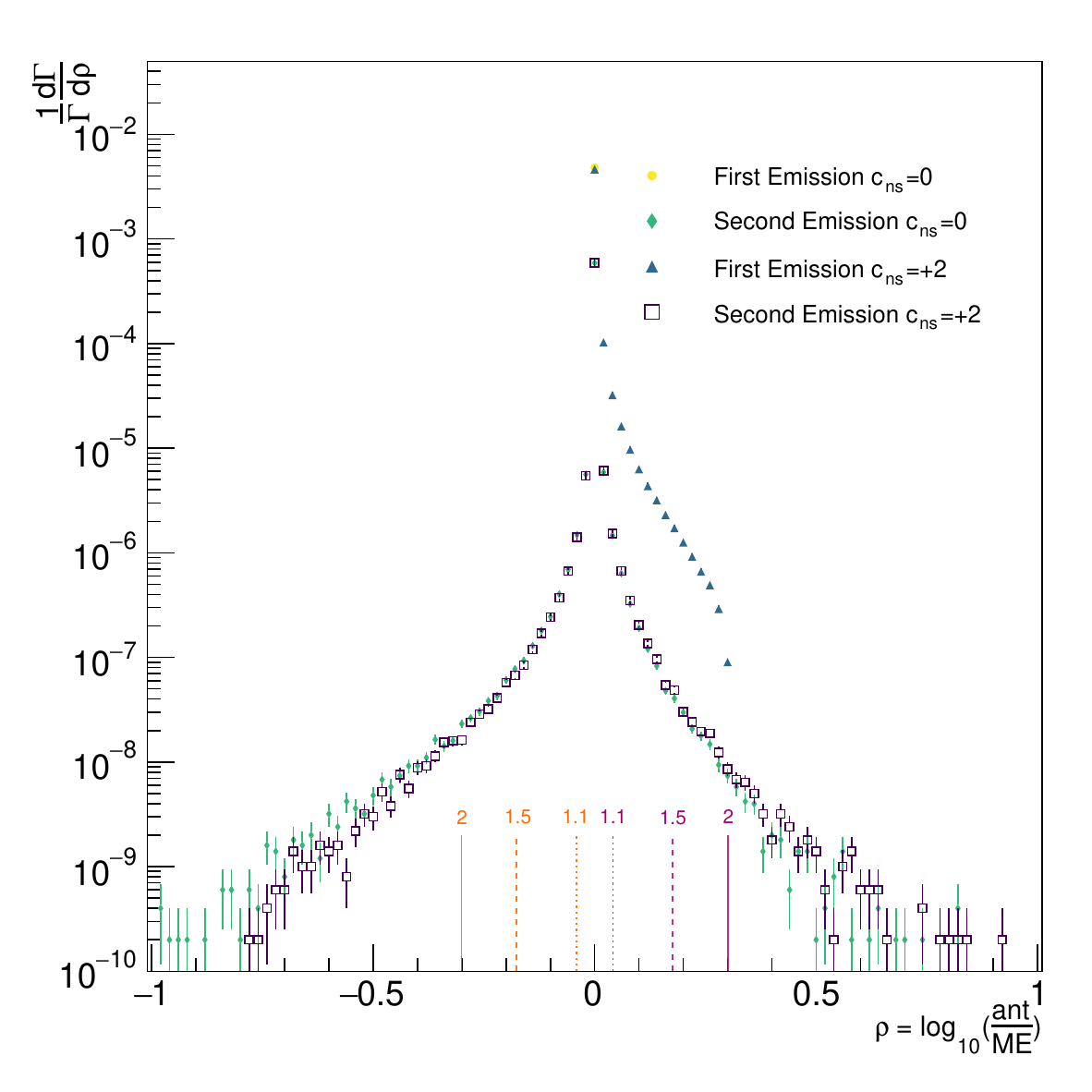}
  \caption{\decay{\jpsi}{\mumu}}
  \label{fig:JpsiMuMuAntOverMec}
\end{subfigure}
\hfill
\begin{subfigure}{0.45\textwidth}
  \centering
  \includegraphics[width=\linewidth]{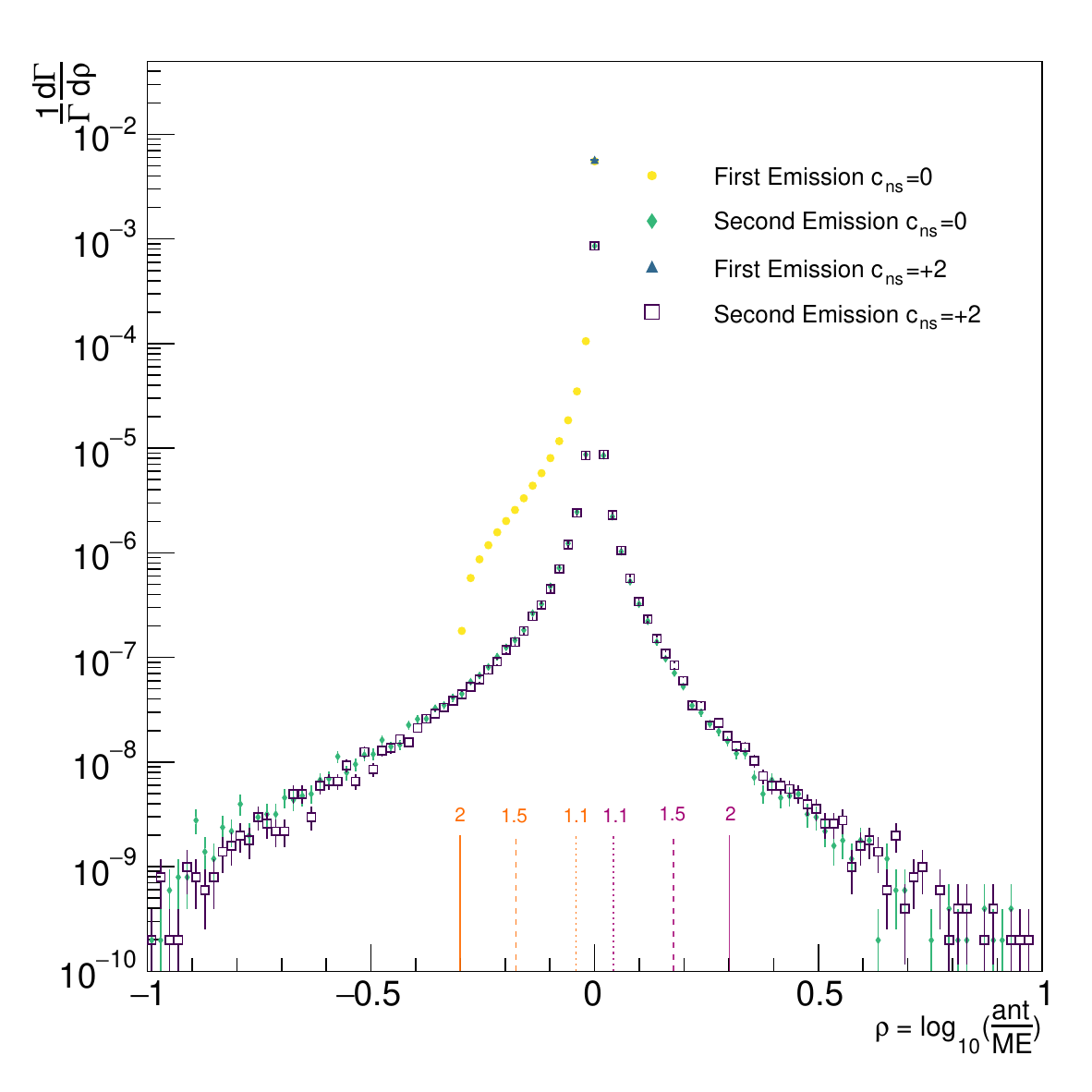}
  \caption{\decay{\Bs}{\mumu}}
  \label{fig:BsmumuMECAntOverMec}
\end{subfigure}
    \caption{Distribution of the base-10 logarithm of $R^{\mathrm{PS}}_{\mathrm{ME}}$ (\eqRef{eq:RPSME}), normalised to the total number of generated events and bin width, for the first and second photon emissions in (left)~\decay{\jpsi}{\mumu} and (right)~\decay{\Bs}{\mumu} decays. 
    Results are presented for $c_\mathrm{NS}=0$ (yellow and green markers) and $c_\mathrm{NS}=2$ (blue and purple markers). 
    Vertical reference lines indicate selected values of the shower-to-matrix-element ratio, corresponding (from left to right) to: a solid orange line for the expanded shower being a factor of $2$ smaller than the tree-level matrix element ($\rho=\log_{10}(1/2)$), a dotted orange line for a $50\%$ deficit ($\rho=\log_{10}(1/1.5)$), a dashed orange line for a $10\%$ deficit ($\rho=\log_{10}(1/1.1)$), a dashed purple line for a $10\%$ excess ($\rho=\log_{10}(1.1)$), a dotted purple line for a $50\%$ excess ($\rho=\log_{10}(1.5)$), and a solid purple line for the expanded shower being a factor of $2$ larger than the matrix element.}
    \label{fig:antOverMec}
\end{figure}

We also compare the probability distribution of $R^\mathrm{PS}_{\mathrm{ME}}$ (as sampled by the ME-corrected shower). 
In singular regions this ratio should be unity, while deviations in hard wide-angle regions signal a non-trivial ME correction. 
\figRef{fig:antOverMec} shows this behaviour for the first two photon emissions in \decay{\jpsi}{\mumu} and \decay{\Bs}{\mumu} decays.

For \decay{\jpsi}{\mumu} decays (\figRef{fig:JpsiMuMuAntOverMec}), with the default $c_\mathrm{NS}=0$, all events fall in the zero bin for the first photon emission (logarithmic $x$ axis, so 0 corresponds to $R^\mathrm{PS}_\mathrm{ME}=1$); 
with $c_\mathrm{NS}=+2$, the shower overestimates the first-order matrix element and therefore has a tail to the right side of zero.
For the second emission, we observe asymmetric broadening: the default antenna more often underestimates than overestimates the matrix element. 
With $c_\mathrm{NS}=+2$, the second-emission distribution becomes only slightly more symmetric because the added constant modestly increases the numerator of \eqRef{eq:RPSME}, but broad tails remain where the genuine second-order ME correction is still required for a correct description.

For \decay{\Bs}{\mumu} decays (\figRef{fig:BsmumuMECAntOverMec}), all events with $c_\mathrm{NS}=+2$ now sit in the zero bin for the first emission, while the baseline (uncorrected) shower has a tail to the left side of zero. 
For the second emission, the pattern is similar as for vector decays: setting $c_\mathrm{NS}=+2$ increases the numerator of \eqRef{eq:RPSME}, but the shower still underestimates the matrix element, so the second-order correction is equally necessary to recover the correct expansion at this order.
Again, the important point is that, regardless of the baseline antenna function chosen, the ME corrections enforce the correct tree-level matrix elements in both decays.

\section{Results and validations}
\label{sec:results}

All results in this section are obtained using the \evtgen generator~\cite{Lange:2001uf,Abudinen:2024wpc,evtgen:2025} to produce events prior to QED radiation.
\evtgen is a simulation package specialised in the decays of heavy hadrons, incorporating a variety of models specific to different decay topologies.
All of the four QED radiation generators we consider are employed as afterburners, acting on an identical set of input events containing undressed particles, thereby enabling a consistent comparison of the results obtained from the various QED radiation generators within a common framework.

We compare the QED radiation generators using a set of representative decay modes of neutral and charged mesons. 
Unless otherwise specified, all decays are simulated in \evtgen employing the phase-space model (\texttt{PHSP}). 
In \secRef{sec:2body}, we consider two-body decay processes, focussing first on neutral meson decays:
\begin{itemize}
\item \decay{\jpsi}{ \ep\en} and \decay{\jpsi}{\mumu}, modelled with \texttt{VLL} (\decay{\rm vector}{\rm lepton\, lepton});
\item \decay{\Bs}{\mumu}; 
\item \decay{\Bz}{\pip\pim}, \decay{\Bz}{\Kp\pim}, and \decay{\Bz}{\Ds\Dm};
\item \decay{\Dz}{\pip\pim}, \decay{\Dz}{\Km\pip}, and  \decay{\Dz}{\Kp\Km}; 
\end{itemize}
and then focussing on charged meson decays:
\begin{itemize}
\item \decay{\Bp}{\pip\piz} and \decay{\Bp}{\Kp\piz};
\item \decay{\Dp}{\pip\piz}. 
\end{itemize}
In \secRef{sec:3body}, we consider three-body decay processes: 
\begin{itemize}
\item \decay{\Bz}{\Dz\Kp\pim};
\item \decay{\Bp}{\pip\piz\piz} and \decay{\Bp}{\Dp\Km\pip};
\item \decay{\Dp}{\pip\piz\piz};
\item \decay{\Dp}{\pip\pip\pim} and \decay{\Dp}{\Kp\pip\pim}, modelled with \texttt{D\_Dalitz}, which implements the Dalitz structures of refs.~\cite{E791:2000vek,FOCUS:2004muk}. 
\end{itemize}
Finally, in \secRef{sec:fixed3}, we present dedicated illustrations of multipole radiation patterns for a specific Born-level phase-space point for \decay{\Dp}{\pip\pip\pim} and \decay{\Bp}{\Dp\Kp\pim} decays.
In the analyses presented herein, for the decay channels specified above, charge conjugation is not implied and charge-conjugate modes are not included. This is simply for technical convenience, not a physics limitation, since QED is invariant under charge conjugation.

To avoid a proliferation of plots, we summarise many results in tables showing the total emitted photon energy ($\sum E_\gamma$) and the fraction of events with radiation energy losses above $1$, $10$, and $100\mev$. 
The values in the tables are colour-coded by the ratio to the \photons generator with first-order ME corrections (\photons~21), which is used throughout as a baseline reference.
The ratios below unity are colour-coded in orange (\colorbox{orange1}{$2-5\%$}, \colorbox{orange2}{$5-10\%$}, \colorbox{orange3}{$10-50\%$}), with bordeaux for \colorbox{orange4}{$<50\%$}. 
The ratios above unity are colour-coded in cyan (\colorbox{cyan1}{$2-5\%$}, \colorbox{cyan2}{$5-10\%$}, \colorbox{cyan3}{$10-100\%$}), with purple for \colorbox{cyan4}{$>100\%$}. 
A ratio is taken as $2\%$ below unity if it is $1/1.02$ relative to \photons~21, and $2\%$ above unity if it is $1.02$.

\subsection{Two-body decays}
\label{sec:2body}

\subsubsection{Two-body decays of neutral hadrons}
Leptonic decays of neutral vector particles such as the $\jpsi$ meson (and $Z$ boson) are particularly well studied and so serve as a strong basis for our first set of comparisons. 
For \decay{\jpsi}{ \ep\en}, we compare baseline \vincia (without MECs), \vincia MEC (with second-order MECs), and baseline \pythia (which includes first-order MECs).

\begin{figure}[tp]
\centering
\begin{subfigure}{0.45\textwidth}
  \centering
  \includegraphics[width=\linewidth]{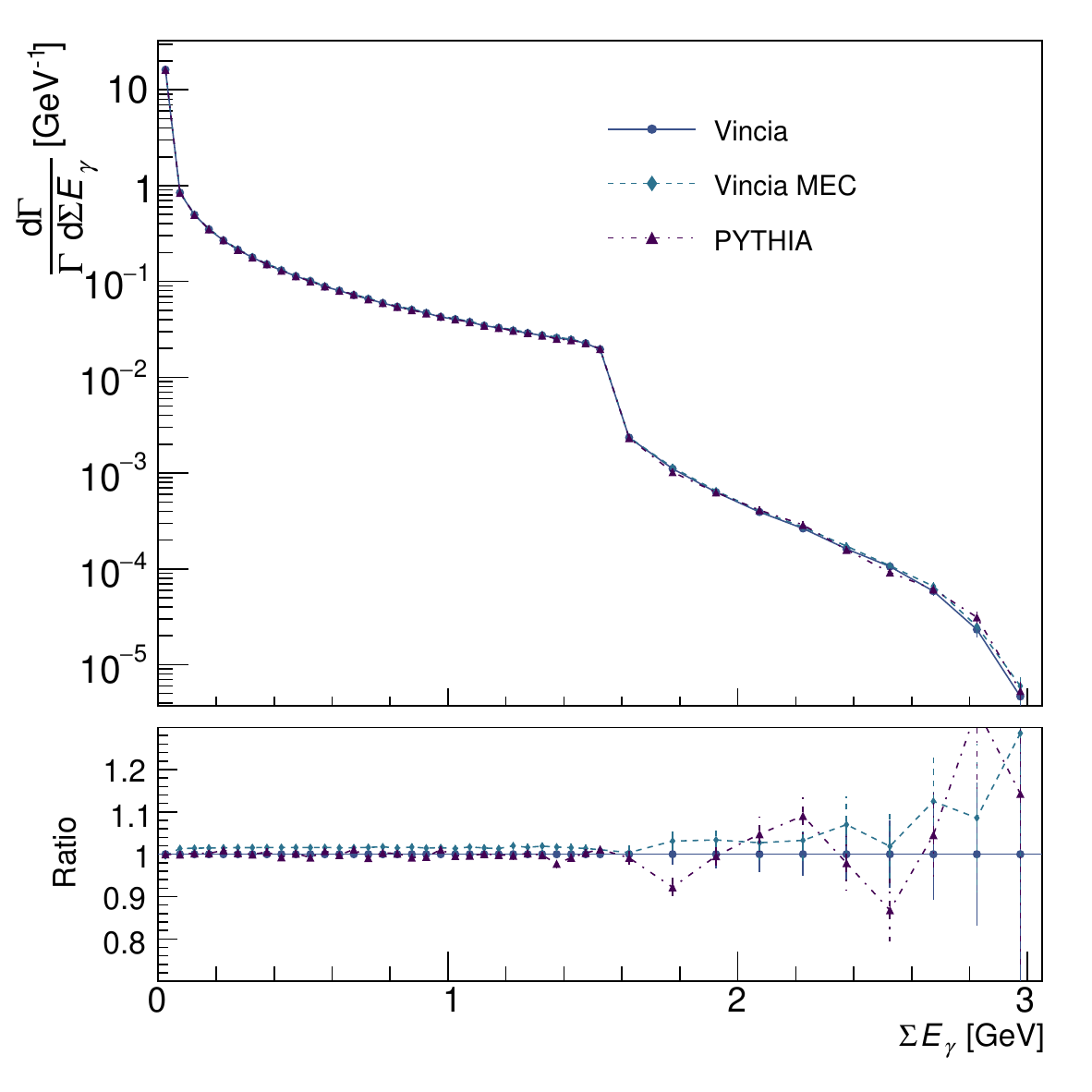}
  \caption{Total photon energy radiated in \mbox{\decay{\jpsi}{\ep\en}} decays for \pythia, \vincia and \vincia~MEC.}
  \label{fig:JpsitoEEVincias}
\end{subfigure}
\hfill
\begin{subfigure}{0.45\textwidth}
  \centering
  \includegraphics[width=\linewidth]{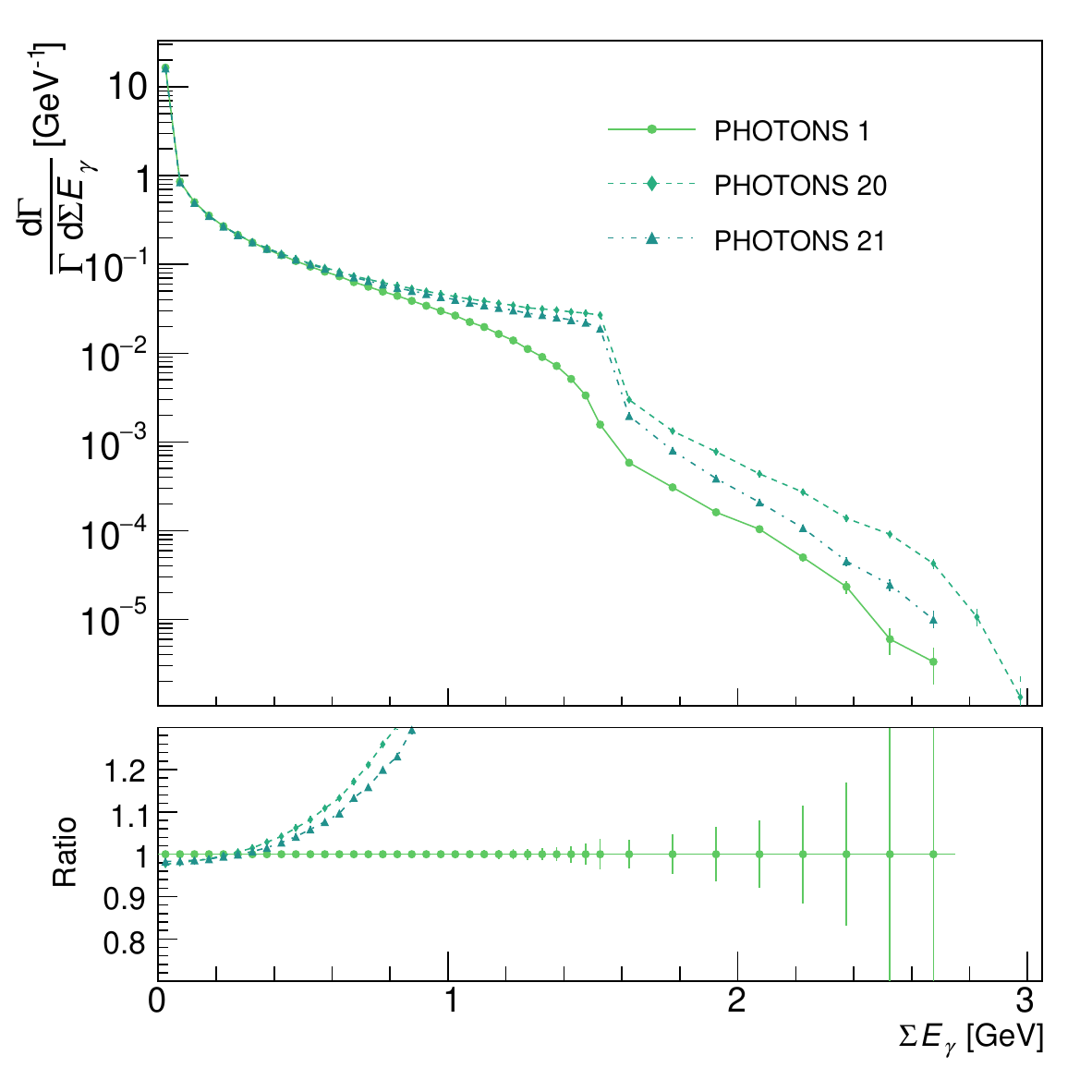}
  \caption{Total photon energy radiated in \mbox{\decay{\jpsi}{\ep\en}} decays for \photons~1, \photons~20 and \photons~21.}
  \label{fig:JpsitoEESherpas}
\end{subfigure}
\begin{subfigure}{0.45\textwidth}
  \centering
  \includegraphics[width=\linewidth]{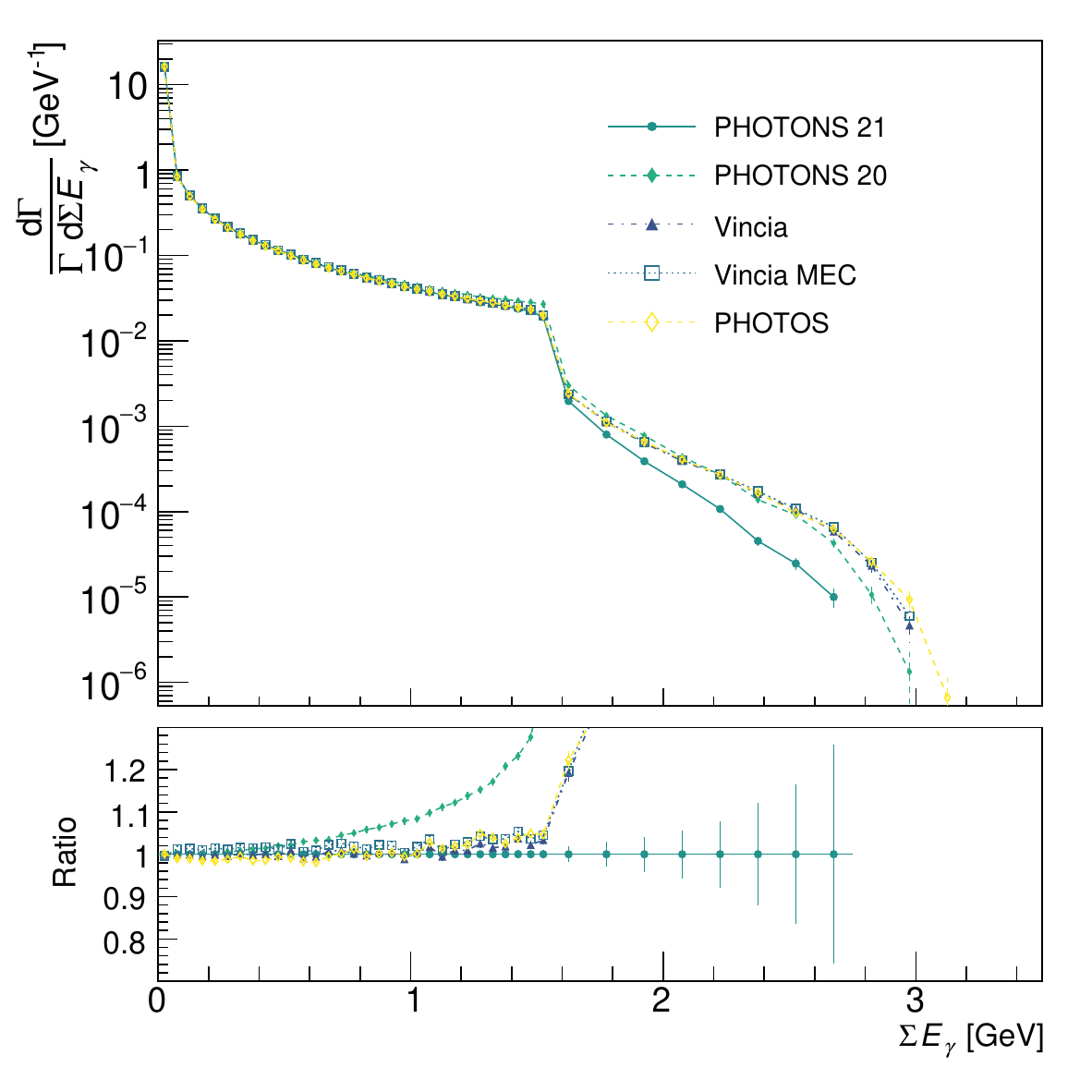}
  \caption{Total photon energy radiated in \mbox{\decay{\jpsi}{\ep\en}} decays for \photons~21, \photons~20, \vincia, \vincia~MEC and \photos.}
  \label{fig:JpsitoEEDeltaEComparisons}
\end{subfigure}
\hfill
\begin{subfigure}{0.45\textwidth}
  \centering
  \includegraphics[width=\linewidth]{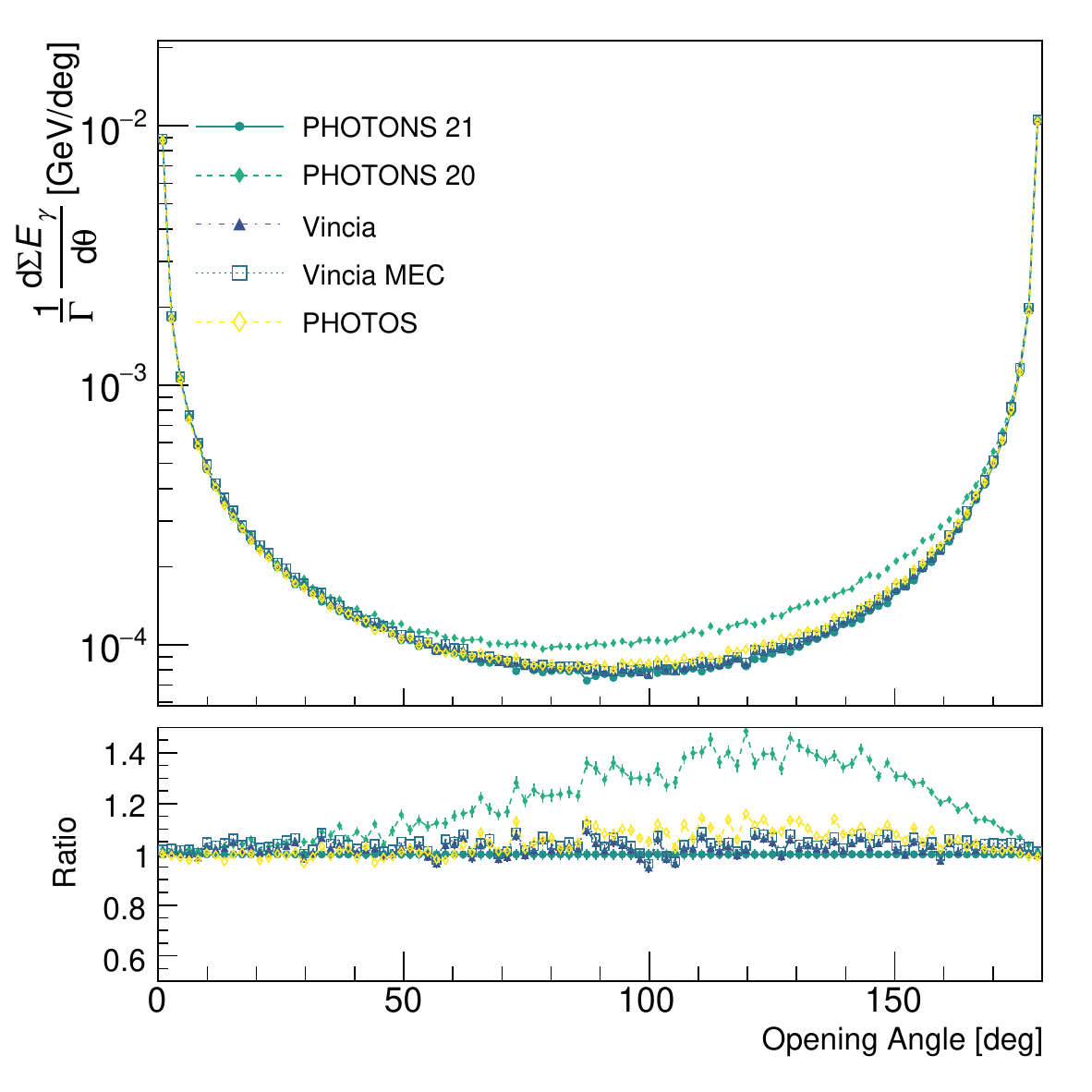}
  \caption{Angular distribution of photon energy in \mbox{\decay{\jpsi}{\ep\en}} decays for \photons~21, \photons~20, \vincia, \vincia~MEC and \photos.}
  \label{fig:JpsitoEEAngularE}
\end{subfigure}
\hfill
    \caption{Total photon energy radiated and angular distribution of photon energy in \mbox{\decay{\jpsi}{\ep\en}} decays for various QED generator options. 
    The opening angle is defined as the angle between the emitted photon and the post-branching final-state electron.}
    \label{fig:JpsitoEEComparisons}
\end{figure}

\FigRef{fig:JpsitoEEVincias} shows the total emitted photon energy. 
For this decay, \vincia without MECs matches \pythia closely. 
This is expected because \pythia includes an explicit first-order ME correction, while, as discussed in \secRef{sec:MEs}, \vincia’s baseline spin-$1/2$ antenna is derived from the \decay{\PZ}{\ellell} matrix element and is therefore identical to the first-order matrix element for this decay. 
\vincia with MEC is slightly higher (consistent with~\figRef{fig:JpsiMuMuAntOverMec} in \secRef{sec:MEs}), which shows that the baseline antenna underestimates the second-emission matrix element, giving a net positive ME correction.
In fig.~\ref{fig:JpsitoEESherpas} we verify that we can reproduce the plot from ref.~\cite{Schonherr:2008av}, where the first-order ME correction mode (\photons~21) sits between the pure scalar QED (\photons~1) and the scalar QED + collinear correction (\photons~20). 
Comparing the three generators (\photons~21, \vincia, and \photos) in~\figRef{fig:JpsitoEEDeltaEComparisons}, we find good agreement in the single-emission region ($\sum E_\gamma \le m_{\jpsi}/2 \sim 1.55\gev$). 
In the tail above $m_{\jpsi}/2$, accessible only to events with multiple emissions, \vincia and \photos broadly agree, while \photons~21 shows a significantly softer multi-emission tail.   

In~\figRef{fig:JpsitoEEAngularE}, we show the angular distribution of the total emitted photon energy. 
Here and in all subsequent angular-distribution plots in this section, the opening angle is defined as the angle between the radiated photon and one of the post-branching final-state daughters~(see figure captions and discussion in \secRef{ssec:phsp_photons}). 
Enhancements in the collinear regions ($0^\circ$ and $180^\circ$) reflect that the electrons are nearly massless, and all generators agree well in this limit. 
There is a slight bias towards $180^\circ$ caused by hard photons, which cause large recoil and are at a wide angle to both the final-state electron and positron. 
At wide angles,  \photons~20 radiates significantly more than any of the other models (consistent with its larger tails in \figRef{fig:JpsitoEEDeltaEComparisons}). 
The rest are much more similar, with  
\photos radiating slightly more and \photons~21 slightly less than \vincia at wide angles.

Table~\ref{tab:Jpsi_ee} shows good agreement between all generators, both on average energy loss and high-energy-loss fractions, with differences greater than $5\%$ only between different \photons configurations. 
Additionally, the second-order correction in \vincia MEC increases the total energy loss slightly, by about $1.5\%$ relative to \pythia and the baseline \vincia, or by about $2.6\%$ relative to \photons~21. 

\begin{table}[tp]
\centering
\begin{tabular}{l c c c c}
\toprule
$J/\psi \to e^+e^-$ & $\langle \sum E_{\gamma}\rangle [\mev]$ & $ f(\sum > 1\mev)$ & $ f(\sum > 10\mev)$ & $ f(\sum > 100\mev)$\\ \midrule 
\photos & \cellcolor[HTML]{FFFFFF} $74.12$ & \cellcolor[HTML]{FFFFFF} $39.34\%$ & \cellcolor[HTML]{FFFFFF} $27.81\%$ & \cellcolor[HTML]{FFFFFF} $14.43\%$\\ \midrule
\photons 1 & \cellcolor{orange2} $57.25$ & \cellcolor{orange1} $38.53\%$ & \cellcolor{orange1} $26.77\%$ & \cellcolor{orange2} $13.07\%$\\
\photons 20 & \cellcolor{cyan2} $78.94$ & \cellcolor[HTML]{FFFFFF} $39.92\%$ & \cellcolor[HTML]{FFFFFF} $28.41\%$ & \cellcolor{cyan1} $14.99\%$\\
\photons 21 & \cellcolor[HTML]{FFFFFF} $73.46$ & \cellcolor[HTML]{FFFFFF} $39.54\%$ & \cellcolor[HTML]{FFFFFF} $27.97\%$ & \cellcolor[HTML]{FFFFFF} $14.49\%$\\ \midrule
\vincia MEC & \cellcolor{cyan1} $75.39$ & \cellcolor[HTML]{FFFFFF} $39.84\%$ & \cellcolor[HTML]{FFFFFF} $28.41\%$ & \cellcolor[HTML]{FFFFFF} $14.75\%$\\
\vincia & \cellcolor[HTML]{FFFFFF} $74.24$ & \cellcolor[HTML]{FFFFFF} $39.35\%$ & \cellcolor[HTML]{FFFFFF} $28.02\%$ & \cellcolor[HTML]{FFFFFF} $14.53\%$\\
\vincia Dipole & \cellcolor[HTML]{FFFFFF} $74.38$ & \cellcolor[HTML]{FFFFFF} $39.37\%$ & \cellcolor[HTML]{FFFFFF} $28.02\%$ & \cellcolor[HTML]{FFFFFF} $14.56\%$\\ \midrule
\pythia & \cellcolor[HTML]{FFFFFF} $74.24$ & \cellcolor[HTML]{FFFFFF} $39.61\%$ & \cellcolor[HTML]{FFFFFF} $28.04\%$ & \cellcolor[HTML]{FFFFFF} $14.55\%$\\ \bottomrule
\end{tabular}
\caption{
Average amounts of photon energy radiated in
 \decay{\jpsi}{\ep\en} decays, and the fractions of events which radiate a total of more than $1$, $10$, and $100\mev$. Colour scheme defined at the start of \secRef{sec:results}.}
\label{tab:Jpsi_ee}
\end{table}

\begin{table}[tp]
\centering
\begin{tabular}{l c c c c}
\toprule
$J/\psi \to\mu^+\mu^-$ & $\langle \sum E_{\gamma}\rangle [\mev]$ & $ f(\sum > 1\mev)$ & $ f(\sum > 10\mev)$ & $ f(\sum > 100\mev)$\\ \midrule 
\photos & \cellcolor{cyan2} $24.70$ & \cellcolor[HTML]{FFFFFF} $15.92\%$ & \cellcolor[HTML]{FFFFFF} $10.59\%$ & \cellcolor{cyan1} $5.07\%$\\ \midrule
\photons 1 & \cellcolor{orange2} $18.87$ & \cellcolor{orange1} $15.40\%$ & \cellcolor{orange1} $10.04\%$ & \cellcolor{orange2} $4.51\%$\\
\photons 20 & \cellcolor{cyan3} $27.39$ & \cellcolor{cyan1} $16.21\%$ & \cellcolor{cyan1} $10.91\%$ & \cellcolor{cyan2} $5.40\%$\\
\photons 21 & \cellcolor[HTML]{FFFFFF} $23.44$ & \cellcolor[HTML]{FFFFFF} $15.80\%$ & \cellcolor[HTML]{FFFFFF} $10.48\%$ & \cellcolor[HTML]{FFFFFF} $4.97\%$\\ \midrule
\vincia MEC & \cellcolor{cyan1} $24.13$ & \cellcolor{cyan1} $16.16\%$ & \cellcolor{cyan1} $10.73\%$ & \cellcolor{cyan1} $5.09\%$\\
\vincia & \cellcolor[HTML]{FFFFFF} $23.76$ & \cellcolor[HTML]{FFFFFF} $15.90\%$ & \cellcolor[HTML]{FFFFFF} $10.56\%$ & \cellcolor[HTML]{FFFFFF} $5.01\%$\\
\vincia Dipole & \cellcolor[HTML]{FFFFFF} $23.80$ & \cellcolor[HTML]{FFFFFF} $15.88\%$ & \cellcolor[HTML]{FFFFFF} $10.55\%$ & \cellcolor[HTML]{FFFFFF} $5.02\%$\\ \midrule
\pythia & \cellcolor[HTML]{FFFFFF} $23.71$ & \cellcolor[HTML]{FFFFFF} $15.90\%$ & \cellcolor[HTML]{FFFFFF} $10.56\%$ & \cellcolor[HTML]{FFFFFF} $5.01\%$\\ \bottomrule
\end{tabular} 
\caption{
Average amounts of photon energy radiated in
 \decay{\jpsi}{\mumu} decays, and the fractions of events which radiate a total of more than $1$, $10$, and $100\mev$. Colour scheme defined at the start of \secRef{sec:results}.}
\label{tab:Jpsi_mumu}
\end{table}

\TabRef{tab:Jpsi_mumu} shows an equivalent summary table for \decay{\jpsi}{\mumu} decays. 
Again, \photons~21, \pythia, and \vincia agree to within $2\%$ of each other, with \vincia MEC exhibiting only slightly larger energy losses, consistent with the net-positive second-order correction illustrated in \figRef{fig:JpsiMuMuAntOverMec}. 
However, we note that the total energy loss predicted by \photos is about $5\%$ greater than that generated by \photons~21.

\begin{figure}[htb]
\centering
\begin{subfigure}{0.45\textwidth}
  \centering
  \includegraphics[width=\linewidth]{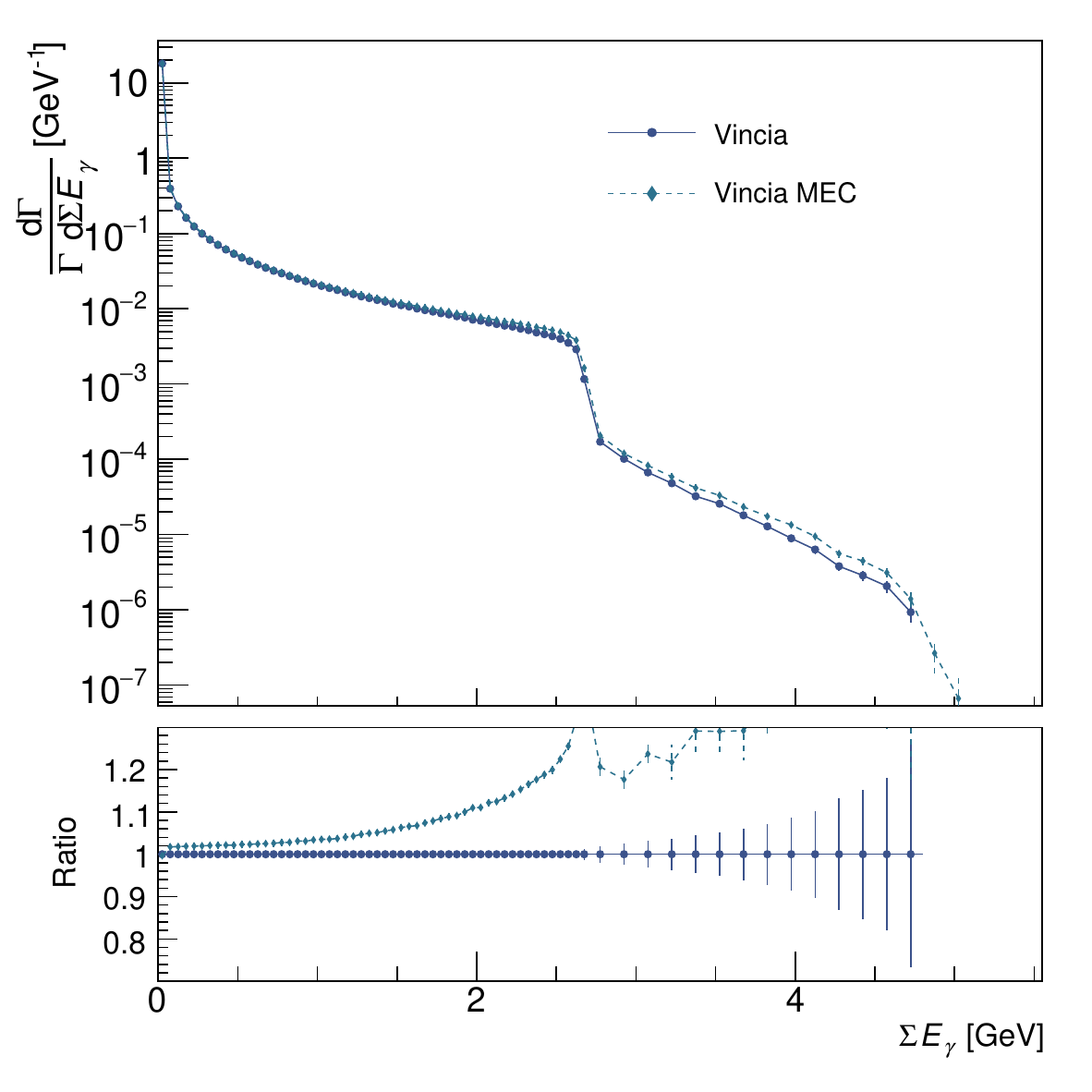}
  \caption{Total photon energy radiated in \mbox{\decay{\Bs}{\mumu}} decays for \vincia and \vincia~MEC.\\}
  \label{fig:BstomumuVincias}
\end{subfigure}
\hfill
\begin{subfigure}{0.45\textwidth}
  \centering
  \includegraphics[width=\linewidth]{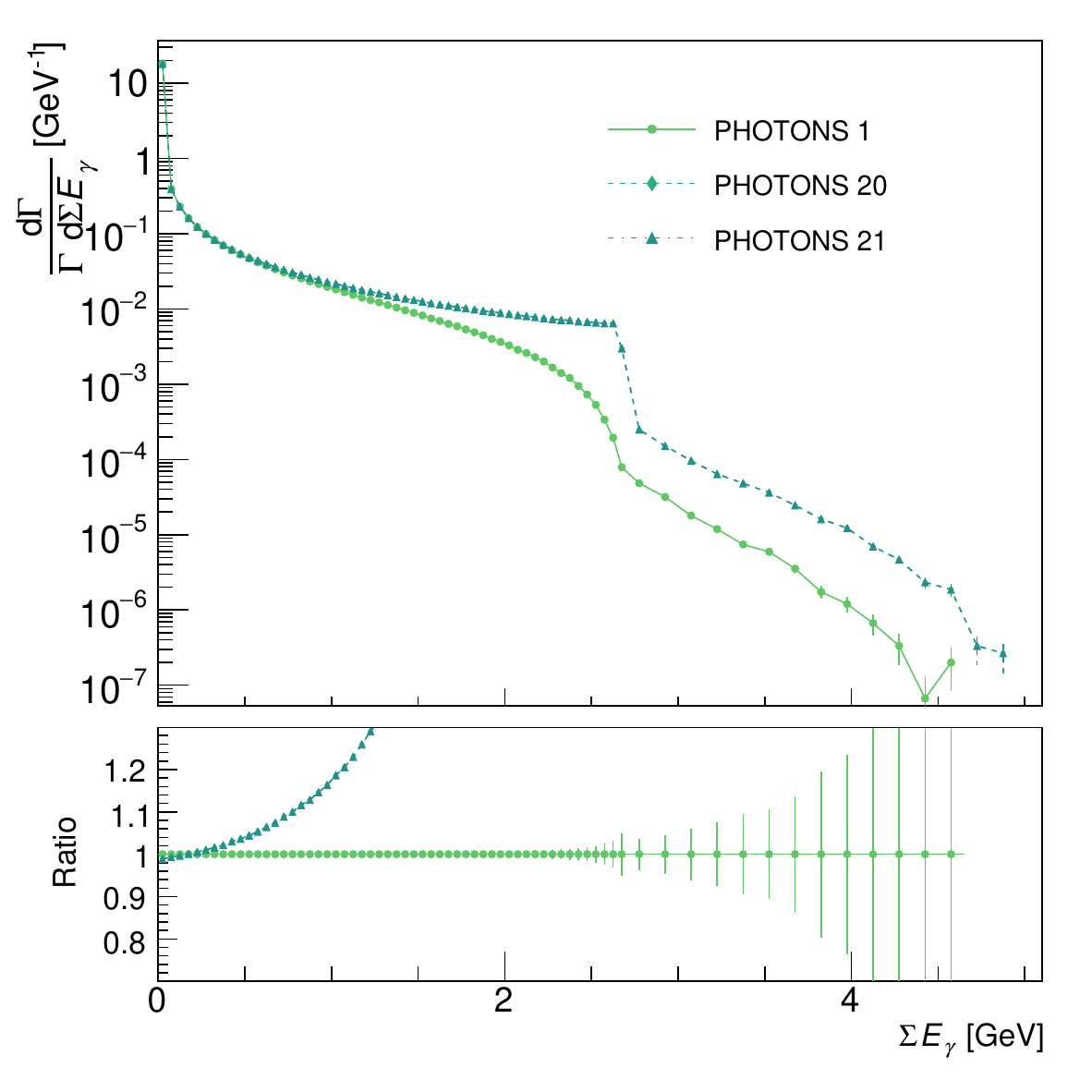}
  \caption{Total photon energy radiated in \mbox{\decay{\Bs}{\mumu}} decays for \photons~1, \photons~20 and \photons~21.\\}
  \label{fig:BstomumuSherpas}
\end{subfigure}
\begin{subfigure}{0.45\textwidth}
  \centering
  \includegraphics[width=\linewidth]{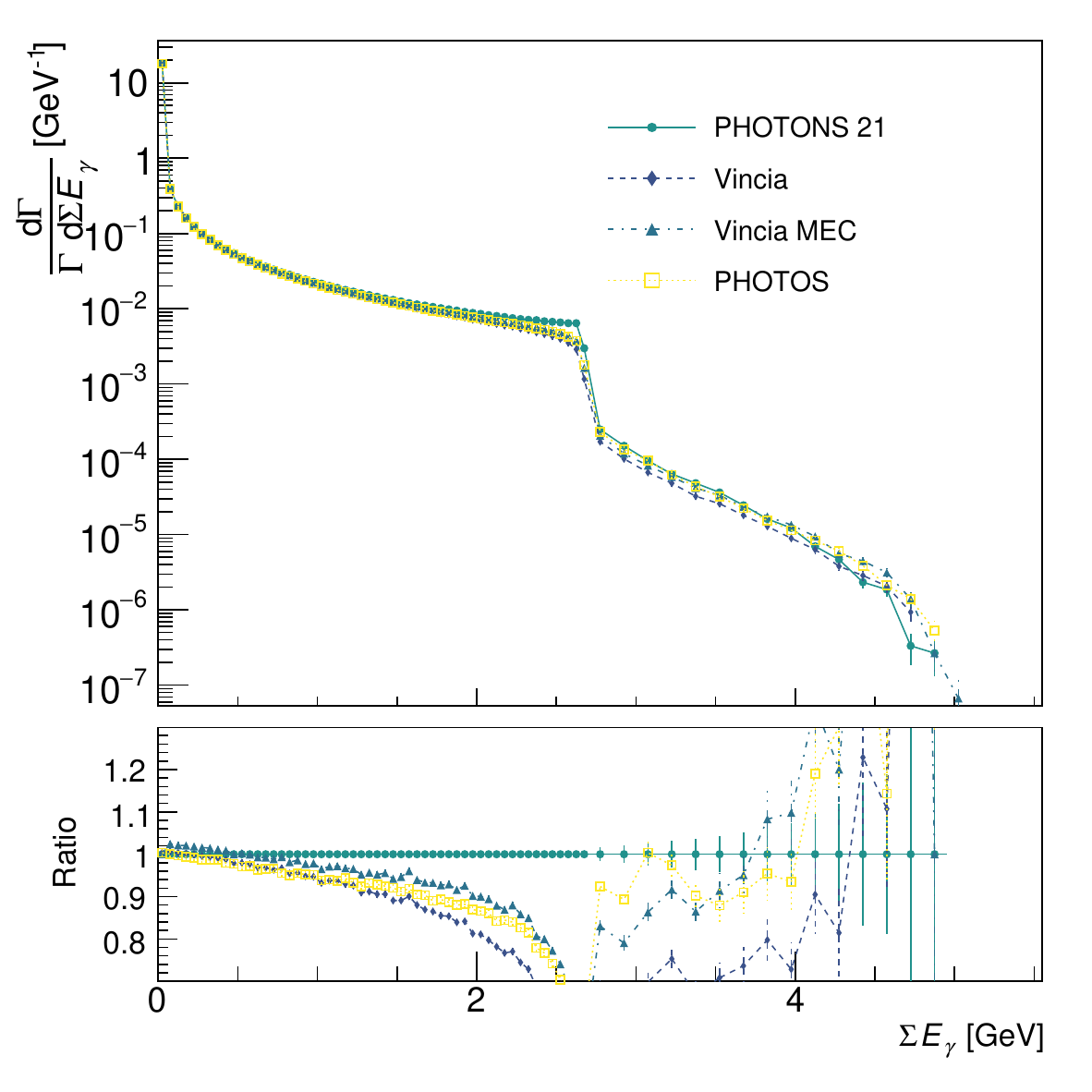}
  \caption{Total photon energy radiated in \mbox{\decay{\Bs}{\mumu}} decays for \photons~21, \vincia, \vincia~MEC and \photos.}
  \label{fig:BstomumuDeltaEComparisons}
\end{subfigure}
\hfill
\begin{subfigure}{0.45\textwidth}
  \centering
  \includegraphics[width=\linewidth]{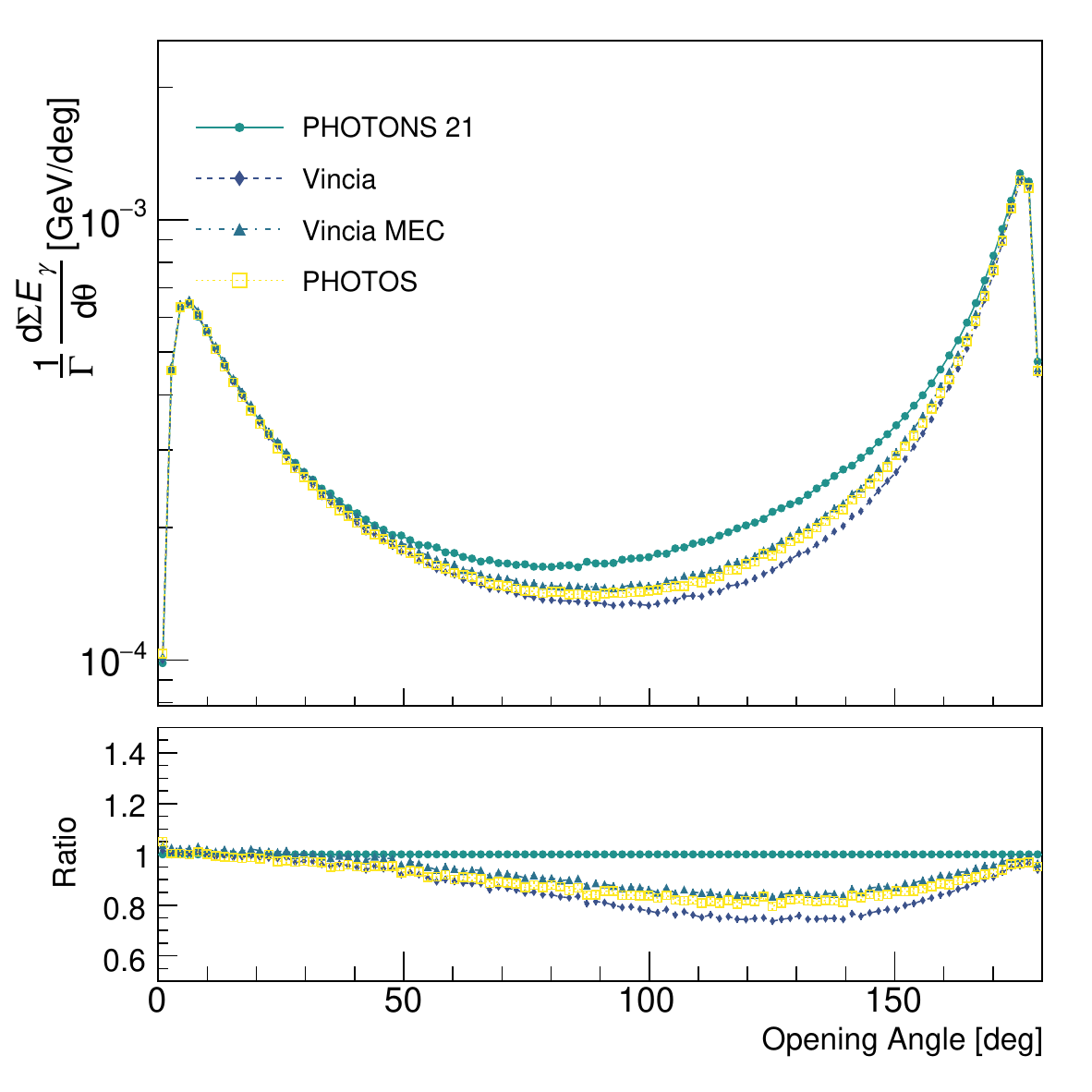}
  \caption{Angular distribution of energy in \mbox{\decay{\Bs}{\mumu}} decays for \photons~21, \vincia, \vincia~MEC and \photos.}
  \label{fig:BstomumuAngularEComparisons}
\end{subfigure}
    \caption{Total photon energy radiated and angular distribution of photon energy in \mbox{\decay{\Bs}{\mumu}} decays for various QED generator options. 
    The opening angle is defined as the angle between the emitted photon and the post-branching final-state negative muon.}
    \label{fig:BstomumuComparisons}
\end{figure}

Moving to \decay{\Bs}{\mumu} decays, \figRef{fig:BstomumuVincias} clearly shows the impact of nonsingular terms in the ME corrections in \vincia, as expected from the studies in \secRef{sec:MEs}.
In \figRef{fig:BstomumuSherpas}, \photons~1 (pure YFS) radiates much less in the tail than \photons~20 and~21, which overlap with each other.
Comparing the three generators in~\figRef{fig:BstomumuDeltaEComparisons}, we find that, unlike the \decay{V^0}{\ellell} case, \photons~21 radiates more than the others in the hard tail.
Similarly to the \decay{V^0}{\ellell} case, \vincia with ME corrections approximately agrees with \photos. 
In the angular energy distributions (\figRef{fig:BstomumuAngularEComparisons}), we now observe the dead-cone effect around the muon directions at $0^\circ$ and $180^\circ$. 
Differences between generators are also clearer: \photons~21 produces more wide-angle radiation and less small-angle radiation than the others. 
The harder tail of \photons~21~(and \photons~20) for \decay{\Bs}{\mumu} decays is also reflected in \tabRef{tab:Bs_mumu}, where all other generators predict somewhat smaller total radiation losses. 
Nevertheless, \vincia~MEC and \photons~21 agree well on the fractions of events that have total energy losses greater than $1$, $10$, and $100\mev$.

\begin{table}[tp]
\centering
\begin{tabular}{l c c c c}
\toprule
$B_s \to \mu^+\mu^-$ & $\langle \sum E_{\gamma}\rangle [\mev]$ & $ f(\sum > 1\mev)$ & $ f(\sum > 10\mev)$ & $ f(\sum > 100\mev)$\\ \midrule 
\photos & \cellcolor{orange2} $51.78$ & \cellcolor[HTML]{FFFFFF} $20.12\%$ & \cellcolor{orange1} $14.06\%$ & \cellcolor{orange1} $7.63\%$\\
\photos MEC & \cellcolor{orange2} $51.80$ & \cellcolor[HTML]{FFFFFF} $20.13\%$ & \cellcolor{orange1} $14.06\%$ & \cellcolor{orange1} $7.63\%$\\ \midrule
\photons 1 & \cellcolor{orange2} $39.49$ & \cellcolor{orange1} $19.53\%$ & \cellcolor{orange2} $13.42\%$ & \cellcolor{orange2} $6.95\%$\\
\photons 20 & \cellcolor[HTML]{FFFFFF} $56.62$ & \cellcolor[HTML]{FFFFFF} $20.41\%$ & \cellcolor[HTML]{FFFFFF} $14.37\%$ & \cellcolor[HTML]{FFFFFF} $7.95\%$\\
\photons 21 & \cellcolor[HTML]{FFFFFF} $56.62$ & \cellcolor[HTML]{FFFFFF} $20.41\%$ & \cellcolor[HTML]{FFFFFF} $14.37\%$ & \cellcolor[HTML]{FFFFFF} $7.95\%$\\ \midrule
\vincia MEC & \cellcolor{orange2} $53.39$ & \cellcolor[HTML]{FFFFFF} $20.59\%$ & \cellcolor[HTML]{FFFFFF} $14.41\%$ & \cellcolor[HTML]{FFFFFF} $7.85\%$\\
\vincia & \cellcolor{orange2} $50.36$ & \cellcolor[HTML]{FFFFFF} $20.12\%$ & \cellcolor{orange1} $14.05\%$ & \cellcolor{orange1} $7.59\%$\\
\vincia Dipole & \cellcolor{orange2} $50.36$ & \cellcolor[HTML]{FFFFFF} $20.11\%$ & \cellcolor{orange1} $14.04\%$ & \cellcolor{orange1} $7.59\%$\\ \bottomrule
\end{tabular}
\caption{
Average amounts of photon energy radiated in
 \decay{\Bs}{\mumu} decays, and the fractions of events which radiate a total of more than $1$, $10$, and $100\mev$. Colour scheme defined at the start of \secRef{sec:results}.}
\label{tab:Bs_mumu}
\end{table}

\begin{figure}[tp]
\centering
\begin{subfigure}{0.45\textwidth}
  \centering
  \includegraphics[width=\linewidth]{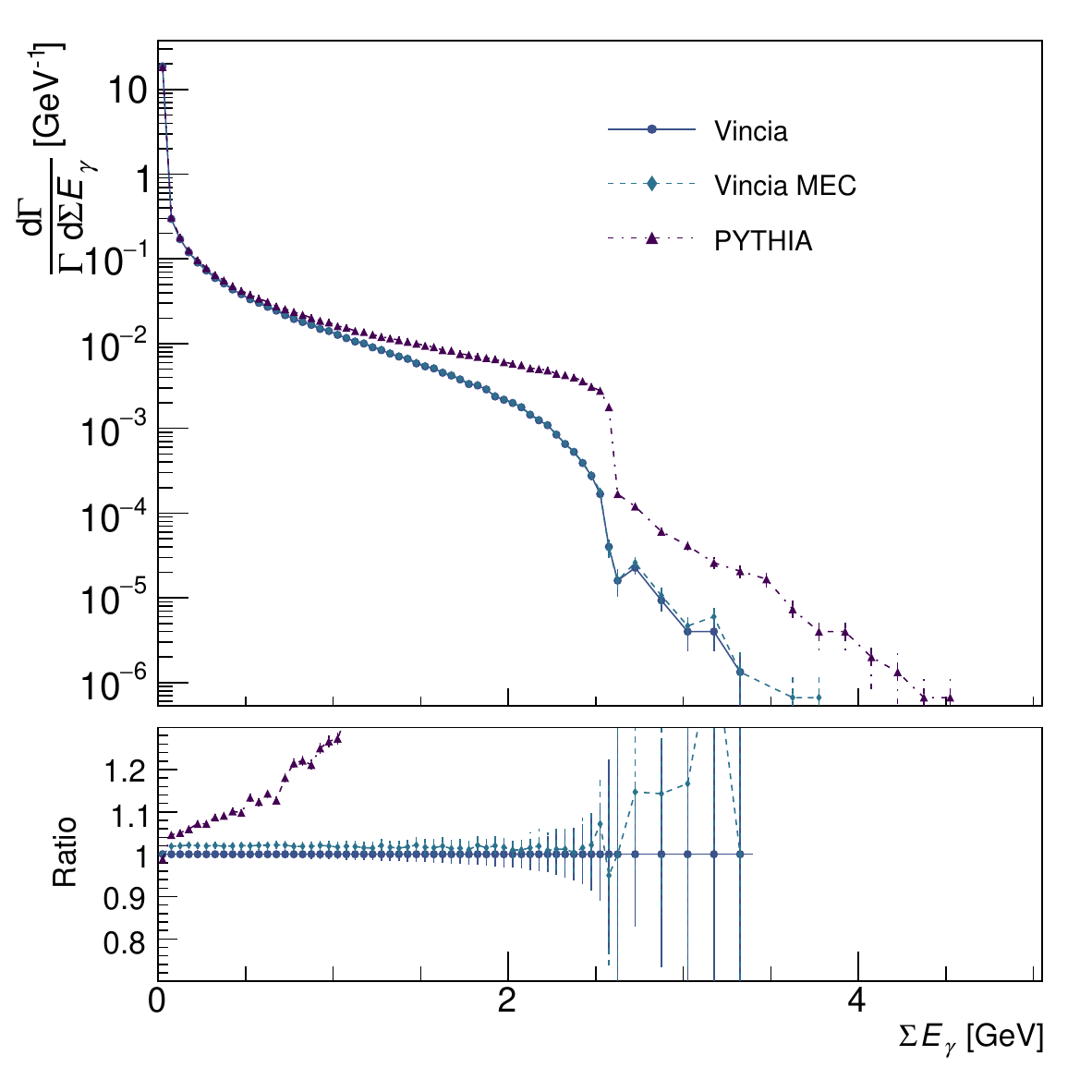}
  \caption{Total photon energy radiated in \mbox{\decay{\Bz}{\Kp\pim}} decays for \pythia, \vincia and \vincia~MEC.\\}
  \label{fig:BdtoKpiVincias}
\end{subfigure}
\hfill
\begin{subfigure}{0.45\textwidth}
  \centering
  \includegraphics[width=\linewidth]{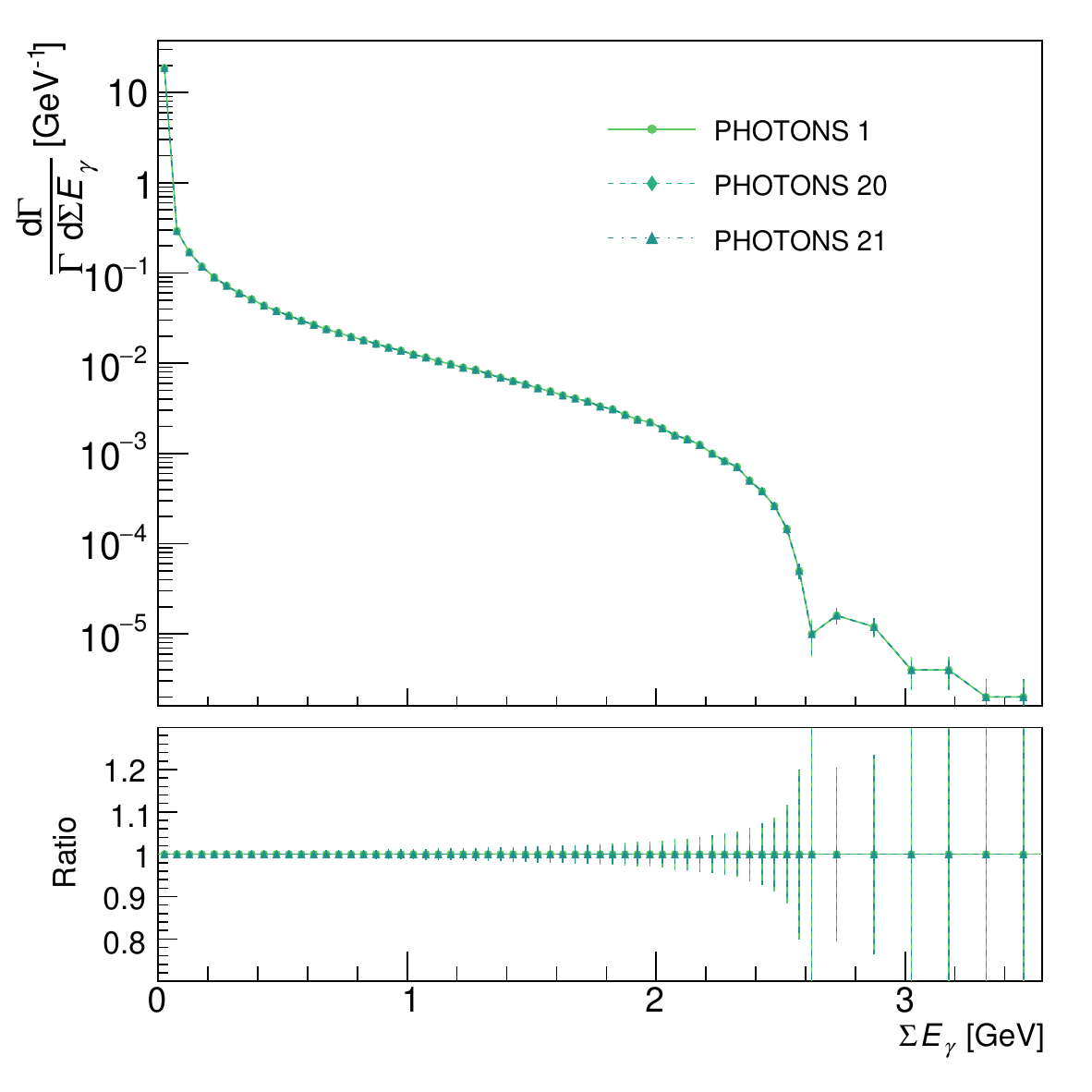}
  \caption{Total photon energy radiated in \mbox{\decay{\Bz}{\Kp\pim}} decays for \photons~1, \photons~20 and \photons~21.\\}
  \label{fig:BdtoKpiSherpas}
\end{subfigure}
\begin{subfigure}{0.45\textwidth}
  \centering
  \includegraphics[width=\linewidth]{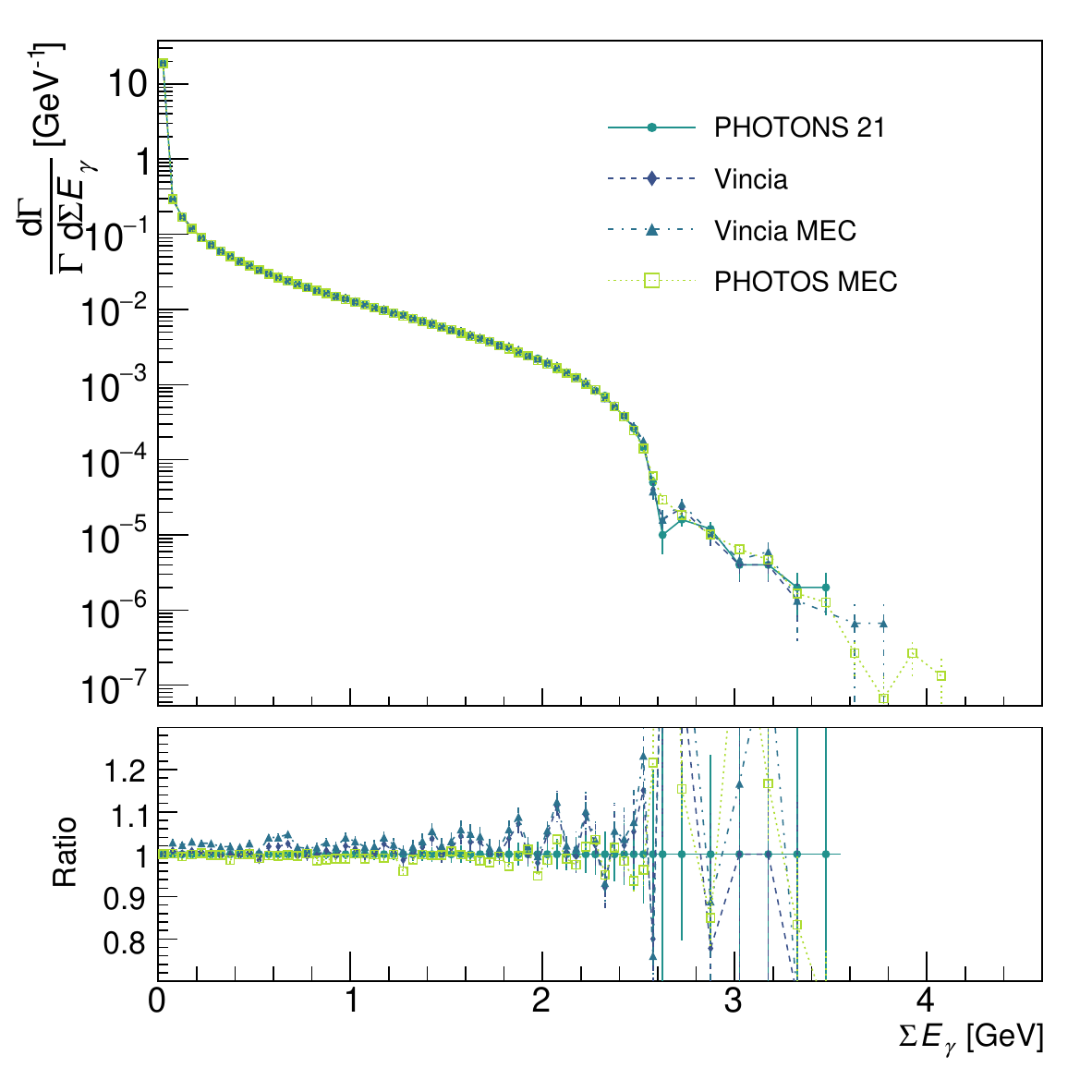}
  \caption{Total photon energy radiated in \mbox{\decay{\Bz}{\Kp\pim}} decays for \photons~21, \vincia, \vincia~MEC and \photos~MEC.}
  \label{fig:BdtoKpiDeltaEComparisons}
\end{subfigure}
\hfill
\begin{subfigure}{0.45\textwidth}
  \centering
  \includegraphics[width=\linewidth]{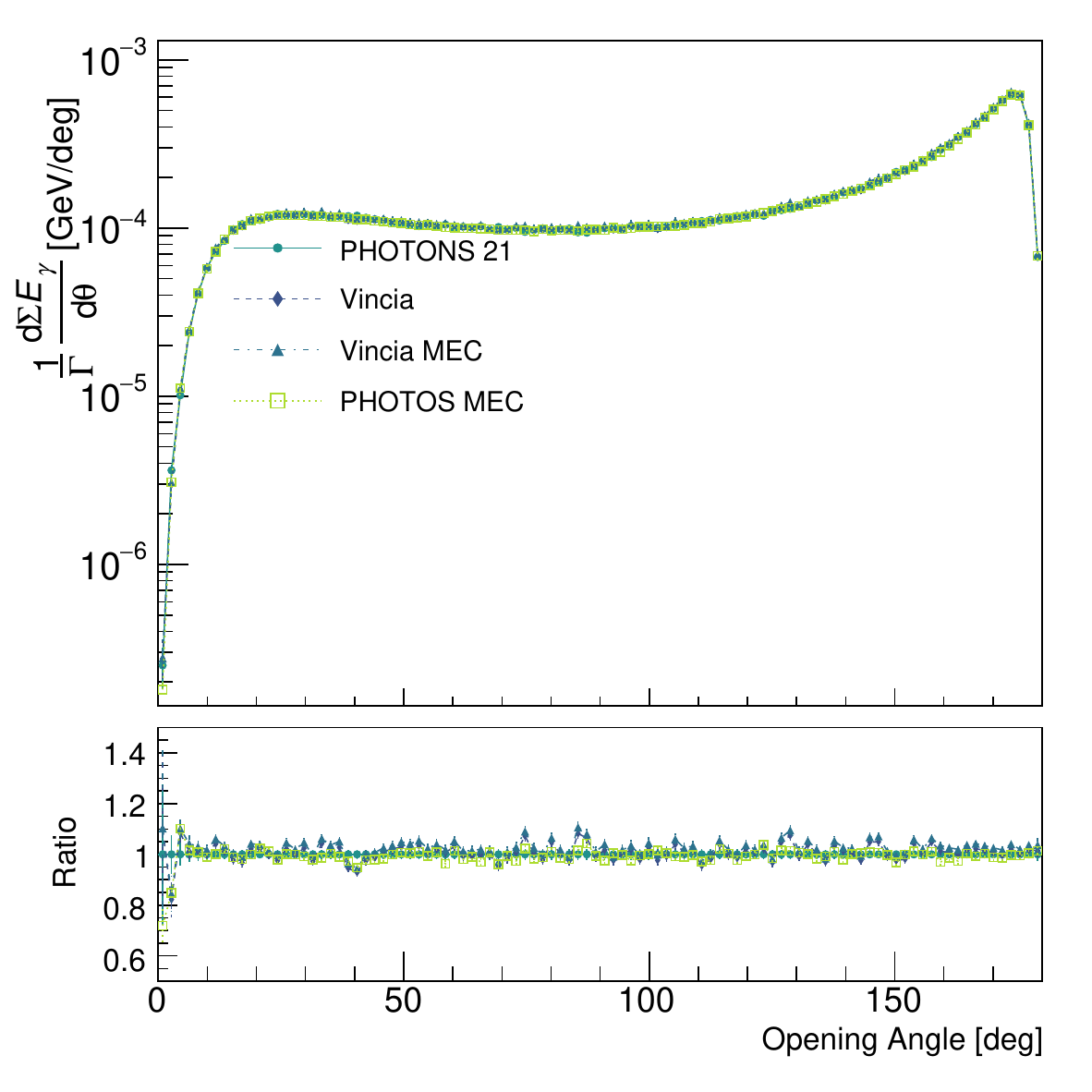}
  \caption{Angular distribution of photon energy in \mbox{\decay{\Bz}{\Kp\pim}} decays for \photons~21, \vincia, \vincia~MEC and \photos~MEC.}
  \label{fig:BdtoKpiAngularEComparisons}
\end{subfigure}
    \caption{Total photon energy radiated and angular distribution of photon energy in \decay{\Bz}{\Kp\pim} decays for various QED generator options. 
    The opening angle is defined as the angle between the emitted photon and the post-branching final-state kaon.}
    \label{fig:BdtoKpiComparisons}
\end{figure}


For purely (pseudo)scalar \decay{\Bz}{\Kp\pim} decays, \figRef{fig:BdtoKpiVincias} shows that the main \pythia shower still treats the kaon and pion as fermions (since it lacks scalar splitting functions). 
This is why, in released \pythia versions, the QED shower is only available for purely leptonic hadron decays. 
As~\figRef{fig:JpsitoEEVincias} shows, \vincia~MEC is around $2\%$ higher than the uncorrected configuration, while having around $0.1\%$ fewer entries in the first bin ($0-50\mev$).
For the three \photons configurations in \figRef{fig:BdtoKpiSherpas}, we used the same random-number seed in each run to show that, for pure (pseudo)scalar decays, switching between these configurations produces no difference.
Comparing the generators in \figRef{fig:BdtoKpiDeltaEComparisons} and \tabRef{tab:Bd_Kpi} shows excellent agreement, with only \vincia~MEC giving slightly higher radiation losses than the others (except \pythia, which, as noted above, is not a realistic model for these decays).
This agreement is reflected in the angular distributions (\figRef{fig:BdtoKpiAngularEComparisons}), where all generators agree very well. 
The visible asymmetry in the dead-cone effect is caused by the difference between the kaon and pion masses.
This pattern repeats overall for the other two-body decays of neutral hadrons considered in this study, summarised in \tabsRef{tab:Bd_DDs}--\ref{tab:D0_pipi}.

\begin{table}[tp]
\centering
\begin{tabular}{l c c c c}
\toprule
$B^0 \to K^+ \pi^-$ & $\langle \sum E_{\gamma}\rangle [\mev]$ & $ f(\sum > 1\mev)$ & $ f(\sum > 10\mev)$ & $ f(\sum > 100\mev)$\\ \midrule 
\photos & \cellcolor[HTML]{FFFFFF} $27.05$ & \cellcolor[HTML]{FFFFFF} $14.51\%$ & \cellcolor[HTML]{FFFFFF} $9.82\%$ & \cellcolor[HTML]{FFFFFF} $4.96\%$\\
\photos MEC & \cellcolor[HTML]{FFFFFF} $27.16$ & \cellcolor[HTML]{FFFFFF} $14.52\%$ & \cellcolor[HTML]{FFFFFF} $9.83\%$ & \cellcolor[HTML]{FFFFFF} $4.97\%$\\ \midrule
\photons 1 & \cellcolor[HTML]{FFFFFF} $27.30$ & \cellcolor[HTML]{FFFFFF} $14.52\%$ & \cellcolor[HTML]{FFFFFF} $9.84\%$ & \cellcolor[HTML]{FFFFFF} $4.99\%$\\
\photons 20 & \cellcolor[HTML]{FFFFFF} $27.30$ & \cellcolor[HTML]{FFFFFF} $14.52\%$ & \cellcolor[HTML]{FFFFFF} $9.84\%$ & \cellcolor[HTML]{FFFFFF} $4.99\%$\\
\photons 21 & \cellcolor[HTML]{FFFFFF} $27.30$ & \cellcolor[HTML]{FFFFFF} $14.52\%$ & \cellcolor[HTML]{FFFFFF} $9.84\%$ & \cellcolor[HTML]{FFFFFF} $4.99\%$\\ \midrule
\vincia MEC & \cellcolor{cyan1} $28.06$ & \cellcolor{cyan1} $14.86\%$ & \cellcolor{cyan1} $10.09\%$ & \cellcolor{cyan1} $5.12\%$\\
\vincia & \cellcolor[HTML]{FFFFFF} $27.54$ & \cellcolor[HTML]{FFFFFF} $14.60\%$ & \cellcolor[HTML]{FFFFFF} $9.91\%$ & \cellcolor[HTML]{FFFFFF} $5.02\%$\\
\vincia Dipole & \cellcolor[HTML]{FFFFFF} $27.54$ & \cellcolor[HTML]{FFFFFF} $14.56\%$ & \cellcolor[HTML]{FFFFFF} $9.89\%$ & \cellcolor[HTML]{FFFFFF} $5.02\%$\\ \midrule
\pythia & \cellcolor{cyan3} $40.19$ & \cellcolor{cyan3} $15.98\%$ & \cellcolor{cyan3} $11.11\%$ & \cellcolor{cyan3} $6.03\%$\\ \bottomrule
\end{tabular} 
\caption{
Average amounts of photon energy radiated in
 \decay{\Bz}{\Kp\pim} decays and the fractions of events which radiate a total of more than $1$, $10$, and $100\mev$. 
 Colour scheme defined at the start of \secRef{sec:results}.}
\label{tab:Bd_Kpi}
\end{table}
\begin{table}[tp]
\centering
\begin{tabular}{l c c c c}
\toprule
$B^0 \to D_s^+D^-$ & $\langle \sum E_{\gamma}\rangle [\mev]$ & $ f(\sum > 1\mev)$ & $ f(\sum > 10\mev)$ & $ f(\sum > 100\mev)$\\ \midrule 
\photos & \cellcolor{orange2} $1.68$ & \cellcolor{orange1} $2.12\%$ & \cellcolor{orange1} $1.28\%$ & \cellcolor{orange2} $0.47\%$\\
\photos MEC & \cellcolor[HTML]{FFFFFF} $1.88$ & \cellcolor[HTML]{FFFFFF} $2.17\%$ & \cellcolor[HTML]{FFFFFF} $1.33\%$ & \cellcolor[HTML]{FFFFFF} $0.51\%$\\ \midrule
\photons 1 & \cellcolor[HTML]{FFFFFF} $1.91$ & \cellcolor[HTML]{FFFFFF} $2.17\%$ & \cellcolor[HTML]{FFFFFF} $1.33\%$ & \cellcolor[HTML]{FFFFFF} $0.52\%$\\
\photons 20 & \cellcolor[HTML]{FFFFFF} $1.91$ & \cellcolor[HTML]{FFFFFF} $2.17\%$ & \cellcolor[HTML]{FFFFFF} $1.33\%$ & \cellcolor[HTML]{FFFFFF} $0.52\%$\\
\photons 21 & \cellcolor[HTML]{FFFFFF} $1.91$ & \cellcolor[HTML]{FFFFFF} $2.17\%$ & \cellcolor[HTML]{FFFFFF} $1.33\%$ & \cellcolor[HTML]{FFFFFF} $0.52\%$\\ \midrule
\vincia MEC & \cellcolor[HTML]{FFFFFF} $1.94$ & \cellcolor{cyan1} $2.27\%$ & \cellcolor{cyan1} $1.37\%$ & \cellcolor{cyan1} $0.53\%$\\
\vincia & \cellcolor[HTML]{FFFFFF} $1.90$ & \cellcolor[HTML]{FFFFFF} $2.18\%$ & \cellcolor[HTML]{FFFFFF} $1.34\%$ & \cellcolor[HTML]{FFFFFF} $0.52\%$\\
\vincia Dipole & \cellcolor[HTML]{FFFFFF} $1.91$ & \cellcolor[HTML]{FFFFFF} $2.18\%$ & \cellcolor[HTML]{FFFFFF} $1.33\%$ & \cellcolor[HTML]{FFFFFF} $0.52\%$\\ \bottomrule
\end{tabular} 
\caption{
Average amounts of photon energy radiated in
 \decay{\Bz}{\Ds\Dm} decays and the fractions of events which radiate a total of more than $1$, $10$, and $100\mev$. 
 Colour scheme defined at the start of \secRef{sec:results}.}
\label{tab:Bd_DDs}
\end{table}


\begin{table}[tp]
\centering
\begin{tabular}{l c c c c}
\toprule
$B^0 \to \pi^+ \pi^-$ & $\langle \sum E_{\gamma}\rangle [\mev]$ & $ f(\sum > 1\mev)$ & $ f(\sum > 10\mev)$ & $ f(\sum > 100\mev)$\\ \midrule 
\photos & \cellcolor[HTML]{FFFFFF} $35.01$ & \cellcolor[HTML]{FFFFFF} $17.95\%$ & \cellcolor[HTML]{FFFFFF} $12.27\%$ & \cellcolor[HTML]{FFFFFF} $6.29\%$\\
\photos MEC & \cellcolor[HTML]{FFFFFF} $35.08$ & \cellcolor[HTML]{FFFFFF} $17.96\%$ & \cellcolor[HTML]{FFFFFF} $12.28\%$ & \cellcolor[HTML]{FFFFFF} $6.29\%$\\ \midrule
\photons 1 & \cellcolor[HTML]{FFFFFF} $35.37$ & \cellcolor[HTML]{FFFFFF} $17.96\%$ & \cellcolor[HTML]{FFFFFF} $12.29\%$ & \cellcolor[HTML]{FFFFFF} $6.32\%$\\
\photons 20 & \cellcolor[HTML]{FFFFFF} $35.37$ & \cellcolor[HTML]{FFFFFF} $17.96\%$ & \cellcolor[HTML]{FFFFFF} $12.29\%$ & \cellcolor[HTML]{FFFFFF} $6.32\%$\\
\photons 21 & \cellcolor[HTML]{FFFFFF} $35.37$ & \cellcolor[HTML]{FFFFFF} $17.96\%$ & \cellcolor[HTML]{FFFFFF} $12.29\%$ & \cellcolor[HTML]{FFFFFF} $6.32\%$\\ \midrule
\vincia MEC & \cellcolor{cyan1} $36.22$ & \cellcolor{cyan1} $18.35\%$ & \cellcolor{cyan1} $12.58\%$ & \cellcolor{cyan1} $6.48\%$\\
\vincia & \cellcolor[HTML]{FFFFFF} $35.55$ & \cellcolor[HTML]{FFFFFF} $18.04\%$ & \cellcolor[HTML]{FFFFFF} $12.36\%$ & \cellcolor[HTML]{FFFFFF} $6.35\%$\\
\vincia Dipole & \cellcolor[HTML]{FFFFFF} $35.57$ & \cellcolor[HTML]{FFFFFF} $18.02\%$ & \cellcolor[HTML]{FFFFFF} $12.35\%$ & \cellcolor[HTML]{FFFFFF} $6.36\%$\\ \bottomrule
\end{tabular}
\caption{
Average amounts of photon energy radiated in
 \decay{\Bz}{\pip\pim} decays and the fractions of events which radiate a total of more than $1$, $10$, and $100\mev$. 
 Colour scheme defined at the start of \secRef{sec:results}.}
\label{tab:Bd_pipi}
\end{table}
\begin{table}[tp]
\centering
\begin{tabular}{l c c c c}
\toprule
$D^0 \to K^+K^-$ & $\langle \sum E_{\gamma}\rangle [\mev]$ & $ f(\sum > 1\mev)$ & $ f(\sum > 10\mev)$ & $ f(\sum > 100\mev)$\\ \midrule 
\photos & \cellcolor{orange2} $1.85$ & \cellcolor[HTML]{FFFFFF} $3.64\%$ & \cellcolor{orange1} $2.06\%$ & \cellcolor{orange2} $0.58\%$\\
\photos MEC & \cellcolor[HTML]{FFFFFF} $1.98$ & \cellcolor[HTML]{FFFFFF} $3.70\%$ & \cellcolor[HTML]{FFFFFF} $2.12\%$ & \cellcolor[HTML]{FFFFFF} $0.63\%$\\ \midrule
\photons 1 & \cellcolor[HTML]{FFFFFF} $1.99$ & \cellcolor[HTML]{FFFFFF} $3.71\%$ & \cellcolor[HTML]{FFFFFF} $2.12\%$ & \cellcolor[HTML]{FFFFFF} $0.63\%$\\
\photons 20 & \cellcolor[HTML]{FFFFFF} $1.99$ & \cellcolor[HTML]{FFFFFF} $3.71\%$ & \cellcolor[HTML]{FFFFFF} $2.12\%$ & \cellcolor[HTML]{FFFFFF} $0.63\%$\\
\photons 21 & \cellcolor[HTML]{FFFFFF} $1.99$ & \cellcolor[HTML]{FFFFFF} $3.71\%$ & \cellcolor[HTML]{FFFFFF} $2.12\%$ & \cellcolor[HTML]{FFFFFF} $0.63\%$\\ \midrule
\vincia MEC & \cellcolor[HTML]{FFFFFF} $2.03$ & \cellcolor[HTML]{FFFFFF} $3.77\%$ & \cellcolor[HTML]{FFFFFF} $2.16\%$ & \cellcolor[HTML]{FFFFFF} $0.64\%$\\
\vincia & \cellcolor[HTML]{FFFFFF} $2.01$ & \cellcolor[HTML]{FFFFFF} $3.72\%$ & \cellcolor[HTML]{FFFFFF} $2.13\%$ & \cellcolor[HTML]{FFFFFF} $0.63\%$\\
\vincia Dipole & \cellcolor[HTML]{FFFFFF} $2.01$ & \cellcolor[HTML]{FFFFFF} $3.72\%$ & \cellcolor[HTML]{FFFFFF} $2.13\%$ & \cellcolor[HTML]{FFFFFF} $0.63\%$\\ \bottomrule
\end{tabular} 
\caption{
Average amounts of photon energy radiated in
 \decay{\Dz}{\Kp\Km} decays and the fractions of events which radiate a total of more than $1$, $10$, and $100\mev$. 
 Colour scheme defined at the start of \secRef{sec:results}.}
\label{tab:D0_KK}
\end{table}


\begin{table}[tp]
\centering
\begin{tabular}{l c c c c}
\toprule
$D^0\to K^+\pi^-$ & $\langle \sum E_{\gamma}\rangle [\mev]$ & $ f(\sum > 1\mev)$ & $ f(\sum > 10\mev)$ & $ f(\sum > 100\mev)$\\ \midrule 
\photos & \cellcolor{orange1} $4.56$ & \cellcolor[HTML]{FFFFFF} $7.03\%$ & \cellcolor[HTML]{FFFFFF} $4.18\%$ & \cellcolor{orange1} $1.40\%$\\
\photos MEC & \cellcolor[HTML]{FFFFFF} $4.68$ & \cellcolor[HTML]{FFFFFF} $7.07\%$ & \cellcolor[HTML]{FFFFFF} $4.21\%$ & \cellcolor[HTML]{FFFFFF} $1.43\%$\\ \midrule
\photons 1 & \cellcolor[HTML]{FFFFFF} $4.70$ & \cellcolor[HTML]{FFFFFF} $7.06\%$ & \cellcolor[HTML]{FFFFFF} $4.21\%$ & \cellcolor[HTML]{FFFFFF} $1.44\%$\\
\photons 20 & \cellcolor[HTML]{FFFFFF} $4.70$ & \cellcolor[HTML]{FFFFFF} $7.06\%$ & \cellcolor[HTML]{FFFFFF} $4.21\%$ & \cellcolor[HTML]{FFFFFF} $1.44\%$\\
\photons 21 & \cellcolor[HTML]{FFFFFF} $4.70$ & \cellcolor[HTML]{FFFFFF} $7.06\%$ & \cellcolor[HTML]{FFFFFF} $4.21\%$ & \cellcolor[HTML]{FFFFFF} $1.44\%$\\ \midrule
\vincia MEC & \cellcolor[HTML]{FFFFFF} $4.79$ & \cellcolor[HTML]{FFFFFF} $7.19\%$ & \cellcolor[HTML]{FFFFFF} $4.29\%$ & \cellcolor{cyan1} $1.47\%$\\
\vincia & \cellcolor[HTML]{FFFFFF} $4.73$ & \cellcolor[HTML]{FFFFFF} $7.10\%$ & \cellcolor[HTML]{FFFFFF} $4.23\%$ & \cellcolor[HTML]{FFFFFF} $1.45\%$\\
\vincia Dipole & \cellcolor[HTML]{FFFFFF} $4.73$ & \cellcolor[HTML]{FFFFFF} $7.10\%$ & \cellcolor[HTML]{FFFFFF} $4.25\%$ & \cellcolor[HTML]{FFFFFF} $1.44\%$\\  \bottomrule
\end{tabular} 
\caption{
Average amounts of photon energy radiated in
 \decay{\Dz}{\Km\pip} decays, and the fractions of events which radiate a total of more than $1$, $10$, and $100\mev$. Colour scheme defined at the start of \secRef{sec:results}.}
\label{tab:D0_Kpi}
\end{table}

\clearpage


\begin{table}[tp]
\centering
\begin{tabular}{l c c c c}
\toprule
$D^0\to\pi^+\pi^-$ & $\langle \sum E_{\gamma}\rangle [\mev]$ & $ f(\sum > 1\mev)$ & $ f(\sum > 10\mev)$ & $ f(\sum > 100\mev)$\\ \midrule 
\photos & \cellcolor[HTML]{FFFFFF} $7.76$ & \cellcolor[HTML]{FFFFFF} $10.43\%$ & \cellcolor[HTML]{FFFFFF} $6.36\%$ & \cellcolor[HTML]{FFFFFF} $2.30\%$\\
\photos MEC & \cellcolor[HTML]{FFFFFF} $7.82$ & \cellcolor[HTML]{FFFFFF} $10.45\%$ & \cellcolor[HTML]{FFFFFF} $6.38\%$ & \cellcolor[HTML]{FFFFFF} $2.32\%$\\ \midrule
\photons 1 & \cellcolor[HTML]{FFFFFF} $7.85$ & \cellcolor[HTML]{FFFFFF} $10.44\%$ & \cellcolor[HTML]{FFFFFF} $6.38\%$ & \cellcolor[HTML]{FFFFFF} $2.33\%$\\
\photons 20 & \cellcolor[HTML]{FFFFFF} $7.85$ & \cellcolor[HTML]{FFFFFF} $10.44\%$ & \cellcolor[HTML]{FFFFFF} $6.38\%$ & \cellcolor[HTML]{FFFFFF} $2.33\%$\\
\photons 21 & \cellcolor[HTML]{FFFFFF} $7.85$ & \cellcolor[HTML]{FFFFFF} $10.44\%$ & \cellcolor[HTML]{FFFFFF} $6.38\%$ & \cellcolor[HTML]{FFFFFF} $2.33\%$\\ \midrule
\vincia MEC & \cellcolor{cyan1} $8.02$ & \cellcolor[HTML]{FFFFFF} $10.62\%$ & \cellcolor[HTML]{FFFFFF} $6.50\%$ & \cellcolor{cyan1} $2.38\%$\\
\vincia & \cellcolor[HTML]{FFFFFF} $7.92$ & \cellcolor[HTML]{FFFFFF} $10.49\%$ & \cellcolor[HTML]{FFFFFF} $6.41\%$ & \cellcolor[HTML]{FFFFFF} $2.35\%$\\
\vincia Dipole & \cellcolor[HTML]{FFFFFF} $7.90$ & \cellcolor[HTML]{FFFFFF} $10.48\%$ & \cellcolor[HTML]{FFFFFF} $6.41\%$ & \cellcolor[HTML]{FFFFFF} $2.34\%$\\ \bottomrule
\end{tabular}
\caption{
Average amounts of photon energy radiated in
\decay{\Dz}{\pip\pim}  decays, and the fractions of events which radiate a total of more than $1$, $10$, and $100\mev$. Colour scheme defined at the start of \secRef{sec:results}.}
\label{tab:D0_pipi}
\end{table}

\subsubsection{Two-body decays of charged hadrons}

In \vincia, the description of radiative corrections in decays of charged initial states is based on its unique ``resonance-final'' antennae, which were initially developed for top-quark decays in~\cite{Brooks:2019xso} and generalised to QED in~\cite{Skands:2020lkd}. In most other generators we are aware of, radiation in decays of charged initial states is treated by considering the initial hadron to be at rest and hence non-radiating, and then including explicit radiation functions only for the final-state decay products. In singular limits, these treatments should be equivalent.

\begin{figure}[tp]
\centering
\begin{subfigure}{0.45\textwidth}
  \centering
  \includegraphics[width=\linewidth]{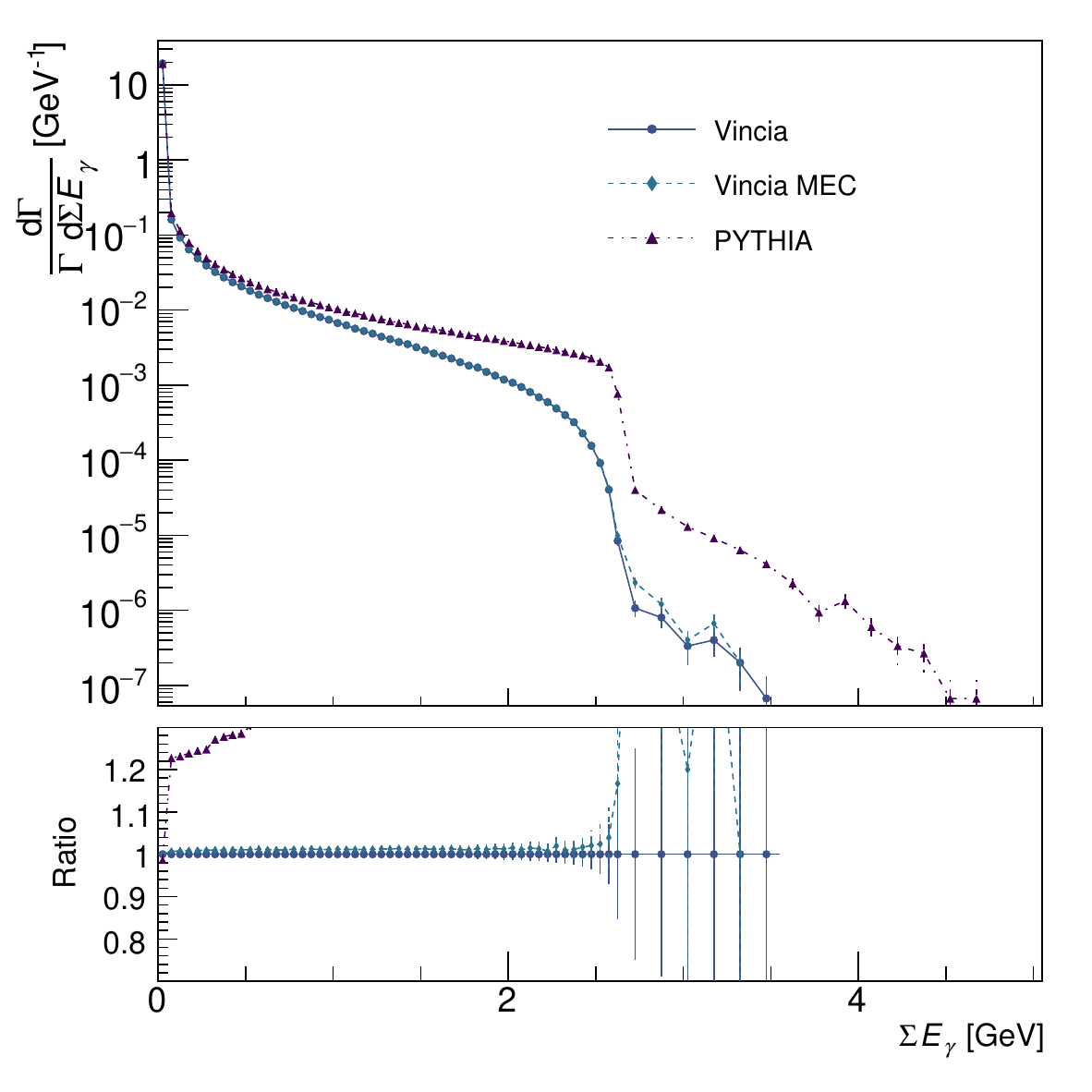}
  \caption{Total photon energy radiated in \mbox{\decay{\Bp}{\pip\piz}} decays for \pythia, \vincia, and \vincia~MEC.\\}
  \label{fig:Butopipi0Vincias}
\end{subfigure}
\hfill
\begin{subfigure}{0.45\textwidth}
  \centering
  \includegraphics[width=\linewidth]{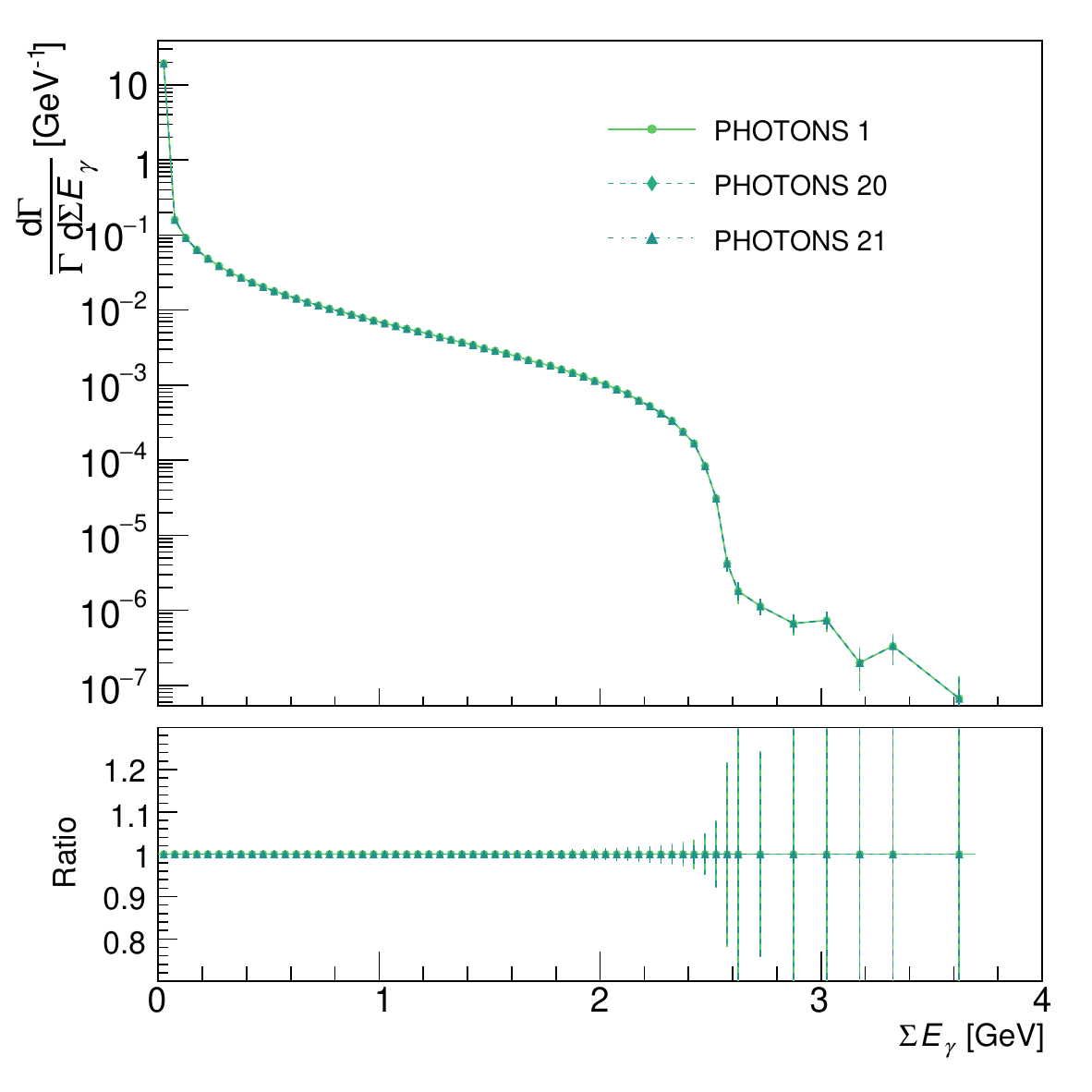}
  \caption{Total photon energy radiated in \mbox{\decay{\Bp}{\pip\piz}} decays for \photons~1, \photons~20, and \photons~21.\\}
  \label{fig:Butopipi0Sherpas}
\end{subfigure}
\begin{subfigure}{0.45\textwidth}
  \centering
  \includegraphics[width=\linewidth]{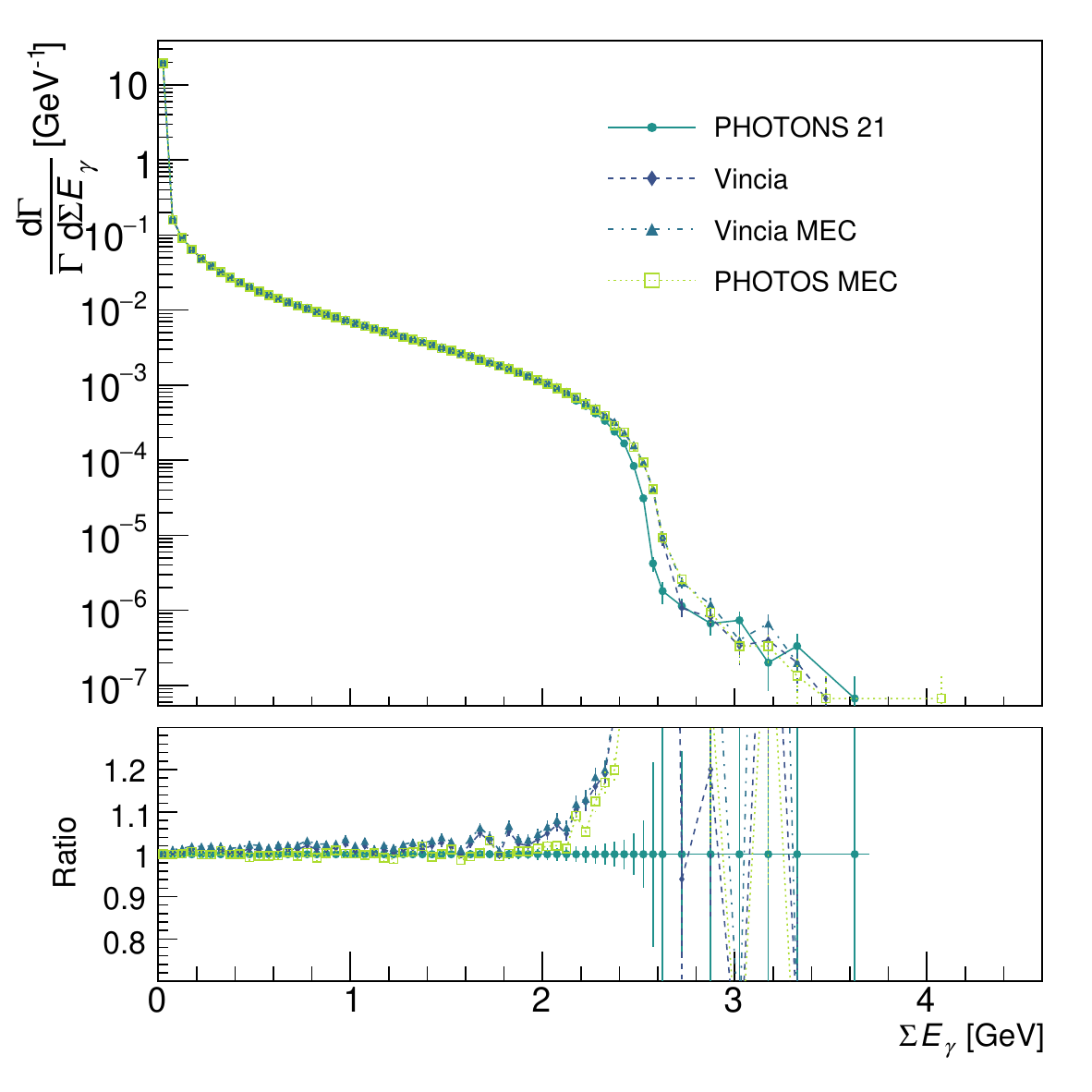}
  \caption{Total photon energy radiated in \mbox{\decay{\Bp}{\pip\piz}} decays for \photons~21, \vincia, \vincia~MEC and \photos~MEC.}
  \label{fig:Butopipi0DeltaEComparisons}
\end{subfigure}
\hfill
\begin{subfigure}{0.45\textwidth}
  \centering
  \includegraphics[width=\linewidth]{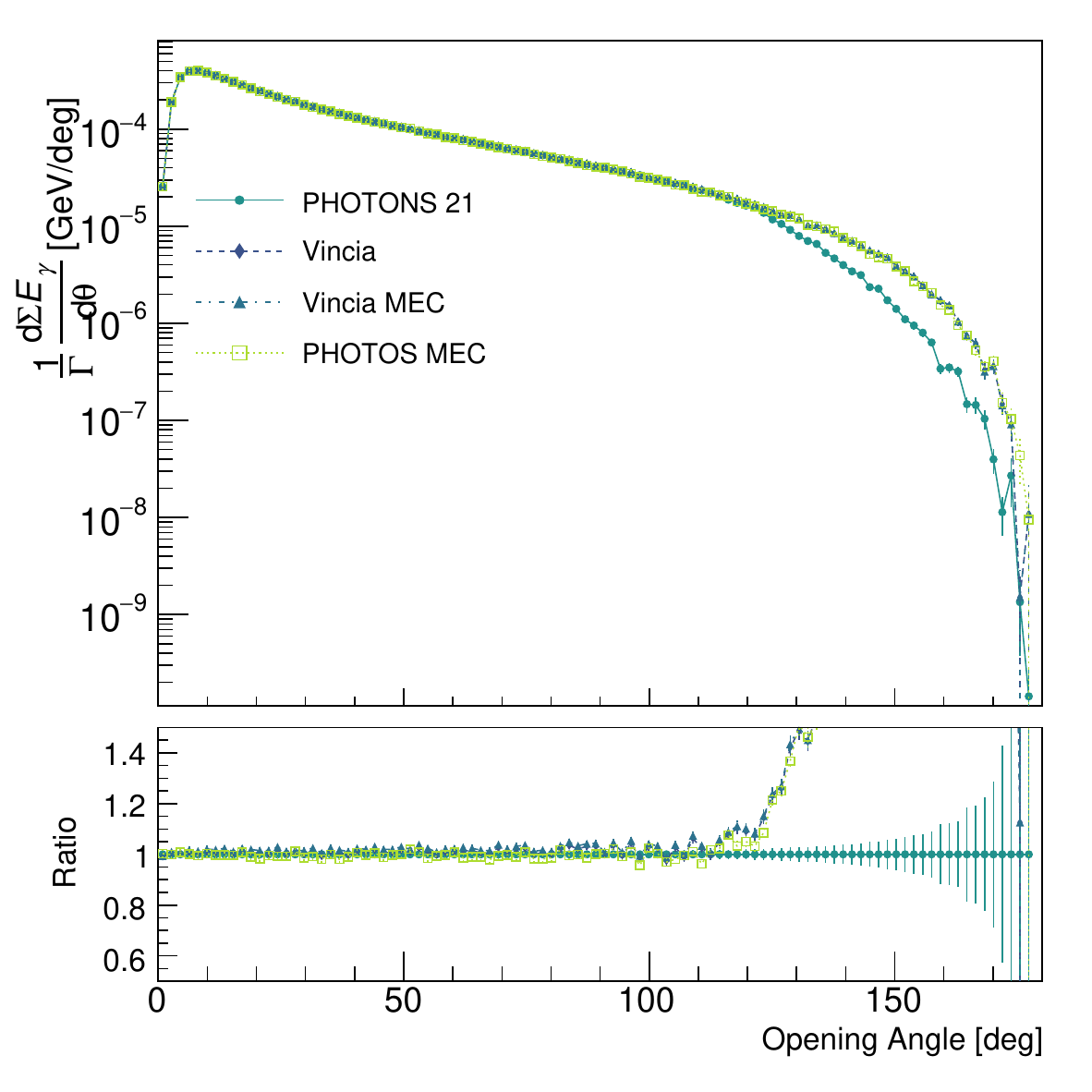}
  \caption{Angular distribution of photon energy in \mbox{\decay{\Bp}{\pip\piz}} decays for \photons~21, \vincia, \vincia~MEC and \photos~MEC.}
  \label{fig:Butopipi0AngularEComparisons}
\end{subfigure}
    \caption{Total photon energy radiated and angular distribution
of photon energy in \decay{\Bp}{\pip\piz} decays for various QED generator options. 
The opening angle is defined as the angle between the emitted photon and the post-branching final-state charged pion.}
    \label{fig:Butopipi0Comparisons}
\end{figure}

Taking \decay{\Bp}{\pip\piz} as our main example, \figRef{fig:Butopipi0Vincias} shows that the two \vincia configurations agree well, while \pythia's fermion-radiator treatment yields a significantly harder radiation spectrum.
As in the case of neutral decays, the ME corrected antennae nearly overlap with the uncorrected antennae. 
In \figRef{fig:Butopipi0Sherpas}, we again used the same random-number seeds to show that the \photons distributions remain unchanged in the three configurations.
Comparing the three generators in \figRef{fig:Butopipi0DeltaEComparisons}, we find good agreement up to the multi-emission tail near $2\gev$, where \photos~MEC and \vincia radiate more than \photons. 
This is consistent with \tabRef{tab:Bu_pipi0}: only \vincia~MEC shows a slightly higher total energy loss, about $400\kev$ more than \photons~21.
In the angular distribution~(\figRef{fig:Butopipi0AngularEComparisons}), the radiation pattern is now highly asymmetric since only one of the decay products is charged. 
The agreement between the generators is very strong in the small-angle region, with \photons~21 emitting less in the wide-angle region than the other generators. 

\noindent

In \tabsRef{tab:B+_Kpi0} and \ref{tab:D+_pipi0}, we summarise the radiation losses for \decay{\Bpm}{\Kpm\piz} and \decay{\Dpm}{\pipm\piz}, respectively, with the same consistent pattern of 
\vincia~MEC radiating slightly more energy on average than the other generators.

\begin{table}[tp]
\centering
\begin{tabular}{l c c c c}
\toprule
$B^+ \to \pi^+\pi^0$ & $\langle \sum E_{\gamma}\rangle [\mev]$ & $ f(\sum > 1\mev)$ & $ f(\sum > 10\mev)$ & $ f(\sum > 100\mev)$\\ \midrule 
\photos & \cellcolor[HTML]{FFFFFF} $14.57$ & \cellcolor[HTML]{FFFFFF} $7.96\%$ & \cellcolor[HTML]{FFFFFF} $5.33\%$ & \cellcolor[HTML]{FFFFFF} $2.67\%$\\
\photos MEC & \cellcolor[HTML]{FFFFFF} $14.60$ & \cellcolor[HTML]{FFFFFF} $7.96\%$ & \cellcolor[HTML]{FFFFFF} $5.33\%$ & \cellcolor[HTML]{FFFFFF} $2.67\%$\\ \midrule
\photons 1 & \cellcolor[HTML]{FFFFFF} $14.53$ & \cellcolor[HTML]{FFFFFF} $7.96\%$ & \cellcolor[HTML]{FFFFFF} $5.33\%$ & \cellcolor[HTML]{FFFFFF} $2.66\%$\\
\photons 20 & \cellcolor[HTML]{FFFFFF} $14.53$ & \cellcolor[HTML]{FFFFFF} $7.96\%$ & \cellcolor[HTML]{FFFFFF} $5.33\%$ & \cellcolor[HTML]{FFFFFF} $2.66\%$\\
\photons 21 & \cellcolor[HTML]{FFFFFF} $14.53$ & \cellcolor[HTML]{FFFFFF} $7.96\%$ & \cellcolor[HTML]{FFFFFF} $5.33\%$ & \cellcolor[HTML]{FFFFFF} $2.66\%$\\ \midrule
\vincia MEC & \cellcolor{cyan1} $14.94$ & \cellcolor[HTML]{FFFFFF} $8.05\%$ & \cellcolor[HTML]{FFFFFF} $5.41\%$ & \cellcolor{cyan1} $2.72\%$\\
\vincia & \cellcolor[HTML]{FFFFFF} $14.78$ & \cellcolor[HTML]{FFFFFF} $8.00\%$ & \cellcolor[HTML]{FFFFFF} $5.37\%$ & \cellcolor[HTML]{FFFFFF} $2.69\%$\\
\vincia Dipole & \cellcolor[HTML]{FFFFFF} $14.78$ & \cellcolor[HTML]{FFFFFF} $8.00\%$ & \cellcolor[HTML]{FFFFFF} $5.37\%$ & \cellcolor[HTML]{FFFFFF} $2.69\%$\\  \bottomrule
\end{tabular} 
\caption{
Average amounts of photon energy radiated in
 \decay{\Bp}{\pip\piz} decays and the fractions of events which radiate a total of more than $1$, $10$, and $100\mev$. 
 Colour scheme defined at the start of \secRef{sec:results}.}
\label{tab:Bu_pipi0}
\end{table}
\begin{table}[tp]
\centering
\begin{tabular}{l c c c c}
\toprule
$B^+\to K^+\pi^0$ & $\langle \sum E_{\gamma}\rangle [\mev]$ & $ f(\sum > 1\mev)$ & $ f(\sum > 10\mev)$ & $ f(\sum > 100\mev)$\\ \midrule 
\photos & \cellcolor[HTML]{FFFFFF} $7.15$ & \cellcolor[HTML]{FFFFFF} $4.28\%$ & \cellcolor[HTML]{FFFFFF} $2.82\%$ & \cellcolor[HTML]{FFFFFF} $1.37\%$\\
\photos MEC & \cellcolor[HTML]{FFFFFF} $7.17$ & \cellcolor[HTML]{FFFFFF} $4.28\%$ & \cellcolor[HTML]{FFFFFF} $2.82\%$ & \cellcolor[HTML]{FFFFFF} $1.37\%$\\ \midrule
\photons 1 & \cellcolor[HTML]{FFFFFF} $7.11$ & \cellcolor[HTML]{FFFFFF} $4.27\%$ & \cellcolor[HTML]{FFFFFF} $2.82\%$ & \cellcolor[HTML]{FFFFFF} $1.37\%$\\
\photons 20 & \cellcolor[HTML]{FFFFFF} $7.11$ & \cellcolor[HTML]{FFFFFF} $4.27\%$ & \cellcolor[HTML]{FFFFFF} $2.82\%$ & \cellcolor[HTML]{FFFFFF} $1.37\%$\\
\photons 21 & \cellcolor[HTML]{FFFFFF} $7.11$ & \cellcolor[HTML]{FFFFFF} $4.27\%$ & \cellcolor[HTML]{FFFFFF} $2.82\%$ & \cellcolor[HTML]{FFFFFF} $1.37\%$\\ \midrule
\vincia MEC & \cellcolor{cyan1} $7.42$ & \cellcolor{cyan1} $4.35\%$ & \cellcolor{cyan1} $2.88\%$ & \cellcolor{cyan1} $1.41\%$\\
\vincia & \cellcolor{cyan1} $7.33$ & \cellcolor[HTML]{FFFFFF} $4.31\%$ & \cellcolor[HTML]{FFFFFF} $2.86\%$ & \cellcolor[HTML]{FFFFFF} $1.40\%$\\
\vincia Dipole & \cellcolor{cyan1} $7.33$ & \cellcolor[HTML]{FFFFFF} $4.29\%$ & \cellcolor[HTML]{FFFFFF} $2.85\%$ & \cellcolor[HTML]{FFFFFF} $1.39\%$\\ \bottomrule
\end{tabular}
\caption{
Average amounts of photon energy radiated in
\decay{\Bp}{\Kp\piz} decays and the fractions of events which radiate a total of more than $1$, $10$, and $100\mev$. 
Colour scheme defined at the start of \secRef{sec:results}.}
\label{tab:B+_Kpi0}
\end{table}
\begin{table}[tp]
\centering
\begin{tabular}{l c c c c}
\toprule
$D^+ \to \pi^+\pi^0$ & $\langle \sum E_{\gamma}\rangle [\mev]$ & $ f(\sum > 1\mev)$ & $ f(\sum > 10\mev)$ & $ f(\sum > 100\mev)$\\ \midrule 
\photos & \cellcolor[HTML]{FFFFFF} $2.95$ & \cellcolor[HTML]{FFFFFF} $4.16\%$ & \cellcolor[HTML]{FFFFFF} $2.49\%$ & \cellcolor[HTML]{FFFFFF} $0.88\%$\\
\photos MEC & \cellcolor[HTML]{FFFFFF} $2.98$ & \cellcolor[HTML]{FFFFFF} $4.16\%$ & \cellcolor[HTML]{FFFFFF} $2.50\%$ & \cellcolor[HTML]{FFFFFF} $0.89\%$\\ \midrule
\photons 1 & \cellcolor[HTML]{FFFFFF} $2.94$ & \cellcolor[HTML]{FFFFFF} $4.16\%$ & \cellcolor[HTML]{FFFFFF} $2.49\%$ & \cellcolor[HTML]{FFFFFF} $0.88\%$\\
\photons 20 & \cellcolor[HTML]{FFFFFF} $2.94$ & \cellcolor[HTML]{FFFFFF} $4.16\%$ & \cellcolor[HTML]{FFFFFF} $2.49\%$ & \cellcolor[HTML]{FFFFFF} $0.88\%$\\
\photons 21 & \cellcolor[HTML]{FFFFFF} $2.94$ & \cellcolor[HTML]{FFFFFF} $4.16\%$ & \cellcolor[HTML]{FFFFFF} $2.49\%$ & \cellcolor[HTML]{FFFFFF} $0.88\%$\\ \midrule
\vincia MEC & \cellcolor{cyan1} $3.03$ & \cellcolor[HTML]{FFFFFF} $4.20\%$ & \cellcolor[HTML]{FFFFFF} $2.53\%$ & \cellcolor{cyan1} $0.90\%$\\
\vincia & \cellcolor{cyan1} $3.00$ & \cellcolor[HTML]{FFFFFF} $4.18\%$ & \cellcolor[HTML]{FFFFFF} $2.51\%$ & \cellcolor[HTML]{FFFFFF} $0.89\%$\\
\vincia Dipole & \cellcolor{cyan1} $3.00$ & \cellcolor[HTML]{FFFFFF} $4.18\%$ & \cellcolor[HTML]{FFFFFF} $2.51\%$ & \cellcolor[HTML]{FFFFFF} $0.89\%$\\ \bottomrule
\end{tabular} 
\caption{
Average amounts of photon energy radiated in
 \decay{\Dp}{\pip\piz} decays and the fractions of events which radiate a total of more than $1$, $10$, and $100\mev$. 
 Colour scheme defined at the start of \secRef{sec:results}.}
\label{tab:D+_pipi0}
\end{table}

\clearpage

\subsection{Three-body decays}
\label{sec:3body}

Three-body decays, and many-body decays generally, differ greatly from two-body decays because the configuration of radiating charges is not fixed as a single dipole of fixed invariant mass. 
Instead, they allow multiple orientations and relative energies of the decay products and possibly multiple charged antennae. 
We consider three examples of three-body decay topologies:
\begin{itemize}
    \item Neutral particle decays to two charged and one neutral particle, e.g. \decay{\Bz}{\Dz\Kp\pim};
    \item Charged particle decays to one charged and two neutral particles, e.g. \decay{\Bp}{\pip\piz\piz} and \decay{\Dp}{\pip\piz\piz};
    \item Charged particle decays to three charged particles, e.g. \decay{\Bp}{\Dp\Km\pip}, \decay{\Dp}{\Kp\pip\pim}, and \decay{\Dp}{\pip\pim\pip}.
\end{itemize}
Only the last case should exhibit genuine multipole effects beyond dipole terms.

In this study, QED radiation is handled by treating Born-level decays as point-like and direct, without any special treatment of sequential decays via intermediate resonances (e.g., $\rho$, $K^*$, $D^*$); 
these appear only as resonances in the Born-level Dalitz distributions generated by \evtgen. 
A more careful treatment of QED radiation sensitive to the presence or absence of such intermediate resonances will be the subject of a follow-up study. 

\subsubsection{\decay{\Bz}{\Dz\Kp\pim} decays}

\begin{figure}[tp]
\centering
\begin{subfigure}{0.45\textwidth}
  \centering
  \includegraphics[width=\linewidth]{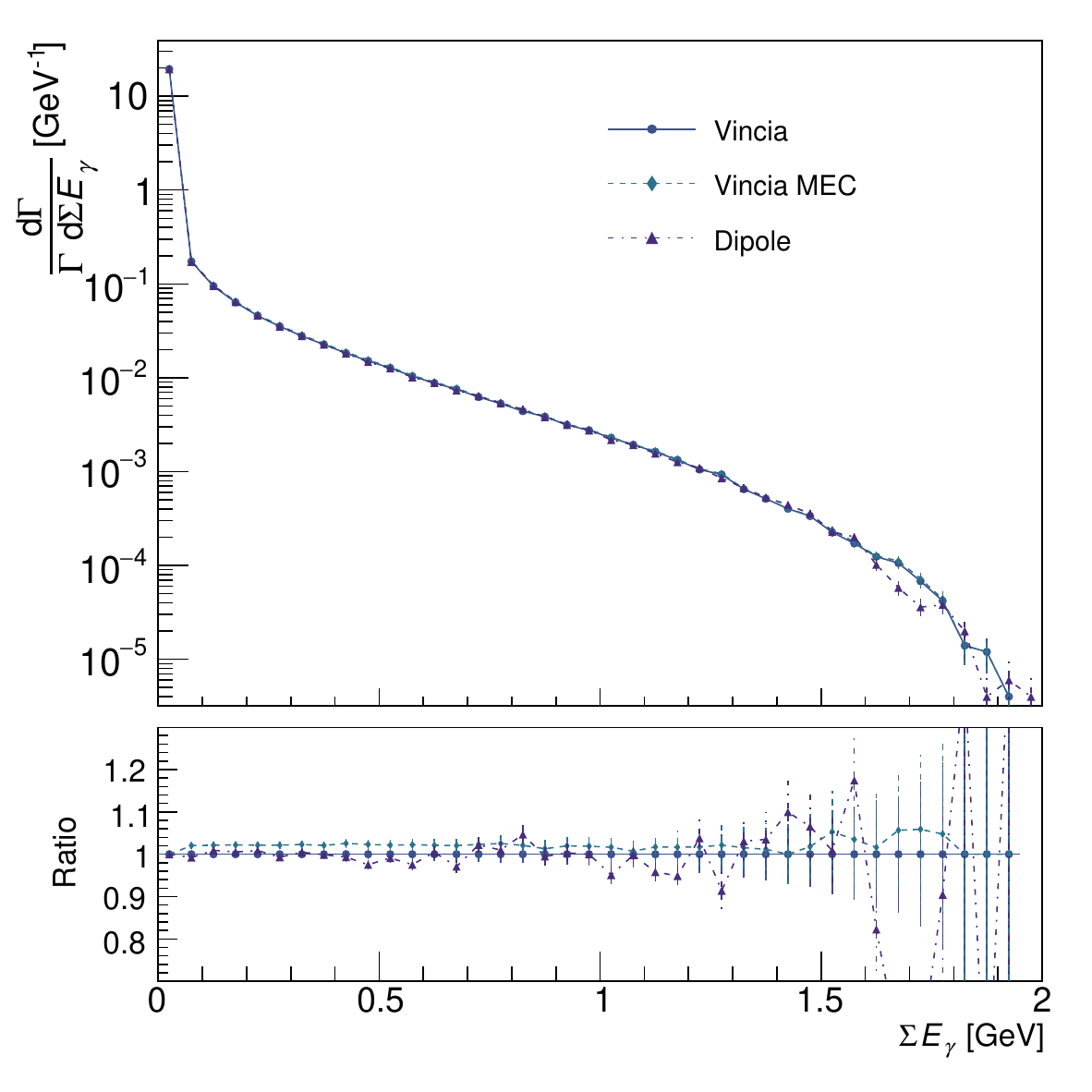}
  \caption{Total photon energy radiated in \mbox{\decay{\Bz}{\Dz\Kp\pim}} decays for \vincia, \vincia~MEC and \vincia's dipole approximation.}
  \label{fig:BdtoD0KpiVincias}
\end{subfigure}
\hfill
\begin{subfigure}{0.45\textwidth}
  \centering
  \includegraphics[width=\linewidth]{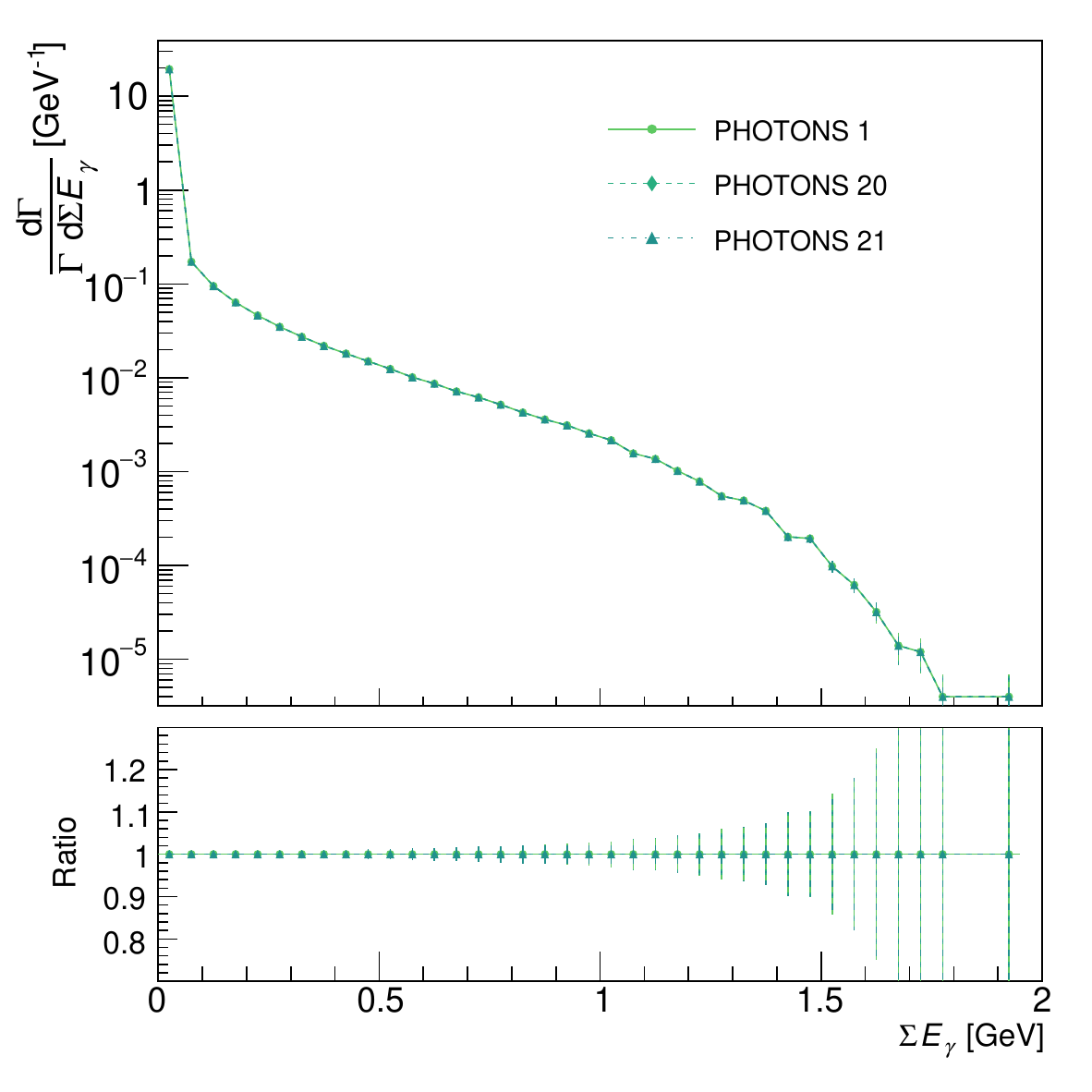}
  \caption{Total photon energy radiated in \mbox{\decay{\Bz}{\Dz\Kp\pim}} decays for \photons~1, \photons~20 and \photons~21.}
  \label{fig:BdtoD0KpiSherpas}
\end{subfigure}
\begin{subfigure}{0.60\textwidth}
  \centering
  \includegraphics[width=0.75\linewidth]{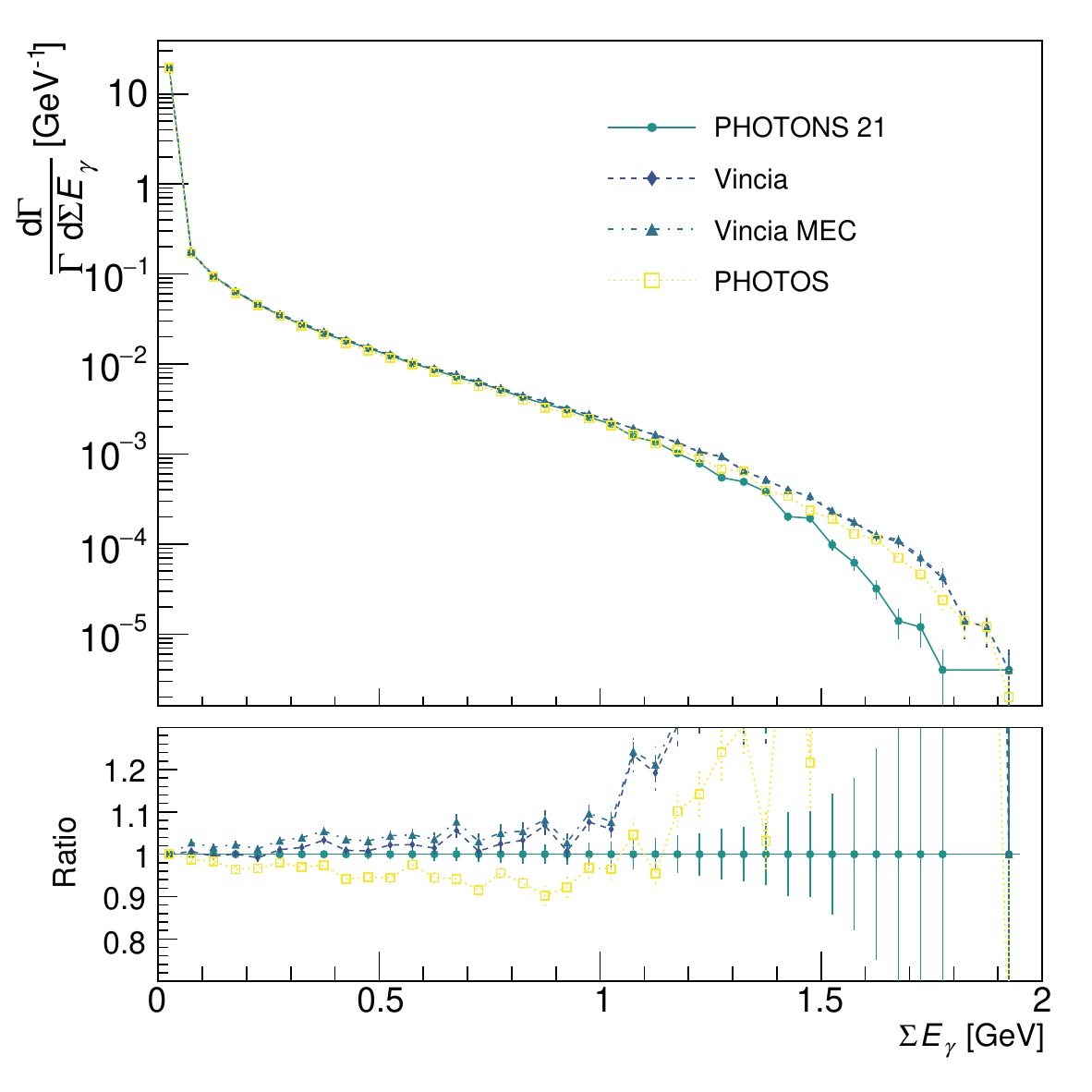}
  \caption{Total photon energy radiated in \mbox{\decay{\Bz}{\Dz\Kp\pim}} decays for \photons~21, \vincia, \vincia~MEC and \photos~MEC.}
  \label{fig:BdtoD0KpiDeltaEComparisons}
\end{subfigure}
\hfill
    \caption{Total photon energy radiated in \decay{\Bz}{\Dz\Kp\pim} decays for various QED generator options.}
    \label{fig:BdtoD0KpiComparisons}
\end{figure}

\FigRef{fig:BdtoD0KpiComparisons} shows the comparisons for \decay{\Bz}{\Dz\Kp\pim} decays.
\FigRef{fig:BdtoD0KpiVincias} shows that \vincia~MEC lies slightly above the uncorrected shower, as in the two-body decays. 
The dipole approximation matches the uncorrected shower within statistical fluctuations, as expected for a decay with only one antenna. 
In~\figRef{fig:BdtoD0KpiSherpas}, all \photons configurations yield again identical plots as in the fully scalar cases. 
Comparing all generators in \figRef{fig:BdtoD0KpiDeltaEComparisons} shows, unlike in the \decay{\Bz}{\Kp\pim} case~(\figRef{fig:BdtoKpiComparisons}), that \vincia radiates more than \photons beyond the first few bins. 

Comparing the summary table for \decay{\Bz}{\Dz\Kp\pim} decays~(\tabRef{tab:Bd_D0Kpi}) with that for \mbox{\decay{\Bz}{\Kp\pim}} decays~(\tabRef{tab:Bd_Kpi}), the average radiated energy differences shift up by one colour band for all \vincia modes, while \photos shifts down by one. 
For the fractions of events with total radiation losses above $1$, $10$, and $100\mev$, the relations between the generators remain nearly the same. 
This similarity to the two-body case is expected: a single fundamental dipole still radiates, now with an invariant mass distributed by the three-body decay into a smaller phase space.

\begin{table}[bp]
\centering
\begin{tabular}{l c c c c}
\toprule
$B^0 \to D^0K^+\pi^-$ & $\langle \sum E_{\gamma}\rangle [\mev]$ & $ f(\sum > 1\mev)$ & $ f(\sum > 10\mev)$ & $ f(\sum > 100\mev)$\\ \midrule 
\photos & \cellcolor{orange1} $7.48$ & \cellcolor[HTML]{FFFFFF} $7.83\%$ & \cellcolor[HTML]{FFFFFF} $4.84\%$ & \cellcolor{orange1} $1.91\%$\\
\photos MEC & \cellcolor{cyan1} $7.85$ & \cellcolor[HTML]{FFFFFF} $7.91\%$ & \cellcolor[HTML]{FFFFFF} $4.93\%$ & \cellcolor[HTML]{FFFFFF} $1.98\%$\\ \midrule
\photons 1 & \cellcolor[HTML]{FFFFFF} $7.67$ & \cellcolor[HTML]{FFFFFF} $7.92\%$ & \cellcolor[HTML]{FFFFFF} $4.93\%$ & \cellcolor[HTML]{FFFFFF} $1.97\%$\\
\photons 20 & \cellcolor[HTML]{FFFFFF} $7.67$ & \cellcolor[HTML]{FFFFFF} $7.92\%$ & \cellcolor[HTML]{FFFFFF} $4.93\%$ & \cellcolor[HTML]{FFFFFF} $1.97\%$\\
\photons 21 & \cellcolor[HTML]{FFFFFF} $7.67$ & \cellcolor[HTML]{FFFFFF} $7.92\%$ & \cellcolor[HTML]{FFFFFF} $4.93\%$ & \cellcolor[HTML]{FFFFFF} $1.97\%$\\ \midrule
\vincia MEC & \cellcolor{cyan2} $8.12$ & \cellcolor{cyan1} $8.10\%$ & \cellcolor{cyan1} $5.07\%$ & \cellcolor{cyan1} $2.04\%$\\
\vincia & \cellcolor{cyan1} $7.95$ & \cellcolor[HTML]{FFFFFF} $7.95\%$ & \cellcolor[HTML]{FFFFFF} $4.96\%$ & \cellcolor[HTML]{FFFFFF} $2.00\%$\\
\vincia Dipole & \cellcolor{cyan1} $7.93$ & \cellcolor[HTML]{FFFFFF} $7.95\%$ & \cellcolor[HTML]{FFFFFF} $4.96\%$ & \cellcolor[HTML]{FFFFFF} $2.00\%$\\ \bottomrule
\end{tabular}
\caption{
Average amounts of photon energy radiated in
\decay{\Bz}{\Dz\Kp\pim} decays, and the fractions of events which radiate a total of more than $1$, $10$, and $100\mev$. Colour scheme defined at the start of \secRef{sec:results}.}
\label{tab:Bd_D0Kpi}
\end{table}

\clearpage

\subsubsection{\decay{\Bp}{\pip\piz\piz} and \decay{\Dz}{\pip\piz\piz} decays}

\begin{figure}[tp]
\centering
\begin{subfigure}{0.45\textwidth}
  \centering
  \includegraphics[width=\linewidth]{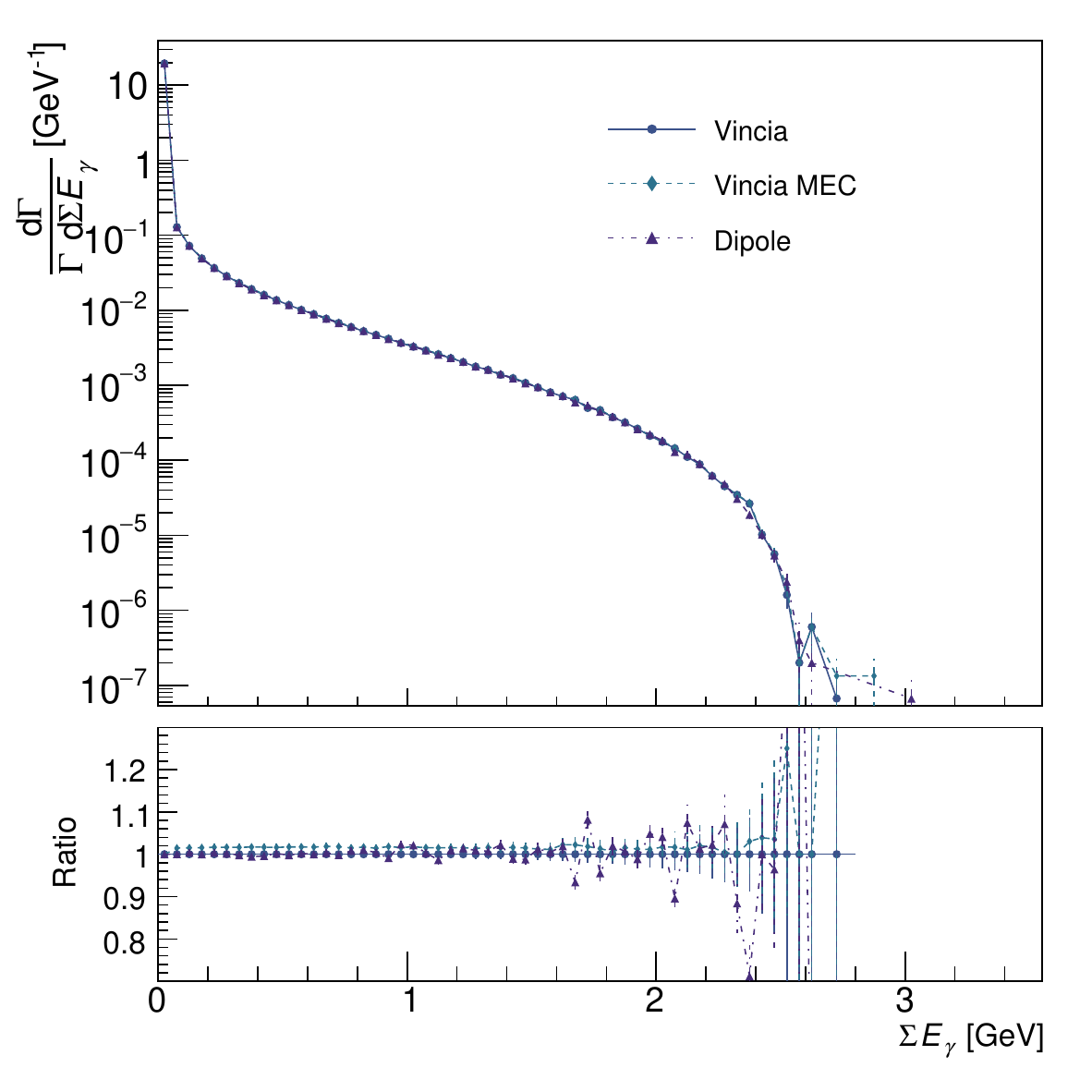}
  \caption{Total photon energy radiated in \mbox{\decay{\Bp}{\pip\piz\piz}} decays for \vincia, \vincia~MEC and \vincia's dipole approximation. }
  \label{fig:Butopi0pi0piVincias}
\end{subfigure}
\hfill
\begin{subfigure}{0.45\textwidth}
  \centering
  \includegraphics[width=\linewidth]{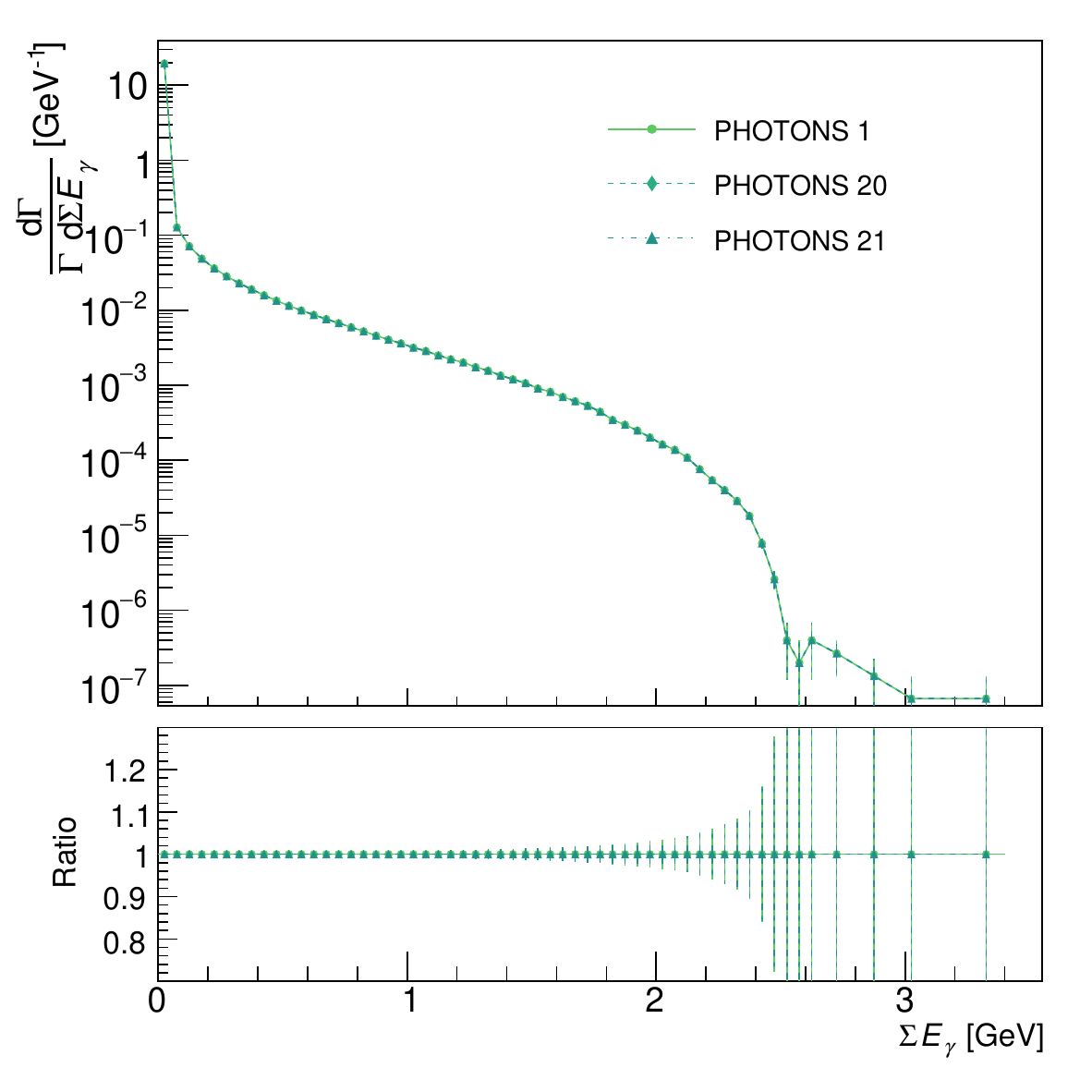}
  \caption{Total photon energy radiated in \mbox{\decay{\Bp}{\pip\piz\piz}} decays for \photons~1, \photons~20 and \photons~21.}
  \label{fig:Butopi0pi0piSherpas}
\end{subfigure}
\begin{subfigure}{0.60\textwidth}
  \centering
  \includegraphics[width=0.75\linewidth]{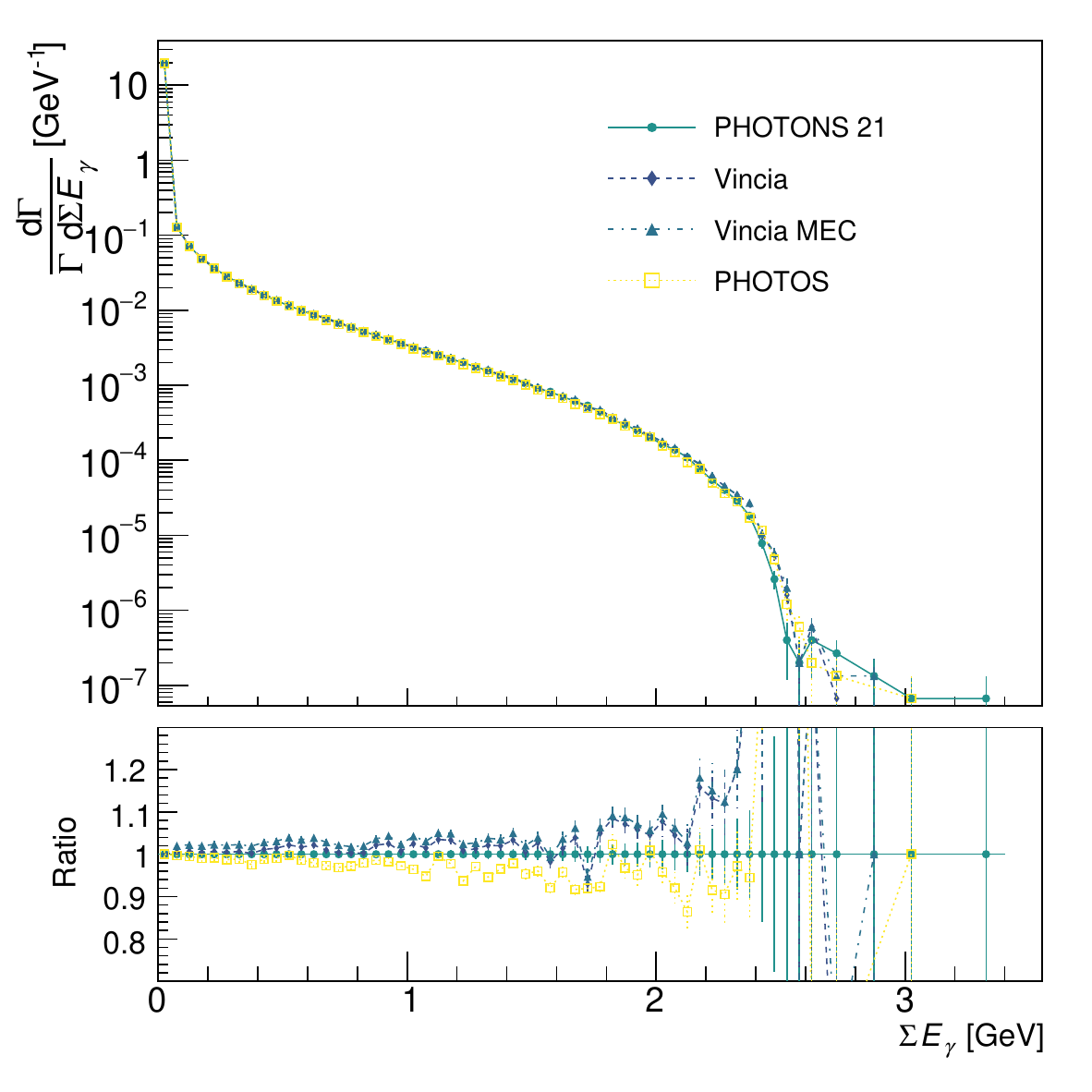}
  \caption{Total photon energy radiated in \mbox{\decay{\Bp}{\pip\piz\piz}} decays for \photons~21, \vincia, \vincia~MEC and \photos~MEC.}
  \label{fig:Butopi0pi0piDeltaEComparisons}
\end{subfigure}
\hfill
    \caption{Total photon energy radiated in \decay{\Bp}{\pip\piz\piz} decays for various QED generator options. }
    \label{fig:Butopi0pi0piComparisons}
\end{figure}

We next consider \decay{X^+}{\pip\piz\piz} decays (with $X$ = \B or \D), again with a single charged dipole but now spanned between the decaying hadron and one final-state particle. 
Comparing the total emitted photon-energy distribution for \decay{\Bp}{\pip\piz\piz} decays in \figRef{fig:Butopi0pi0piComparisons} with that for the two-body decay with one fewer neutral pion~(\figRef{fig:Butopipi0Comparisons}), we observe qualitatively similar generator behaviour, but slightly different tails:
for large energy losses, \vincia sits slightly above \photons, while \photos (as for \decay{\Bz}{\Dz\Kp\pim} decays in \figRef{fig:BdtoD0KpiDeltaEComparisons}) decays) sits below it. 
However, the summary in \tabRef{tab:B+topi+pi0pi0} shows that these differences are not particularly significant.

Analogous results for \decay{\Dz}{\pip\piz\piz} decays are summarised in \tabRef{tab:D+topi+pi0pi0}, now with the somewhat lower total energies in \photos showing up in orange shading.
However, since the total radiated energy is quite small for this decay, the absolute difference is less than $100\kev$ on average. 

\begin{table}[hp]
\centering
\begin{tabular}{l c c c c}
\toprule
$B^+ \to \pi^+\pi^0\pi^0$ & $\langle \sum E_{\gamma}\rangle [\mev]$ & $ f(\sum > 1\mev)$ & $ f(\sum > 10\mev)$ & $ f(\sum > 100\mev)$\\ \midrule 
\photos & \cellcolor[HTML]{FFFFFF} $7.88$ & \cellcolor[HTML]{FFFFFF} $6.07\%$ & \cellcolor[HTML]{FFFFFF} $3.89\%$ & \cellcolor[HTML]{FFFFFF} $1.73\%$\\
\photos MEC & \cellcolor[HTML]{FFFFFF} $8.06$ & \cellcolor[HTML]{FFFFFF} $6.10\%$ & \cellcolor[HTML]{FFFFFF} $3.92\%$ & \cellcolor[HTML]{FFFFFF} $1.75\%$\\ \midrule
\photons 1 & \cellcolor[HTML]{FFFFFF} $8.03$ & \cellcolor[HTML]{FFFFFF} $6.09\%$ & \cellcolor[HTML]{FFFFFF} $3.91\%$ & \cellcolor[HTML]{FFFFFF} $1.75\%$\\
\photons 20 & \cellcolor[HTML]{FFFFFF} $8.03$ & \cellcolor[HTML]{FFFFFF} $6.09\%$ & \cellcolor[HTML]{FFFFFF} $3.91\%$ & \cellcolor[HTML]{FFFFFF} $1.75\%$\\
\photons 21 & \cellcolor[HTML]{FFFFFF} $8.03$ & \cellcolor[HTML]{FFFFFF} $6.09\%$ & \cellcolor[HTML]{FFFFFF} $3.91\%$ & \cellcolor[HTML]{FFFFFF} $1.75\%$\\ \midrule
\vincia MEC & \cellcolor{cyan1} $8.26$ & \cellcolor[HTML]{FFFFFF} $6.20\%$ & \cellcolor{cyan1} $4.00\%$ & \cellcolor{cyan1} $1.79\%$\\
\vincia & \cellcolor[HTML]{FFFFFF} $8.13$ & \cellcolor[HTML]{FFFFFF} $6.12\%$ & \cellcolor[HTML]{FFFFFF} $3.94\%$ & \cellcolor[HTML]{FFFFFF} $1.77\%$\\
\vincia Dipole & \cellcolor[HTML]{FFFFFF} $8.15$ & \cellcolor[HTML]{FFFFFF} $6.13\%$ & \cellcolor[HTML]{FFFFFF} $3.95\%$ & \cellcolor[HTML]{FFFFFF} $1.77\%$\\ \bottomrule
\end{tabular}
\caption{
Average amounts of photon energy radiated in
\decay{\Bp}{\pip\piz\piz} decays and the fractions of events which radiate a total of more than $1$, $10$, and $100\mev$. 
Colour scheme defined at the start of \secRef{sec:results}.}
\label{tab:B+topi+pi0pi0}
\end{table}

\begin{table}[tp]
\centering
\begin{tabular}{l c c c c}
\toprule
$D^+ \to \pi^+\pi^0\pi^0$ & $\langle \sum E_{\gamma}\rangle [\mev]$ & $ f(\sum > 1\mev)$ & $ f(\sum > 10\mev)$ & $ f(\sum > 100\mev)$\\ \midrule 
\photos & \cellcolor{orange2} $1.34$ & \cellcolor[HTML]{FFFFFF} $2.75\%$ & \cellcolor{orange1} $1.53\%$ & \cellcolor{orange2} $0.41\%$\\
\photos MEC & \cellcolor[HTML]{FFFFFF} $1.41$ & \cellcolor[HTML]{FFFFFF} $2.78\%$ & \cellcolor[HTML]{FFFFFF} $1.56\%$ & \cellcolor[HTML]{FFFFFF} $0.43\%$\\ \midrule
\photons 1 & \cellcolor[HTML]{FFFFFF} $1.41$ & \cellcolor[HTML]{FFFFFF} $2.78\%$ & \cellcolor[HTML]{FFFFFF} $1.56\%$ & \cellcolor[HTML]{FFFFFF} $0.43\%$\\
\photons 20 & \cellcolor[HTML]{FFFFFF} $1.41$ & \cellcolor[HTML]{FFFFFF} $2.78\%$ & \cellcolor[HTML]{FFFFFF} $1.56\%$ & \cellcolor[HTML]{FFFFFF} $0.43\%$\\
\photons 21 & \cellcolor[HTML]{FFFFFF} $1.41$ & \cellcolor[HTML]{FFFFFF} $2.78\%$ & \cellcolor[HTML]{FFFFFF} $1.56\%$ & \cellcolor[HTML]{FFFFFF} $0.43\%$\\ \midrule
\vincia MEC & \cellcolor{cyan1} $1.45$ & \cellcolor[HTML]{FFFFFF} $2.83\%$ & \cellcolor{cyan1} $1.60\%$ & \cellcolor{cyan1} $0.44\%$\\
\vincia & \cellcolor[HTML]{FFFFFF} $1.43$ & \cellcolor[HTML]{FFFFFF} $2.80\%$ & \cellcolor[HTML]{FFFFFF} $1.58\%$ & \cellcolor[HTML]{FFFFFF} $0.44\%$\\
\vincia Dipole & \cellcolor[HTML]{FFFFFF} $1.43$ & \cellcolor[HTML]{FFFFFF} $2.80\%$ & \cellcolor[HTML]{FFFFFF} $1.57\%$ & \cellcolor[HTML]{FFFFFF} $0.44\%$\\ \bottomrule
\end{tabular}
\caption{
Average amounts of photon energy radiated in
\decay{\Dp}{\pip\piz\piz} decays and the fractions of events which radiate a total of more than $1$, $10$, and $100\mev$. 
Colour scheme defined at the start of \secRef{sec:results}.}
\label{tab:D+topi+pi0pi0}
\end{table}

\subsubsection{\decay{\Bp}{\Dp\Km\pip}, \decay{\Dp}{\Kp\pip\pim}, and \decay{\Dp}{\pip\pip\pim}}

For fully charged three-body topologies such as \decay{\Bp}{\Dp\Km\pip} decays, genuine QED multipole effects are now possible. 
As discussed in \secsRef{sec:generators} and~\ref{sec:vincia}, these effects are taken into account in the YFS formalism in the soft limit and in \vincia's (default) multipole shower mode, while \vincia's dipole mode approximates them in terms of a direct sum of two (non-interfering) dipoles, identified based on a principle of maximal screening~\cite{Skands:2020lkd}.

\begin{figure}[tp]
\centering
\begin{subfigure}{0.45\textwidth}
  \centering
  \includegraphics[width=\linewidth]{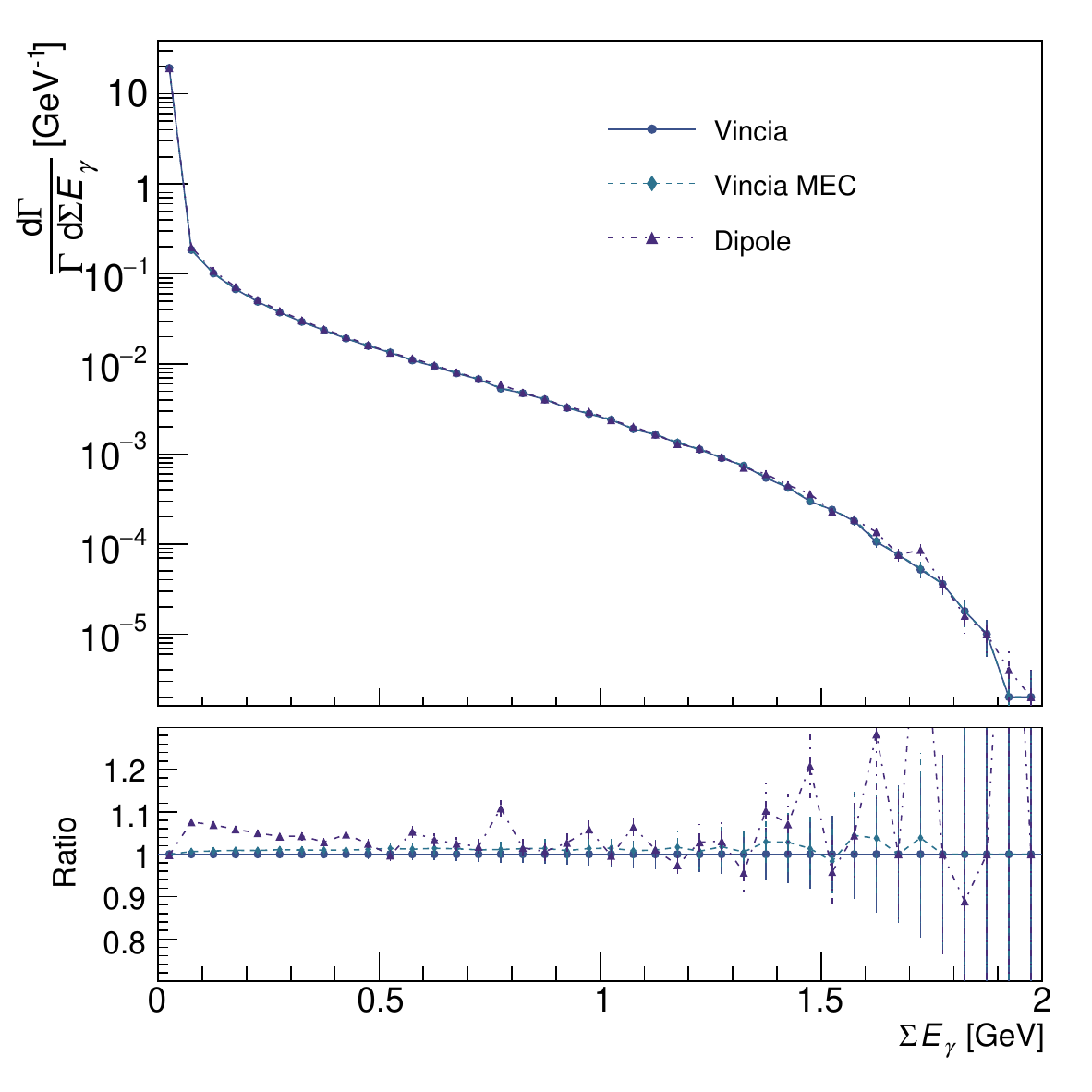}
  \caption{Total photon energy radiated in \mbox{\decay{\Bp}{\Dp\Km\pip}} decays for \vincia, \vincia~MEC and \vincia's dipole approximation.}
  \label{fig:ButoDKpiVincias}
\end{subfigure}
\hfill
\begin{subfigure}{0.45\textwidth}
  \centering
  \includegraphics[width=\linewidth]{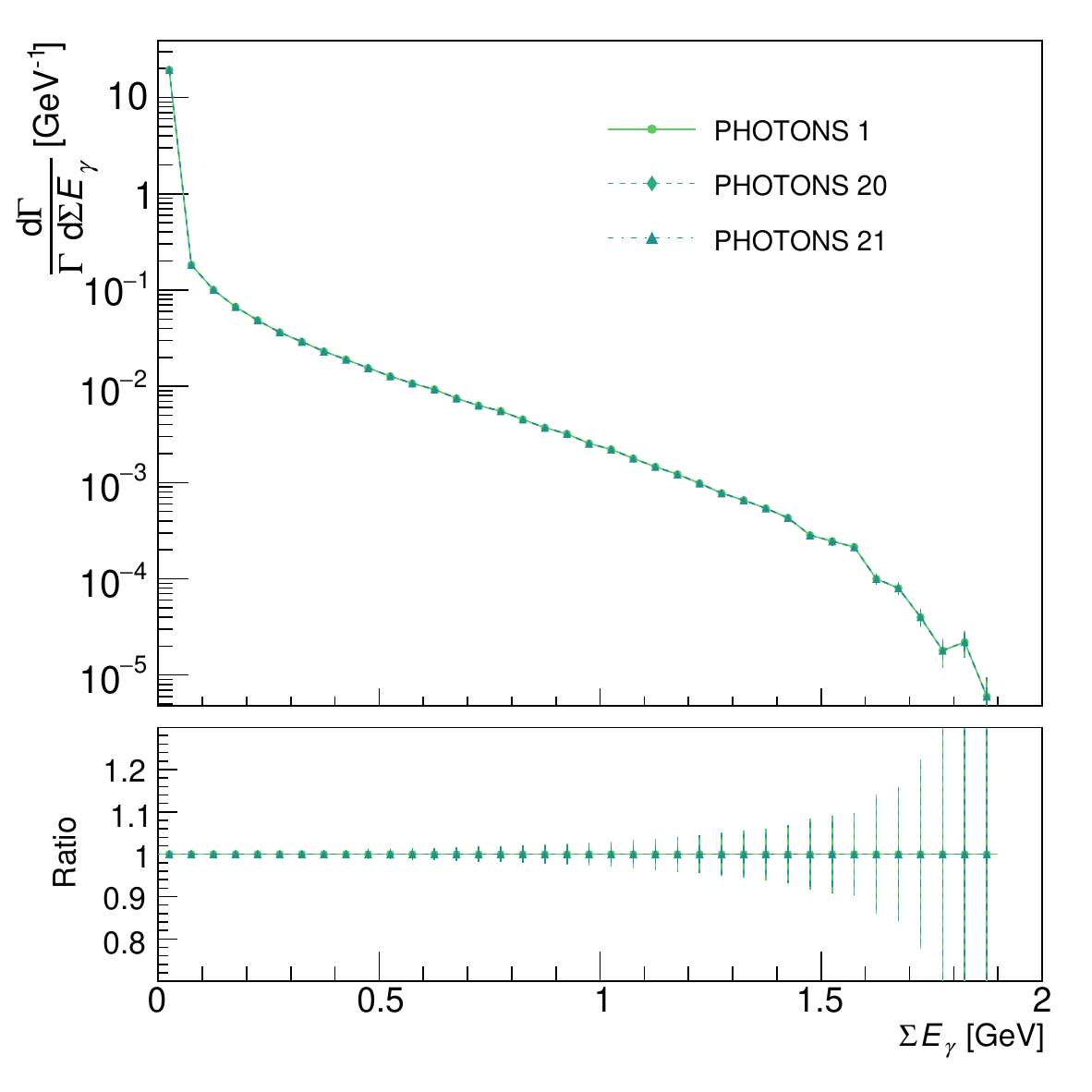}
  \caption{Total photon energy radiated in \mbox{\decay{\Bp}{\Dp\Km\pip}} decays for \photons~1, \photons~20 and \photons~21.}
  \label{fig:ButoDKpiSherpas}
\end{subfigure}
\begin{subfigure}{0.60\textwidth}
  \centering
  \includegraphics[width=0.75\textwidth]{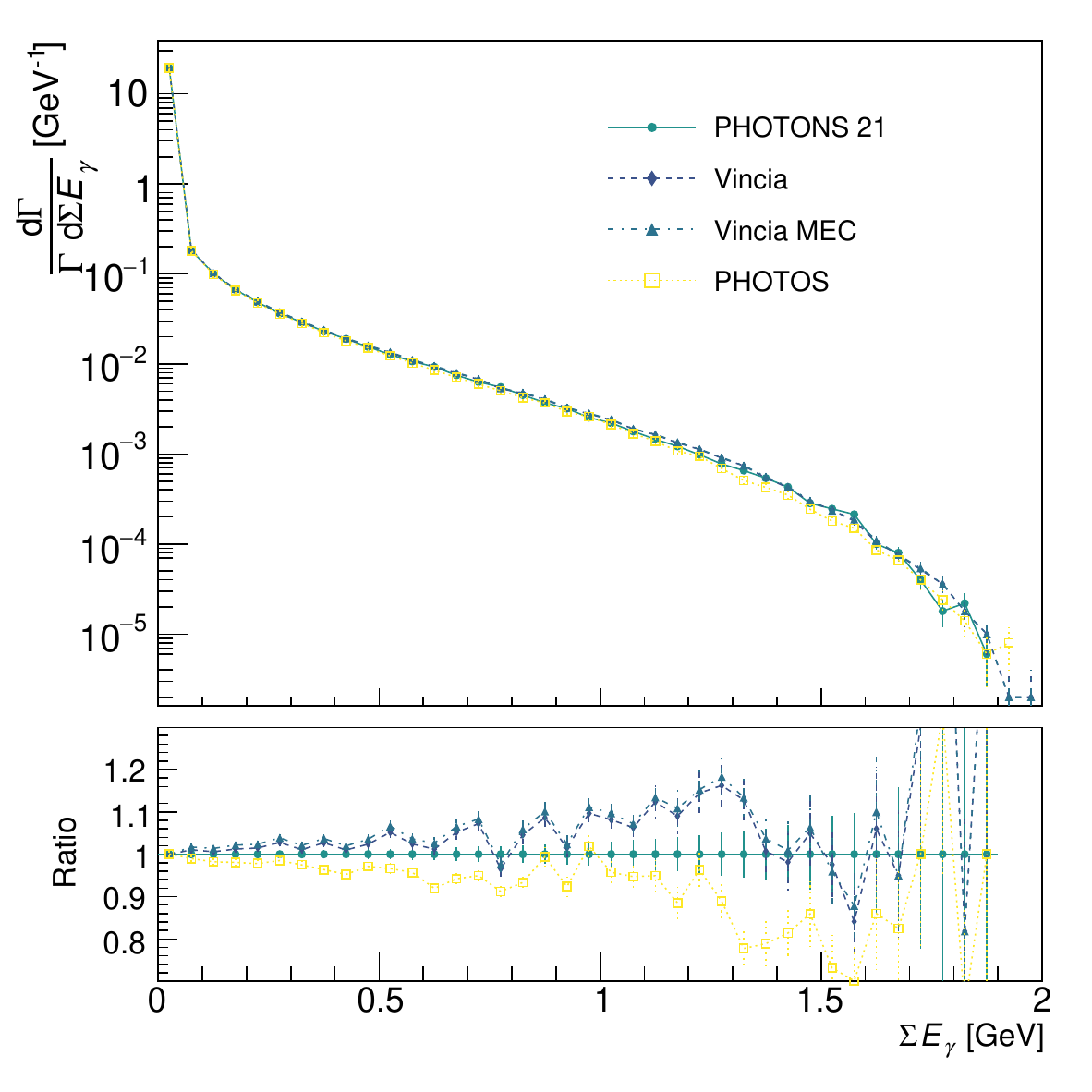}
  \caption{Total photon energy radiated in \mbox{\decay{\Bp}{\Dp\Km\pip}} decays for various QED generator options.}
  \label{fig:ButoDKpiDeltaEComparisons}
\end{subfigure}
\hfill
    \caption{
    Total photon energy radiated in \decay{\Bp}{\Dp\Km\pip} decays for various QED generator options.
    }
    \label{fig:ButoDKpiComparisons}
\end{figure}

\FigRef{fig:ButoDKpiVincias} shows that \vincia with and without MECs exhibit the same behaviour found before, with the ME corrected shower slightly above the uncorrected one. 
\vincia's dipole approximation sits consistently above the other \vincia modes, especially in the low $\sum E_\gamma$ region. 
This leads us to conclude that the multipole corrections are mostly negative.

This behaviour is consistent across the three considered fully-charged three-body decays, summarised in \tabsRef{tab:B+2D+K+pi-}, \ref{tab:D+2K+pi+pi-}, and \ref{tab:D+2pi+pi+pi-}. 
These tables also show that the relation between \photos and \photons~21 is relatively sporadic: \photos under-radiates (\tabRef{tab:B+2D+K+pi-}), over-radiates (\tabRef{tab:D+2K+pi+pi-}), or radiates within our 2\% bound (\tabRef{tab:D+2pi+pi+pi-}). 
The amount of this radiation seems to correlate negatively with the relative size of the phase space.

\begin{table}[p]
\centering
\begin{tabular}{l c c c c}
\toprule
$B^+ \to D^-K^+\pi^+$ & $\langle \sum E_{\gamma}\rangle [\mev]$ & $ f(\sum > 1\mev)$ & $ f(\sum > 10\mev)$ & $ f(\sum > 100\mev)$\\ \midrule 
\photos & \cellcolor{orange1} $7.88$ & \cellcolor[HTML]{FFFFFF} $8.31\%$ & \cellcolor[HTML]{FFFFFF} $5.14\%$ & \cellcolor{orange1} $2.02\%$\\
\photos MEC & \cellcolor[HTML]{FFFFFF} $8.30$ & \cellcolor[HTML]{FFFFFF} $8.40\%$ & \cellcolor[HTML]{FFFFFF} $5.24\%$ & \cellcolor[HTML]{FFFFFF} $2.10\%$\\ \midrule
\photons 1 & \cellcolor[HTML]{FFFFFF} $8.17$ & \cellcolor[HTML]{FFFFFF} $8.38\%$ & \cellcolor[HTML]{FFFFFF} $5.22\%$ & \cellcolor[HTML]{FFFFFF} $2.08\%$\\
\photons 20 & \cellcolor[HTML]{FFFFFF} $8.17$ & \cellcolor[HTML]{FFFFFF} $8.38\%$ & \cellcolor[HTML]{FFFFFF} $5.22\%$ & \cellcolor[HTML]{FFFFFF} $2.08\%$\\
\photons 21 & \cellcolor[HTML]{FFFFFF} $8.17$ & \cellcolor[HTML]{FFFFFF} $8.38\%$ & \cellcolor[HTML]{FFFFFF} $5.22\%$ & \cellcolor[HTML]{FFFFFF} $2.08\%$\\ \midrule
\vincia MEC & \cellcolor{cyan1} $8.47$ & \cellcolor[HTML]{FFFFFF} $8.54\%$ & \cellcolor{cyan1} $5.32\%$ & \cellcolor{cyan1} $2.14\%$\\
\vincia & \cellcolor{cyan1} $8.39$ & \cellcolor[HTML]{FFFFFF} $8.50\%$ & \cellcolor[HTML]{FFFFFF} $5.29\%$ & \cellcolor[HTML]{FFFFFF} $2.12\%$\\
\vincia Dipole & \cellcolor{cyan2} $8.80$ & \cellcolor{cyan3} $9.48\%$ & \cellcolor{cyan2} $5.71\%$ & \cellcolor{cyan2} $2.23\%$\\ \bottomrule
\end{tabular}
\caption{
Average amounts of photon energy radiated in
\decay{\Bp}{\Dm\Kp\pip} decays and the fractions of events which radiate a total of more than $1$, $10$, and $100\mev$. 
Colour scheme defined at the start of \secRef{sec:results}.}
\label{tab:B+2D+K+pi-}
\end{table}

\begin{table}[tp]
\centering
\begin{tabular}{l c c c c}
\toprule
$D^+\to K^+\pi^+\pi^-$ & $\langle \sum E_{\gamma}\rangle [\mev]$ & $ f(\sum > 1\mev)$ & $ f(\sum > 10\mev)$ & $ f(\sum > 100\mev)$\\ \midrule 
\photos & \cellcolor{cyan1} $2.52$ & \cellcolor[HTML]{FFFFFF} $5.55\%$ & \cellcolor[HTML]{FFFFFF} $3.02\%$ & \cellcolor{cyan1} $0.77\%$\\
\photos MEC & \cellcolor{cyan3} $2.67$ & \cellcolor[HTML]{FFFFFF} $5.63\%$ & \cellcolor{cyan1} $3.11\%$ & \cellcolor{cyan3} $0.82\%$\\ \midrule
\photons 1 & \cellcolor[HTML]{FFFFFF} $2.40$ & \cellcolor[HTML]{FFFFFF} $5.56\%$ & \cellcolor[HTML]{FFFFFF} $3.03\%$ & \cellcolor[HTML]{FFFFFF} $0.73\%$\\
\photons 20 & \cellcolor[HTML]{FFFFFF} $2.40$ & \cellcolor[HTML]{FFFFFF} $5.56\%$ & \cellcolor[HTML]{FFFFFF} $3.03\%$ & \cellcolor[HTML]{FFFFFF} $0.73\%$\\
\photons 21 & \cellcolor[HTML]{FFFFFF} $2.40$ & \cellcolor[HTML]{FFFFFF} $5.56\%$ & \cellcolor[HTML]{FFFFFF} $3.03\%$ & \cellcolor[HTML]{FFFFFF} $0.73\%$\\ \midrule
\vincia MEC & \cellcolor{cyan1} $2.52$ & \cellcolor{cyan1} $5.67\%$ & \cellcolor[HTML]{FFFFFF} $3.09\%$ & \cellcolor{cyan1} $0.77\%$\\
\vincia & \cellcolor{cyan1} $2.52$ & \cellcolor{cyan1} $5.67\%$ & \cellcolor[HTML]{FFFFFF} $3.09\%$ & \cellcolor{cyan1} $0.77\%$\\
\vincia Dipole & \cellcolor{cyan3} $3.01$ & \cellcolor{cyan3} $7.22\%$ & \cellcolor{cyan3} $3.78\%$ & \cellcolor{cyan3} $0.90\%$\\ \bottomrule
\end{tabular}
\caption{
Average amounts of photon energy radiated in \decay{\Dp}{\Kp\pip\pim} decays and the fractions of events which radiate a total of more than $1$, $10$, and $100\mev$. 
Colour scheme defined at the start of \secRef{sec:results}.}
\label{tab:D+2K+pi+pi-}
\end{table}

\clearpage

\begin{table}[tp]
\centering
\begin{tabular}{l c c c c}
\toprule
$D^+\to\pi^+\pi^+\pi^-$ & $\langle \sum E_{\gamma}\rangle [\mev]$ & $ f(\sum > 1\mev)$ & $ f(\sum > 10\mev)$ & $ f(\sum > 100\mev)$\\ \midrule 
\photos & \cellcolor[HTML]{FFFFFF} $4.33$ & \cellcolor[HTML]{FFFFFF} $8.18\%$ & \cellcolor[HTML]{FFFFFF} $4.65\%$ & \cellcolor[HTML]{FFFFFF} $1.33\%$\\
\photos MEC & \cellcolor{cyan2} $4.52$ & \cellcolor[HTML]{FFFFFF} $8.28\%$ & \cellcolor[HTML]{FFFFFF} $4.74\%$ & \cellcolor{cyan1} $1.39\%$\\ \midrule
\photons 1 & \cellcolor[HTML]{FFFFFF} $4.30$ & \cellcolor[HTML]{FFFFFF} $8.22\%$ & \cellcolor[HTML]{FFFFFF} $4.68\%$ & \cellcolor[HTML]{FFFFFF} $1.33\%$\\
\photons 20 & \cellcolor[HTML]{FFFFFF} $4.30$ & \cellcolor[HTML]{FFFFFF} $8.22\%$ & \cellcolor[HTML]{FFFFFF} $4.68\%$ & \cellcolor[HTML]{FFFFFF} $1.33\%$\\
\photons 21 & \cellcolor[HTML]{FFFFFF} $4.30$ & \cellcolor[HTML]{FFFFFF} $8.22\%$ & \cellcolor[HTML]{FFFFFF} $4.68\%$ & \cellcolor[HTML]{FFFFFF} $1.33\%$\\ \midrule
\vincia MEC & \cellcolor[HTML]{FFFFFF} $4.33$ & \cellcolor[HTML]{FFFFFF} $8.35\%$ & \cellcolor[HTML]{FFFFFF} $4.71\%$ & \cellcolor[HTML]{FFFFFF} $1.32\%$\\
\vincia & \cellcolor[HTML]{FFFFFF} $4.30$ & \cellcolor[HTML]{FFFFFF} $8.31\%$ & \cellcolor[HTML]{FFFFFF} $4.68\%$ & \cellcolor[HTML]{FFFFFF} $1.31\%$\\
\vincia Dipole & \cellcolor{cyan3} $4.96$ & \cellcolor{cyan3} $9.82\%$ & \cellcolor{cyan3} $5.39\%$ & \cellcolor{cyan3} $1.50\%$\\ \bottomrule
\end{tabular}
\caption{
Average amounts of photon energy radiated in  \decay{\Dp}{\pip\pip\pim} decays and the fractions of events which radiate a total of more than $1$, $10$, and $100\mev$. 
Colour scheme defined at the start of \secRef{sec:results}.}
\label{tab:D+2pi+pi+pi-}
\end{table}

\subsection{A fixed three-body multipole configuration}

\label{sec:fixed3}

\begin{figure}[tp]
    \centering
    \includegraphics[width=0.5\linewidth]{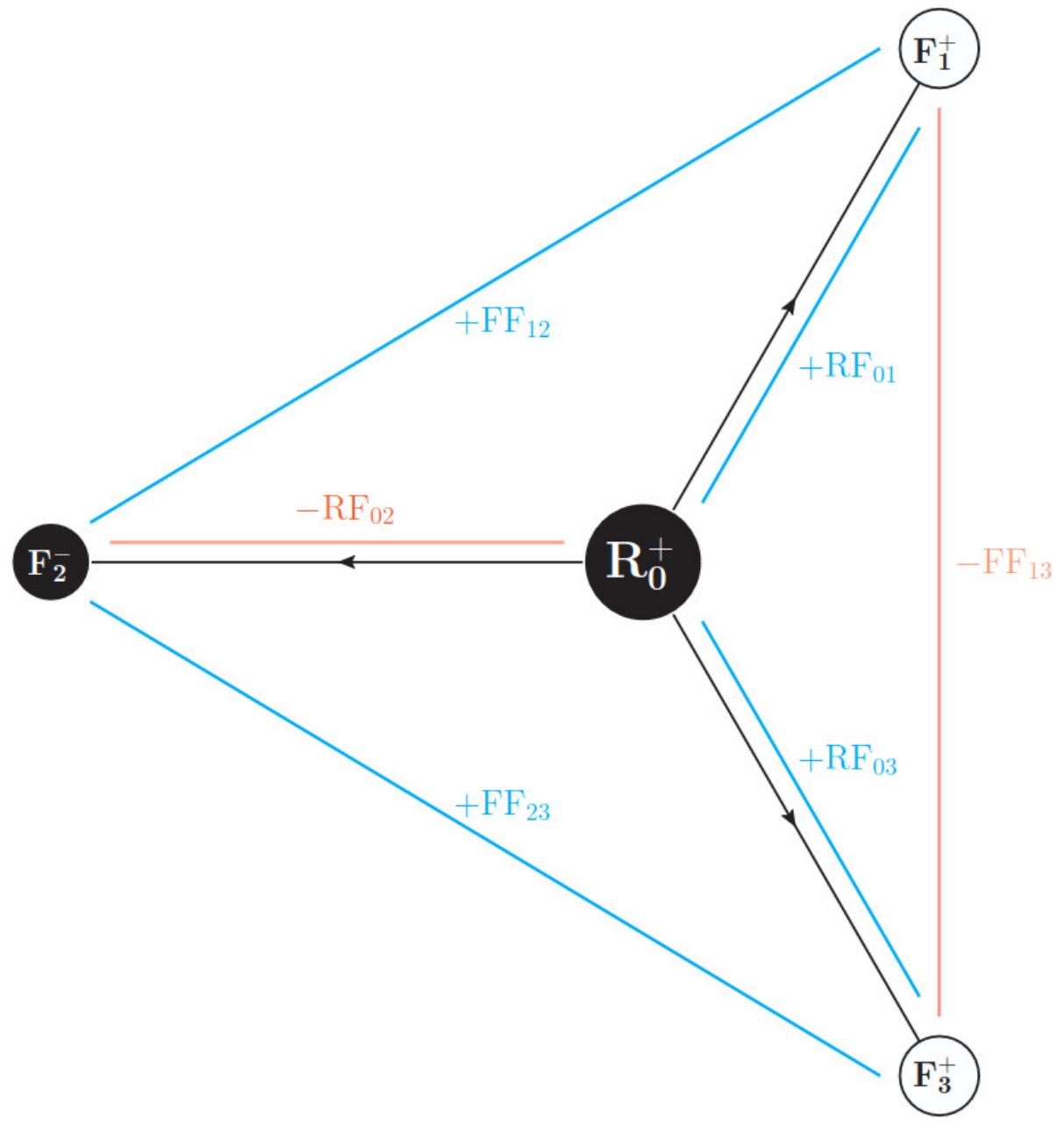}
    \caption{The Mercedes configuration used for the fixed decay of a charged mother into three charged daughters. The positive-sign (constructive) antennae are labelled in light blue and the negative-sign (destructive) ones in light red.
    ${\rm R}$ stands for resonance (initial state), while ${\rm F}$ stands for final state.  }
    \label{fig:mercedesConfiguration}
\end{figure}

We inspect the angular distributions of photons in decays to more than two bodies by considering a useful fixed phase-space point referred to as ``Mercedes'' configuration, illustrated in \figRef{fig:mercedesConfiguration}.
This configuration serves as direct  Born-level input to \pythia, and is then dressed with QED radiation showered with \vincia.  
This configuration should maximise multipole effects, highlighting differences between \vincia's multipole and dipole modes. 
 
\begin{figure}[t]
\centering
\begin{subfigure}{0.45\textwidth}
  \centering
  \includegraphics[width=\linewidth]{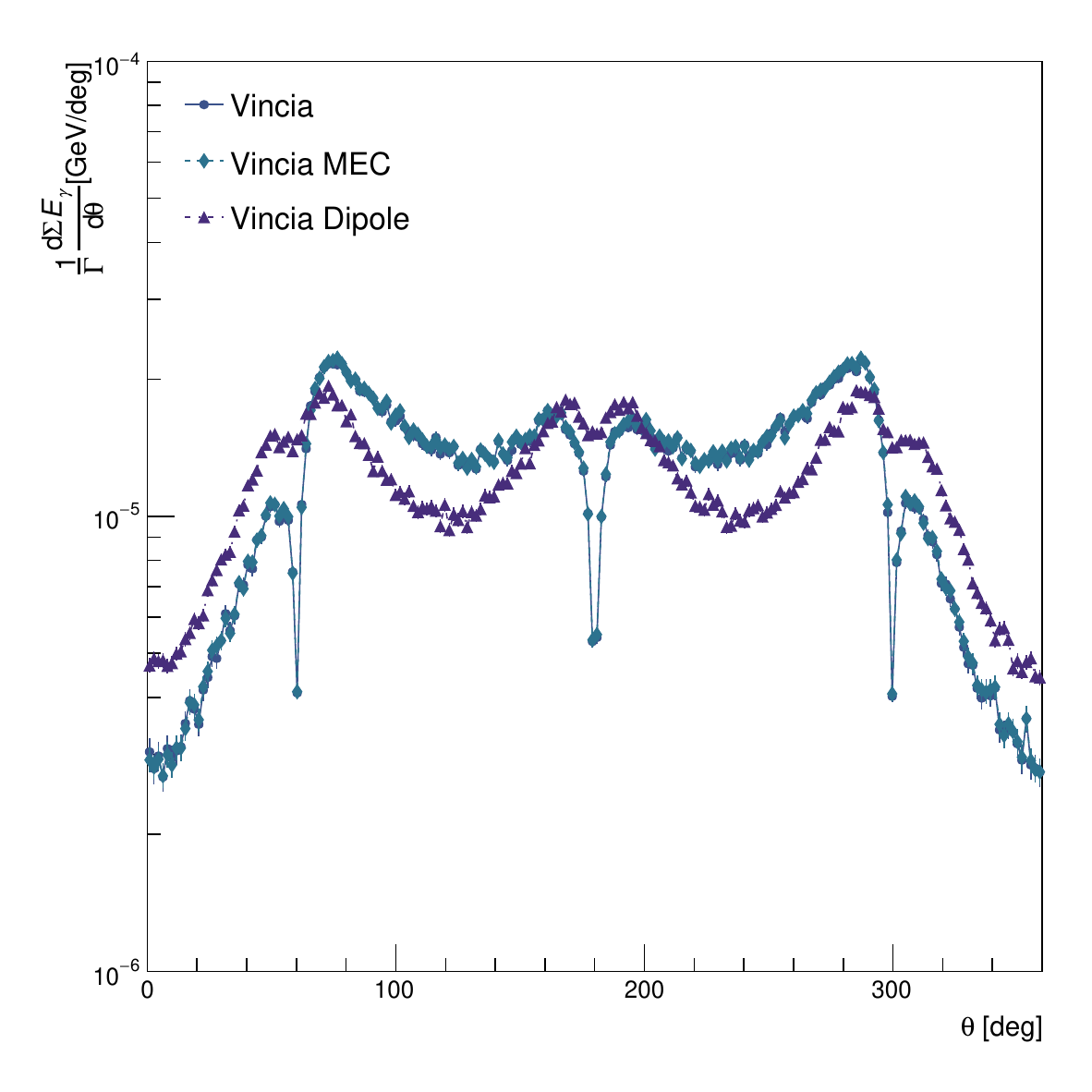}
  \caption{Angular distribution of photon energy for the first emission in \vincia, \vincia~MEC and \vincia's dipole approximation.}
  \label{fig:D+_3piAngular}
\end{subfigure}
\hfill
\begin{subfigure}{0.45\textwidth}
  \centering
  \includegraphics[width=\linewidth]{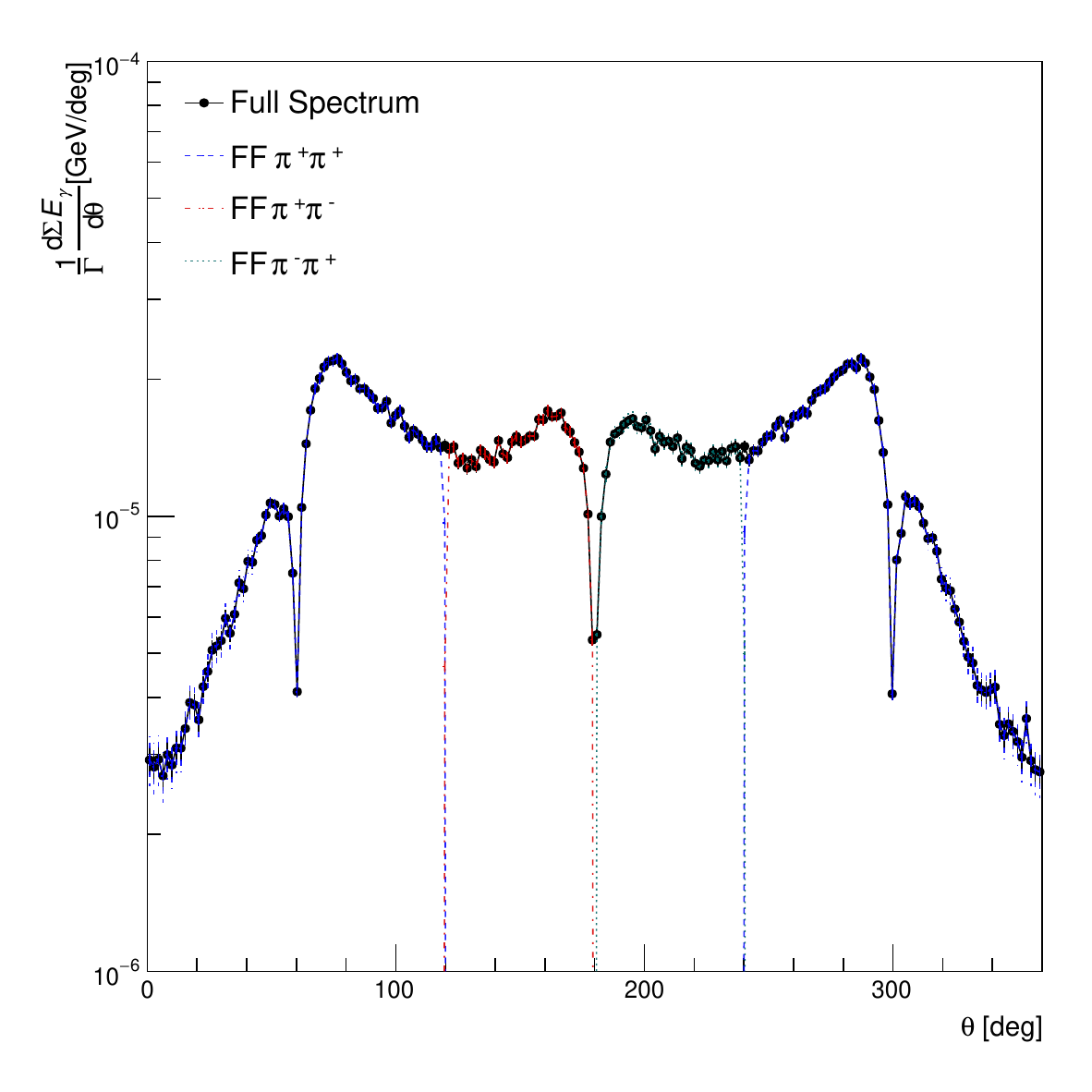}
  \caption{Contributions from each ${\rm FF}$ sector to the angular distribution of energy for the first emission in \vincia~MEC.}
  \label{fig:D+_3piAngularAntennae}
\end{subfigure}
    \caption{Angular distribution of radiated photon energy for the first emission in \decay{\Dp}{\pip\pip\pim} decays in a fixed Mercedes configuration~(see \figRef{fig:mercedesConfiguration}) with the $\pim$ at $180^\circ$ and the $\pip$'s at $60^\circ$ and $300^\circ$. 
    Left: comparison of different configurations in \vincia. 
    Right: antenna contributions in each sector for \vincia~MEC.
    The $\pip\pip$ antenna is shown in blue, the $\pip\pim$ antenna with the $\pip$ at $60^\circ$ in red and the $\pip\pim$ antenna with the $\pip$ at $300^\circ$ in green.}
    \label{fig:D+_3piAngularComparisons}
\end{figure}

We first consider the case where all three daughter particles have the same mass, using \decay{\Dp}{\pip\pip\pim} decays.  \FigRef{fig:D+_3piAngular} shows the angular distribution of the total emitted photon energy for \vincia's default (multipole), ME-corrected, and dipole (uncorrected) shower. 
In both multipole modes, we find sharp dips at $60^\circ$, $180^\circ$, and $240^\circ$, corresponding to the dead cones of the three pions. 
However, in the dipole approximation, these dips are much less pronounced. 
We attribute this to radiation from the ``other'' dipole in the system, which, in the dipole approximation, is allowed to radiate into the gap. 
In the multipole treatment, the phase-space sectorisation prevents this, as positive and negative multipole terms are explicitly summed into a single kernel in each phase-space point.

To illustrate the sectorisation in \vincia's multipole mode,  \figRef{fig:D+_3piAngularAntennae} shows the relative contributions of each of the three final-final QED sectors to the full emission pattern for \vincia~MEC \footnote{The resonance-final~(RF) sectors do not contribute for this phase-space point.}. 
The $\pip\pip$ sector accounts for all of the radiation in the angular regions between $0^\circ$ and $120^\circ$ and between $240^\circ$  and $360^\circ$. 
At $120^\circ$, there is a very sharp transition to one of the $\pip\pim$ sectors, which at $180^\circ$ (the location of the negative pion) switches to the other $\pip\pim$ sector.
These borders are particularly sharp because of the high degree of symmetry of the Mercedes configuration with identical final-state masses. 
The underlying sectorisation is based on the size of the individual (positive or negative) antenna contributions, but in this case it corresponds closely to the opening angle.

Another interesting feature of \figRef{fig:D+_3piAngular} is that it exhibits a change in sign of the multipole corrections. In the angular region in-between the two positive pions (from 300$^\circ$ to 60$^\circ$), the multipole distribution is below the dipole approximation, reflecting predominantly negative (destructive) multipole corrections. 
While in the complementary angular region between 60$^\circ$ and 300$^\circ$, the multipole distribution is generally higher than the dipole approximation, reflecting predominantly positive (constructive) multipole corrections.
The lack of the full multipole structure in the dipole case is also visible in the region from $60^\circ$ to $180^\circ$ (and, identically, $180^\circ$ to $300^\circ$), where the peaks near 60$^\circ$ and 180$^\circ$ have similar heights in the dipole case but not in the multipole case.

Turning now to the case of different daughter masses, we consider \decay{\Bp}{\Dp\Kp\pim} decays, again in a Mercedes configuration, with four-momenta: 
\begin{equation*}
\begin{split}
    p_{\Bp} =&
    \begin{pmatrix}
        5.279, \phantom{-}0.000, \phantom{-}0.000, \phantom{-}0.000
    \end{pmatrix}^\mathsf{T}\gev\,,\\
p_{\Dp} =&
    \begin{pmatrix}
        2.348, \phantom{-}0.710, -1.230, \phantom{-}0.000
    \end{pmatrix}^\mathsf{T}\gev\,,\\
p_{\Kp} =&
    \begin{pmatrix}
        1.504, \phantom{-}0.710, \phantom{-}1.230, \phantom{-}0.000
    \end{pmatrix}^\mathsf{T}\gev\,,\\
p_{\pim} =&
    \begin{pmatrix}
        1.427, -1.421, \phantom{-}0.000, \phantom{-}0.000
    \end{pmatrix}^\mathsf{T}\gev\,.\\
\end{split}
\end{equation*}

\begin{figure}[tp]
\centering
\begin{subfigure}{0.45\textwidth}
  \centering
  \includegraphics[width=\linewidth]{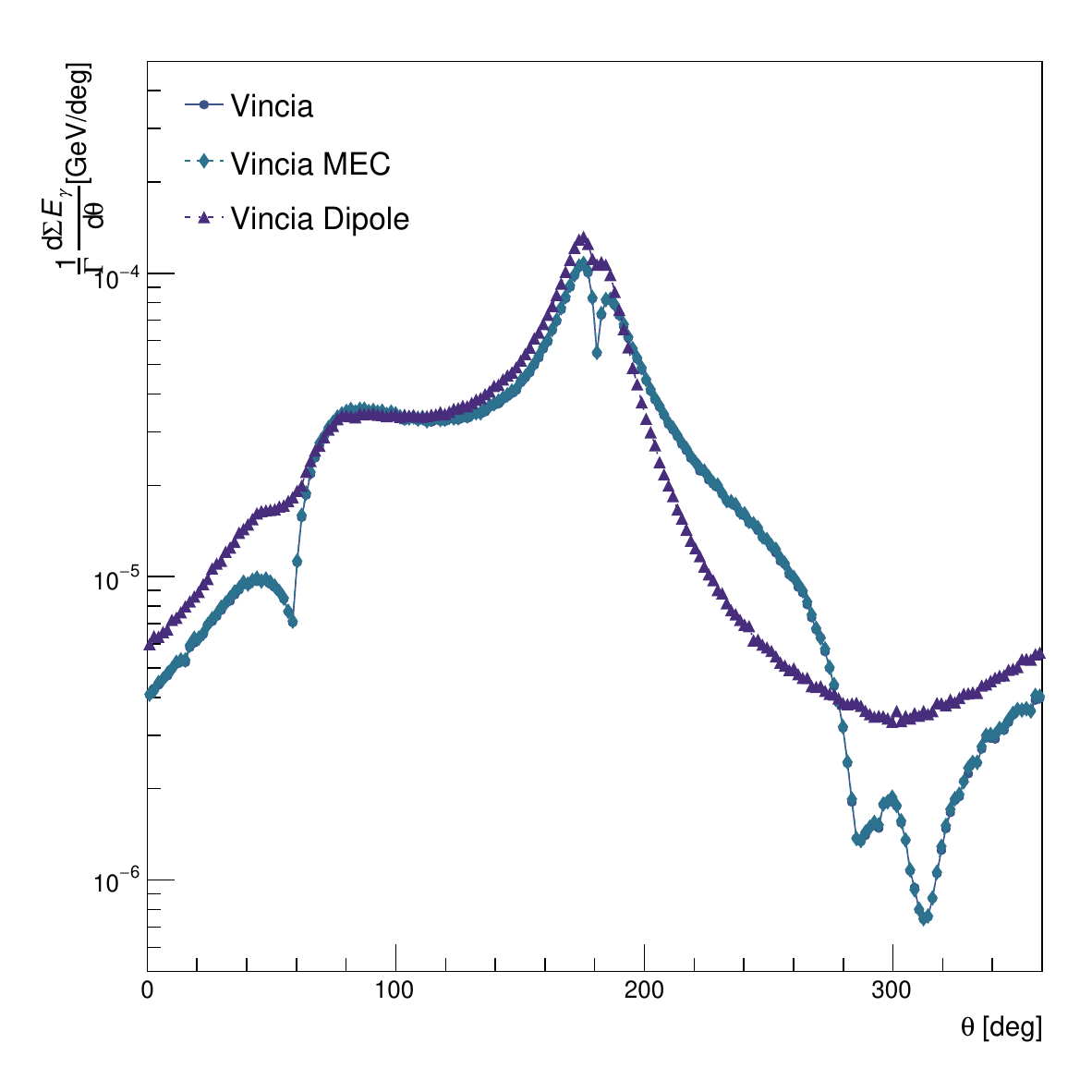}
  \caption{Angular distribution of photon energy for the first emission in \vincia, \vincia~MEC and \vincia's dipole approximation.}
  \label{fig:Bu_DKpiAngular}
\end{subfigure}
\hfill
\begin{subfigure}{0.45\textwidth}
  \centering
  \includegraphics[width=\linewidth]{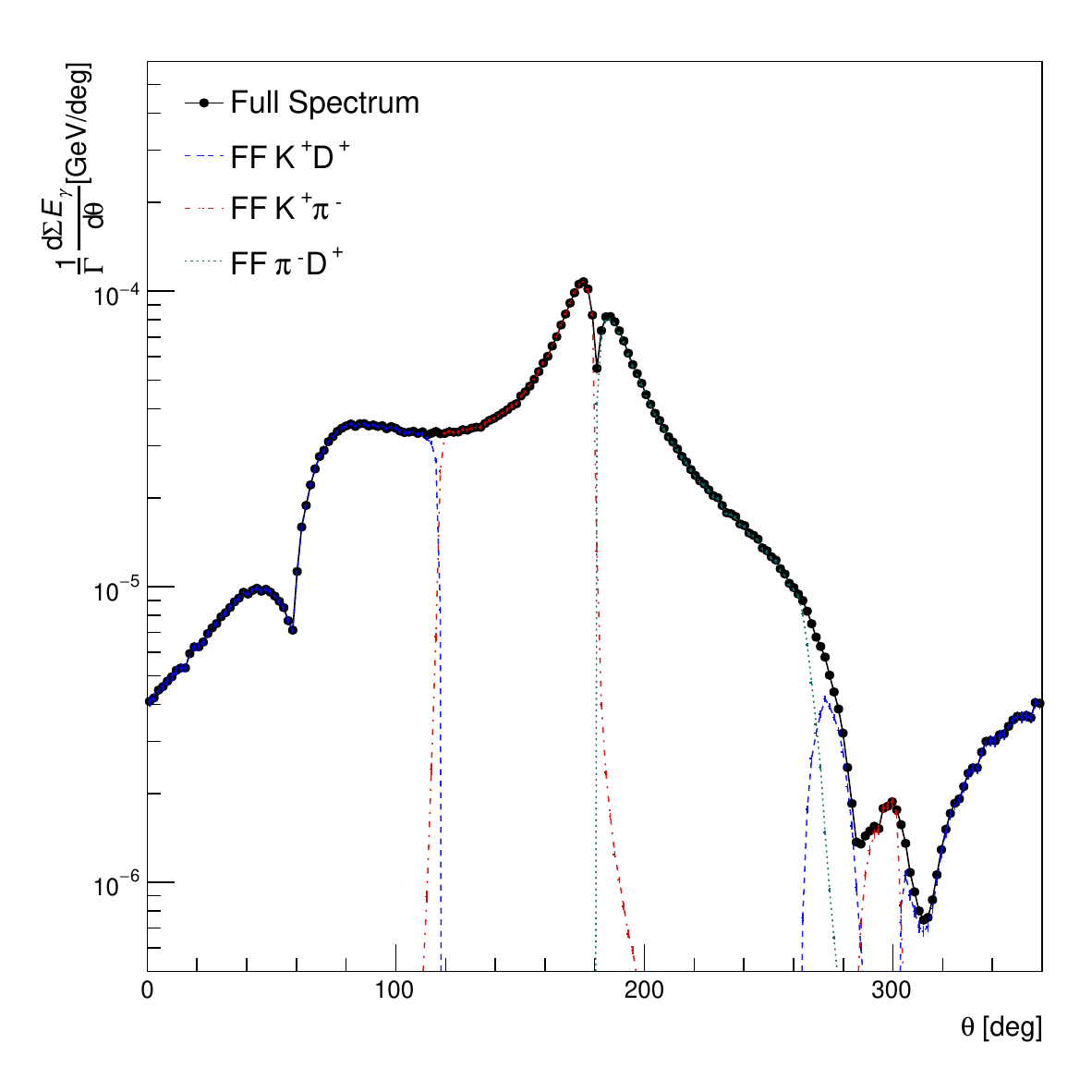}
  \caption{Contributions from individual antennae in each sector for the first emission in \vincia~MEC.\\}
  \label{fig:Bu_DKpiAngularAntennae}
\end{subfigure}
    \caption{Angular distribution of radiated photon energy for the first emission in \decay{\Bp}{\Dp\Kp\pim} decays in a fixed Mercedes configuration~(see \figRef{fig:mercedesConfiguration}). 
    The $\pim$ sits at $180^\circ$, the $\Kp$ at $60^\circ$ and the $\Dp$ at~$300^\circ$. 
    Left: distribution for different modes in \vincia. 
    Right: antenna contributions in each sector for \vincia~MEC. 
    The $\Dp\Kp$ antenna is shown in blue, the $\Kp\pim$ antenna in red and the $\Dp\pim$ antenna in green.}
    \label{fig:Bu_DKpiAngularComparisons}
\end{figure}

\FigRef{fig:Bu_DKpiAngularComparisons} shows features similar to the equal-mass case: 
the asymmetry in the peaks around the mass-suppressed dead-cone regions of the pion ($180^\circ$) and kaon ($60^\circ$), and much shallower dips in those regions for the dipole approximation. 
Interestingly, in the multipole treatment (including the case with MECs), we find a small relative enhancement at $300^\circ$ (where the $D$ sits), with an asymmetric dip on either side. 
Inspecting the sector contribution plot shown in \figRef{fig:Bu_DKpiAngularAntennae},
the region around $300^\circ$ is characterised by quite a lot of transitions between different sectors, and that the enhancement at $300^\circ$ is associated with a transition between radiation from the $\Kp\Dp$ sectors to radiation from the $\Kp\pim$ sector. 
Naively, we would expect that radiation from this antenna would be vetoed by the sectorisation in this region, the curves of the $\Dp\Kp$ and $\Dp\pip$ antennae show. 
However, here the mass suppression caused by being so close to the \D~meson is large enough that the $\Dp\Kp$ and $\Dp\pip$ antennae are not sufficiently large to veto the radiation, leading to a small local enhancement. 

The fact that the ME-corrected and -uncorrected distributions agree implies that we do not regard this as in principle problematic; in multibody decays, different sectorisations produce different ``histories''; different definitions of which radiative phase-space points are mapped to which Born-level phase-space points. 
Here, we artificially consider only a single specific Born-level phase-space point. 
Our sectorisation then defines which hypersurface in the Born+$\gamma$ phase space that point is mapped onto. 
Sector boundaries correspond to discontinuities in that map; but since the full matrix element is smooth, the ME-corrected distribution considered as a function defined on the full Born+$\gamma$ phase space will also be smooth.

Another notable feature across both~\figsRef{fig:D+_3piAngularAntennae} and~\ref{fig:Bu_DKpiAngularAntennae} is that the only contributing sectors are final-final ones. 
This is due to all the resonance-final sectors being vetoed in the sectorisation process in the Mercedes configuration.

In summary, in the dipole approximation, non-singular contributions from other dipole antennae in the event can partly ``fill in'' the dead cones of the most suppressed massive radiators, while the multipole shower (with or without MECs) captures the full set of soft interference effects as well as (all of) the quasi-collinear limits. 

\section{Performance}
\label{sec:performance}
To evaluate the computational efficiency of the QED radiation generators we have considered in this work, we benchmark both CPU execution time and peak memory consumption through the \evtgen generator framework. 
The generation of events is conducted using \evtgen release 3.0.0, which provides interfaces to \photos release 3.64, \sherpa release 3.0.1 (with physics consistency verified against release 2.2.15), and \pythia 8.315. The software was compiled within a standardised LHC Computing Grid environment (\texttt{LCG\_107}, platform \texttt{x86\_64-el9-clang16-opt}) distributed via the Cern Virtual Machine File System (CVMFS)~\cite{Blomer:2011zz}. 

We evaluate benchmarking metrics for two operational regimes: single-threaded execution to establish baseline single-core performance, and multi-threaded processing to quantify scaling and speed-up behaviour enabled by recent multi-threading adaptations in \evtgen~\cite{Abudinen:2024wpc,evtgen:2025}. 
We perform benchmark tests for four representative decay topologies: \mbox{\decay{\jpsi}{\ep\en}}, \mbox{\decay{\jpsi}{\mumu}}, \mbox{\decay{\Bz}{\Kp\pim}}, and \mbox{\decay{\Dp}{\Kp\pip\pim}} decays. 
The wall-clock time is evaluated using \texttt{chrone} by measuring the elapsed time required to simulate $10^7$ events in \evtgen, with final-state radiation~(FSR) simulation enabled or disabled. 
When FSR is enabled, the individual QED plugins are benchmarked separately.
We assess the memory usage by measuring the peak resident set size (RSS) recorded during the generation of the $10^7$ events. 
All execution runs were performed on a dedicated benchmarking host equipped with an Intel Core i9-10980XE processor (18 physical cores, 36 logical threads, base clock frequency 3.00~GHz, maximum boost 4.80~GHz).

\begin{figure}[tp]
\centering
\begin{subfigure}{0.49\textwidth}
  \centering
  \includegraphics[width=\linewidth]{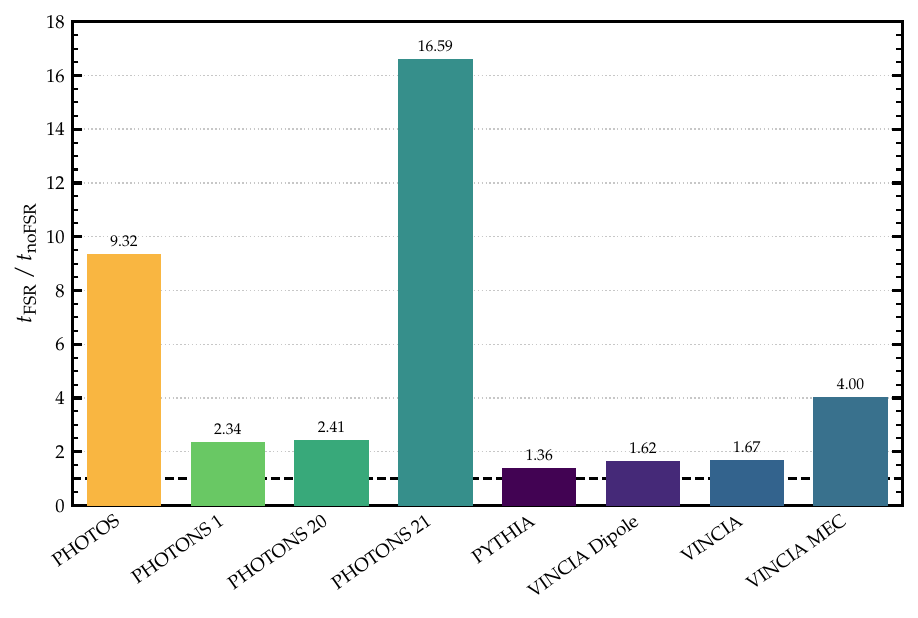}
  \caption{\decay{\jpsi}{\ep\en}\\}
  \label{fig:timing_Jpsi_ee}
\end{subfigure}
\hfill
\begin{subfigure}{0.49\textwidth}
  \centering
  \includegraphics[width=\linewidth]{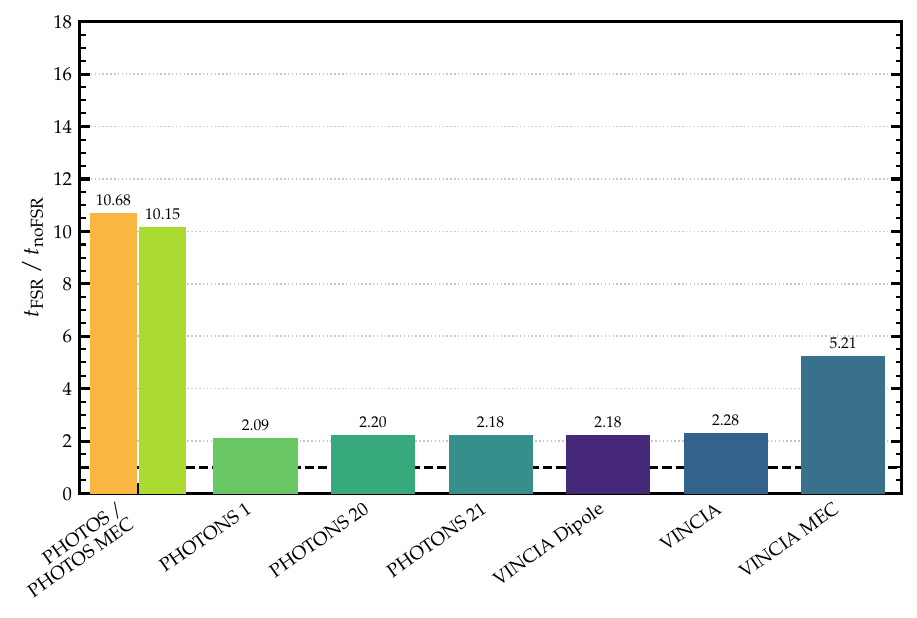}
  \caption{\decay{\Bz}{\Kp\pim}}
  \label{fig:timing_Bd_Kpi}
\end{subfigure}
\vspace{0.3cm}

\begin{subfigure}{0.49\textwidth}
  \centering
  \includegraphics[width=\linewidth]{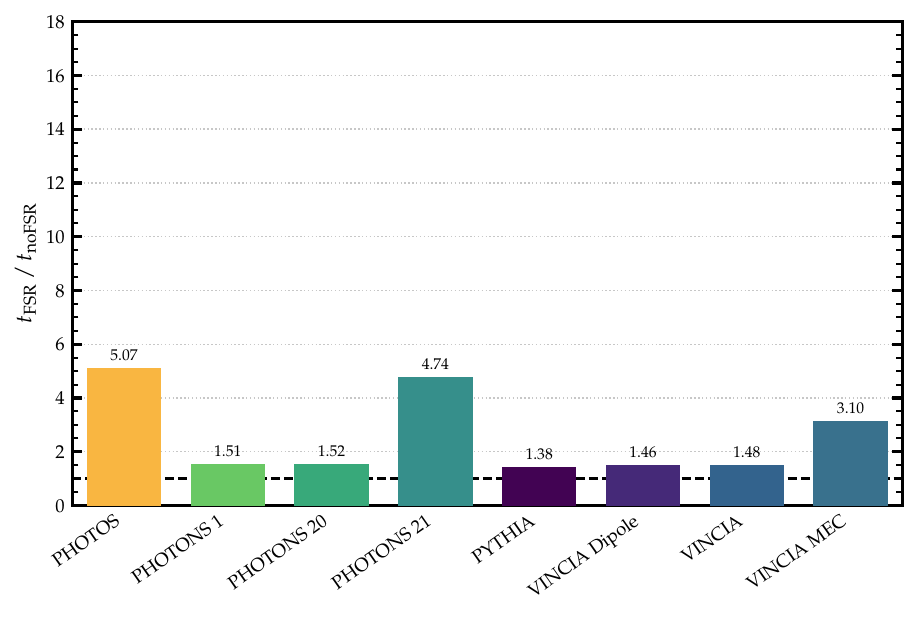}
  \caption{\decay{\jpsi}{\mumu}\\}
  \label{fig:timing_Jpsi_mumu}
\end{subfigure}
\hfill
\begin{subfigure}{0.49\textwidth}
  \centering
  \includegraphics[width=\linewidth]{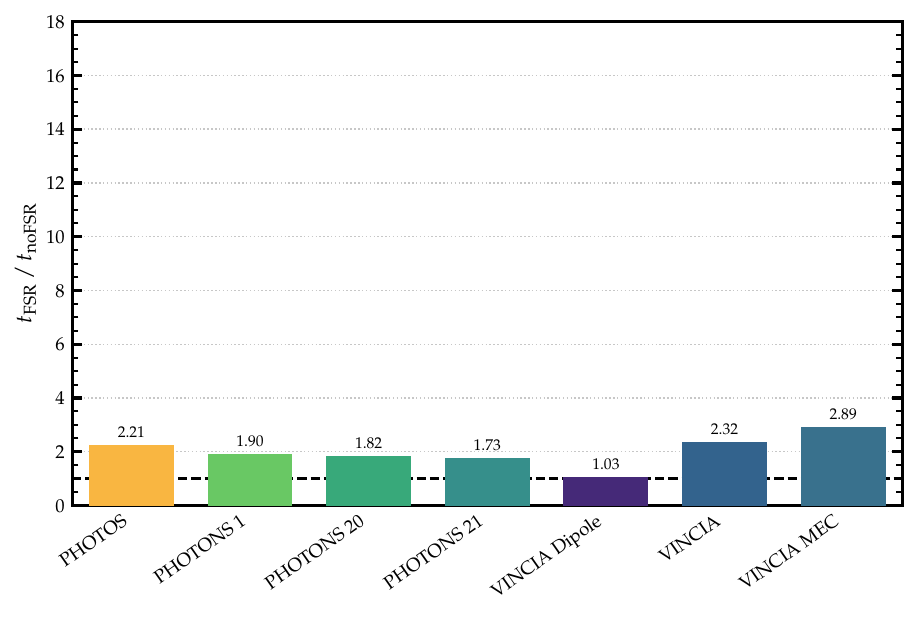}
  \caption{\decay{\Dp}{\Kp\pip\pim}}
  \label{fig:timing_Dp_Kppipi}
\end{subfigure}
    \caption{Overhead factor $t/t_{\mathrm{noFSR}}$ for single-thread execution.}
    \label{fig:timing}
\end{figure}

\TabRef{tab:timing} reports the elapsed time per event and the corresponding ratio of the runtime with FSR simulation relative to the baseline configuration without FSR, for single-threaded execution. The runtime ratios are also presented in~\figRef{fig:timing}.
\TabRef{tab:memory} reports the results obtained for a single-threaded execution. 
The difference in peak RSS between runs performed with and without FSR simulation is presented in \figRef{tab:memory}.

Overall, we find that the FSR simulation using \vincia achieves run times comparable to the fastest available plug-ins. 
When ME corrections are enabled, the execution time increases by approximately a factor of two.
Nevertheless, the resulting generator performance remains competitive. 
With respect to memory usage, enabling \vincia increases the memory footprint of the \evtgen simulation by only about $10\%$.

\begin{figure}[tp]
\centering
\begin{subfigure}{0.49\textwidth}
  \centering
  \includegraphics[width=\linewidth]{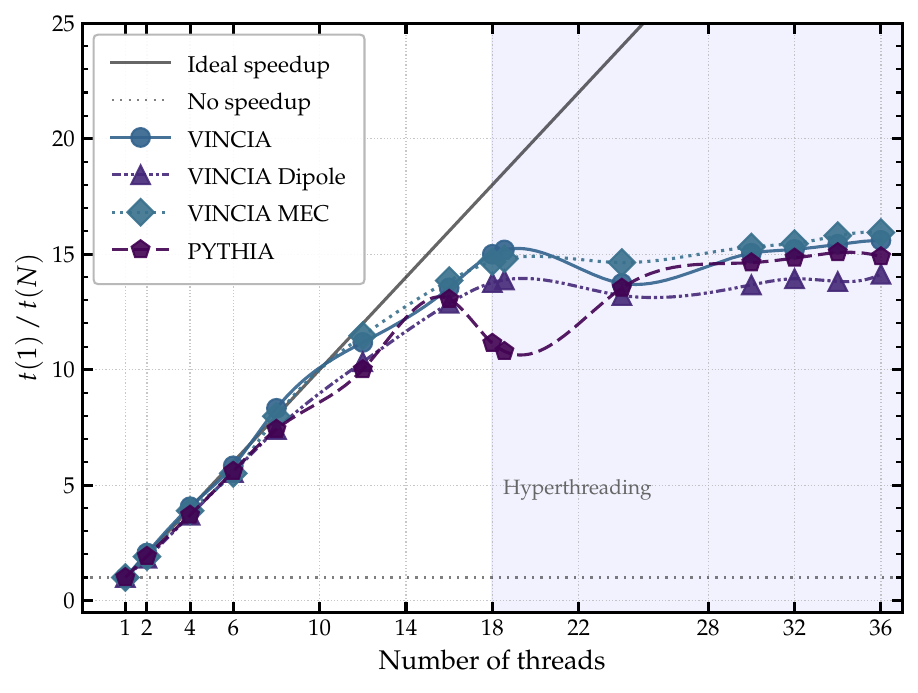}
  \caption{\vincia}
  \label{fig:speedup_Jpsi_ee_a}
\end{subfigure}
\hfill
\begin{subfigure}{0.49\textwidth}
  \centering
  \includegraphics[width=\linewidth]{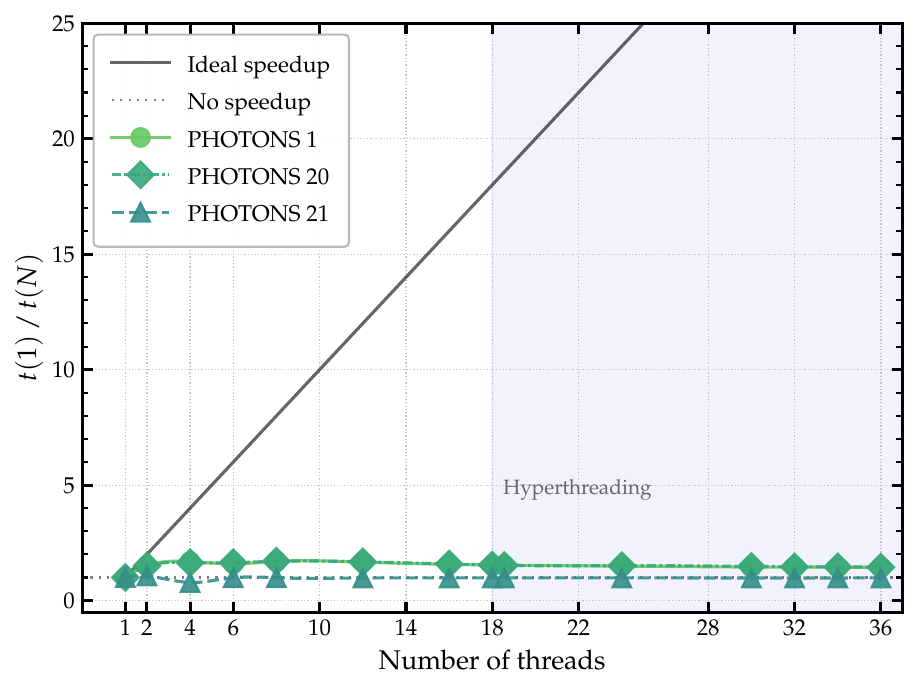}
  \caption{\photons}
  \label{fig:speedup_Jpsi_ee_b}
\end{subfigure}
\vspace{0.3cm}

\begin{subfigure}{0.49\textwidth}
  \centering
  \includegraphics[width=\linewidth]{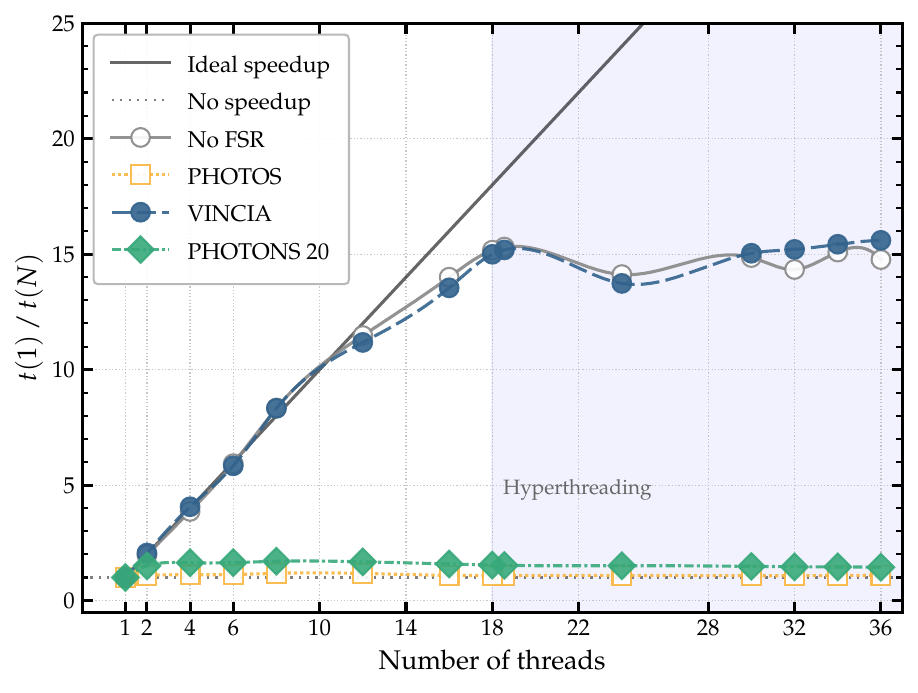}
  \caption{FSR without MEC vs. no FSR}
  \label{fig:speedup_Jpsi_ee_c}
\end{subfigure}
\hfill
\begin{subfigure}{0.49\textwidth}
  \centering
  \includegraphics[width=\linewidth]{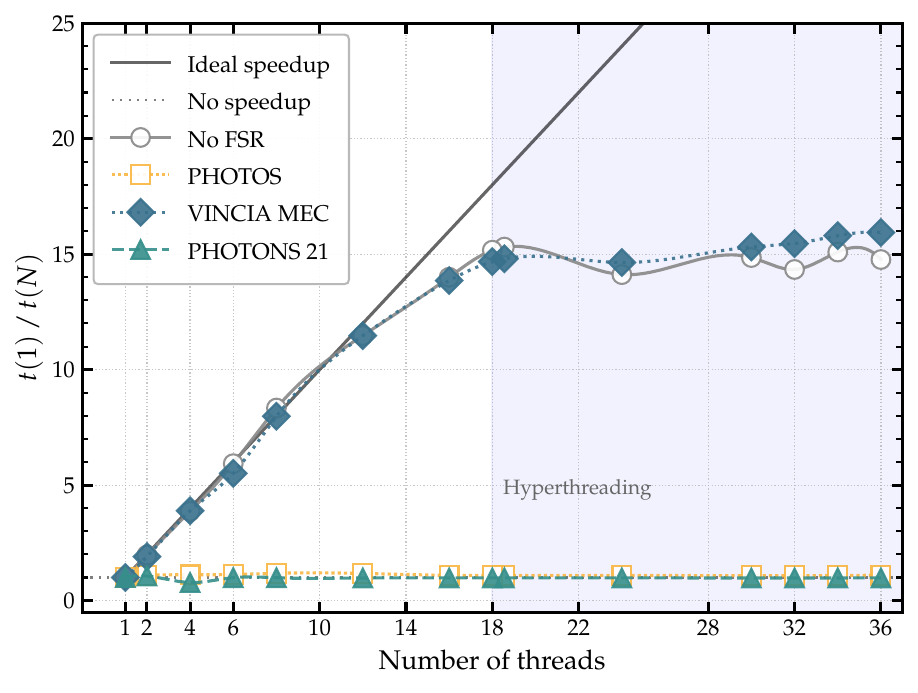}
  \caption{FSR with MEC (where available) vs. no FSR}
  \label{fig:speedup_Jpsi_ee_d}
\end{subfigure}
    \caption{Speed-up curves for \decay{\jpsi}{\ep\en} decays.}
    \label{fig:speedup_Jpsi_ee}
\end{figure}

\begin{figure}[tp]
\centering
\begin{subfigure}{0.49\textwidth}
  \centering
  \includegraphics[width=\linewidth]{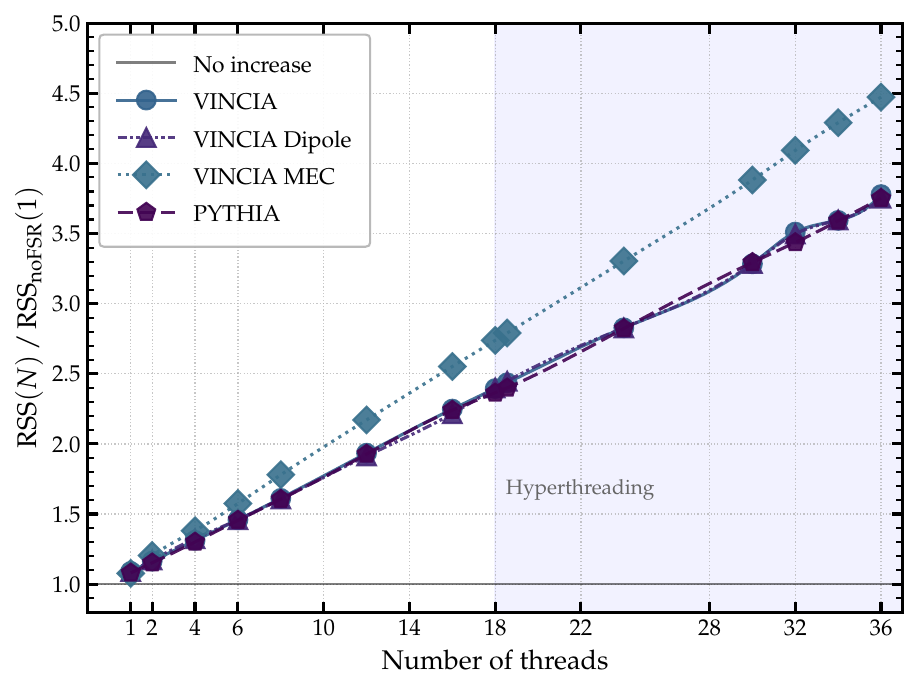}
  \caption{\vincia}
  \label{fig:memory_Jpsi_ee_a}
\end{subfigure}
\hfill
\begin{subfigure}{0.49\textwidth}
  \centering
  \includegraphics[width=\linewidth]{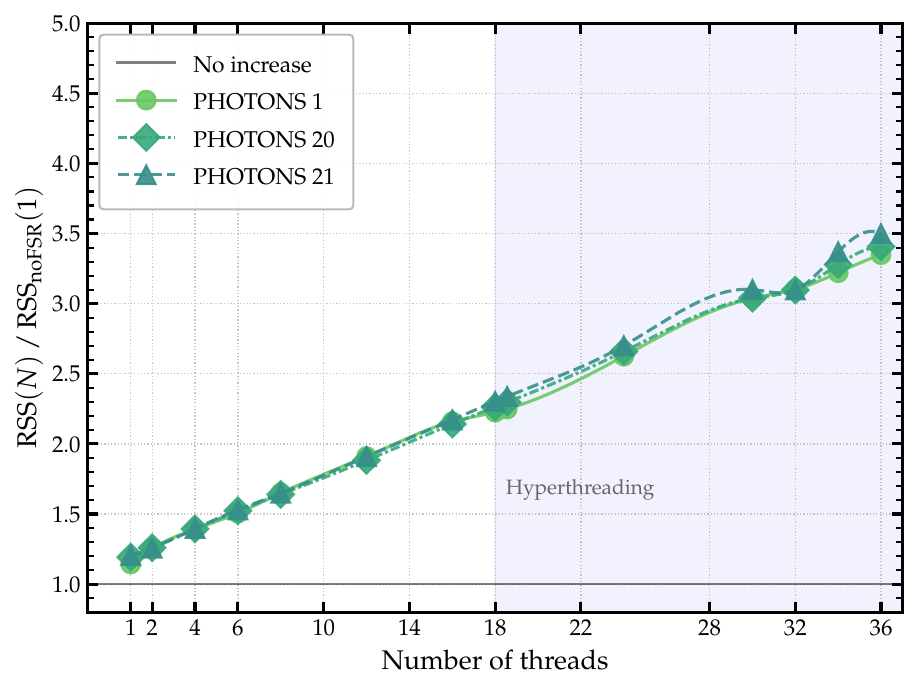}
  \caption{\photons}
  \label{fig:memory_Jpsi_ee_b}
\end{subfigure}
\vspace{0.3cm}

\begin{subfigure}{0.49\textwidth}
  \centering
  \includegraphics[width=\linewidth]{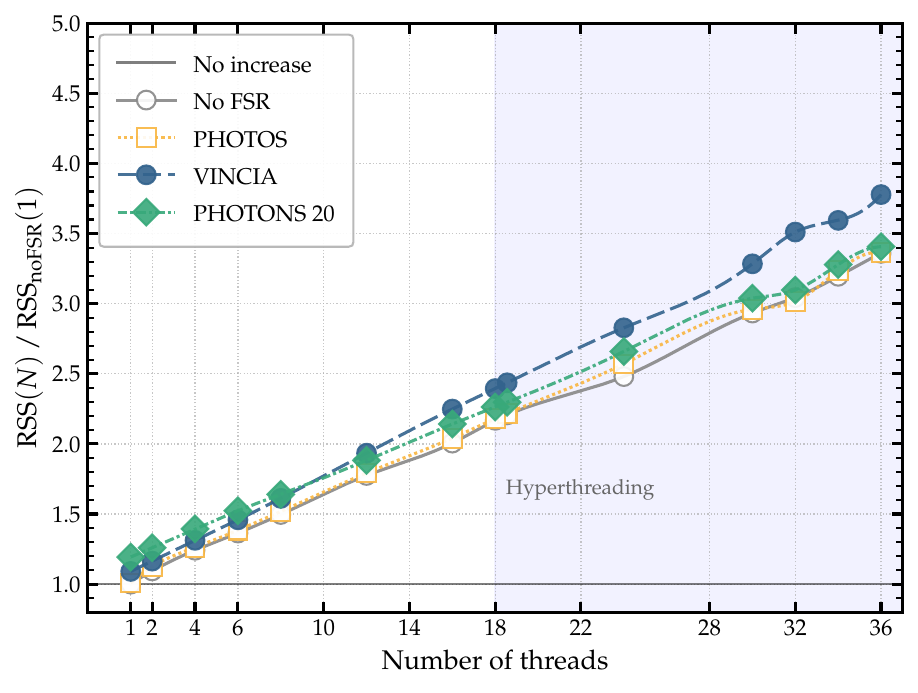}
  \caption{FSR without MEC vs. no FSR}
  \label{fig:memory_Jpsi_ee_c}
\end{subfigure}
\hfill
\begin{subfigure}{0.49\textwidth}
  \centering
  \includegraphics[width=\linewidth]{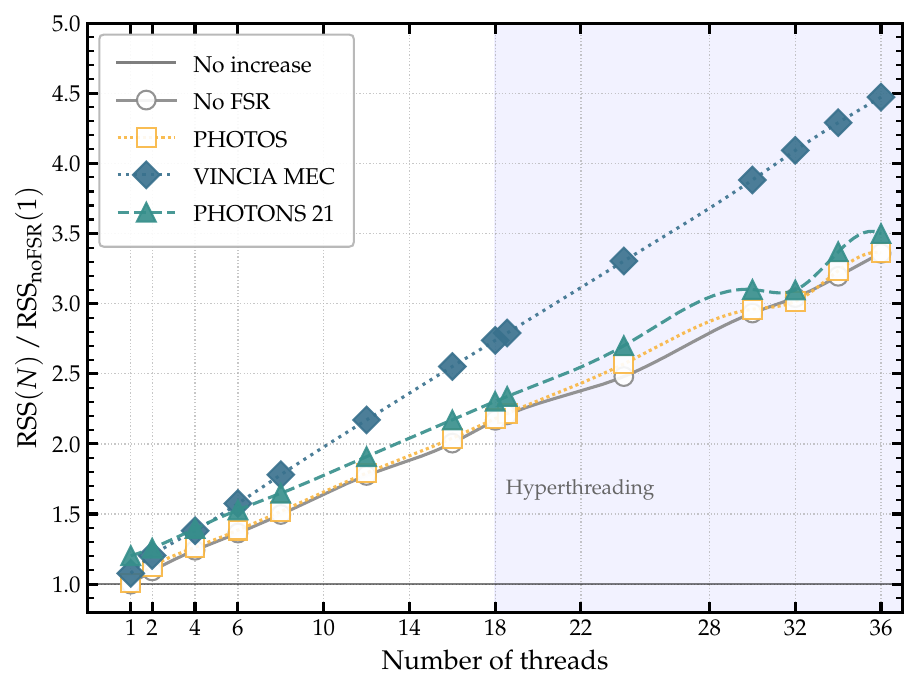}
  \caption{FSR with MEC (where available) vs. no FSR}
  \label{fig:memory_Jpsi_ee_d}
\end{subfigure}
    \caption{Maximum RSS memory consumption relative to single-thread execution for \mbox{\decay{\jpsi}{\ep\en}}~decays.}
    \label{fig:memory_Jpsi_ee_mt}
\end{figure}

Multi-threaded tests use the \cpp \texttt{std} threading library, with one \evtgen instance created per thread. 
In each test, the $10^7$ events are evenly distributed between the threads. For the \decay{\jpsi}{\ep\en} mode, \figRef{fig:speedup_Jpsi_ee} shows the speed-up, defined as the ratio of single-threaded to multi-threaded wall-clock time, while \figRef{fig:memory_Jpsi_ee_mt} shows the corresponding RSS memory consumption normalised to the single-threaded value.
The figures corresponding to the other modes are collected in \appRef{app:performance}. 
Although some performance differences are observed among the various decay modes, the overall behaviour is consistent with that found for the \decay{\jpsi}{\ep\en} channel.

The \photos and \photons libraries are not thread-safe.
Therefore, their invocations are serialised via a \texttt{thread lock} inside \evtgen. 
For \photos, both the event-record conversion and the shower call must be locked, whereas for \photons only the shower call requires serialisation. 
As a result, multi-threading yields negligible speed-up for these generators, as threads stall waiting for the lock.

When \vincia or \pythia are used, \evtgen creates one \pythia instance per thread. 
Since these instances are thread-local, no serialisation is required, and a near-ideal linear speed-up is achieved. 
The cost is a $\sim\!10$--$30\%$ increase in memory per thread compared to the case without FSR, reflecting the \pythia initialisation per thread. 
This overhead is modest and well within the per-core cache capacity of modern processors. 
As expected, the linear scaling does not extend to the hyper-threaded regime, where physical cores are shared between logical threads.

In general, \vincia performs competitively, even with ME corrections, while significantly reducing the CPU time of FSR simulation in \evtgen compared to the current default configuration. 
In multithreaded-threaded runs, \vincia enables near-ideal \evtgen speed-up with a small memory increase, provided it operates in the normally multithreaded regime (no hyper-threading).
With hyper-threading, the speed-up is still significant but less linear and more saturated.

\begin{table}[htbp]
  \centering
  \begin{tabular}{l c c c c c c c c }
  \toprule
  FSR option & \multicolumn{2}{c}{\decay{\jpsi}{\ep\en}} & \multicolumn{2}{c}{\decay{\jpsi}{\mumu}} & \multicolumn{2}{c}{\decay{\Bz}{\Kp\pim}} & \multicolumn{2}{c}{\decay{\Dp}{\Kp\pip\pim}} \\
   & ms/ev & \multicolumn{1}{c}{$t/t_{\mathrm{noFSR}}$} & ms/ev & $t/t_{\mathrm{noFSR}}$ & ms/ev & $t/t_{\mathrm{noFSR}}$ & ms/ev & $t/t_{\mathrm{noFSR}}$ \\
  \midrule
  No FSR & 0.0112 & \phantom{1}1.00 & 0.0112 & 1.00 & 0.0046 & \phantom{1}1.00 & 0.0462 & 1.00 \\
  \midrule
  \photos & 0.1043 & \phantom{1}9.32 & 0.0567 & 5.07 & 0.0493 & 10.68 & 0.1195 & 2.59 \\
  \photos MEC & --- & --- & --- & --- & 0.0468 & 10.15 & --- & --- \\
  \photons~1 & 0.0262 & \phantom{1}2.34 & 0.0169 & 1.51 & 0.0096 & \phantom{1}2.09 & 0.0877 & 1.90 \\
  \photons~20 & 0.0269 & \phantom{1}2.41 & 0.0170 & 1.52 & 0.0101 & \phantom{1}2.20 & 0.0841 & 1.82 \\
  \photons~21 & 0.1855 & 16.59 & 0.0531 & 4.74 & 0.0101 & \phantom{1}2.18 & 0.0798 & 1.73 \\
  \pythia & 0.0153 & \phantom{0}1.36 & 0.0155 & 1.38 & --- & --- & --- & --- \\
  \vincia Dipole & 0.0181 & \phantom{1}1.62 & 0.0164 & 1.46 & 0.0101 & \phantom{1}2.18 & 0.0476 & 1.03 \\  
  \vincia & 0.0187 & \phantom{1}1.67 & 0.0166 & 1.48 & 0.0105 & \phantom{1}2.28 & 0.1075 & 2.32 \\
  \vincia MEC & 0.0447 & \phantom{1}4.00 & 0.0347 & 3.10 & 0.0240 & \phantom{1}5.21 & 0.1334 & 2.89 \\
  \bottomrule
  \end{tabular}
    \caption{Absolute timing (ms per event) and overhead factor $t/t_{\mathrm{noFSR}}$ for single-thread execution. 
    Missing entries indicate the configuration is not applicable for that decay topology.}
  \label{tab:timing}
\end{table}

\begin{table}[htbp]
  \centering
  \begin{tabular}{lcccc}
  \toprule
  FSR option & \decay{\jpsi}{\ep\en} & \decay{\jpsi}{\mumu} & \decay{\Bz}{\Kp\pim} & \decay{\Dp}{\Kp\pip\pim} \\
  \midrule
  No FSR         & 255.7 & 255.0 & 255.8 & 257.7 \\\midrule
  \photos        & 257.8 & 258.3 & 256.1 & --- \\
  \photos MEC    & ---   & ---   & 256.3 & --- \\\midrule
  \photons~1     & 292.5 & 305.5 & 302.0 & 306.6 \\
  \photons~20    & 305.0 & 306.3 & 303.7 & 306.5 \\
  \photons~21    & 307.6 & 307.4 & 304.1 & 306.5 \\\midrule
  \pythia        & 276.1 & 275.6 & ---   & --- \\\midrule
  \vincia Dipole & 278.8 & 277.6 & 275.9 & 278.1 \\  
  \vincia        & 278.8 & 277.1 & 274.5 & 278.4 \\
  \vincia MEC    & 275.3 & 275.9 & 272.1 & 288.3 \\
  \bottomrule
  \end{tabular}
    \caption{Maximum resident set size memory (max RSS, in MB) for single-thread execution ($10^7$ events). 
    Missing entries indicate the configuration is not applicable for that decay topology.}
  \label{tab:memory}
\end{table}

\begin{figure}[tp]
\centering
\begin{subfigure}{0.49\textwidth}
  \centering
  \includegraphics[width=\linewidth]{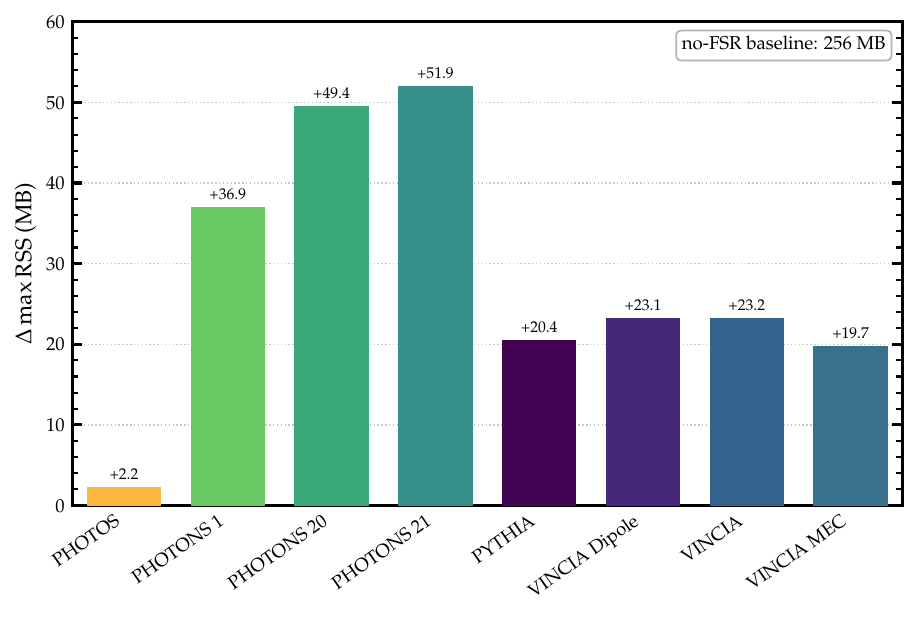}
  \caption{\decay{\jpsi}{\ep\en}\\}
  \label{fig:memory_Jpsi_ee}
\end{subfigure}
\hfill
\begin{subfigure}{0.49\textwidth}
  \centering
  \includegraphics[width=\linewidth]{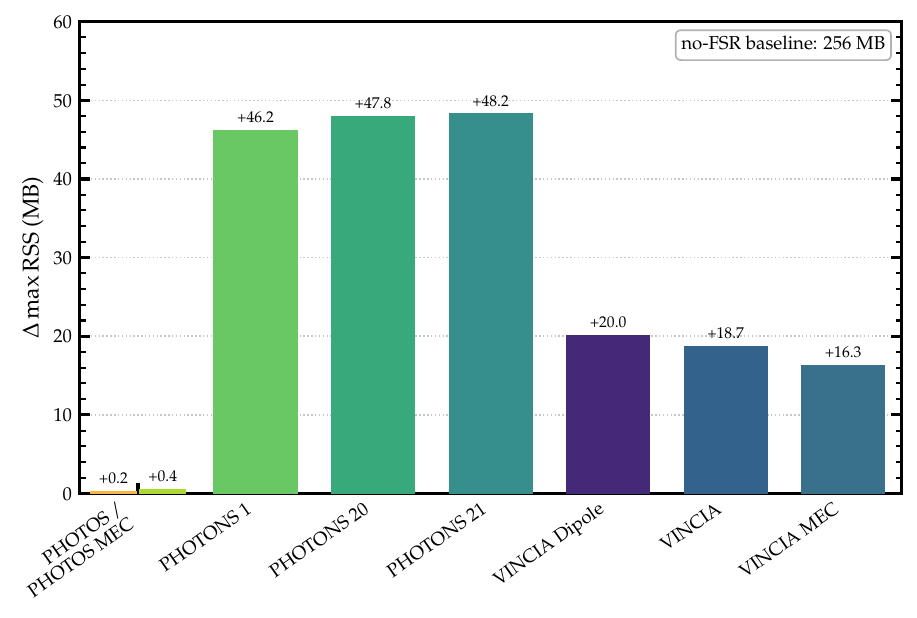}
  \caption{\decay{\Bz}{\Kp\pim}}
  \label{fig:memory_Bd_Kpi}
\end{subfigure}
\vspace{0.5cm}

\begin{subfigure}{0.49\textwidth}
  \centering
  \includegraphics[width=\linewidth]{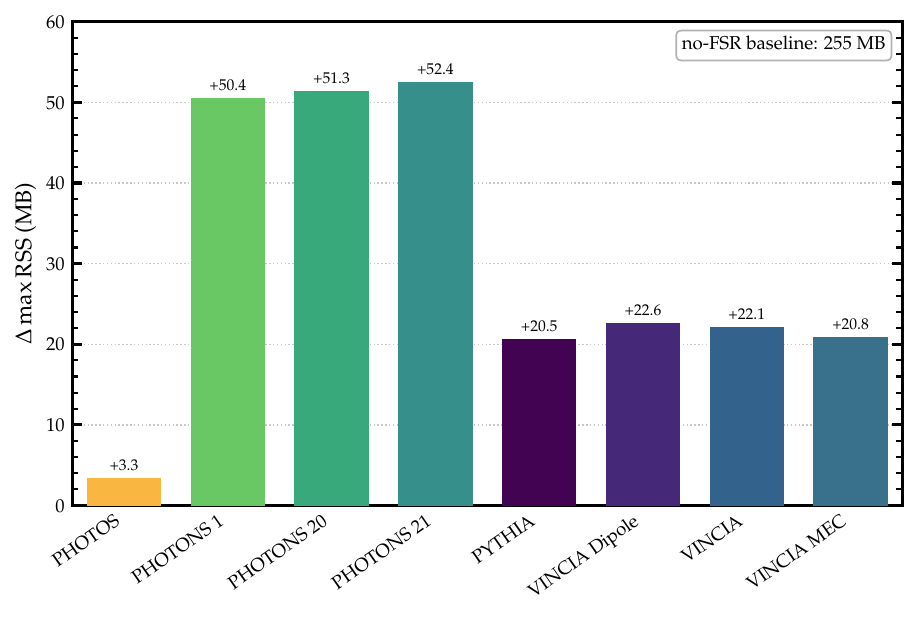}
  \caption{\decay{\jpsi}{\mumu}\\}
  \label{fig:memory_Jpsi_mumu}
\end{subfigure}
\hfill
\begin{subfigure}{0.49\textwidth}
  \centering
  \includegraphics[width=\linewidth]{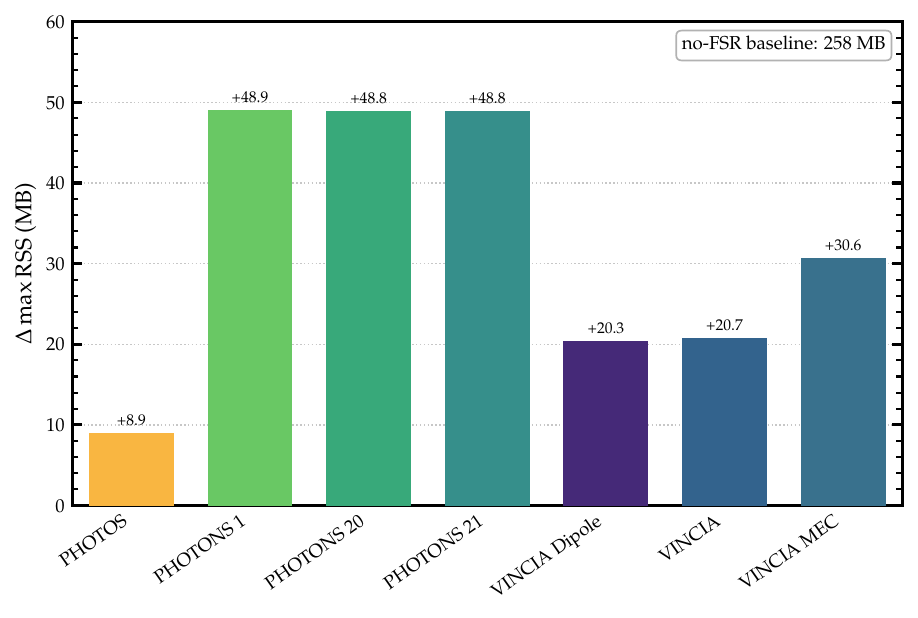}
  \caption{\decay{\Dp}{\Kp\pip\pim}}
  \label{fig:memory_Dp_Kppipi}
\end{subfigure}
    \caption{Difference in RSS memory between simulation with and without FSR simulated for single-threaded execution. }
    \label{fig:memory}
\end{figure}

\clearpage

\section{Summary and outlook}
\label{sec:summary}

Precise descriptions of hadron decays require a reliable treatment of the QED radiation that accompanies them. 
We have presented a new framework for QED antenna showers in hadron decays, implemented in the \vincia shower module~\cite{Brooks:2020upa} of \pythia~8~\cite{Bierlich:2022pfr}. 

Our antenna functions include full mass dependence and spin-dependent (quasi-)collinear terms for
spin-$0$, spin-$1/2$, and spin-$1$ emitters (spin-$3/2$ ones are so far treated the same as spin-$1/2$ ones). 
For soft photon emissions, our antenna functions reproduce the full coherent
multipole eikonal structure~\cite{Kleiss:2017iir,Skands:2020lkd} with the same fidelity as the YFS formalism~\cite{Yennie:1961ad}. 
Unlike YFS, however, they also reproduce the correct DGLAP kernels in collinear limits. 
By default, the full soft multipole pattern is included via phase-space sectorisation~\cite{Lopez-Villarejo:2011pwr,Skands:2020lkd}; 
a faster dipole option based on maximal screening is also available.

To improve the accuracy of the shower, in particular for semi-hard emissions (which can be important since they can be associated with significant photon energy losses), we have developed a semi-automated tool chain for iterated tree-level ME corrections (MECs). 
Starting from an effective Lagrangian for generic hadronic degrees of freedom, implemented in \textsc{FeynRules}, we use \textsc{MadGraph} to produce a library of plugins that \textsc{Vincia} calls at run time. 
By default, this library currently includes MECs for up to two photon emissions, for decays involving only
(pseudo)scalars and for decays of (pseudo)scalars and vectors to lepton pairs,
see app.~\ref{app:MEs}. 
Since the ME correction factor is the ratio of the full
matrix element to the shower kernel, the corrected result does not depend on the
choice of baseline antenna function. 
We illustrated this by varying the
nonsingular term of the antenna function for \decay{\jpsi}{\mumu} and
\decay{\Bs}{\mumu} decays.

We also advocated $\cos\theta_\mathrm{dead} = \beta$ as an alternative definition of the dead-cone angle, which reduces to the conventional $1/\gamma$ for relativistic emitters, but remains somewhat more meaningful for the only mildly relativistic decay products that are typical in many hadron decays, such as the \D~meson in \decay{\PB}{\PD\PK}.

In \secRef{sec:IR}, we emphasised that the number of emitted photons is not an infrared safe quantity; it depends entirely on the infrared cutoff and on the variable it is imposed on, and both differ between generators. 
Only infrared-safe quantities, defined with respect to a common notion of a resolved photon, can be meaningfully compared, with cutoffs chosen sufficiently low to resolve the full phase space of such photons. 
In this context, we studied the effect of varying the cutoff values in various generators and observed an
unexplained suppression of the total emitted photon energy in \photos~\cite{Barberio:1990ms,Barberio:1993qi,Golonka:2005pn,Nanava:2006vv,davidson2015photos} for
\decay{\jpsi}{\ep\en} with cutoffs below $10\,$\kev.

Using \textsc{EvtGen}~\cite{Lange:2001uf,Abudinen:2024wpc,evtgen:2025} to provide identical Born-level events, we compared \vincia to \photos and \sherpa's \photonspp module~\cite{Schonherr:2008av} for a range of two- and three-body decays.
\pythia's internal QED shower was also included for some of these comparisons, although, by construction, it treats all radiators as spin-$1/2$ particles. 
For the total radiated energy, and for the fractions of events losing more than $1$, $10$, and $100\,$\mev,
\textsc{Vincia} typically agrees with \photons with first-order MECs to within a few percent. 
The second-order MECs appear to consistently increase the average
energy loss, by about $1$--$3\%$.

For decays to three charged particles, the dipole approximation overestimates the average energy loss of the multipole shower by $5$--$20\%$. 
This indicates that the
multipole corrections are predominantly destructive (at least relative to the maximal-screening dipole approximation). 
In a fixed Mercedes configuration~(see \secRef{sec:fixed3}), the dipole approximation also partly fills in the dead cones,
since the ``other'' dipoles are allowed to radiate into them. 

Regarding computational performance, \vincia achieves single-core run times comparable to the currently available final-state radiation (FSR) plug-ins (\sherpa's \photons module and \photos), with ME corrections increasing the execution time by approximately a factor of two while remaining competitive overall. 
The memory footprint of \evtgen increases only by $\sim\!10\%$ when \vincia is enabled. 
In multi-threaded operation, the \photos and \photons libraries require serialisation of their calls via a thread lock, yielding negligible speed-up with increasing thread count. 
In contrast, \vincia and \pythia can be used as thread-local instances, achieving a near-ideal linear speed-up at the cost of a $\sim\!10$--$30\%$ increase in memory per thread relative to the no-FSR baseline, which is well within the per-core cache capacity of modern processors.

Several extensions are open for future work. 
These include radiation from spin-$3/2$ baryons, MECs for decays
involving vector mesons and baryons, hadronic form factors, vector-meson-dominance contributions, exact one-loop corrections, and an
interleaved treatment of QED radiation with hadron decays. 
The latter of these would build on generalising \vincia's interleaved resonance decays~\cite{Brooks:2021kji} to
intermediate hadronic resonances (such as intermediate \Prho or $\PK^{*}$ states), so that
short-lived states radiate only down to scales of order their widths. 
The resulting modification can induce subtle shifts in reconstructed invariant-mass distributions, relative to the conventional (non-interleaved) treatments of resonance decays.

The present framework, for pseudoscalars, will be included in the next \pythia release (tentatively 8.319). 

\subsection*{Acknowledgements}

We are grateful to M.~Kreps, T.~Latham, and J.~Back for their valuable assistance in reviewing the \evtgen-\vincia interface code and for their insightful discussions. 
We also gratefully acknowledge fruitful discussions related to this work with U.~Egede, G.~Valencia, M.~Schoenherr, and P.~Richardson.
Regarding the technical aspects of multithreading, we also acknowledge valuable discussions and technical support provided by the Research Software Engineers at the University of Warwick, Heather Ratcliffe and Christopher Brady.
FA is supported by UKRI FLF grant MR/W009048/1.
PS is supported by ARC grant DP230103014. 
GM is supported by the Commonwealth through an Australian Government Research Training Program Scholarship [DOI: https://doi.org/10.82133/C42F-K220].
This work was also supported by the Monash-Warwick Alliance in Particle Physics~(MWAPP). 

In formulating the summary and outlook section (and only that section), we made partial use of suggestions from Claude AI. All statements were explicitly edited by the authors, and we take full responsibility for the statements made in that section.

\appendix

\section{Implementation of the QED shower algorithm}

\label{app:QEDshower}

A description of the antenna formalism underlying \vincia’s QED shower is given in \secRef{sec:vincia}. 
The shower is embedded in \pythia and can be accessed via the generic shower models of \pythia (base classes \texttt{ShowerModel}, \texttt{TimeShower}, and \texttt{SpaceShower}) in three main ways:
\begin{enumerate}
\item As part of the standard interleaved QCD+QED perturbative evolution of a partonic hard-scattering event, via \vincia's main \texttt{prepare()}, \texttt{pTnext()}, and \texttt{branch()} methods;
\item As a continuation of the evolution of hadronic collisions, after adding beam remnants but before hadronisation and particle decays, via \texttt{showerQEDafterRemnants()};
\item As a plugin for QED showering in hadron decays via the \texttt{showerQED()} method, callable within \pythia at hadron decay generation or directly from external programmes such as \evtgen.
\end{enumerate}
The primary QED radiation algorithm for hadron decays is \pythia’s \texttt{TimeShower}-inherited \texttt{showerQED()} method, with the signature:
\begin{center}
\texttt{showerQED(int iBeg, int iEnd, Event\& event, double pTmax, int nBranchMax);}
\end{center}

\begin{figure}[H]
    \centering
    \includegraphics[width=0.75\linewidth]{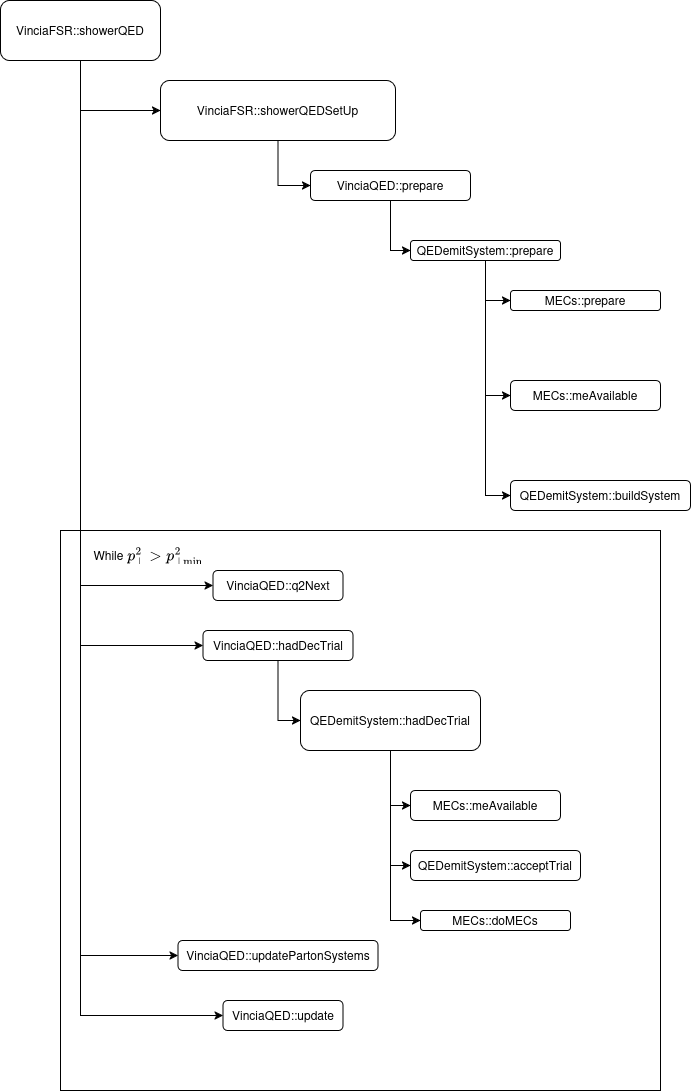}
    \caption{A diagram showing the key function calls in \texttt{showerQED} for emitting a photon with matrix-element corrections. 
    Arrows coming out of the bottom of a block show a function called by the block.}
    \label{fig:codeFlow}
\end{figure}

\FigRef{fig:codeFlow} shows an overview of the functions in the \texttt{showerQED} algorithm.
As a first step, the procedure calls a preparation routine called \texttt{VinciaQED::prepare}, which discriminates between hadronic decays and configurations involving leptons and quarks. 
This distinction is required because, in the case of bare partonic constituents, additional interactions (such as QCD) may need to be interleaved. 
Prepare also determines whether the initial system provided has MECs (\texttt{MECs::prepare} and \texttt{MECs::meAvailable}) and builds the internal systems (\texttt{buildSystem}).  
The \texttt{buildSystem}\footnote{In all operational modes, this is the stage at which QED emissions are constructed} function includes different implementations\footnote{Here different implementations refers to if-else control rather than providing multiple definitions for the method.} for the multipole and dipole shower configurations. 
The choice between multipole and dipole  configurations is controlled by the user. 

In the dipole configuration, the \texttt{buildSystem} function aggregates all charges into a single vector of vectors, grouped according to their absolute value and sign.  
If the charges correspond to the initial state, their signs are inverted (reflecting their effective role in the antenna picture, cf. \EqRef{eq:QEDAntenna}).  
Subsequently, the Hungarian algorithm is applied, which guarantees a matching between charges of opposite sign and equal magnitude.  
These pairs are then stored in a vector that represents the complete collection of dipoles.
In the multipole case\footnote{The multipole shower is also invoked in dipole mode after the dipole routine whenever residual fractional electric charges ($1/3$, $2/3$) remain. This implementation detail, however, is not relevant for the hadronic decays considered in this work.}, the \texttt{buildSystem} function constructs a matrix of dipoles that enumerates all possible charge combinations.

As a next step, the shower evolution (\texttt{showerQED}) is initiated at a scale set by the mass of the parent particle. 
The algorithm now enters a loop which starts with \texttt{VinciaQED::q2Next} choosing a scale for the next emission\footnote{q2Next also determines whether the chosen $\pTs$ scale is associated with a photon emission, photon splitting or photon conversion. Though here we only concern ourselves with emissions.}.  
At each iteration, the loop goes over all possible charge-pair configurations.
In the dipole case, the loop runs over the charged pairs obtained from the Hungarian algorithm~\cite{kuhn1955hungarian,munkres1957algorithms,jonker1987shortest}.  
In the multipole case, the loop instead runs over the elements of the matrix of dipoles, which is not endowed with any particular ordering.  
 
For each antenna, the algorithm generates a trial emission and selects as the physical emission the trial with the largest transverse momentum \pT, designating it as the kinematic dipole.  
It then proceeds to the \texttt{QEDemitSystem::hadDecTrial} step, which checks the availability of MECs for the current system if requested\footnote{In case the last successful trial happened to be for example a photon splitting}. 
Finally, \texttt{QEDemitSystem::acceptTrial} accepts or rejects the trial emission based on the antenna function value and, in the multipole case, a sector check that avoids double counting.
When the ME correction mode is enabled, the splitting kernel is constructed using the matrix elements provided by the \madgraph plugins.
Otherwise, the standard antenna function is employed. 
In the ME correction mode, the uncorrected antenna function is used for the phase-space sectorisation, and the subsequent accept/reject step for photon emission is performed on the basis of the exact matrix elements. 
If the event is accepted, a photon is generated and the event record is updated accordingly (\texttt{QEDemitSystem::updatePartonSystems} and \texttt{VinciaQED::update}). 

In the next iteration, the $\pTs$ value starts at the scale of the last trial emission, regardless of whether it was accepted or not.
This pattern continues iteratively. 
If, after a given iteration, the $\pTs$ values of all trials lie below the cutoff, then the algorithm terminates.

\section{Matrix elements for corrections}
\label{app:MEs}

To implement the ME corrections~(see~\secRef{sec:mecs}), we specify the complete set of particles and encode their interactions in terms of Lagrangian densities in \textsc{Mathematica} syntax~\cite{Mathematica}. 
In this way, the particle content and the structure of the interaction terms are fully defined. 
This model definition is then passed to \feynrules~\cite{Degrande:2011ua,Alloul:2013bka}. 
Within \feynrules we formulate an effective field theory (EFT) in terms of generic hadronic degrees of freedom, characterised primarily by their electric charges and spins. 
\feynrules expands the EFT Lagrangian densities, systematically enumerates all admissible vertex topologies implied by the interaction terms, and writes them into a universal \textsc{UFO} file~\cite{Degrande:2011ua}.

The resulting \textsc{UFO} model is imported into, together with a set of run cards that specify the physical processes to be studied (including, for example, the decay chain structure and QED perturbative order). 
\madgraph~\cite{Alwall:2011uj,Alwall:2014hca} is used to construct all combinations of interaction vertices consistent with the specified process, generating all contributing Feynman diagrams. 
For each diagram, \madgraph produces numerical routines corresponding to the given amplitude, which can be accessed from within \vincia via a dedicated run-time plugin interface to \madgraph libraries.

For each specified process, \madgraph generates a dedicated \cpp plugin. 
This plugin contains routines that iterate over all individual amplitudes, compute them, and assemble the total squared matrix element. 
The plugin output is used as input to \vincia, where both the pre- and post-branching matrix elements are evaluated. 
Their ratio is the ME-corrected antenna function. 

In our setup, the antenna functions are computed using tree-level matrix elements obtained from the combined \feynrules + \madgraph framework.  
These matrix elements are then employed to define the antenna radiation pattern, in place of traditional dipole or multipole approximations.
The full suite of generic matrix elements currently included for corrections in \vincia is listed below. 
For all modes, tree-level matrix elements for up to two-photon emissions are included. 
We list (pseudo)scalar particles as $S$, vectors as $V$, and leptons by their species. 
Particles sharing the same charge type and spin but different masses are distinguished with a subscript;
The subscript labels distinct mass states with the same charge type and spin (Lorentz structure) and carries no implied particle ordering.
For example, in the \decay{\Bp}{\Dp\Km\pip} decay the \D~meson would be labelled $S^\pm_2$ while in the \decay{\Dp}{\pip\pip\pim} decay the \D~meson would be labelled $S^\pm_1$.
This convention allows matrix elements to be reused across equivalent spin-charge 
topologies that differ only in their mass assignments, e.g \decay{\Bz}{\Dp\pim} and \decay{\Bz}{\Kp\pim}, making the library more compact.

\begin{enumerate}  
 \item Neutral vector to leptons:
 \begin{itemize}
  \item \decay{V^0}{\ep\en};
  \item \decay{V^0}{\mumu}.
 \end{itemize}

\item Neutral scalar to leptons:
\begin{itemize}
\item \decay{S_1^0}{\ep\en};
\item \decay{S_1^0}{\mumu}.
\end{itemize}

\item Charged scalar to leptons and (anti-)neutrinos:
\begin{itemize}
    \item \decay{S_1^+}{\ep\neue};
    \item \decay{S_1^+}{\mup \neum};
    \item \decay{S_1^-}{\en \neueb};
    \item \decay{S_1^-}{\mun \neumb}.
\end{itemize}

\item Neutral scalar to two charged scalars:
\begin{itemize}
    \item \decay{S_1^0}{S_1^+ S_1^-} (\eg, \decay{\Bz}{\Kp\Km}, \decay{\Dz}{\pip\pim}, \decay{\KS}{\pip\pim});
    \item \decay{S_1^0}{S_1^+ S_2^-}  (\eg, \decay{\Dzb}{\Kp\pim}, \decay{\Bzb}{\Dp\Km}).
\end{itemize}

\item Neutral scalar to two charged scalars and a neutral scalar:
\begin{itemize}
    \item \decay{S_1^0}{S_2^0 S_1^+ S_2^-}(\eg, \decay{\Bz}{\Dzb\Kp\pim}, \decay{\Dzb}{\Kzb\Kp\pim});
    \item \decay{S_1^0}{S_2^0 S_1^+ S_1^-} (\eg, \decay{\Bz}{\Dzb\pip\pim}, \decay{\Bz}{\piz\pip\pim}, \decay{\KL}{\piz\pip\pim}).
\end{itemize}

\item Charged scalar to a charged scalar and a neutral scalar:
\begin{itemize}
    \item \decay{S_1^{\pm}}{S_2^{\pm} S_1^0}(\eg, \decay{\Bpm}{\pipm\piz}).
\end{itemize}

\item Charged scalar to three charged scalars:
\begin{itemize}
    \item \decay{S_1^{\pm}}{S_2^{\pm} S_2^{\pm} S_2^{\mp}} (\eg, \decay{\Dpm}{\pipm\pipm\pimp});
    \item \decay{S_1^{\pm}}{S_3^{\pm} S_2^{\pm} S_2^{\mp}} (\eg, \decay{\Dpm}{\Kpm\pipm\pimp});
    \item \decay{S_1^{\pm}}{S_4^{\mp} S_3^{\pm} S_2^{\pm}} (\eg, \decay{\Bpm}{\Dmp\Kpm\pipm}).
\end{itemize}

\item Charged scalar to one charged scalar and two neutral scalars:
\begin{itemize}
    \item \decay{S_1^{\pm}}{S_2^{\pm} S_1^0 S_1^0} (\eg \decay{\Dpm}{\pipm\piz\piz}).
\end{itemize}

\end{enumerate}

\section{Deadcone derivations}

We provide complete derivations of the expressions used in \secRef{sec:deadConeEffect}: upper-bound on $y^*$ provided in eq.~\eqRef{eq:yStarMax} and eq.~\eqRef{eq:yStarMaxmK}. 
We start with the definition of the massive antenna function and set it to zero:
\begin{equation*}
    \frac{s_{ik}}{s_{i\gamma}s_{s_{\gamma k}}} - \frac{m_I^2}{s_{i\gamma}^2} - \frac{m_K^2}{s_{\gamma k}^2} = 0\,.
\end{equation*}
We define $r = \frac{s_{jk}}{s_{ij}}$ and rearrange to obtain,
\begin{equation*}
\begin{split}
    \frac{s_{ik}}{s_{i\gamma}s_{\gamma k}} - \frac{m_I^2}{s_{i\gamma}^2} - \frac{m_K^2}{s_{\gamma k}^2} &= 0\,,\\
    \frac{s_{ik}s_{\gamma k}}{s_{i\gamma}} - \frac{m_I^2s_{\gamma k}^2}{s_{i\gamma}^2} - m_K^2 &= 0\,,\\
    s_{ik}r-r^2m_I^2-m_K^2&=0\,.
\end{split}
\end{equation*}
Here, we can solve for $r$ to obtain
\begin{equation*}
\begin{split}
    r &= \frac{s_{ik}\pm\sqrt{s_{ik}^2-4m_I^2m_K^2}}{2m_I^2}\,.\\
\end{split}
\end{equation*}
Using the relationship $s_{IK} = s_{i\gamma} +s_{\gamma k} + s_{ik}$ leads to,
\begin{equation*}
\begin{split}
    r= \frac{s_{IK} - s_{i\gamma} - s_{\gamma k}\pm\sqrt{(s_{IK} - s_{i\gamma} - s_{\gamma k})^2-4m_I^2m_K^2}}{2m_I^2}\,.
\end{split}
\end{equation*}
Using the relationships $E_j = \frac{s_{i\gamma}+s_{\gamma k}+2m_\gamma^2}{2m_{IK}}$ and $m_{IK}^2 = s_{IK}+m_I^2+m_K^2$ gives,
\begin{equation*}
\begin{split}
    r&= \frac{s_{IK} - s_{i\gamma} - s_{\gamma k}\pm\sqrt{(s_{IK} - s_{i\gamma} - s_{\gamma k})^2-4m_I^2m_K^2}}{2m_I^2}\,,\\
    &= \frac{s_{IK} - 2E_\gamma m_{IK} \pm\sqrt{(s_{IK} - 2E_\gamma m_{IK})^2-4m_I^2m_K^2}}{2m_I^2}\,,\\
    &= \frac{m_{IK}^2(1-x_\gamma)-m_I^2 - m_K^2 \pm\sqrt{(m_{IK}^2(1-x_\gamma)-m_I^2 - m_K^2)^2-4m_I^2m_K^2}}{2m_I^2}\,.
\end{split}
\end{equation*}
For the sake of readability we will define $\alpha = m_{IK}^2(1-x_\gamma)-m_I^2 - m_K^2$, leading to:
\begin{equation*}
\begin{split}
    r&= \frac{\alpha \pm\sqrt{\alpha^2-4m_I^2m_K^2}}{2m_I^2}\,,\\
    &= \frac{\alpha \pm \alpha\sqrt{1-\frac{4m_I^2m_K^2}{\alpha^2}}}{2m_I^2}\,.
\end{split}
\end{equation*}
We proceed taking the the logarithm of both sides and separating out common factors,
\begin{equation*}
\begin{split}
    \ln\left(r\right)&= \ln\left(\frac{\alpha \pm \alpha\sqrt{1-\frac{4m_I^2m_K^2}{\alpha^2}}}{2m_I^2}\right)\,,\\
    &= \ln\left(\frac{\alpha}{m_I^2}\right)+\ln\left(1 \pm \sqrt{1-\frac{4m_I^2m_K^2}{\alpha^2}}\right) - \ln\left(2\right).\\
\end{split}
\end{equation*}
Now we substitute $\alpha$ back in and take the positive solution to arrive at eq.~\eqRef{eq:yStarMaxmK}.
\begin{equation*}
\begin{split}
y^*\leq\ &\frac12 \ln\left(\frac{m_{IK}^2(1-x_\gamma)-m_I^2 - m_K^2}{m_I^2}\right)+ \\ &\frac12 \ln\left(1 + \sqrt{1-\frac{4m_I^2m_K^2}{(m_{IK}^2(1-x_\gamma)-m_I^2 - m_K^2)^2}}\right) - \frac12 \ln\left(2\right).
\end{split}
\end{equation*}

From this \eqRef{eq:yStarMax} follows clearly from taking $m_K=0$,
\begin{equation*}
y^*\leq\ \frac12 \ln\left(\frac{m_{IK}^2(1-x_\gamma)-m_I^2}{m_I^2}\right).
\end{equation*}

The definition of $y^*$ provided in eq.~\eqRef{eq:yStarRec} is fully laid out as
\begin{equation*}
\begin{split}
     y^* &= \frac{1}{2}\ln(\frac{s_{jk}}{s_{ij}})\,,\\
     &=\frac{1}{2}\ln(\frac{2E_jE_k-E_j|p_k|\cos{\theta_{jk}}}{2E_jE_i-E_j|p_i|\cos{\theta_{ij}}})\,,\\
     &=\frac{1}{2}\ln(\frac{E_k(1-\beta_k\cos{\theta_{jk}})}{E_i(1-\beta_i\cos{\theta_{ij}})})\,,\\
     &=\frac{1}{2}\ln(\frac{E_k}{E_i})+\frac{1}{2}\ln(\frac{1-\beta_k\cos{\theta_{jk}}}{1-\beta_i\cos{\theta_{ij}}})\,.\\
\end{split}   
\end{equation*}

\clearpage

\section{Additional studies on performance}
\label{app:performance}

\begin{figure}[htb]
\centering
\begin{subfigure}{0.49\textwidth}
  \centering
  \includegraphics[width=\linewidth]{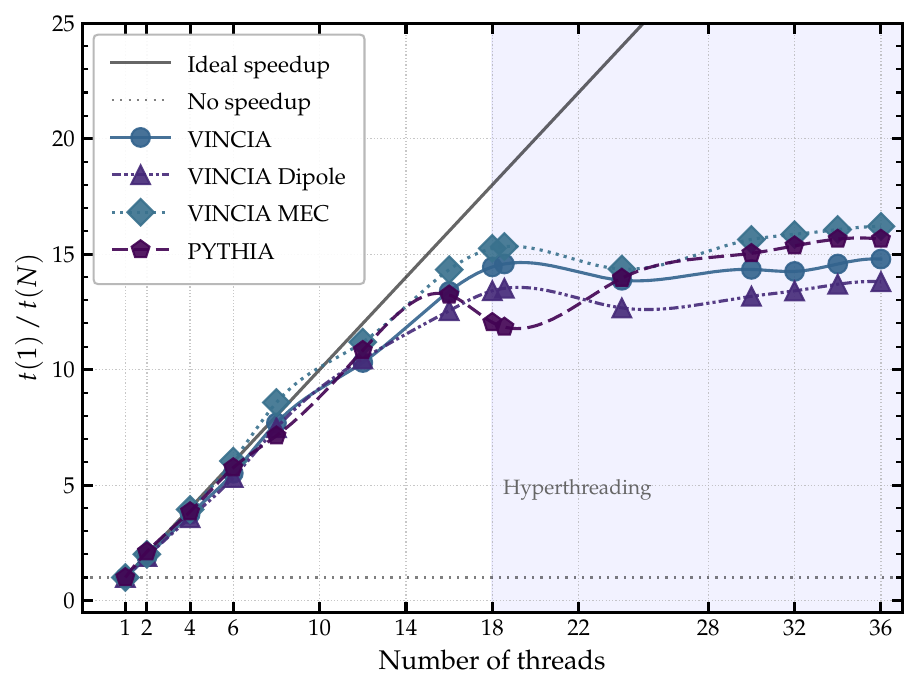}
  \caption{\vincia}
  \label{fig:speedup_Jpsi_mumu_a}
\end{subfigure}
\hfill
\begin{subfigure}{0.49\textwidth}
  \centering
  \includegraphics[width=\linewidth]{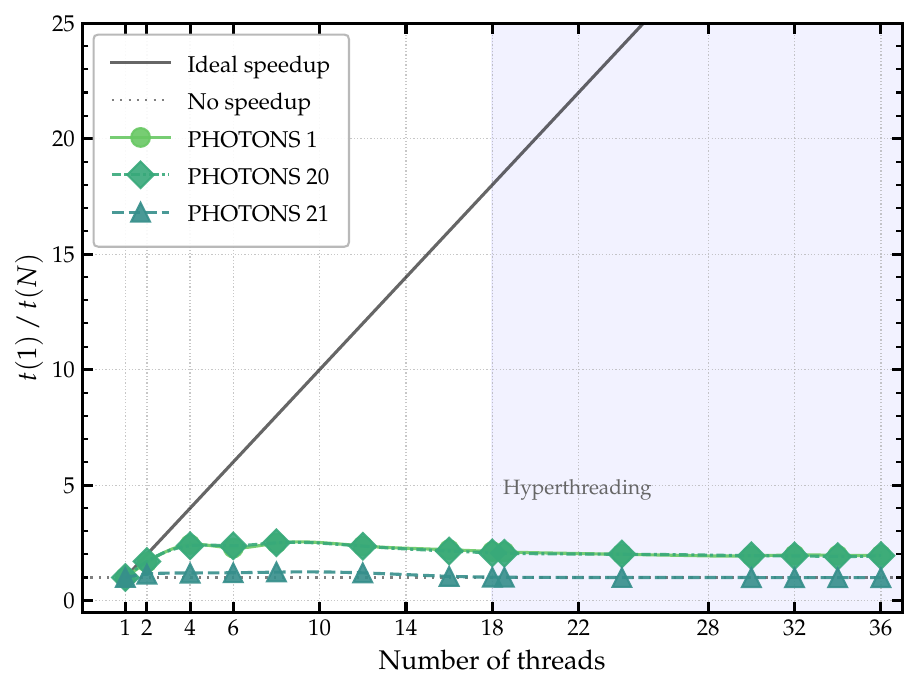}
  \caption{\photons}
  \label{fig:speedup_Jpsi_mumu_b}
\end{subfigure}
\vspace{0.3cm}

\begin{subfigure}{0.49\textwidth}
  \centering
  \includegraphics[width=\linewidth]{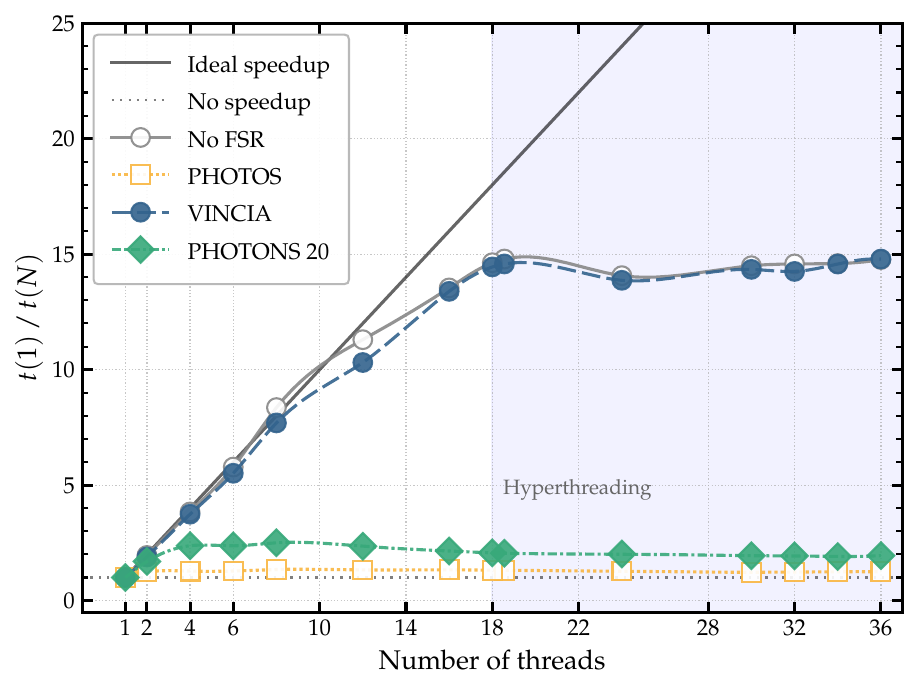}
  \caption{FSR without MEC vs. no FSR}
  \label{fig:speedup_Jpsi_mumu_c}
\end{subfigure}
\hfill
\begin{subfigure}{0.49\textwidth}
  \centering
  \includegraphics[width=\linewidth]{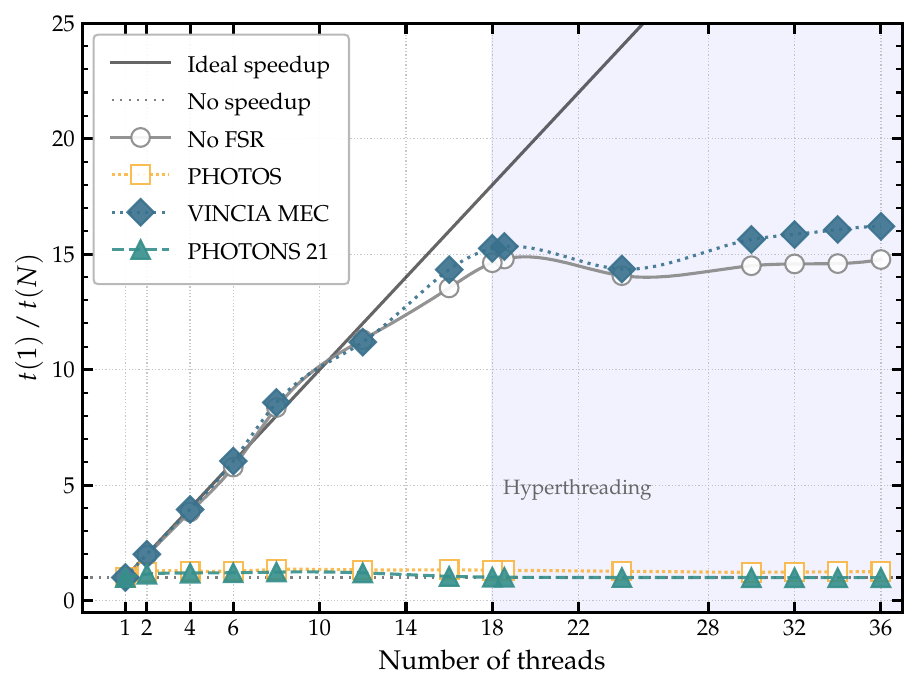}
  \caption{FSR with MEC (where available) vs. no FSR}
  \label{fig:speedup_Jpsi_mumu_d}
\end{subfigure}
    \caption{Speed-up curves for \mbox{\decay{\jpsi}{\mumu}}~decays.}
    \label{fig:speedup_Jpsi_mumu}
\end{figure}

\FigsRef{fig:speedup_Jpsi_mumu}--\ref{fig:memory_Dp_Kppipi} show the speed-up and relative RSS memory increase for the \mbox{\decay{\jpsi}{\mumu}}, \mbox{\decay{\Bd}{\Kp\pim}} and \mbox{\decay{\Dp}{\Kp\pip\pim}} decay modes (see details in \secRef{sec:performance}). 

\begin{figure}[htb]
\centering
\begin{subfigure}{0.49\textwidth}
  \centering
  \includegraphics[width=\linewidth]{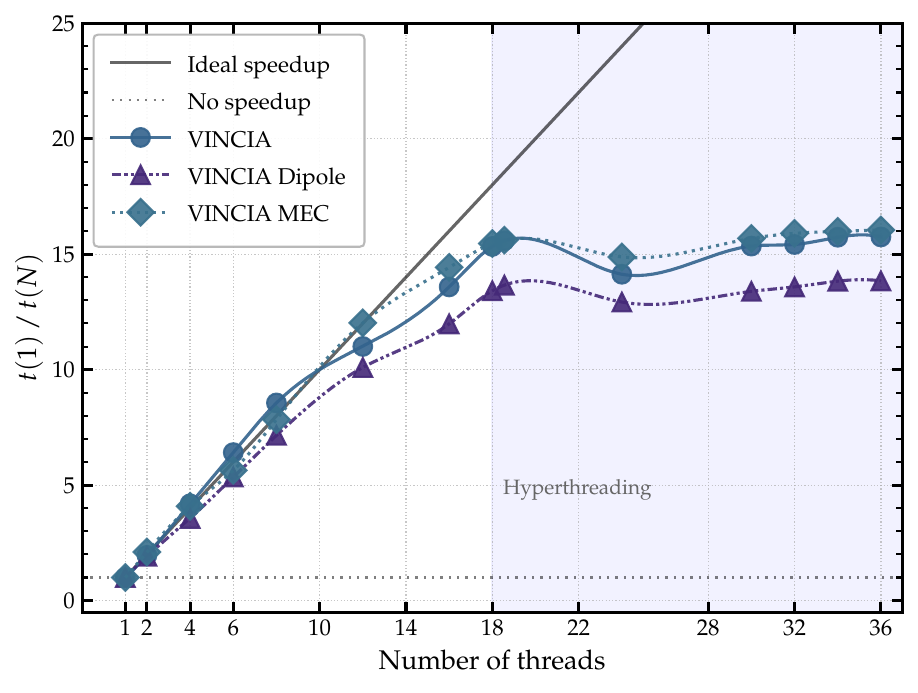}
  \caption{\vincia}
  \label{fig:speedup_Bd_Kpi_a}
\end{subfigure}
\hfill
\begin{subfigure}{0.49\textwidth}
  \centering
  \includegraphics[width=\linewidth]{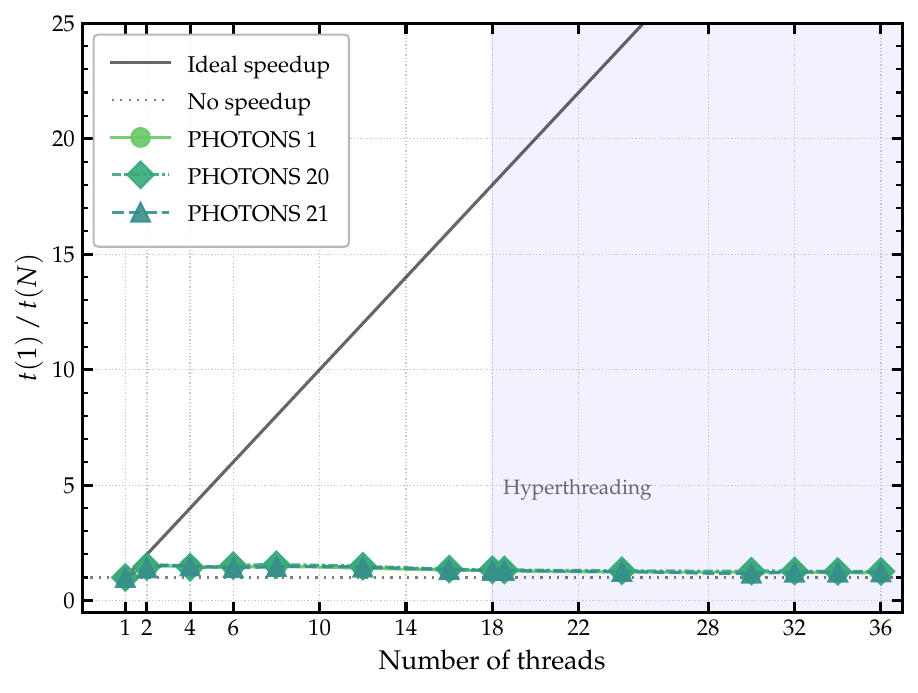}
  \caption{\photons}
  \label{fig:speedup_Bd_Kpi_b}
\end{subfigure}
\vspace{0.3cm}

\begin{subfigure}{0.49\textwidth}
  \centering
  \includegraphics[width=\linewidth]{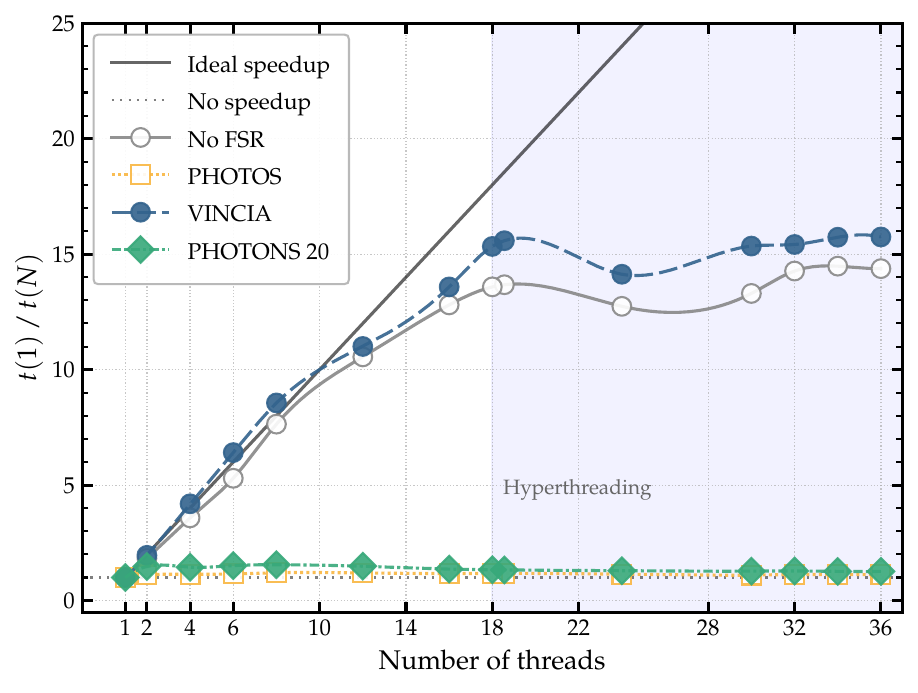}
  \caption{FSR without MEC vs. no FSR}
  \label{fig:speedup_Bd_Kpi_c}
\end{subfigure}
\hfill
\begin{subfigure}{0.49\textwidth}
  \centering
  \includegraphics[width=\linewidth]{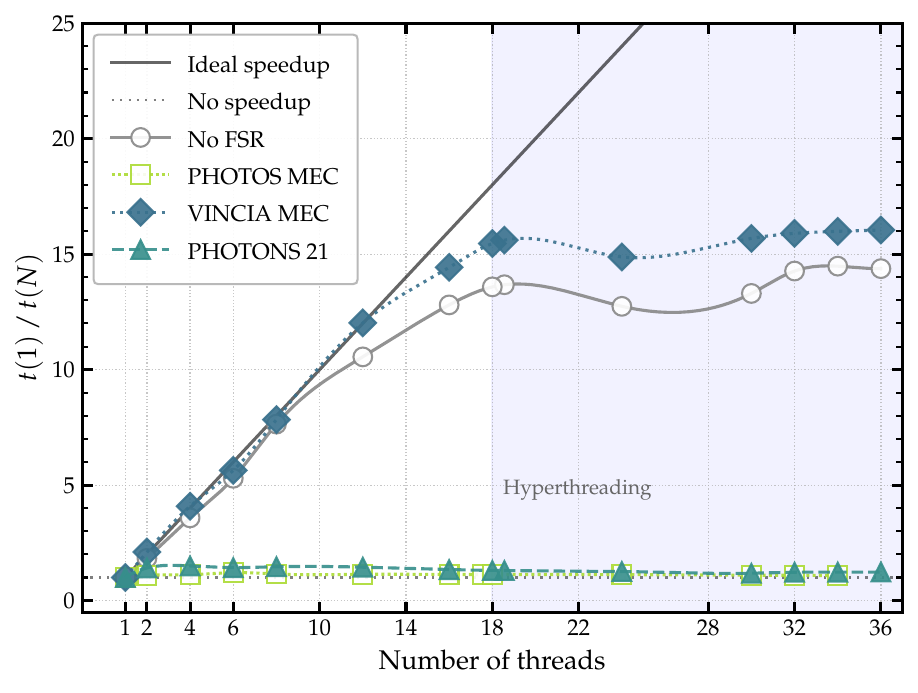}
  \caption{FSR with MEC (where available) vs. no FSR}
  \label{fig:speedup_Bd_Kpi_d}
\end{subfigure}
    \caption{Speed-up curves for \mbox{\decay{\Bz}{\Kp\pim}}~decays.}
    \label{fig:speedup_Bd_Kpi}
\end{figure}

\begin{figure}[htb]
\centering
\begin{subfigure}{0.49\textwidth}
  \centering
  \includegraphics[width=\linewidth]{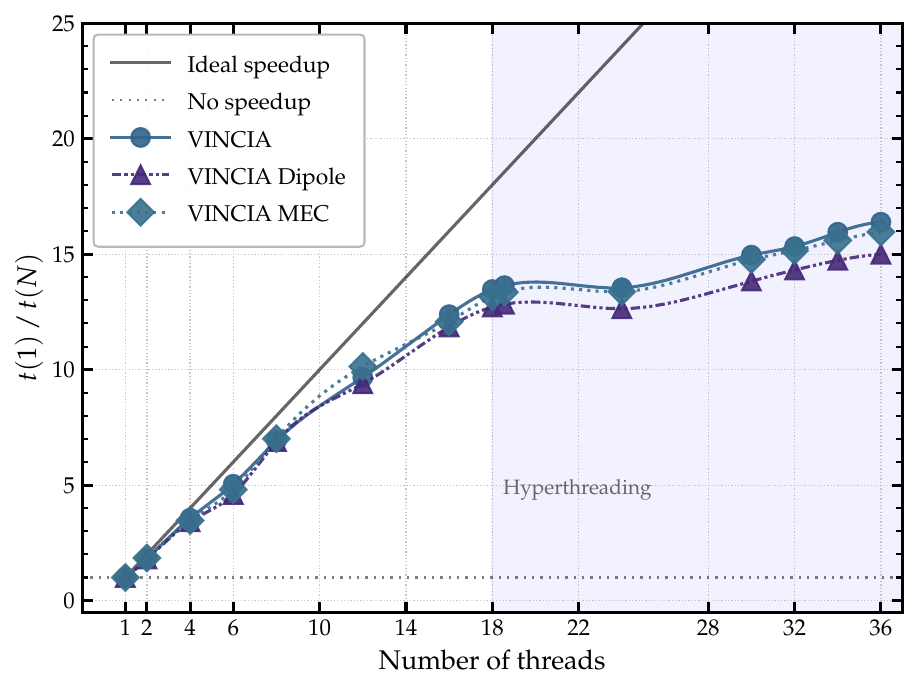}
  \caption{\vincia}
  \label{fig:speedup_Dp_Kppipi_a}
\end{subfigure}
\hfill
\begin{subfigure}{0.49\textwidth}
  \centering
  \includegraphics[width=\linewidth]{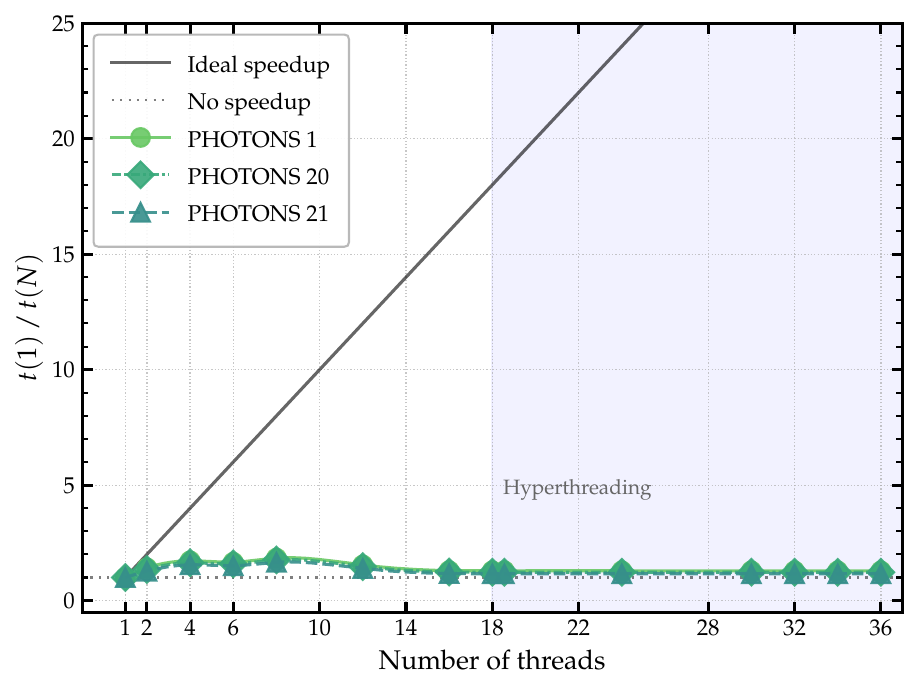}
  \caption{\photons}
  \label{fig:speedup_Dp_Kppipi_b}
\end{subfigure}
\vspace{0.3cm}

\begin{subfigure}{0.49\textwidth}
  \centering
  \includegraphics[width=\linewidth]{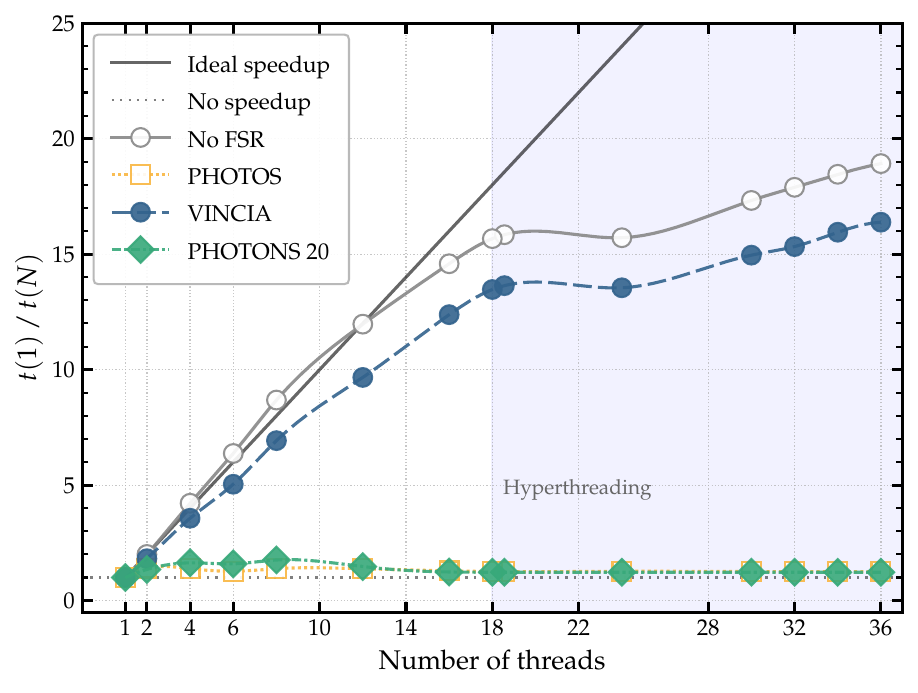}
  \caption{FSR without MEC vs. no FSR}
  \label{fig:speedup_Dp_Kppipi_c}
\end{subfigure}
\hfill
\begin{subfigure}{0.49\textwidth}
  \centering
  \includegraphics[width=\linewidth]{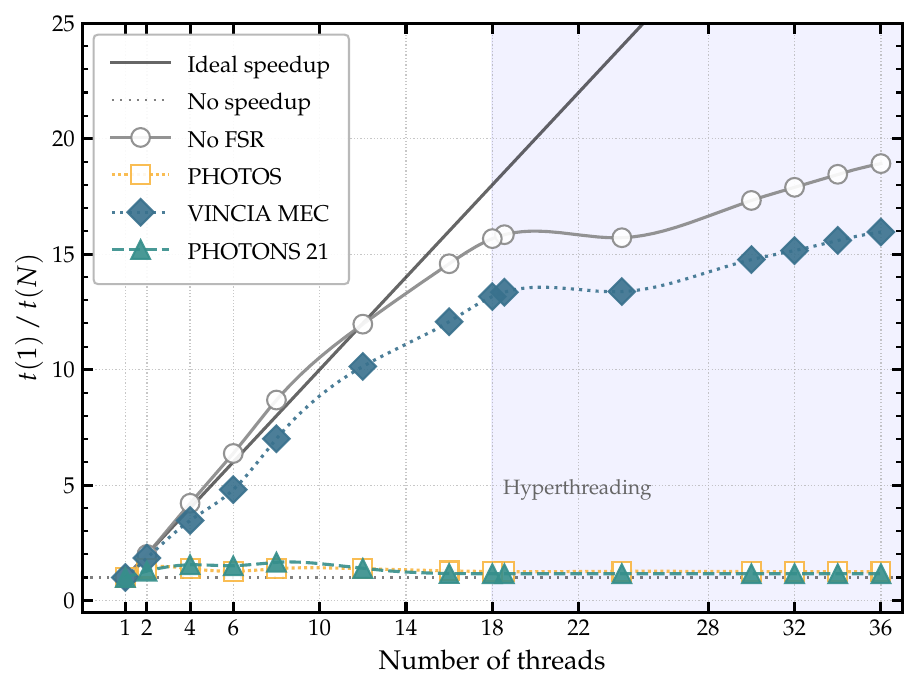}
  \caption{FSR with MEC (where available) vs. no FSR}
  \label{fig:speedup_Dp_Kppipi_d}
\end{subfigure}
    \caption{Speed-up curves for \mbox{\decay{\Dp}{\Kp\pip\pim}}~decays.}
    \label{fig:speedup_Dp_Kppipi}
\end{figure}

\begin{figure}[htb]
\centering
\begin{subfigure}{0.49\textwidth}
  \centering
  \includegraphics[width=\linewidth]{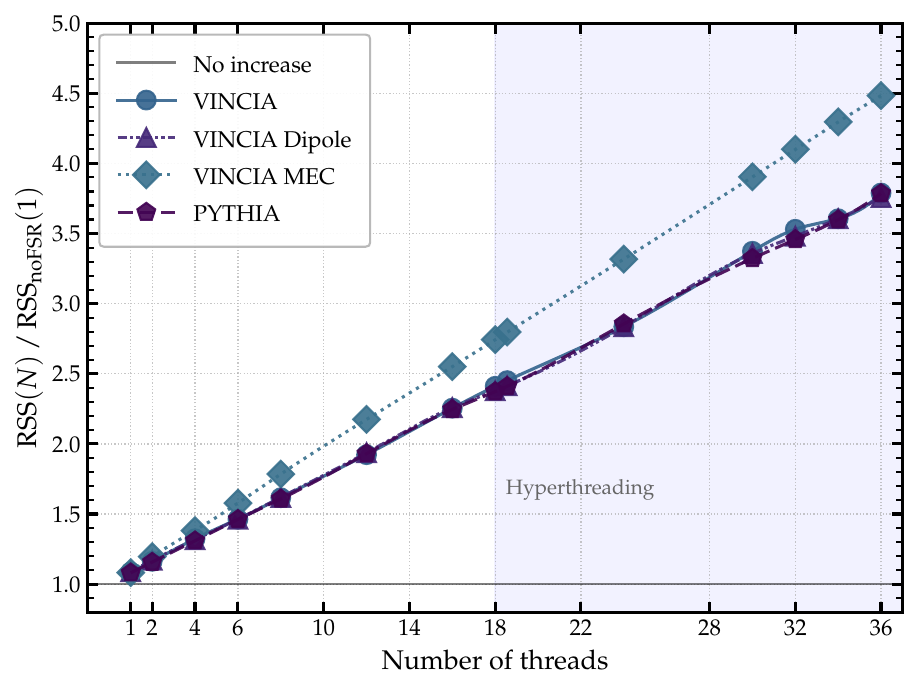}
  \caption{\vincia}
  \label{fig:memory_Jpsi_mumu_a}
\end{subfigure}
\hfill
\begin{subfigure}{0.49\textwidth}
  \centering
  \includegraphics[width=\linewidth]{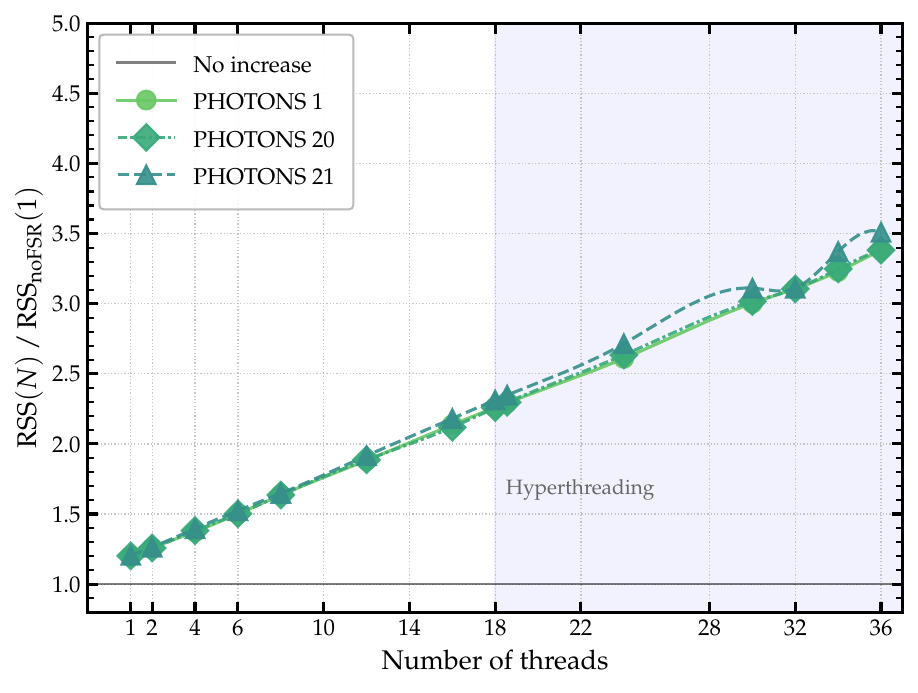}
  \caption{\photons}
  \label{fig:memory_Jpsi_mumu_b}
\end{subfigure}
\vspace{0.3cm}

\begin{subfigure}{0.49\textwidth}
  \centering
  \includegraphics[width=\linewidth]{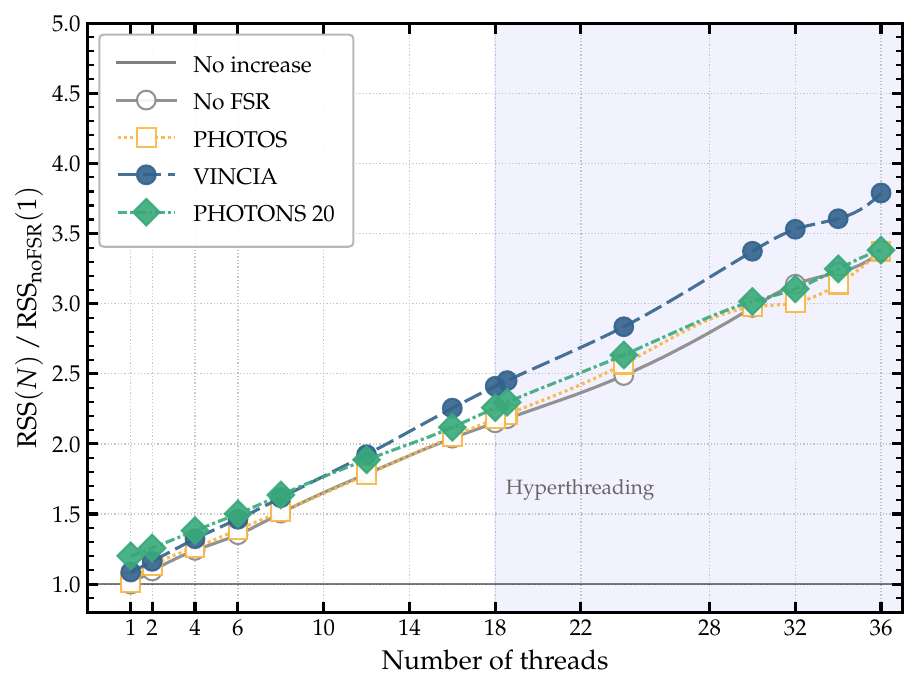}
  \caption{FSR without MEC vs. no FSR}
  \label{fig:memory_Jpsi_mumu_c}
\end{subfigure}
\hfill
\begin{subfigure}{0.49\textwidth}
  \centering
  \includegraphics[width=\linewidth]{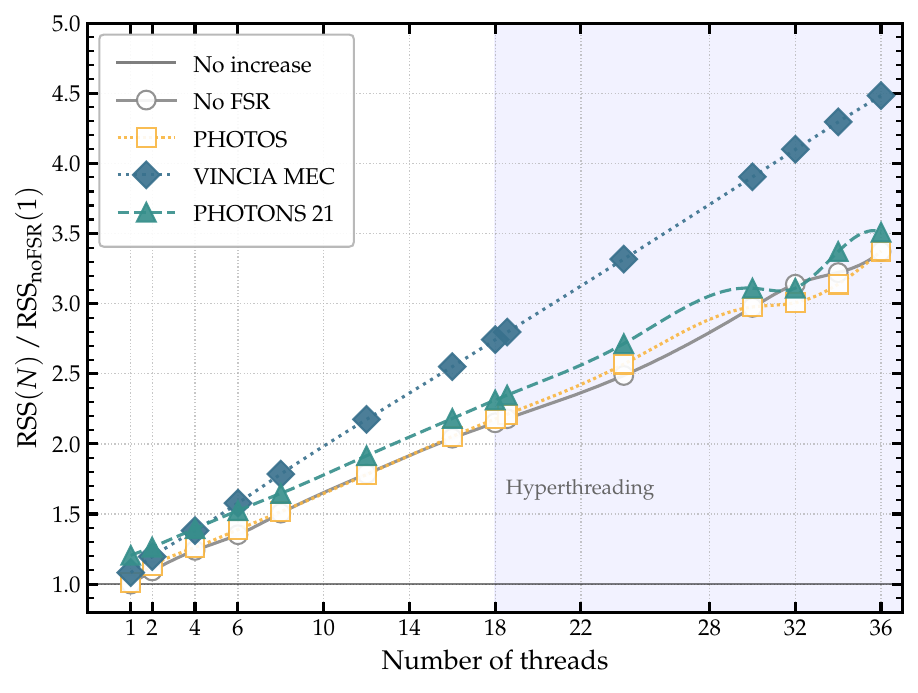}
  \caption{FSR with MEC (where available) vs. no FSR}
  \label{fig:memory_Jpsi_mumu_d}
\end{subfigure}
    \caption{Maximum RSS memory consumption relative to single-thread execution for \mbox{\decay{\jpsi}{\mumu}}~decays.}
    \label{fig:memory_Jpsi_mumu_mt}
\end{figure}

\begin{figure}[htb]
\centering
\begin{subfigure}{0.49\textwidth}
  \centering
  \includegraphics[width=\linewidth]{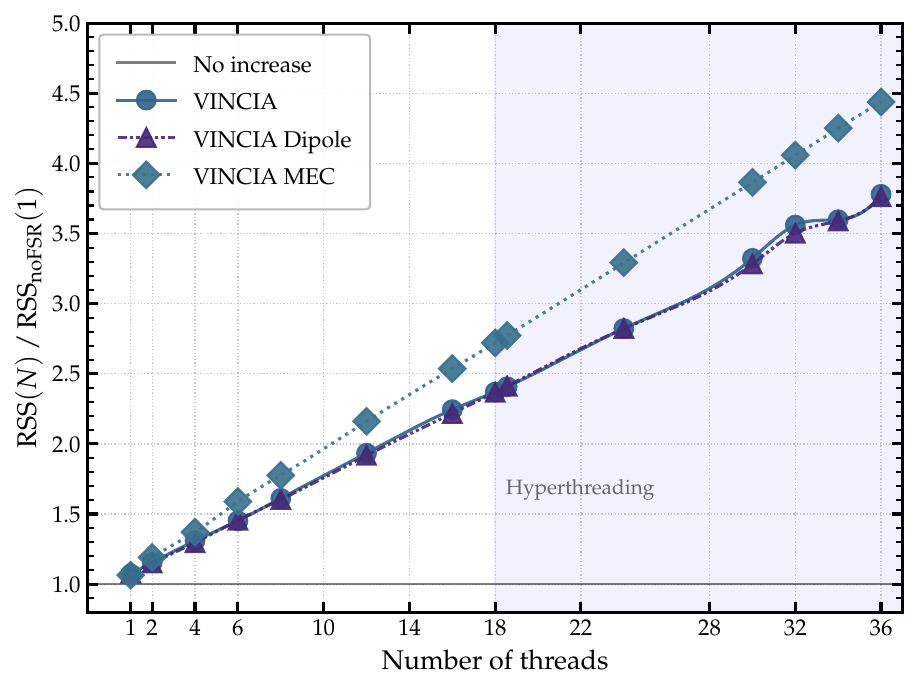}
  \caption{\vincia}
  \label{fig:memory_Bd_Kpi_a}
\end{subfigure}
\hfill
\begin{subfigure}{0.49\textwidth}
  \centering
  \includegraphics[width=\linewidth]{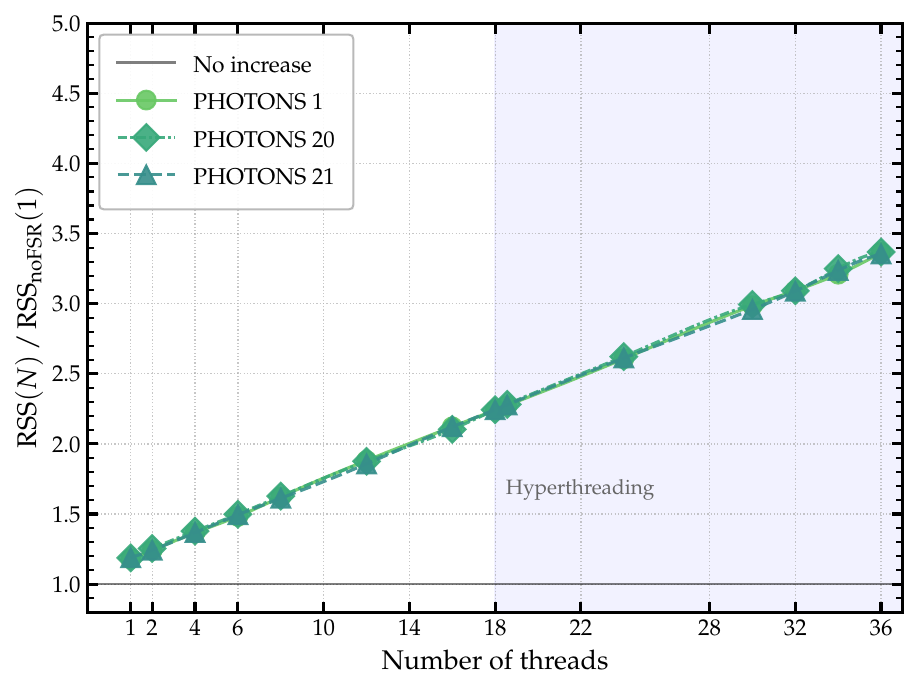}
  \caption{\photons}
  \label{fig:memory_Bd_Kpi_b_mt}
\end{subfigure}
\vspace{0.3cm}

\begin{subfigure}{0.49\textwidth}
  \centering
  \includegraphics[width=\linewidth]{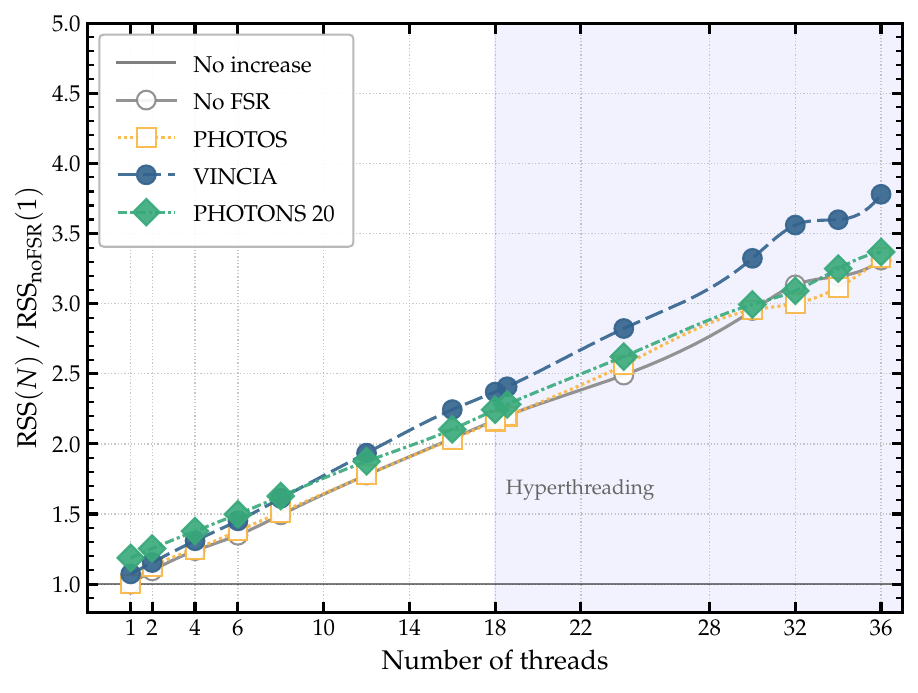}
  \caption{FSR without MEC vs. no FSR}
  \label{fig:memory_Bd_Kpi_c}
\end{subfigure}
\hfill
\begin{subfigure}{0.49\textwidth}
  \centering
  \includegraphics[width=\linewidth]{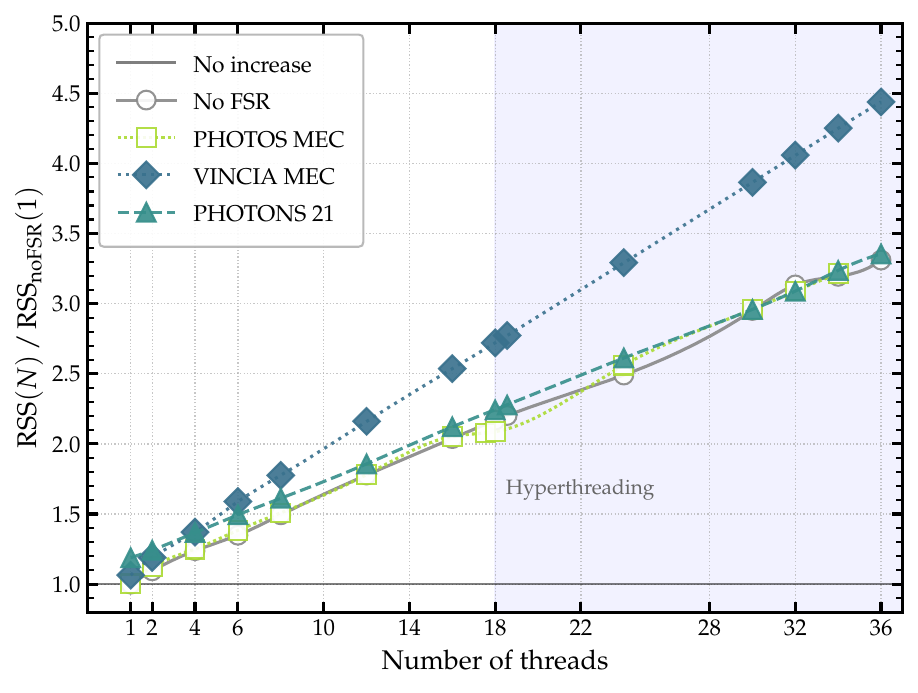}
  \caption{FSR with MEC (where available) vs. no FSR}
  \label{fig:memory_Bd_Kpi_d}
\end{subfigure}
    \caption{Maximum RSS memory consumption relative to single-thread execution for \mbox{\decay{\Bd}{\Kp\pim}}~decays.}
    \label{fig:memory_Bd_Kpi_mt}
\end{figure}

\begin{figure}[htb]
\centering
\begin{subfigure}{0.49\textwidth}
  \centering
  \includegraphics[width=\linewidth]{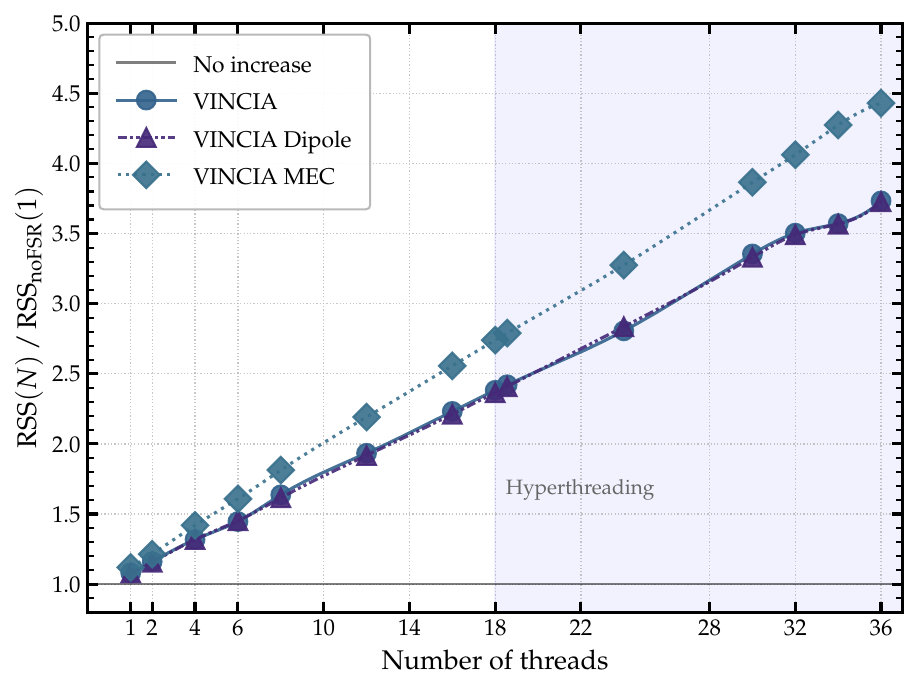}
  \caption{\vincia}
  \label{fig:memory_Dp_Kppipi_a}
\end{subfigure}
\hfill
\begin{subfigure}{0.49\textwidth}
  \centering
  \includegraphics[width=\linewidth]{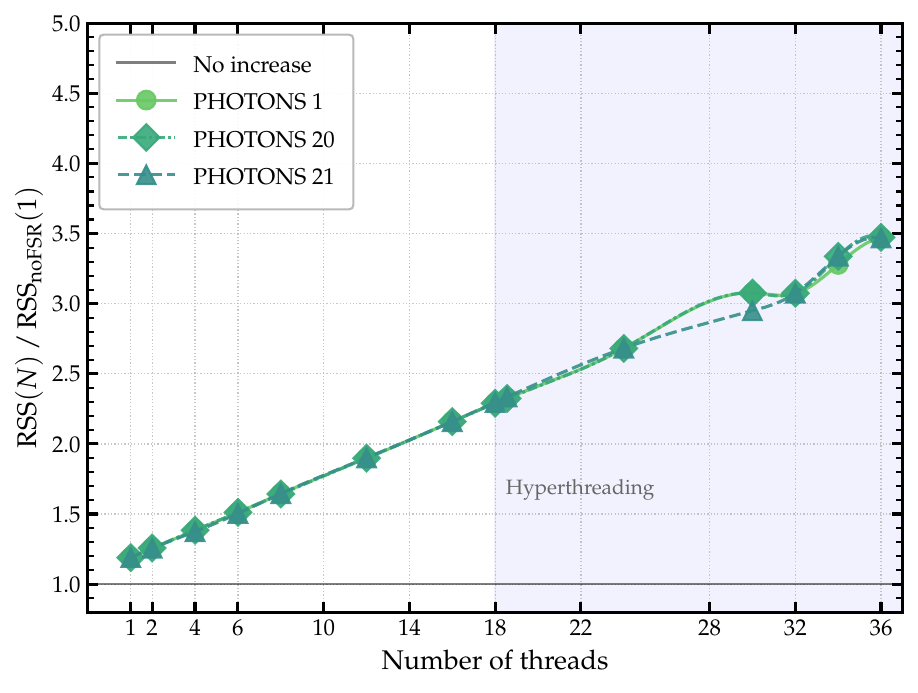}
  \caption{\photons}
  \label{fig:memory_Dp_Kppipi_b}
\end{subfigure}
\vspace{0.3cm}

\begin{subfigure}{0.49\textwidth}
  \centering
  \includegraphics[width=\linewidth]{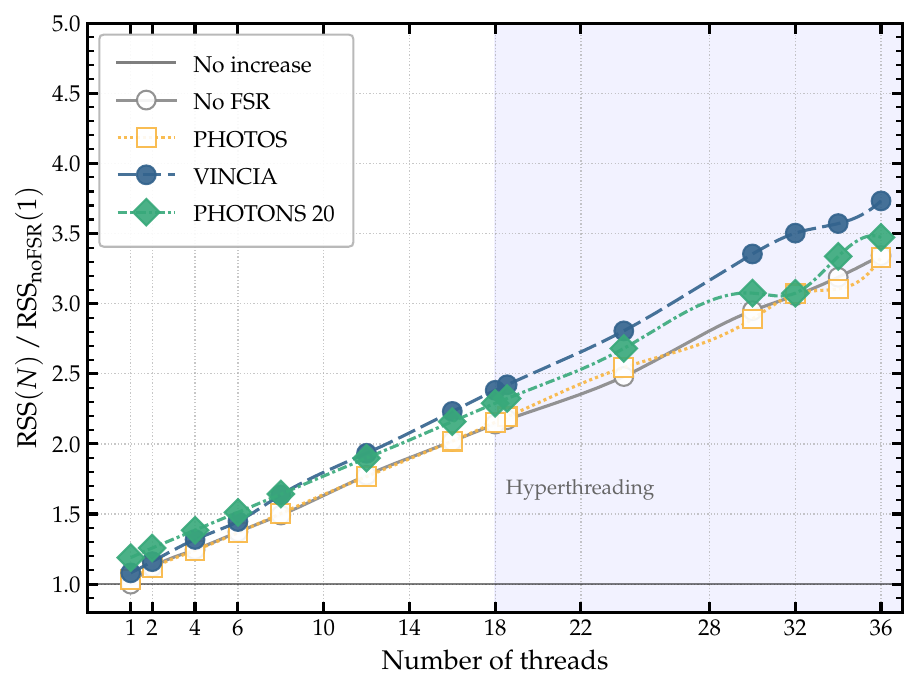}
  \caption{FSR without MEC vs. no FSR}
  \label{fig:memory_Dp_Kppipi_c}
\end{subfigure}
\hfill
\begin{subfigure}{0.49\textwidth}
  \centering
  \includegraphics[width=\linewidth]{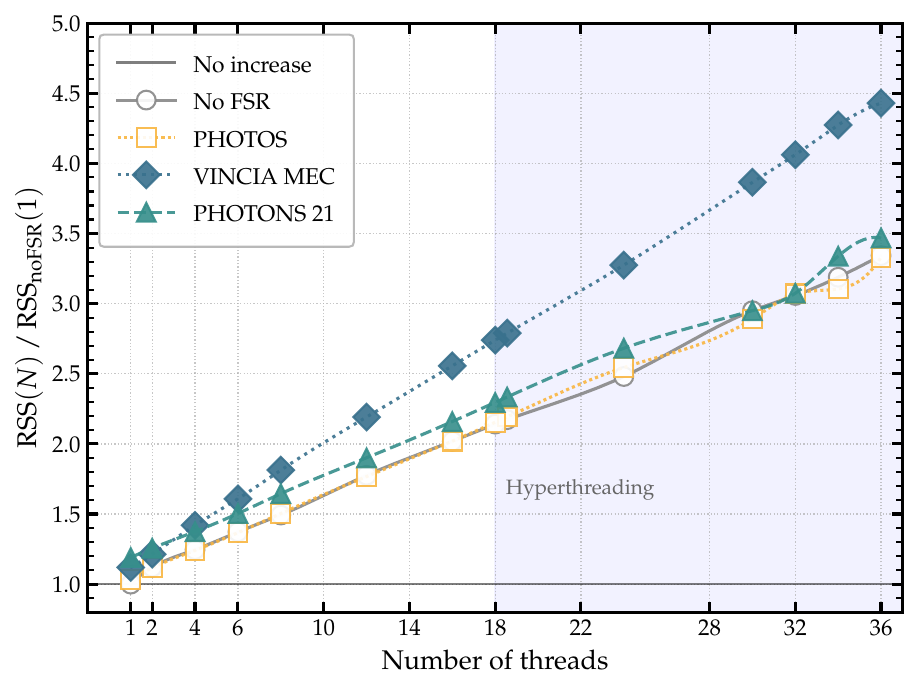}
  \caption{FSR with MEC (where available) vs. no FSR}
  \label{fig:memory_Dp_Kppipi_d}
\end{subfigure}
    \caption{Maximum RSS memory consumption relative to single-thread execution for \mbox{\decay{\Dp}{\Kp\pip\pim}}~decays.}
    \label{fig:memory_Dp_Kppipi_mt}
\end{figure}

\clearpage

\addcontentsline{toc}{section}{References}
\bibliographystyle{LHCb}
\bibliography{main}

\end{document}